\documentclass[
  superscriptaddress,
  reprint,
  amsmath,amssymb,
  aps,
  prd,
  nofootinbib,
  floatfix
]{revtex4-2}

\usepackage{graphicx}
\usepackage{dcolumn}   
\usepackage{bm}
\usepackage[table]{xcolor}
\usepackage{colortbl}
\usepackage{float}
\usepackage{afterpage}
\usepackage{microtype}
\usepackage[linesnumbered,ruled]{algorithm2e}
\usepackage{url}
\usepackage{multirow,makecell}
\usepackage{booktabs}
\usepackage[caption=false]{subfig}

\usepackage{hyperref}
\hypersetup{
  hidelinks,
  colorlinks=true,
  citecolor=teal,
  linkcolor=teal,
  urlcolor=teal
}

\renewcommand\thesection{\Roman{section}}
\renewcommand\thesubsection{\Alph{subsection}}

\makeatletter
\renewcommand\p@subsection{\thesection.}
\renewcommand\p@subsubsection{\thesection.\thesubsection.}
\makeatother

\usepackage[colorinlistoftodos]{todonotes}

\usepackage{lipsum}

\usepackage{parskip}
\begin{document}

\title{Signal selection and model-independent extraction of pionless charged-current muon neutrino cross section using double-differential kinematic imbalance observables on carbon and oxygen with the T2K experiment}

\newcommand{\INSTHD}{\affiliation{University Autonoma Madrid, Department of Theoretical Physics, 28049 Madrid, Spain}}
\newcommand{\INSTFE}{\affiliation{Boston University, Department of Physics, Boston, Massachusetts, U.S.A.}}
\newcommand{\INSTD}{\affiliation{University of British Columbia, Department of Physics and Astronomy, Vancouver, British Columbia, Canada}}
\newcommand{\INSTGA}{\affiliation{University of California, Irvine, Department of Physics and Astronomy, Irvine, California, U.S.A.}}
\newcommand{\INSTI}{\affiliation{IRFU, CEA, Universit\'e Paris-Saclay, F-91191 Gif-sur-Yvette, France}}
\newcommand{\INSTGB}{\affiliation{University of Colorado at Boulder, Department of Physics, Boulder, Colorado, U.S.A.}}
\newcommand{\INSTFH}{\affiliation{Duke University, Department of Physics, Durham, North Carolina, U.S.A.}}
\newcommand{\INSTEF}{\affiliation{ETH Zurich, Institute for Particle Physics and Astrophysics, Zurich, Switzerland}}
\newcommand{\INSTIG}{\affiliation{VNU University of Science, Vietnam National University, Hanoi, Vietnam}}
\newcommand{\INSTIE}{\affiliation{CERN European Organization for Nuclear Research, CH-1211 Gen\'eve 23, Switzerland}}
\newcommand{\INSTEG}{\affiliation{University of Geneva, Section de Physique, DPNC, Geneva, Switzerland}}
\newcommand{\INSTHJ}{\affiliation{University of Glasgow, School of Physics and Astronomy, Glasgow, United Kingdom}}
\newcommand{\INSTJG}{\affiliation{Ghent University, Department of Physics and Astronomy, Proeftuinstraat 86, B-9000 Gent, Belgium}}
\newcommand{\INSTDG}{\affiliation{H. Niewodniczanski Institute of Nuclear Physics PAN, Cracow, Poland}}
\newcommand{\INSTCB}{\affiliation{High Energy Accelerator Research Organization (KEK), Tsukuba, Ibaraki, Japan}}
\newcommand{\INSTIB}{\affiliation{University of Houston, Department of Physics, Houston, Texas, U.S.A.}}
\newcommand{\INSTED}{\affiliation{Institut de Fisica d'Altes Energies (IFAE) - The Barcelona Institute of Science and Technology, Campus UAB, Bellaterra (Barcelona) Spain}}
\newcommand{\INSTJC}{\affiliation{Institut f\"ur Physik, Johannes Gutenberg-Universit\"at Mainz, Staudingerweg 7, 55128 Mainz, Germany}}
\newcommand{\INSTHH}{\affiliation{Institute For Interdisciplinary Research in Science and Education (IFIRSE), ICISE, Quy Nhon, Vietnam}}
\newcommand{\INSTEI}{\affiliation{Imperial College London, Department of Physics, London, United Kingdom}}
\newcommand{\INSTGF}{\affiliation{INFN Sezione di Bari and Universit\`a e Politecnico di Bari, Dipartimento Interuniversitario di Fisica, Bari, Italy}}
\newcommand{\INSTBE}{\affiliation{INFN Sezione di Napoli and Universit\`a di Napoli, Dipartimento di Fisica, Napoli, Italy}}
\newcommand{\INSTBF}{\affiliation{INFN Sezione di Padova and Universit\`a di Padova, Dipartimento di Fisica, Padova, Italy}}
\newcommand{\INSTBD}{\affiliation{INFN Sezione di Roma and Universit\`a di Roma ``La Sapienza'', Roma, Italy}}
\newcommand{\INSTEB}{\affiliation{Institute for Nuclear Research of the Russian Academy of Sciences, Moscow, Russia}}
\newcommand{\INSTHI}{\affiliation{International Centre of Physics, Institute of Physics (IOP), Vietnam Academy of Science and Technology (VAST), 10 Dao Tan, Ba Dinh, Hanoi, Vietnam}}
\newcommand{\INSTJD}{\affiliation{ILANCE, CNRS – University of Tokyo International Research Laboratory, Kashiwa, Chiba 277-8582, Japan}}
\newcommand{\INSTHA}{\affiliation{Kavli Institute for the Physics and Mathematics of the Universe (WPI), The University of Tokyo Institutes for Advanced Study, University of Tokyo, Kashiwa, Chiba, Japan}}
\newcommand{\INSTID}{\affiliation{Keio University, Department of Physics, Kanagawa, Japan}}
\newcommand{\INSTIF}{\affiliation{King's College London, Department of Physics, Strand, London WC2R 2LS, United Kingdom}}
\newcommand{\INSTCC}{\affiliation{Kobe University, Kobe, Japan}}
\newcommand{\INSTCD}{\affiliation{Kyoto University, Department of Physics, Kyoto, Japan}}
\newcommand{\INSTEJ}{\affiliation{Lancaster University, School of Physics and Astronomy, Lancaster, United Kingdom}}
\newcommand{\INSTII}{\affiliation{Lawrence Berkeley National Laboratory, Berkeley, California, U.S.A.}}
\newcommand{\INSTBA}{\affiliation{Ecole Polytechnique, IN2P3-CNRS, Laboratoire Leprince-Ringuet, Palaiseau, France}}
\newcommand{\INSTFC}{\affiliation{University of Liverpool, Department of Physics, Liverpool, United Kingdom}}
\newcommand{\INSTFI}{\affiliation{Louisiana State University, Department of Physics and Astronomy, Baton Rouge, Louisiana, U.S.A.}}
\newcommand{\INSTIH}{\affiliation{Joint Institute for Nuclear Research, Dubna, Moscow Region, Russia}}
\newcommand{\INSTHB}{\affiliation{Michigan State University, Department of Physics and Astronomy,  East Lansing, Michigan, U.S.A.}}
\newcommand{\INSTCE}{\affiliation{Miyagi University of Education, Department of Physics, Sendai, Japan}}
\newcommand{\INSTDF}{\affiliation{National Centre for Nuclear Research, Warsaw, Poland}}
\newcommand{\INSTFJ}{\affiliation{State University of New York at Stony Brook, Department of Physics and Astronomy, Stony Brook, New York, U.S.A.}}
\newcommand{\INSTEH}{\affiliation{STFC, Rutherford Appleton Laboratory, Harwell Oxford,  and  Daresbury Laboratory, Warrington, United Kingdom}}
\newcommand{\INSTGJ}{\affiliation{Okayama University, Department of Physics, Okayama, Japan}}
\newcommand{\INSTCF}{\affiliation{Osaka Metropolitan University, Department of Physics, Osaka, Japan}}
\newcommand{\INSTGG}{\affiliation{Oxford University, Department of Physics, Oxford, United Kingdom}}
\newcommand{\INSTIC}{\affiliation{University of Pennsylvania, Department of Physics and Astronomy,  Philadelphia, Pennsylvania, U.S.A.}}
\newcommand{\INSTGC}{\affiliation{University of Pittsburgh, Department of Physics and Astronomy, Pittsburgh, Pennsylvania, U.S.A.}}
\newcommand{\INSTGD}{\affiliation{University of Rochester, Department of Physics and Astronomy, Rochester, New York, U.S.A.}}
\newcommand{\INSTHC}{\affiliation{Royal Holloway University of London, Department of Physics, Egham, Surrey, United Kingdom}}
\newcommand{\INSTBC}{\affiliation{RWTH Aachen University, III. Physikalisches Institut, Aachen, Germany}}
\newcommand{\INSTJF}{\affiliation{School of Physics and Astronomy, University of Minnesota, Minneapolis, Minnesota, U.S.A.}}
\newcommand{\INSTJB}{\affiliation{Departamento de F\'isica At\'omica, Molecular y Nuclear, Universidad de Sevilla, 41080 Sevilla, Spain}}
\newcommand{\INSTFB}{\affiliation{University of Sheffield, School of Mathematical and Physical Sciences, Sheffield, United Kingdom}}
\newcommand{\INSTDI}{\affiliation{University of Silesia, Institute of Physics, Katowice, Poland}}
\newcommand{\INSTIA}{\affiliation{SLAC National Accelerator Laboratory, Stanford University, Menlo Park, California, U.S.A.}}
\newcommand{\INSTBB}{\affiliation{Sorbonne Universit\'e, CNRS/IN2P3, Laboratoire de Physique Nucl\'eaire et de Hautes Energies (LPNHE), Paris, France}}
\newcommand{\INSTJE}{\affiliation{South Dakota School of Mines and Technology, 501 East Saint Joseph Street, Rapid City, SD 57701, United States}}
\newcommand{\INSTCH}{\affiliation{University of Tokyo, Department of Physics, Tokyo, Japan}}
\newcommand{\INSTBJ}{\affiliation{University of Tokyo, Institute for Cosmic Ray Research, Kamioka Observatory, Kamioka, Japan}}
\newcommand{\INSTCG}{\affiliation{University of Tokyo, Institute for Cosmic Ray Research, Research Center for Cosmic Neutrinos, Kashiwa, Japan}}
\newcommand{\INSTHF}{\affiliation{Institute of Science Tokyo, Department of Physics, Tokyo}}
\newcommand{\INSTGI}{\affiliation{Tokyo Metropolitan University, Department of Physics, Tokyo, Japan}}
\newcommand{\INSTHG}{\affiliation{Tokyo University of Science, Faculty of Science and Technology, Department of Physics, Noda, Chiba, Japan}}
\newcommand{\INSTB}{\affiliation{TRIUMF, Vancouver, British Columbia, Canada}}
\newcommand{\INSTJH}{\affiliation{University of Toyama, Faculty of Science, Toyama, Japan}}
\newcommand{\INSTJJ}{\affiliation{University of Tsukuba, Faculty of Pure and Applied Sciences, Tsukuba, Ibaraki, Japan}}
\newcommand{\INSTDJ}{\affiliation{University of Warsaw, Faculty of Physics, Warsaw, Poland}}
\newcommand{\INSTDH}{\affiliation{Warsaw University of Technology, Institute of Radioelectronics and Multimedia Technology, Warsaw, Poland}}
\newcommand{\INSTIJ}{\affiliation{Tohoku University, Faculty of Science, Department of Physics, Miyagi, Japan}}
\newcommand{\INSTFD}{\affiliation{University of Warwick, Department of Physics, Coventry, United Kingdom}}
\newcommand{\INSTEA}{\affiliation{Wroclaw University, Faculty of Physics and Astronomy, Wroclaw, Poland}}
\newcommand{\INSTHE}{\affiliation{Yokohama National University, Department of Physics, Yokohama, Japan}}
\newcommand{\INSTH}{\affiliation{York University, Department of Physics and Astronomy, Toronto, Ontario, Canada}}

\INSTHD
\INSTFE
\INSTD
\INSTGA
\INSTI
\INSTGB
\INSTFH
\INSTEF
\INSTIG
\INSTIE
\INSTEG
\INSTHJ
\INSTJG
\INSTDG
\INSTCB
\INSTIB
\INSTED
\INSTJC
\INSTHH
\INSTEI
\INSTGF
\INSTBE
\INSTBF
\INSTBD
\INSTEB
\INSTHI
\INSTJD
\INSTHA
\INSTID
\INSTIF
\INSTCC
\INSTCD
\INSTEJ
\INSTII
\INSTBA
\INSTFC
\INSTFI
\INSTIH
\INSTHB
\INSTCE
\INSTDF
\INSTFJ
\INSTEH
\INSTGJ
\INSTCF
\INSTGG
\INSTIC
\INSTGC
\INSTGD
\INSTHC
\INSTBC
\INSTJF
\INSTJB
\INSTFB
\INSTDI
\INSTIA
\INSTBB
\INSTJE
\INSTCH
\INSTBJ
\INSTCG
\INSTHF
\INSTGI
\INSTHG
\INSTB
\INSTJH
\INSTJJ
\INSTDJ
\INSTDH
\INSTIJ
\INSTFD
\INSTEA
\INSTHE
\INSTH

\author{K.\,Abe}\INSTBJ
\author{S.\,Abe}\INSTBJ
\author{H.\,Adhikary}\INSTDJ
\author{R.\,Akutsu}\INSTCB
\author{H.\,Alarakia-Charles}\INSTEJ
\author{Y.I.\,Alj Hakim}\INSTFB
\author{S.\,Alonso Monsalve}\INSTEF
\author{L.\,Anthony}\INSTEI
\author{S.\,Aoki}\INSTCC
\author{K.A.\,Apte}\INSTEI
% ...

\collaboration{The T2K Collaboration}\noaffiliation
    
\begin{abstract}
\noindent We present the first joint measurement of muon neutrino CC$0\pi Np$ interactions on carbon and oxygen targets, in two double-differential kinematic imbalance (KI) observable spaces, $\delta p_{T}$–$\delta \alpha_{T}$ and $p_{N}$–$\cos\theta_{\mu}$. The measurement employs the ND280 detector of the T2K experiment and includes a detailed description of the event selection used to define signal and control regions, the evaluation of systematic uncertainties, and the signal extraction procedure, together with validation studies supporting a robust cross-section measurement. The results of this analysis indicate that current neutrino–nucleus interaction models do not adequately describe the data, and demonstrate the strong discriminating power of KI observables. This measurement highlights the need for improved theoretical nuclear modeling within neutrino interaction generators to achieve increased precision in neutrino oscillation measurements.

\end{abstract}

\maketitle

\section{Introduction}

The primary goal of current and future accelerator-based neutrino experiments is to achieve precision measurements of neutrino oscillation parameters. The dominant systematic uncertainty in current oscillation measurements~\cite{T2K:2025yoy,NOvA:2025tmb,T2K:2025wet} arises from the modeling of neutrino–nucleus interactions. In particular, the so-called “nuclear effects”\cite{Coloma:2013tba, Mosel:2016cwa, NuSTEC:2017hzk} distort the reconstructed neutrino energy spectrum in ways that can bias the measurements of neutrino oscillation parameters. These effects originate from three main sources: (1) the initial state of nucleons within the nucleus; (2) correlations between the interacting nucleon and the surrounding nuclear medium; and (3) final-state interactions (FSIs) of outgoing particles as they propagate through the nucleus. Together, these processes significantly impact the observed final-state kinematics, complicating the interpretation of interaction channels. A detailed understanding of these contributions is therefore essential for improving the precision and reliability of oscillation measurements in current experiments such as T2K\cite{T2K:2011qtm, T2K:2023smv} and NOvA~\cite{NOvA:2007rmc, NOvA:2021nfi}, as well as in next-generation experiments including Hyper-Kamiokande~\cite{Hyper-Kamiokande:2018ofw} and DUNE~\cite{DUNE:2015lol}.

This analysis aims to probe the nuclear effects by measuring the differential cross section of muon neutrino charged-current interactions jointly on carbon and oxygen targets using the T2K ND280 detector. The signal looks at charged-current quasi-elastic-like (CCQE-like) interactions containing one muon, no pions, and at least one reconstructed proton originating from the neutrino interaction vertex. Throughout the paper this channel will be referred to as $\nu_{\mu}$ CC$0\pi Np$. The cross section is measured as functions of kinematic imbalance (KI) observables~\cite{Lu:2015tcr, Lu:2019nmf}, which exploit correlations between the outgoing lepton and hadrons and provide enhanced sensitivity to nuclear effects. In addition, double-differential measurements are performed in these observables to further exploit their correlations and help to lift degeneracies among different nuclear effect contributions.

This article is structured as follows: Sec.~\ref{sec:exp_setup} provides an overview of the experimental setup, including the neutrino beam and the T2K ND280 detector. Sec.~\ref{sec:analysis_observables} introduces the analysis observables. Sec.~\ref{sec:eve_sele} describes the event selection applied to the neutrino interaction samples used in this cross-section measurement. Sec.~\ref{sec:xsec_extra} outlines the cross-section extraction method and its validation. The results are presented in Sec.~\ref{sec:result}, and Sec.~\ref{sec:model_comp} compares the results with predictions from various neutrino-nucleus interaction generators. Finally, Sec.~\ref{sec:conclu} summarizes the conclusions of this measurement.
 
\section{Experimental Setup}
\label{sec:exp_setup}

Tokai-to-Kamioka (T2K) is a long-baseline, accelerator-based neutrino oscillation experiment located in Japan. The experiment consists of a neutrino beamline (Sec.~\ref{subsec:nu_beam}) and a near detector complex at the Japan Proton Accelerator Research Complex (J-PARC) in Tokai, which includes the INGRID~\cite{Abe:2011xv}, WAGASCI-BabyMIND~\cite{T2K-WB-1}, and ND280 detectors. The far detector, Super-Kamiokande~\cite{Super-Kamiokande:2002weg}, is situated 295~km away in Kamioka. The measurement presented in this article utilizes the neutrino beam data collected specifically at the ND280 (Sec.~\ref{subsec:ND280}), with further details regarding the dataset provided in Sec.~\ref{subsec:data_MC}.

\subsection{Neutrino Beam}
\label{subsec:nu_beam}

The T2K neutrino beam is produced by extracting a 30~GeV proton beam from the J-PARC Main Ring (MR) and directing it onto a monolithic graphite target. The collision of protons with the target produces secondary particles, primarily charged mesons (pions and kaons). Three magnetic horns are positioned around and downstream of the target to select mesons of a specific charge by altering the horn current polarity. This configuration allows for the generation of a neutrino- (antineutrino-) enhanced beam by primarily selecting positively (negatively) charged mesons~\cite{T2K:2011qtm}.

After passing through the magnetic horns, the charged particles propagate through a 90~m long decay tunnel filled with helium gas. Within this tunnel, the charged mesons decay, predominantly producing muon neutrinos via pion decay in flight: $\pi^{\pm} \rightarrow \mu^{\pm} + \nu_{\mu}(\bar{\nu}_{\mu})$. At the end of the tunnel, a composite particle flux is formed, containing neutrinos alongside other decay products (such as muons and electrons) and remaining undecayed mesons. A graphite beam dump is located at the end of the tunnel to absorb the undecayed mesons and low-energy muons. Muons with energies exceeding 5~GeV penetrate the beam dump and are measured by the muon monitor (MUMON)~\cite{Suzuki:2012ova}. These energetic muons are subsequently stopped by the Earth before reaching the near detector complex.

\begin{figure}[!htbp]
\centering
\includegraphics[width=0.45\textwidth]{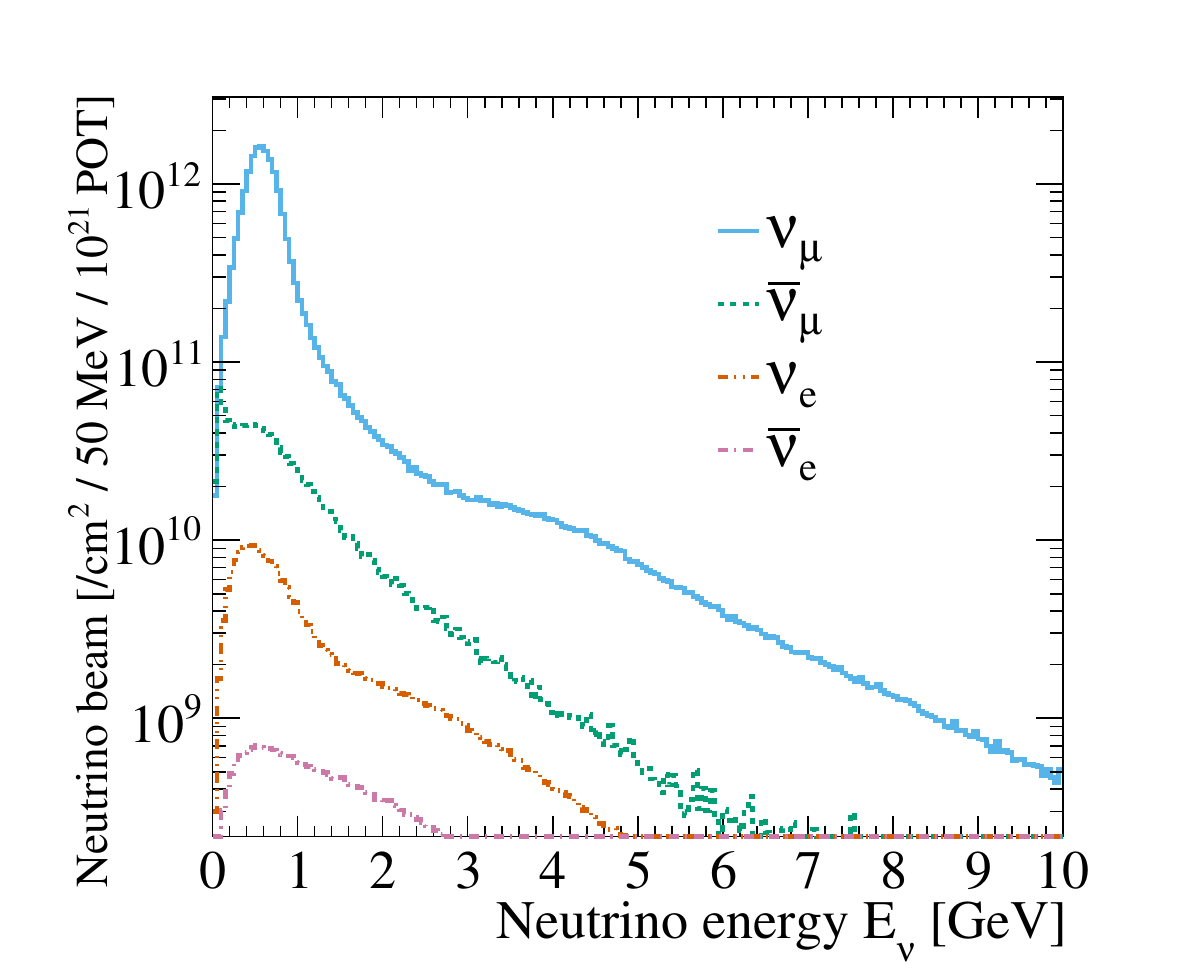}
\caption{Simulated neutrino beam spectra at ND280 in neutrino-enhanced beam mode. The spectra are shown for different neutrino types ($\nu_{\mu}$, $\bar{\nu}_{\mu}$, $\nu_{e}$ and $\bar{\nu}_{e}$) separately. The simulation method is described in Sec.~\ref{subsec:data_MC}.}
\label{fig:nu_flux_ND280}
\end{figure}

The resulting neutrino beam is measured by the T2K near and far detectors. Both ND280 and Super-Kamiokande are positioned at an off-axis angle of $2.5^{\circ}$ relative to the beam direction. This off-axis configuration is designed to produce a narrower neutrino energy spectrum~\cite{E899:1995bzq} and maximize the flux around 0.6~GeV~\cite{T2K:2012bge}, which corresponds to the first oscillation maximum at Super-Kamiokande. Fig.~\ref{fig:nu_flux_ND280} illustrates the simulated neutrino beam spectra at ND280 in neutrino-enhanced mode.

\subsection{The ND280 Detector}
\label{subsec:ND280}

\begin{figure}[!htbp]
\centering
\includegraphics[width=0.45\textwidth]{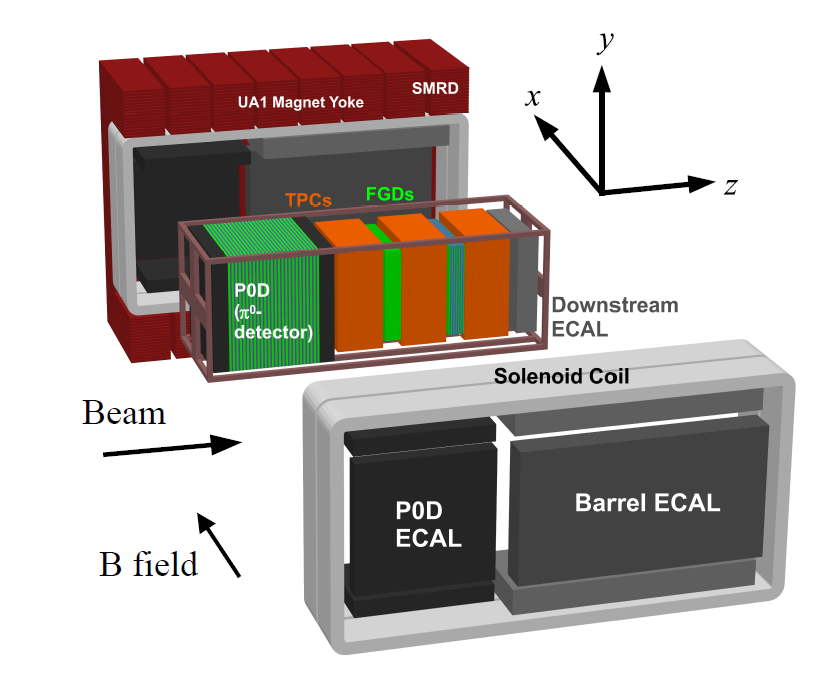}
\caption{Sketch of the ND280 detector in an exploded view. A supporting basket holds the $\pi^{0}$ detector (P{\O}D) as well as the FGDs and TPCs that make up the ND280 tracker. The ECALs surround the basket and the SMRD is in the air gaps of the magnet yoke. The reference coordinate system, the magnetic field (B field) direction and the neutrino beam direction are marked in the sketch as well.}
\label{fig:ND280_sketch}
\end{figure}

The ND280 detector is a key component of the T2K near detector complex, located in an underground hall $\sim 280$~m downstream of the neutrino production target. Its primary function is to provide in-situ measurements of the neutrino beam, which are used to tune experimental simulations and constrain systematic uncertainties in T2K's oscillation analyses. Additionally, the detector has excellent tracking capabilities that allow for dedicated neutrino cross-section measurements. ND280 is a magnetized detector comprising several sub-detectors, as illustrated in Fig.~\ref{fig:ND280_sketch}. The external dimensions of the detector, including the magnet, are $5.6 \times 6.1 \times 7.6$~m$^{3}$ (width $\times$ height $\times$ length). The UA1 magnet~\cite{T2K:2011qtm} provides a $0.2$~T magnetic field oriented along the $X$-axis (refer to Fig.~\ref{fig:ND280_sketch} for the coordinate system). The magnet yoke is instrumented with layers of plastic scintillator, forming the Side Muon Range Detector (SMRD)~\cite{Aoki:2013swe}. Within the magnet enclosure, several electromagnetic calorimeters (ECALs)~\cite{T2KUK:2013wkh} surround the inner detectors. These consist of barrel ECALs and a downstream ECAL, primarily used to distinguish between track-like and electromagnetic shower-like objects. Each ECAL is constructed from alternating layers of plastic scintillator and lead.
 
The ND280 inner detector suite comprises a $\pi^{0}$ detector (P{\O}D,~\cite{Assylbekov:2011sh}), two Fine-Grained scintillator Detectors (FGDs,~\cite{T2KND280FGD:2012umz}), and three Time Projection Chambers (TPCs,~\cite{T2KND280TPC:2010nnd}). The P{\O}D is positioned furthest upstream and consists of alternating layers of water bags, brass sheets, and $X$–$Y$ scintillator planes which are composed of triangular-shaped scintillator bars. In 2023, the ND280 detector was upgraded~\cite{T2K:2019bbb}, which involved the removal of the P{\O}D. However, as this measurement utilizes neutrino beam data collected prior to the upgrade, only the pre-upgrade configuration is described here.

The FGDs serve as the primary targets for measuring neutrino-nucleus interactions. The upstream module, FGD1, consists of 15 polystyrene scintillator modules. Each module measures $186.4 \times 186.4 \times 2.02$~cm$^3$ and is structured as two scintillator layers oriented in the $X$ and $Y$ directions. Each layer contains 192 square bars, each 9.6~mm wide and 2~m long, with readouts at one end. The downstream module, FGD2, has seven similar scintillator modules but interleaved with six passive water layers. This design enables the measurement of neutrino interactions on both carbon and oxygen, the latter being the primary target material for the Super-Kamiokande far detector.

The TPCs are used to reconstruct and identify charged particles produced by neutrino interactions within the FGDs. The three TPCs are positioned upstream of FGD1, between the two FGDs, and downstream of FGD2. Each TPC contains a field cage filled with a gas mixture of Ar:CF$_{4}$:iC$_{4}$H$_{10}$ (95:3:2). The TPC readout anode walls are instrumented with 12 MicroMegas modules arranged in two columns. Each MicroMegas module has an active area of $336 \times 353$~mm$^2$ segmented into 1728 rectangular pads (48 rows by 36 columns), enabling 3D reconstruction of the charged particle tracks traversing the TPCs.

\section{Analysis Observables}
\label{sec:analysis_observables}

This analysis measures the cross sections with the KI observables developed in recent years~\cite{Lu:2015tcr, Lu:2019nmf}. These observables exploit the correlation between final-state leptons and hadrons and offer enhanced sensitivity to underlying nuclear effects. Such observables typically include transverse kinematic imbalance (TKI) variables, which are built upon the conservation of transverse momentum perpendicular to the incident neutrino direction, where any observed imbalance between the lepton and hadronic systems directly probes the initial nuclear state, multi-nucleon correlation and FSI. Furthermore, the ``generalized'' kinematic imbalance (GKI) observables have been developed to incorporate longitudinal information~\cite{MicroBooNE:2023krv}, offering additional sensitivity to nuclear dynamics. Unlike TKI, longitudinal observables require an estimate of the neutrino energy, which is typically reconstructed from final-state kinematics under the quasi-elastic (QE) assumption~\cite{Furmanski:2016wqo}. For the $\nu_\mu$ CC$0\pi Np$ channel, these KI observables are reconstructed using the outgoing muon and the leading proton, defined as the proton with the highest momentum. 

\begin{figure}[htpb!]
\centering
\includegraphics[width=0.3\textwidth]{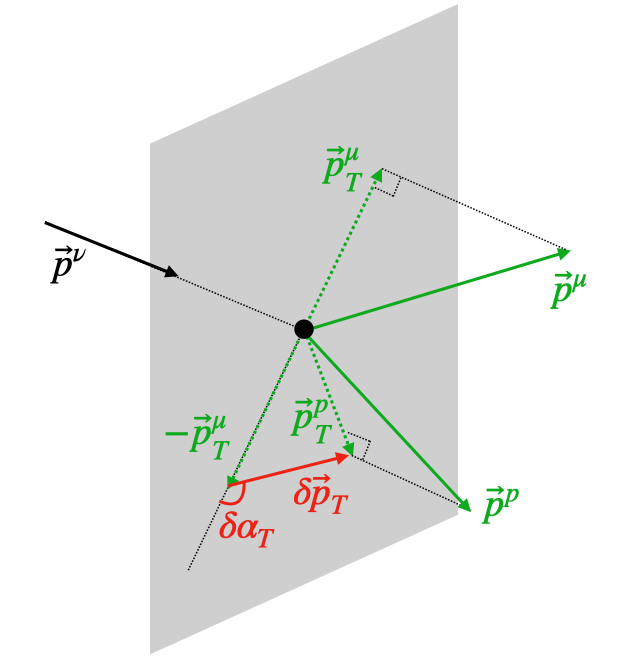}
\caption{Schematic illustration of kinematic imbalance observables $\delta p_{T}$ and $\delta \alpha_{T}$ (shown in red). The observables are defined on the plane transverse to the incident neutrino direction $\vec{p_{T}^{\nu}}$. The central black dot represents the target nucleus. The outgoing muon and leading proton momentum vectors are represented by $\vec{p_{T}^{\mu}}$ and $\vec{p_{T}^{p}}$ (shown in green), respectively. Plot adapted from Ref.~\cite{Lu:2015tcr}.}
\label{fig:TKI_sketch}
\end{figure}

In this measurement, we employ three specific KI observables:
\begin{itemize}
\item $\delta p_{T}$~\cite{Lu:2015tcr}: The magnitude of the total transverse momentum imbalance vector $\vec{\delta p_{T}}$, obtained by summing the transverse muon and leading proton momentum vectors. It is defined as
\begin{equation}
\delta p_{T} = \left| \vec{\delta p_{T}} \right| = \left| \vec{p_{T}^{\mu}}  + \vec{p_{T}^{p}} \right|,
\label{eq:dpt}
\end{equation}
where $\vec{p_{T}^{\mu}}$ and $\vec{p_{T}^{p}}$ are the muon and leading proton momentum vectors projected onto the transverse plane with respect to the incident neutrino direction.

\item $\delta \alpha_{T}$~\cite{Lu:2015tcr}: The angle between the total transverse momentum imbalance vector $\vec{\delta p_{T}}$ and the reversed muon transverse momentum vector $-\vec{p_{T}^{\mu}}$, also known as the ``transverse boosting angle''. It is given by
\begin{equation}
\delta \alpha_{T} = \arccos \left( \frac{-\vec{p_{T}^{\mu}} \cdot \vec{\delta p_{T}}}{\left| \vec{p_{T}^{\mu}} \right| \left| \vec{\delta p_{T}} \right|} \right).
\end{equation}
The schematic illustrations of $\delta p_{T}$ and $\delta \alpha_{T}$ are shown in Fig.~\ref{fig:TKI_sketch}.

\item $p_{N}$~\cite{Furmanski:2016wqo,Lu:2019nmf}: The inferred initial momentum of the target nucleon, computed as
\begin{equation}
p_{N} = \sqrt{\delta p_{T}^{2} + p_{L}^{2}},
\end{equation}
where $\delta p_{T}$ is the magnitude of total transverse momentum imbalance defined in Eq.~\ref{eq:dpt} and $p_{L}$ is the magnitude of initial longitudinal momentum of the nucleon, calculated as
\begin{equation}
\begin{split}
p_{L} = &\frac{1}{2} (M_{N} + p_{L}^{\mu} + p_{L}^{p} - E_{\mu} - E_{p}) \\
        &- \frac{1}{2} \left( \frac{\delta p_{T}^{2} + M_{N'}^{2}}{M_{N} + p_{L}^{\mu} + p_{L}^{p} - E_{\mu} - E_{p}} \right).
\end{split}
\end{equation}
Here, $M_{N'}$ denotes the residual nucleus mass, expressed as $M_{N'} = M_{N} - M_{p} + \left< \epsilon \right>_{p}$. The term $\left< \epsilon \right>_{p}$, the proton mean excitation energy, takes the fixed value 26.1~MeV for carbon and 23.0~MeV for oxygen~\cite{Bodek:2018lmc}. $M_{N}$, $E_{\mu}$, and $E_{p}$ are the target nucleus mass, muon energy, and leading proton energy, respectively.
\end{itemize}
As introduced above, the computation of the KI observables requires defining the plane perpendicular to the incident neutrino direction and projecting the muon and leading-proton momentum vectors onto this plane. Since the neutrino direction cannot be determined on an event-by-event basis, this measurement assumes it to be aligned with the $z$~-axis of the ND280 coordinate system (Fig.~\ref{fig:ND280_sketch}). Any deviation between the true neutrino direction and this assumption is treated as a source of detector-related systematic uncertainties and is included in the evaluation described in Sec.~\ref{subsubsec:syst_det}. From simulation, the maximum deviation is within 2$^{\circ}$, which has a small contribution to the total detector systematic uncertainties.

Nuclear effects contribute to KI observables in distinct and often overlapping ways. To decouple these correlations and isolate specific nuclear processes, it is beneficial to measure cross sections simultaneously across multiple KI dimensions. This analysis presents double-differential cross-section measurements for two primary pairs of observables: $\delta p_{T}$-$\delta \alpha_{T}$ and $p_{N}$--$\cos\theta_{\mu}$, where $\cos\theta_{\mu}$ represents the muon scattering angle with respect to the incident neutrino direction.

In the $\delta p_{T}$-$\delta \alpha_{T}$ phase space, the $\delta p_{T}$ distribution is primarily governed by the nucleon initial motion from CCQE interactions, resulting in a characteristic peak at values corresponding to the transverse component of the initial-state nucleon momentum. This peak is further broadened by additional nuclear effects, most notably final-state interactions (FSI) and multi-nucleon correlations. FSI typically reduces the kinetic energy of outgoing hadrons, manifesting in two distinct ways within the $\delta p_{T}$ spectrum: first, for nucleons that remain above the detection threshold, FSI increases the observed momentum imbalance, populating the distribution at larger $\delta p_{T}$ values; second, FSI can decelerate protons below the tracking threshold, thereby suppressing the $\delta p_{T}$ peak. Multi-nucleon correlations similarly introduce larger momentum imbalances, contributing primarily to the high-$\delta p_{T}$ region. Complementing these features, the $\delta \alpha_{T}$ variable serves as a sensitive probe of FSI, which shifts the distribution toward larger angles. Consequently, in the 2D $\delta p_{T}$--$\delta \alpha_{T}$ space, the initial nucleon motion predominantly shapes the $\delta p_{T}$ peak at small $\delta \alpha_{T}$, while the high-$\delta p_{T}$ region becomes progressively populated by FSI and multi-nucleon correlations as $\delta \alpha_{T}$ increases.

In $p_{N}$--$\cos\theta_{\mu}$ space, the muon scattering angle acts as an effective proxy for discriminating between various interaction channels. Large scattering angles are predominantly populated by CCQE interactions. Conversely, forward-scattering regions contain substantial contributions from multi-nucleon correlations and pion-absorption channels via FSI. The $p_{N}$ distribution functions similarly to $\delta p_{T}$, where the peak region characterizes the initial-state nucleon motion and the high-momentum tail provides sensitivity to FSI and non-QE contributions. Consequently, this two-dimensional space facilitates the separation of CCQE-dominated events from those influenced by FSI and non-QE processes.

\section{Event Selection}
\label{sec:eve_sele}

\subsection{Data and Monte Carlo Samples}
\label{subsec:data_MC}

This measurement utilizes T2K neutrino data collected between 2010 and 2017, totaling $11.61 \times 10^{20}$ protons on target (POT). These data correspond to periods of neutrino-mode operation (forward horn current, FHC), which produces a predominantly muon-neutrino beam.

For the Monte Carlo (MC) samples, the neutrino beam is simulated using FLUKA v4.2.1~\cite{Ferrari:2005zk, Bohlen:2014buj}, with tuning based on hadron production measurements from the NA61/SHINE experiment using a T2K replica target~\cite{NA61SHINE:2016nlf, NA61SHINE:2018rhe}. The simulation framework also incorporates JNUBEAM, a package based on GEANT3~\cite{Brun:1994aa} and GCALOR~\cite{Zeitnitz:1992vw}. Specifically, FLUKA models the proton beam interactions with the graphite target and the subsequent secondary particle production, while JNUBEAM simulates the propagation and decay of these particles within the decay tunnel.

Neutrino-nucleus interactions within the ND280 detector are simulated using NEUT v5.6.4.1~\cite{Hayato:2009zz, Hayato:2021heg}. CCQE interactions are modeled via a spectral function (SF) approach~\cite{Benhar:1994hw,Chakrani:2023htw}, while multi-nucleon interactions, specifically two-particle two-hole (2p2h) processes, are described by the Valencia model~\cite{Nieves:2011pp,Gran:2013kda}. Resonant meson production (CCRES), primarily pions, is simulated via the Rein-Sehgal model~\cite{Rein:1980wg}, updated with an improved nucleon axial form factor~\cite{Graczyk:2014dpa,Graczyk:2007bc} and the inclusion of final-state lepton mass effects~\cite{Kuzmin:2003ji,Berger:2007rq,Graczyk:2007xk}. For higher neutrino energies where the invariant hadronic mass $W$ exceeds 1.4~GeV, charged-current deep inelastic scattering (CCDIS) is modeled using PYTHIA~\cite{SJOSTRAND2020106910} with GRV98 parton distribution functions and Bodek-Yang corrections~\cite{Bodek:2005de}. Below this threshold, a custom NEUT model for multi-pion production is employed. FSIs within the nucleus are simulated using the NEUT intra-nuclear cascade (INC) model~\cite{Serber:1947zza,Salcedo:1987md,Hayato:2009zz}.

The propagation of final-state particles through the detector material is modeled with GEANT4~\cite{GEANT4:2002zbu, Allison:2006ve, Allison:2016lfl} using the \texttt{QGSP\_BERT} physics list. The detector response, including the conversion of energy depositions into electronic signals, is handled by dedicated T2K software~\cite{T2K:2015sqm}. The total exposure of the MC samples is $217.07 \times 10^{20}$~POT, approximately 19 times the statistics of the experimental data.

\subsection{Signal Sample Definition}
\label{subsec:sig_def}

\begin{figure}[!htbp]
\centering
\includegraphics[width=0.45\textwidth]{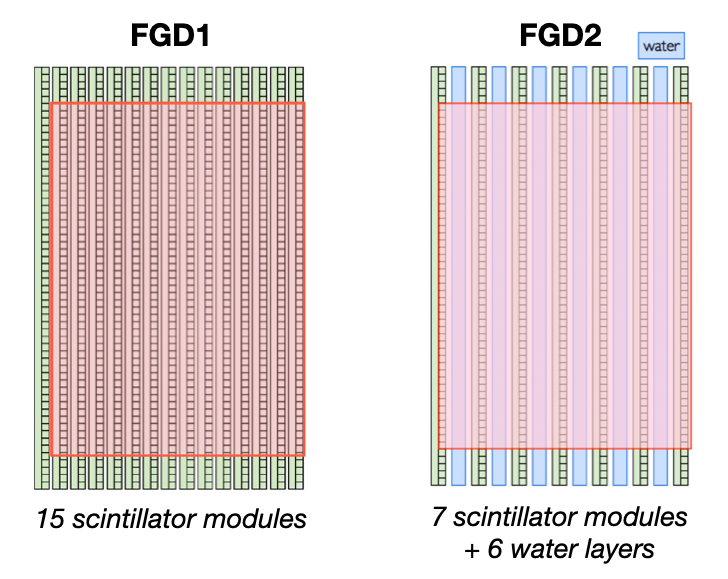}
\caption{Sketches of the FGD1 (left) and FGD2 (right) fiducial volumes (FVs). The FVs are marked as red regions. Plot adapted from Ref.~\cite{T2K:2020jav}.}
\label{fig:fgd_fv}
\end{figure}

In this measurement, we select $\nu_{\mu}$ CC$0\pi Np$ interactions which occurred on the carbon and oxygen targets in the ND280 FGD1 and FGD2 detector fiducial volumes (FV, as sketched in Fig.~\ref{fig:fgd_fv}). The FGD1 FV excludes the first scintillator module along the $z$-axis and the five outermost scintillator layers along the $x$ and $y$ axes. Similarly, the FGD2 FV excludes the first layer of the most upstream scintillator module and the five outermost layers along the $x$ and $y$ axes. With this FV definition, the total mass of carbon target is $\sim$~1230\,kg and that of oxygen is $\sim$~398\,kg. The carbon mass is distributed between two FGDs, with FGD1 contributing $\sim$~861\,kg and FGD2 contributing $\sim$~369\,kg. The oxygen mass is predominantly contained within FGD2.

\begin{figure}[!htbp]
\centering
\includegraphics[width=0.45\textwidth]{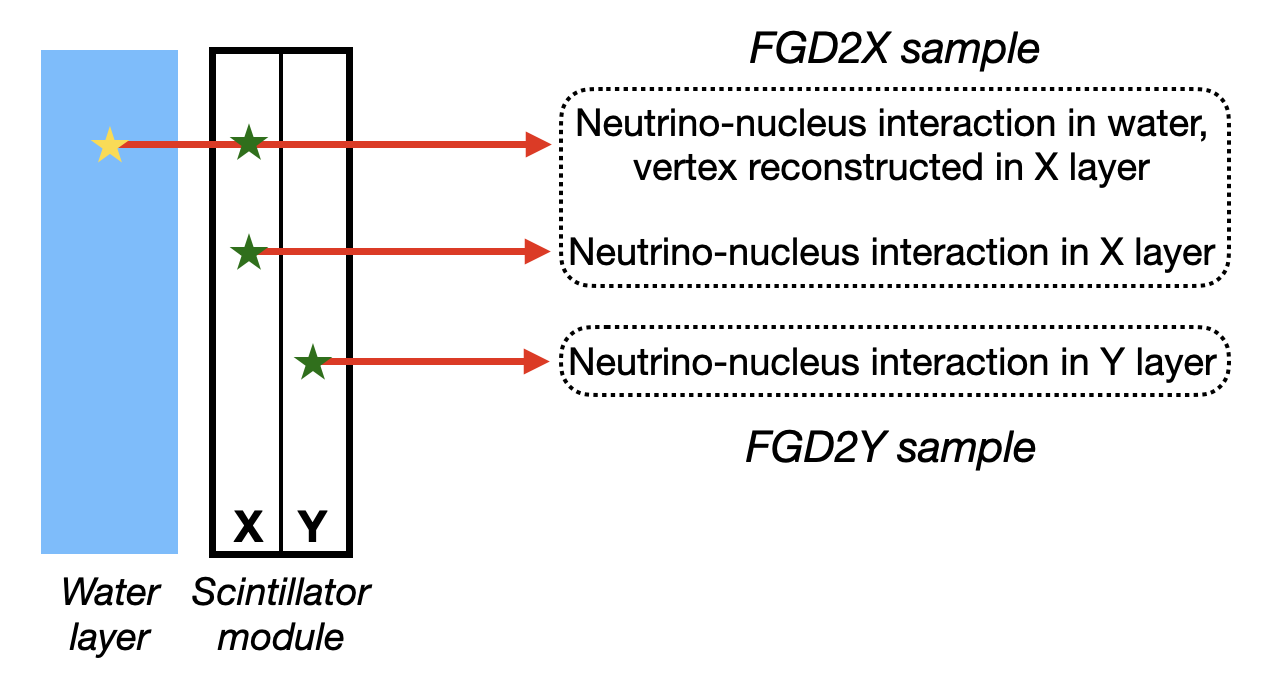}
\caption{Sketch of the FGD2X/Y sample scheme. The green stars indicate the reconstructed interaction vertices in the scintillator modules and the yellow star indicates the true interaction vertex in the water layer.}
\label{fig:fgd2_sample}
\end{figure}

The signal samples are selected separately for the FGD1 and FGD2 detectors. Within FGD2, the sample is further categorized into ``FGD2X'' and ``FGD2Y'' groups based on the reconstructed position of the selected neutrino interaction vertices. The distributions of these selected vertex positions can be found in Appendix~\ref{sec:dis_vtx_pos_FGD2}. This separation is intended to better isolate interactions on carbon and oxygen targets. Fig.~\ref{fig:fgd2_sample} illustrates the FGD2X/Y sample division. As described in Sec.~\ref{subsec:ND280}, each FGD scintillator module consists of layers oriented in the $X$ and $Y$ directions. In the signal selection process (Sec.~\ref{subsec:sig_selection}), we identify the leading primary particle traveling in the forward direction. When vertices are reconstructed in an $X$ layer, the interaction may have occurred either in the preceding water layer (passive region) or within the $X$ layer itself. Conversely, when vertices are reconstructed in a $Y$ layer, the majority of interactions occur within that same layer. Consequently, neutrino-oxygen interactions are preferentially selected in the oxygen-enhanced FGD2X sample, while neutrino-carbon interactions with higher purity are retained in the carbon-enhanced FGD2Y sample. Tab.~\ref{tab:sig_target_frac} summarizes the fractions of $\nu_{\mu}$ CC$0\pi Np$ signal interactions on carbon and oxygen targets for each sample, as determined from MC simulations.

\begin{table}[!htbp]
\centering
\caption{Fractions of $\nu_{\mu}$ CC$0\pi Np$ interactions on carbon and oxygen targets within the selected signal samples. The selection is described in Sec.~\ref{subsec:sig_sample} and the values are estimated based on MC (Sec.~\ref{subsec:data_MC}). The sum of carbon and oxygen fractions in each category is less than 100\% due to background contamination and signal interactions on other nuclei within the detector materials (e.g., titanium, silicon, and nitrogen which compose the glue and optical fibers of the FGDs~\cite{T2KND280FGD:2012umz}).}
\label{tab:sig_target_frac}
\renewcommand{\arraystretch}{1.2}
\begin{tabular}{c|c|c}
\hline
\hline
Category & $\nu_{\mu}$ CC$0\pi Np$ on carbon & $\nu_{\mu}$ CC$0\pi Np$ on oxygen \\
\hline
FGD1 & 74.5\% & 3.9\% \\
FGD2X & 33.0\% & 37.5\% \\
FGD2Y & 52.9\% & 15.9\% \\
\hline
\hline
\end{tabular}
\end{table}

The signal sample is further divided into three sub-samples based on where the selected muons and leading protons are detected after their production. The main purpose of the sub-sample division is to correctly assign detector systematic uncertainties (Sec.~\ref{subsubsec:syst_det}), as different sub-samples rely on different reconstruction and selection tools. The signal sub-sample scheme is summarized below, following the same definitions as the previous T2K ND280 $\nu_{\mu}$ CC$0\pi$ measurements~\cite{T2K:2020jav}:
\begin{itemize}
\item \textbf{TPC $\mu$ + TPC $p$}: Both the selected muon and the leading proton exit the FGD and enter the downstream TPC. This sub-sample is restricted to events with exactly one reconstructed proton.

\item \textbf{TPC $\mu$ + FGD $p$ (+ $Np$)}: The selected muon enters the downstream TPC, while the leading proton stops within the FGD. This sub-sample allows for the selection of multiple protons.

\item \textbf{FGD $\mu$ + TPC $p$ (+ $Np$)}: The selected muon either stops in the FGD or escapes and stops in the barrel ECAL, while the leading proton enters the downstream TPC. This sub-sample allows for the selection of multiple protons.
\end{itemize}

\subsection{Signal Selection}
\label{subsec:sig_selection}

The signal selection identifies the $\nu_{\mu}$ CC$0\pi Np$ interactions originating from the FGD1 and FGD2 FVs. The selection algorithm follows the same structure as one developed for previous T2K measurement~\cite{T2K:2020sbd}. The basic steps are summarized below:
\begin{enumerate}
\item \textbf{Leading track selection}: The highest-momentum track entering TPC2 (TPC3) for the FGD1 (FGD2) samples is selected as the primary muon or leading proton candidate. Here, “primary” denotes a track originating directly from the neutrino interaction vertex. In the TPC $\mu$ + TPC $p$ and TPC $\mu$ + FGD $p$ (+ $Np$) sub-samples, the primary muon candidate is identified, while in the FGD $\mu$ + TPC $p$ (+ $Np$) sub-sample, the primary leading proton candidate is selected. Particle identification (PID) in the TPC is applied to distinguish muon and proton candidates based on measurements of the local energy loss ($dE/dx$). In addition, the track is required to originate within the FV of FGD1 or FGD2.

\item \textbf{Primary vertex identification}: The primary neutrino interaction vertex is defined as the starting point of the leading track selected in the previous step.

\item \textbf{Additional track selection}: Additional primary track candidates are identified if they originate within a specified proximity to the primary vertex (within 50~mm in the $XY$-plane and 30~mm along the $z$-axis) and satisfy either TPC PID or, for tracks not entering the TPC, FGD-based muon or proton PID. The FGD PID relies on the analysis of the local $dE/dx$ profile along the track, including the high $dE/dx$ region near the track endpoint, which helps identify stopping protons or muons. In the TPC $\mu$ + TPC $p$ and TPC $\mu$ + FGD $p$ (+ $Np$) sub-samples, the additional primary tracks are required to be proton candidates only. In the FGD $\mu$ + TPC $p$ (+ $Np$) sub-sample, the remaining primary tracks must include one muon candidate and any number of proton candidates. Events with tracks not identified as muons or protons are rejected.

\item \textbf{Veto cuts}: Two key veto criteria are applied to suppress the dominant backgrounds in the signal selection. These backgrounds primarily arise from neutrino interactions that produce unreconstructed pions, which can mimic the signal topology. Such interactions are predominantly due to CCRES or CCDIS processes, as well as FSI that generate pions in the final state. The first veto rejects events containing Michel electrons produced by the decay of short charged pions in the FGDs. These electrons are identified as time-delayed signals occurring more than 100 ns after the primary vertex time. The second veto removes events containing $\pi^{0}$ production, where the decayed photons enter the surrounding ECALs and initiate electromagnetic showers. The veto requires at least one electromagnetic shower object to be identified in any ECAL.

\end{enumerate}

\subsection{Signal Phase Space Constraint}
\label{subsec:ps_cons}

\begin{figure*}
\centering
\subfloat[Momentum]
{\includegraphics[width=0.4\textwidth]{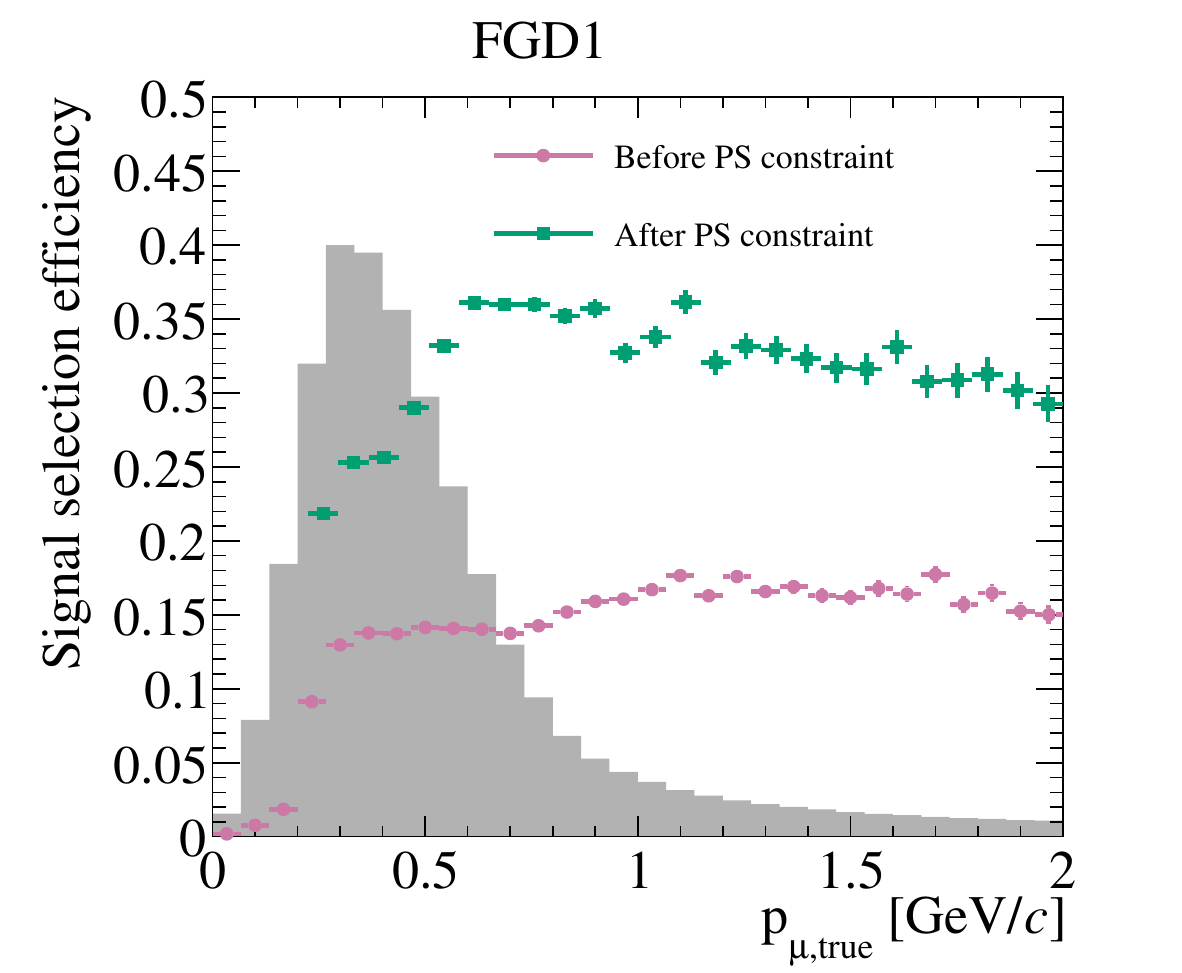}
\includegraphics[width=0.4\textwidth]{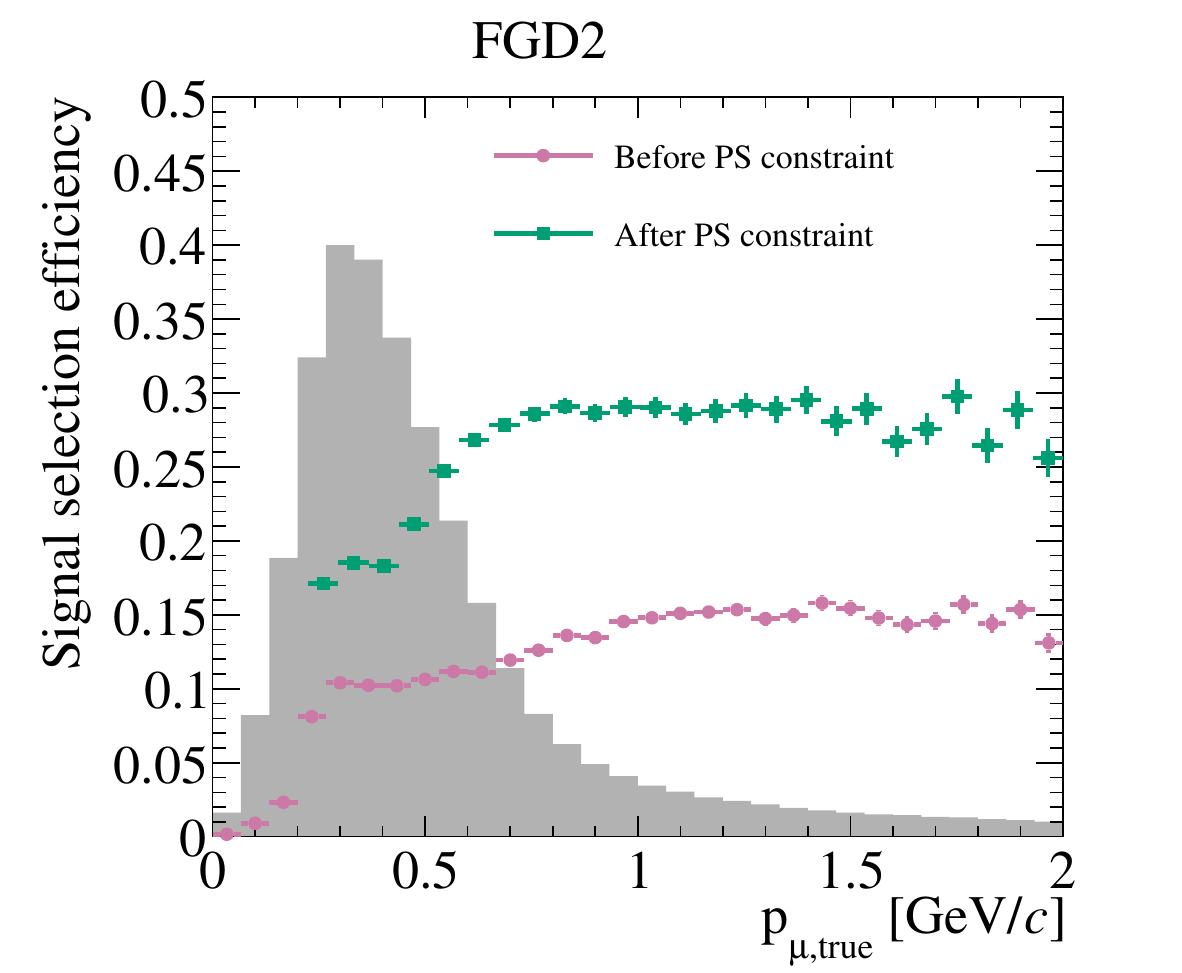}}
\\
\subfloat[Direction]
{\includegraphics[width=0.4\textwidth]{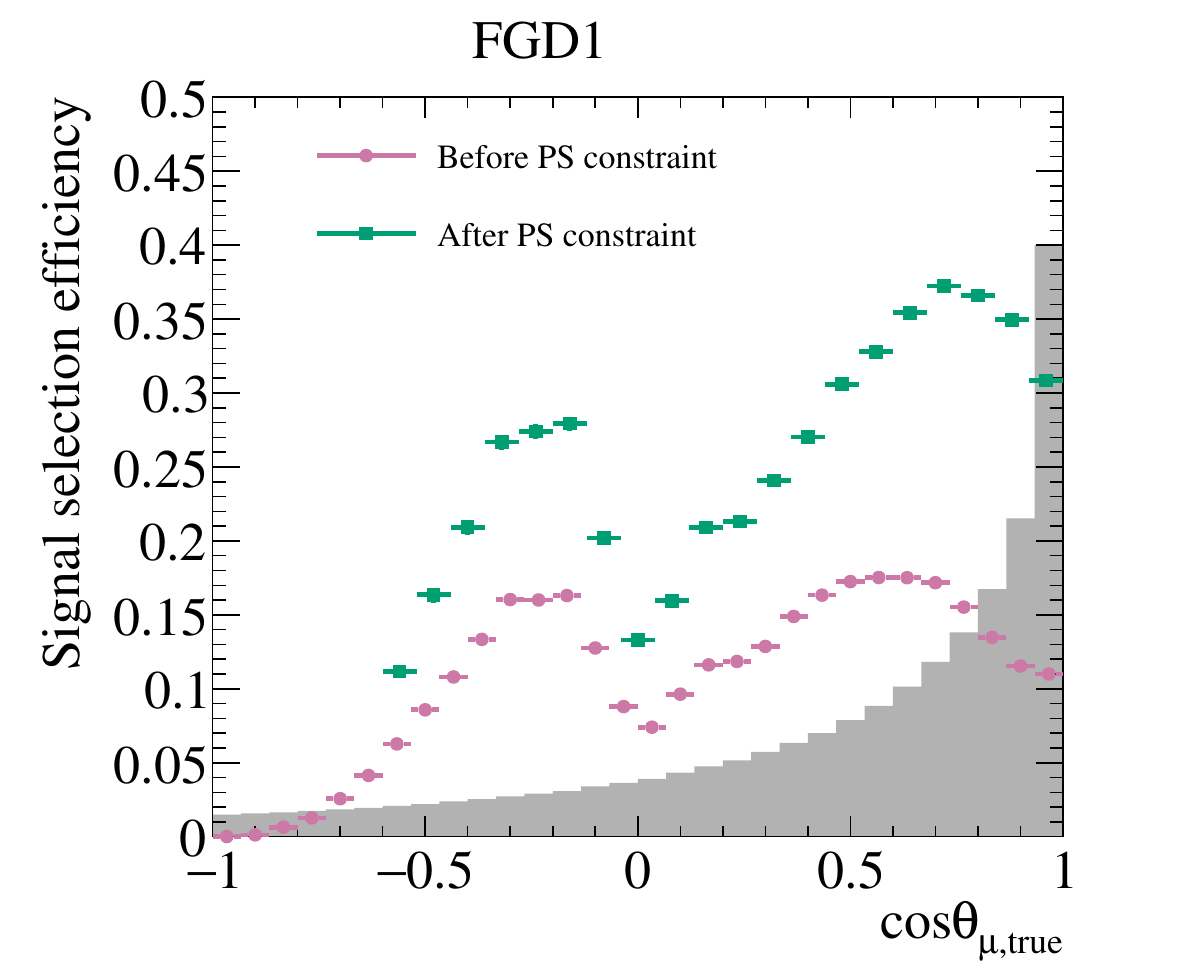}
\includegraphics[width=0.4\textwidth]{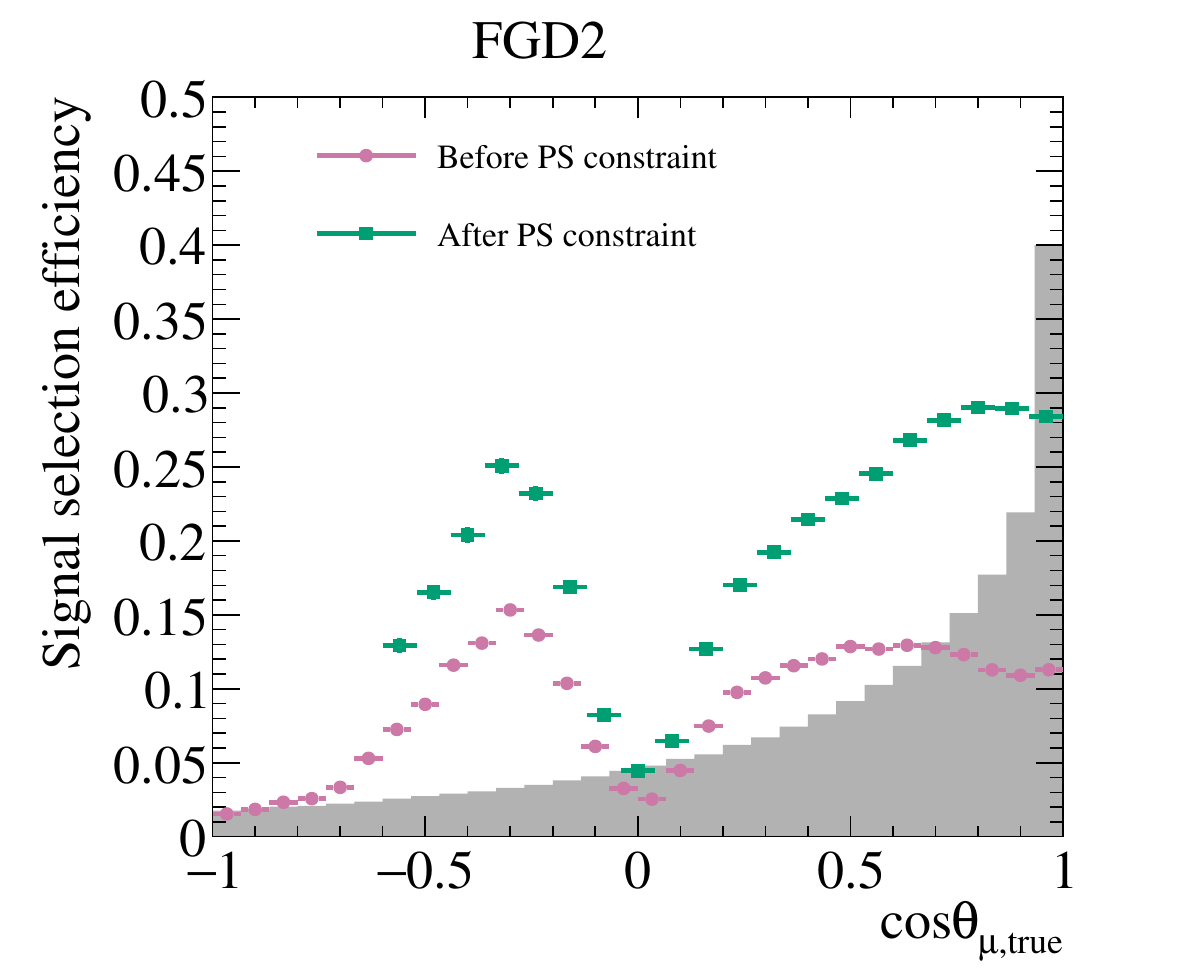}}
\caption{The $\nu_{\mu}$ CC$0\pi Np$ signal selection efficiency as a function of the true primary muon (a) momentum $p_{\mu,\text{true}}$ and (b) direction $\cos\theta_{\mu,\text{true}}$ for FGD1 and FGD2. Efficiencies are shown before (pink) and after (green) the phase space (PS) constraints (Tab.~\ref{tab:ps_cons}). The background gray histograms illustrate the shape of the total $\nu_{\mu}$ CC$0\pi Np$ event rate (in arbitrary unit) predicted by the NEUT v5.6.4.1 MC used in this measurement.}
\label{fig:sig_seleff_mu}
\end{figure*}

\begin{figure*}
\centering
\subfloat[Momentum]
{\includegraphics[width=0.4\textwidth]{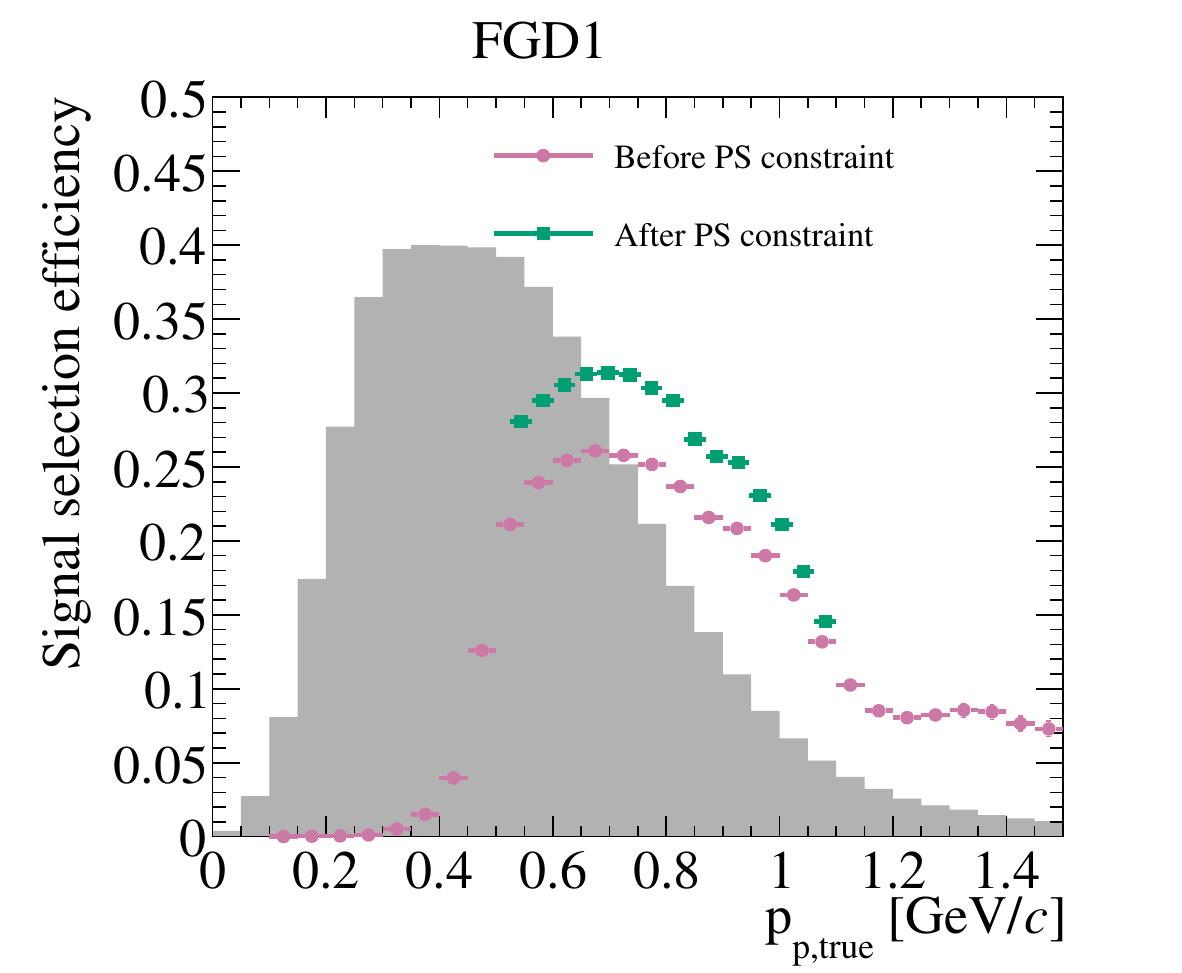}
\includegraphics[width=0.4\textwidth]{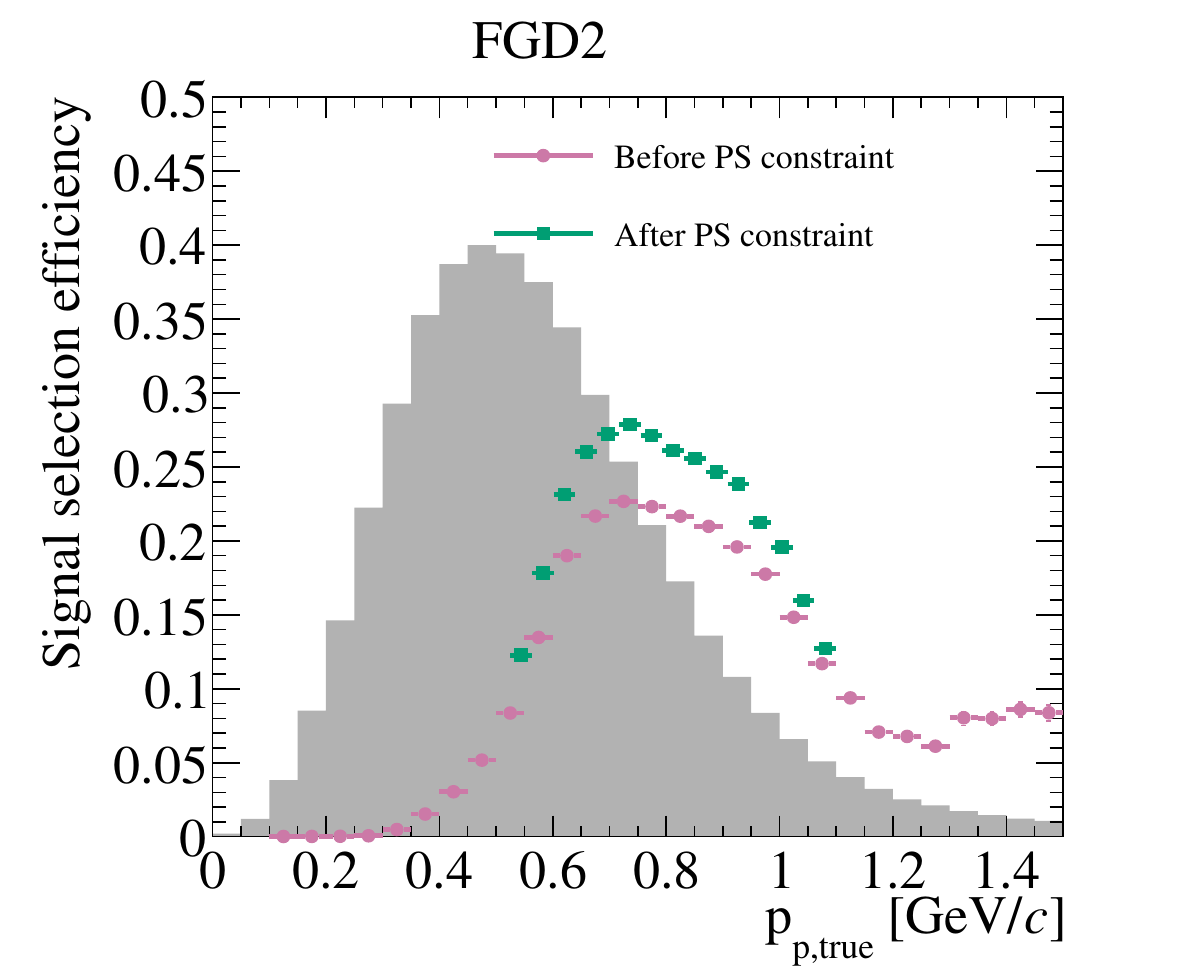}}
\\
\subfloat[Direction]
{\includegraphics[width=0.4\textwidth]{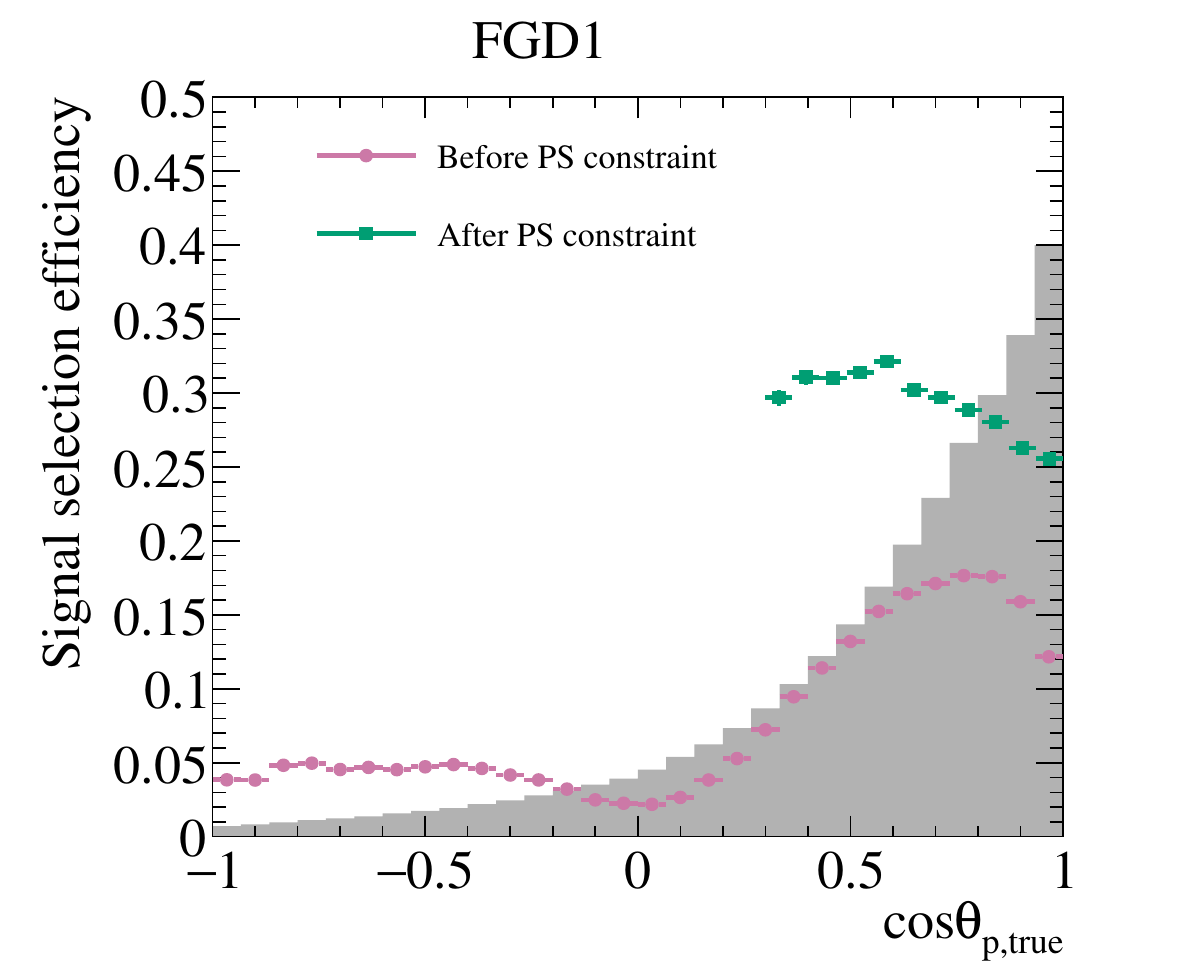}
\includegraphics[width=0.4\textwidth]{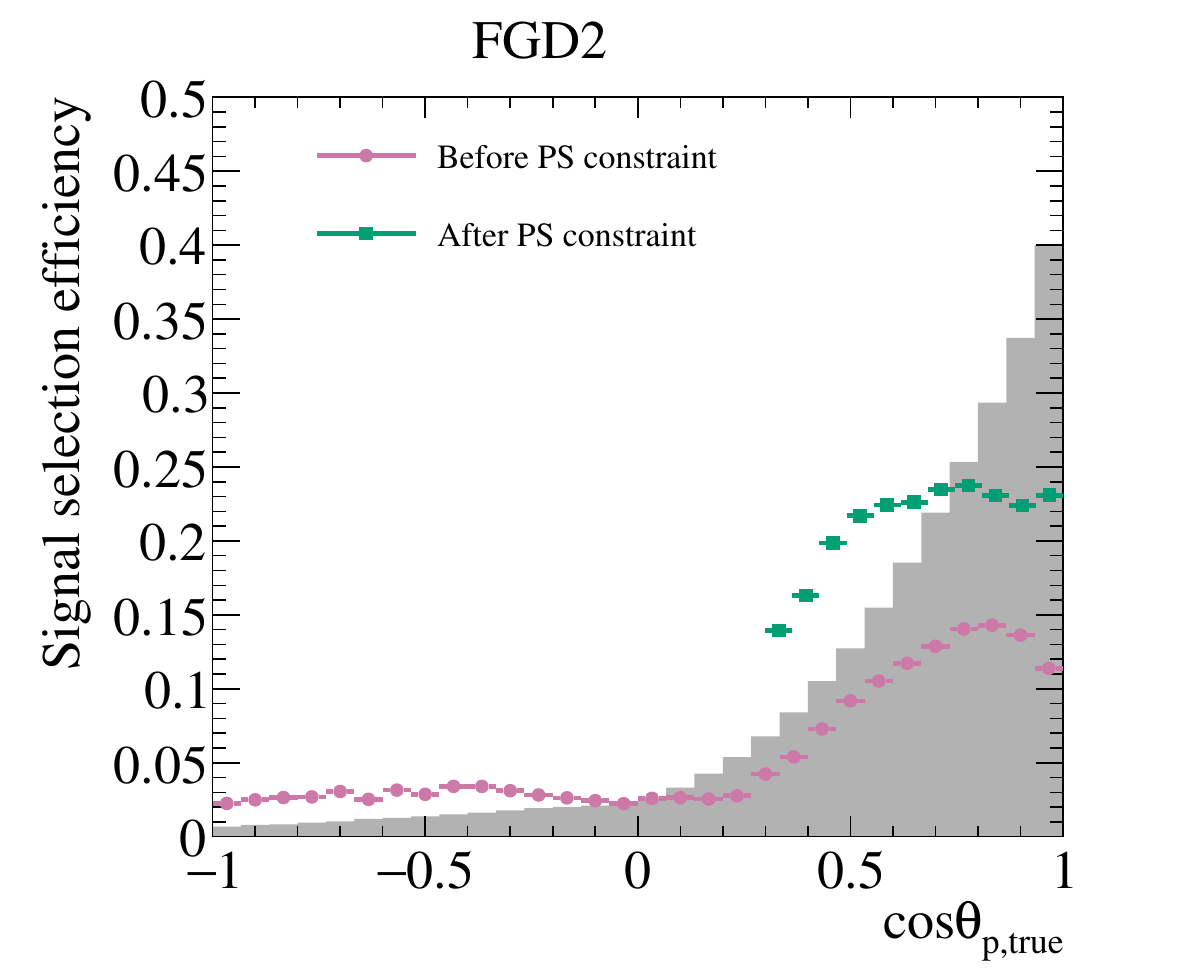}}
\caption{The $\nu_{\mu}$ CC$0\pi Np$ signal selection efficiency as a function of the true primary leading proton (a) momentum $p_{p,\text{true}}$ and (b) direction $\cos\theta_{p,\text{true}}$ for FGD1 and FGD2. Efficiencies are shown before (pink) and after (green) the phase space (PS) constraints (Tab.~\ref{tab:ps_cons}). The background gray histograms illustrate the shape of the total $\nu_{\mu}$ CC$0\pi Np$ event rate (in arbitrary unit) predicted by the NEUT v5.6.4.1 MC used in this measurement.}
\label{fig:sig_seleff_pr}
\end{figure*}

In this measurement, the phase space constraint limits the selected muon and leading proton kinematics to the region where the detector has good acceptance and the related detector systematics are well understood. To determine the exact phase space cut, the $\nu_{\mu}$ CC$0\pi Np$ signal selection efficiency is computed from the simulation as a function of the muon or leading proton momentum and scattering angle ($\cos\theta$). The efficiencies are shown as functions of muon kinematics (Fig.~\ref{fig:sig_seleff_mu}) and leading proton kinematics (Fig.~\ref{fig:sig_seleff_pr}). 

In the muon momentum space (Fig.~\ref{fig:sig_seleff_mu} (a)), the efficiency decreases significantly for momenta below $\sim$~0.2–0.25 GeV/c, as this approaches the tracking threshold of both FGDs. In the muon scattering angle space (Fig.~\ref{fig:sig_seleff_mu} (b)), a reduction in efficiency is observed in the transverse region (approximately $-0.1 < \cos\theta_{\mu} < 0.15$) and in the backward region ($\cos\theta_{\mu} < -0.6$). The decrease in the vertical region is mainly due to the low acceptance of the FGDs at these angles, since reconstructed tracks are required to pass through at least 4 scintillation layers. The reduction in the backward region is primarily attributed to the limited time resolution of the FGDs for identifying backward-going tracks.

In the proton momentum space (Fig.~\ref{fig:sig_seleff_pr} (a)), the efficiency decreases above $\sim$~1.1–1.2 GeV/c, where most protons escape into the forward TPCs. The TPC proton PID performance is limited in this momentum region because the proton $dE/dx$ is similar to that of electrons~\cite{T2KND280TPC:2010nnd}. In addition, towards lower proton momenta ($p_{p,true} < 0.8~ \text{GeV}/c$), the efficiency in FGD2 is lower than that in FGD1 due to the interleaved inactive water layers in FGD2, which lead to poorer reconstruction performance for shorter proton tracks that stop within FGD2. In the proton scattering angle space (Fig.~\ref{fig:sig_seleff_pr} (b)), the efficiency decreases for large angle and backward-going protons ($\cos\theta_{p} < 0.3$), as most protons in this angular region stop in the FGDs, which have limited acceptance for reconstructing these tracks, similar to the behavior observed for muons.

Based on these observations, Tab.~\ref{tab:ps_cons} summarizes the phase space constraint applied to the selected muon and leading proton kinematics used in this measurement. The same constraints are applied to both the FGD1 and FGD2 samples. A cutoff on the muon momentum at 10 GeV/c is applied because event statistics above this momentum are very limited, and the associated detector systematic uncertainties can not be reliably evaluated. In addition, the vertical muon region ($-0.1 < \cos\theta_{\mu} < 0.15$) is retained in this measurement despite its low efficiency. This choice is motivated by the preference to avoid phase space discontinuities. Moreover, the selected statistics in this region are very small compared to other regions (Fig.~\ref{fig:sig_samp_dis} (d)), resulting in a small impact on the overall analysis. 

\begin{table}[!htbp]
\centering
\caption{Phase space constraint applied on the selected muon and leading proton kinematics used by this measurement.}
\label{tab:ps_cons}
\renewcommand{\arraystretch}{1.2}
\begin{tabular}{c|c|c}
\hline
\hline
Particle & Momentum & Direction \\
\hline
Muon & $0.225 < p_{\mu} < 10$ GeV/c & $\cos\theta_{\mu} > -0.6$ \\
Leading proton & $0.525 < p_{p} < 1$ GeV/c & $\cos\theta_{p} > 0.3$ \\
\hline
\hline
\end{tabular}
\end{table}

\begin{figure}
\centering
\subfloat[Carbon signal sample]
{\includegraphics[width=0.45\textwidth]{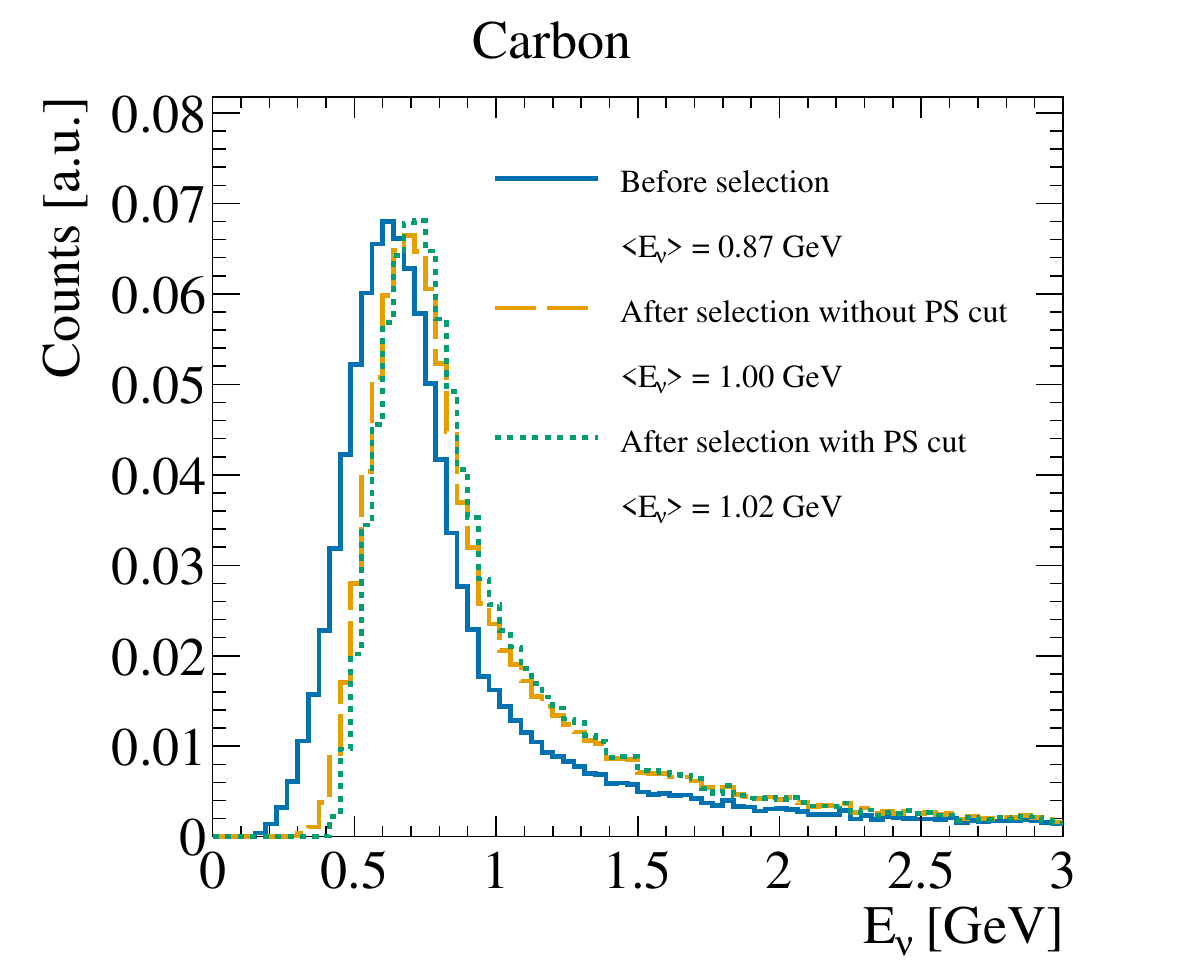}}
\\
\subfloat[Oxygen signal sample]
{\includegraphics[width=0.45\textwidth]{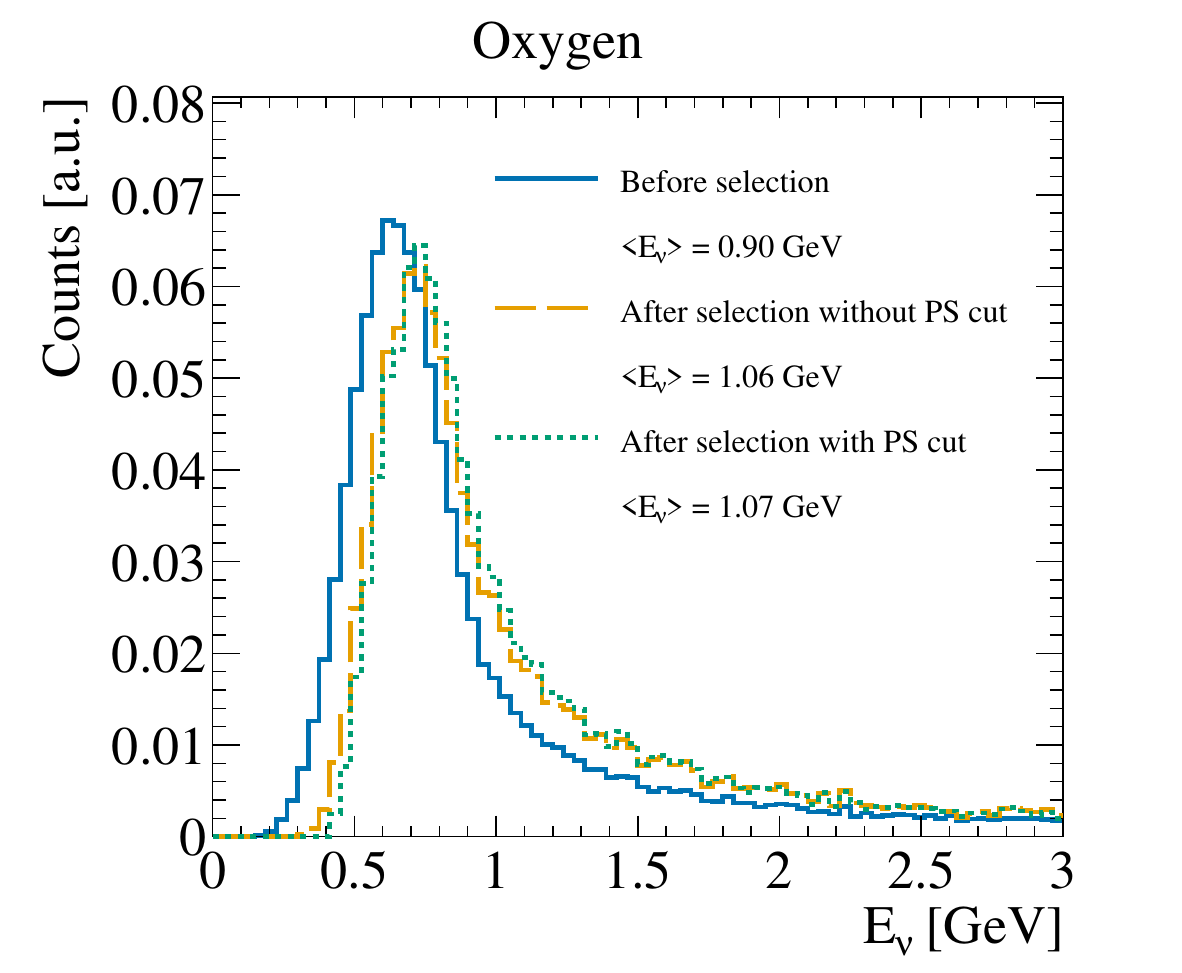}}
\caption{The $\nu_{\mu}$ CC$0\pi Np$ signal distributions from MC as a function of the true incident neutrino energy. Signal interactions on the carbon target (a) and the oxygen target (b) are shown separately. For each target, the signal distributions are shown before selection (sky blue), after selection without the phase-space (PS) cut (orange), and after selection with the PS cut (bluish green). The mean neutrino energy $\langle E_{\nu} \rangle$ of each distribution is also shown. All three distributions are normalized to the same integral to facilitate shape comparison.}
\label{fig:sig_nue}
\end{figure}

To inspect the impact of the selection and the phase-space constraint on the neutrino energy coverage, Fig.~\ref{fig:sig_nue} shows the signal distributions as a function of the true incident neutrino energy. In general, the neutrino energy range after the signal selection and phase-space constraint covers most of the original distribution, except in the low-energy region ($<$ 0.4~GeV), where the threshold is raised by $\sim$ 0.1~GeV owing to the detector's reconstruction and selection limitations for short proton tracks. The selection has a larger impact than the phase-space constraint, since requiring tracks to be reconstructed in the detector already imposes a minimum momentum. Nevertheless, the neutrino peak region ($\sim$ 0.6~GeV), where most signal interactions are located, is well covered by the selected samples. Furthermore, the impact is very similar for the carbon and oxygen targets, and no significant target dependence of the neutrino energy coverage is observed.

\subsection{Selected Signal Sample}
\label{subsec:sig_sample}

\begin{figure*}
\centering
\subfloat[$\delta p_{T}$]
{\includegraphics[width=0.33\linewidth]{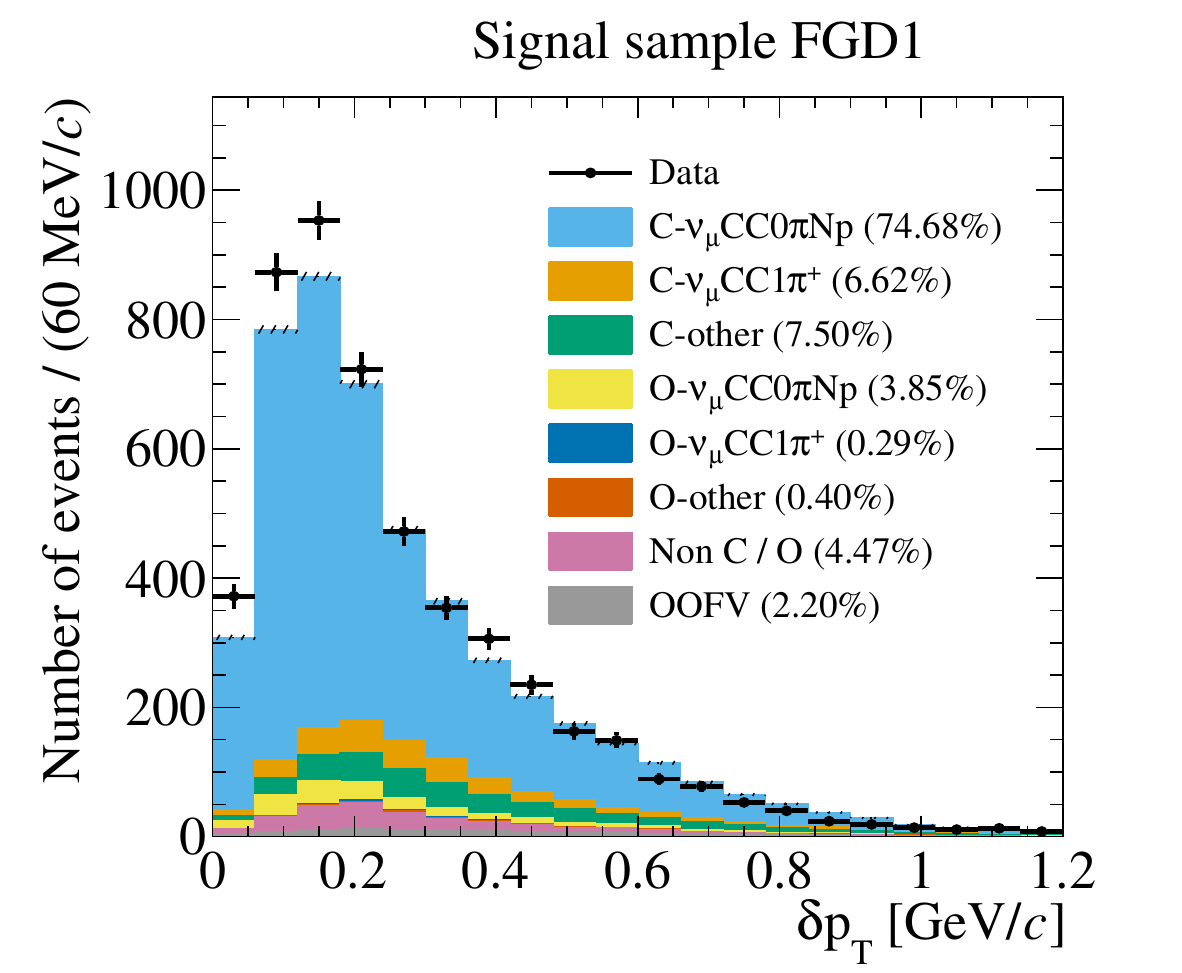}
\includegraphics[width=0.33\linewidth]{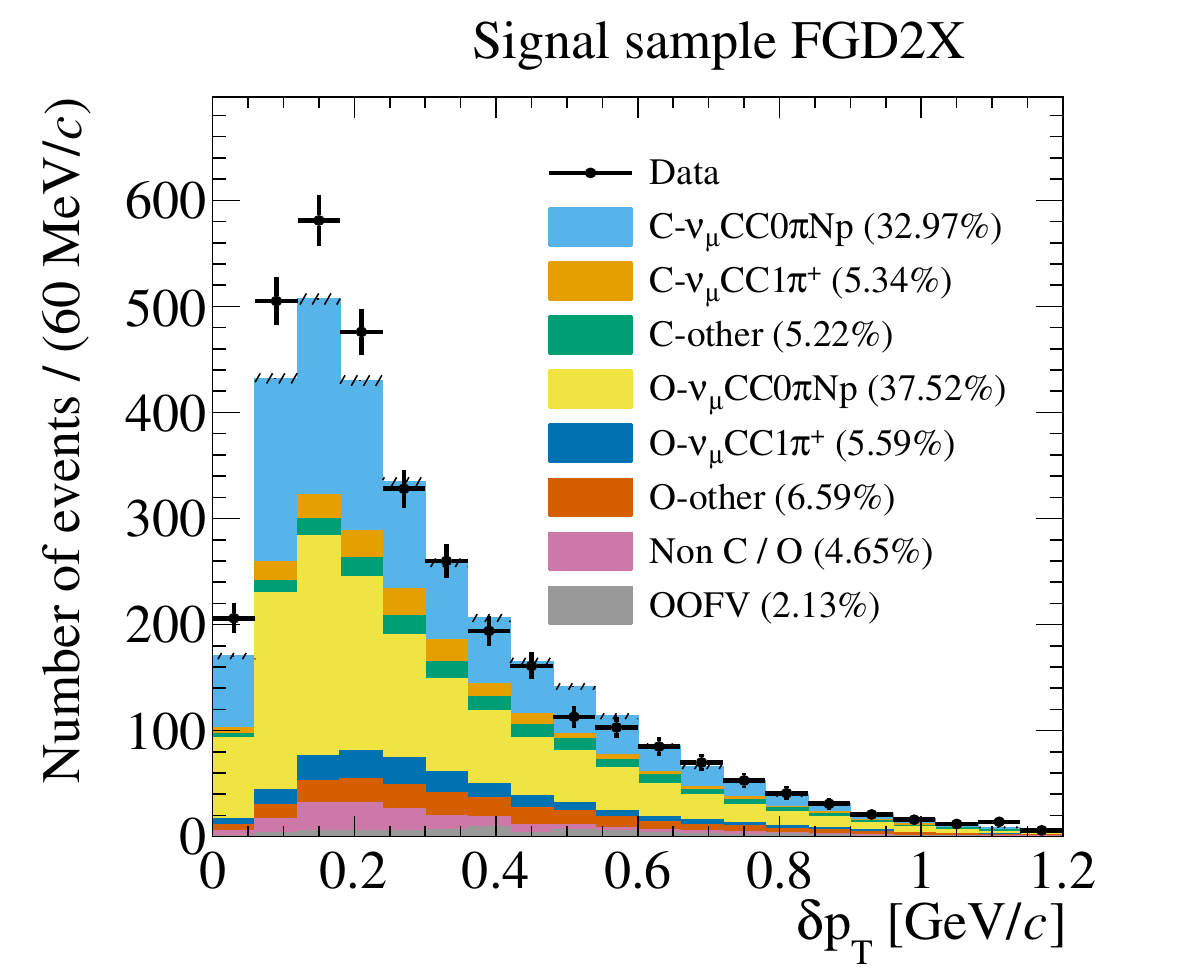}
\includegraphics[width=0.33\linewidth]{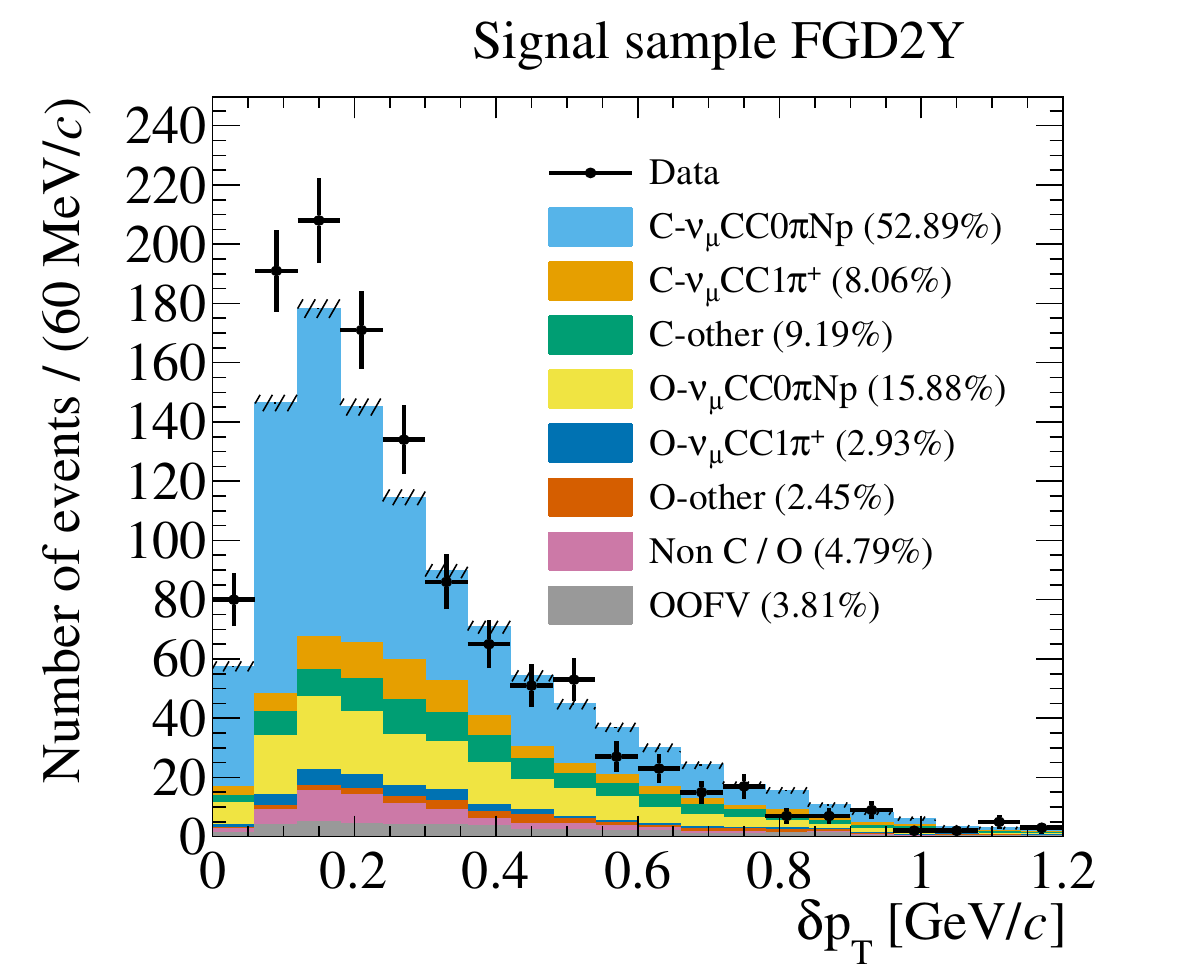}}
\\
\subfloat[$\delta \alpha_{T}$]
{\includegraphics[width=0.33\linewidth]{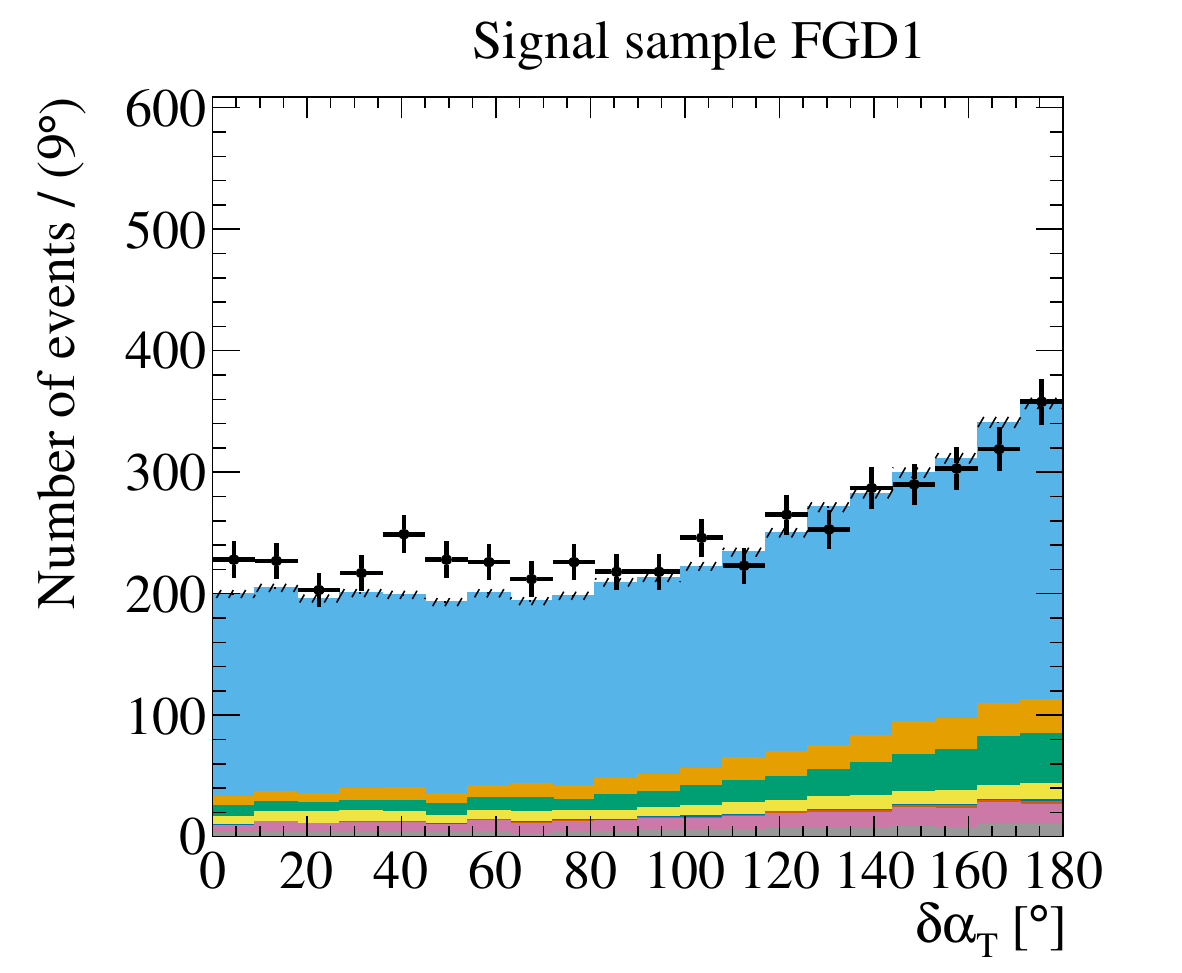}
\includegraphics[width=0.33\linewidth]{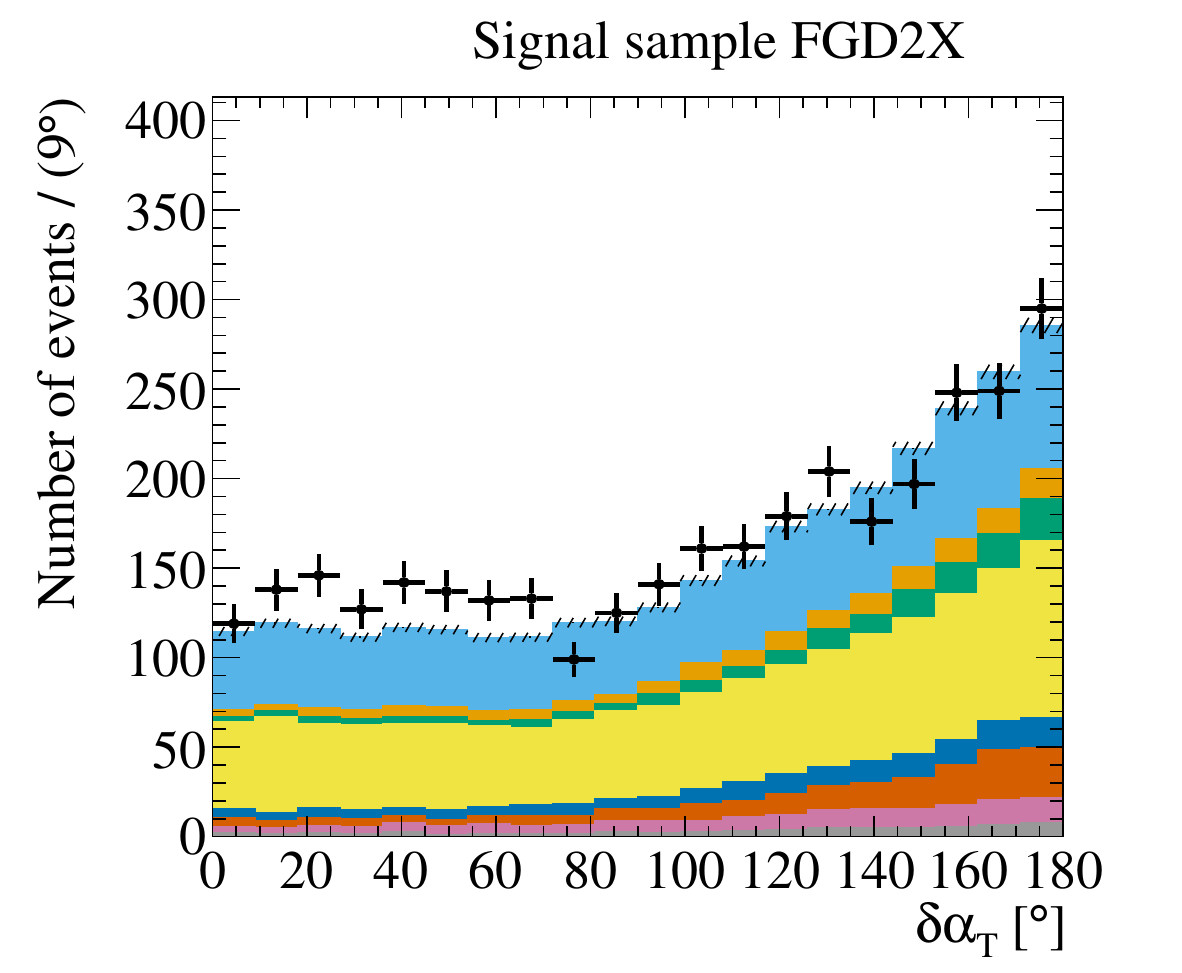}
\includegraphics[width=0.33\linewidth]{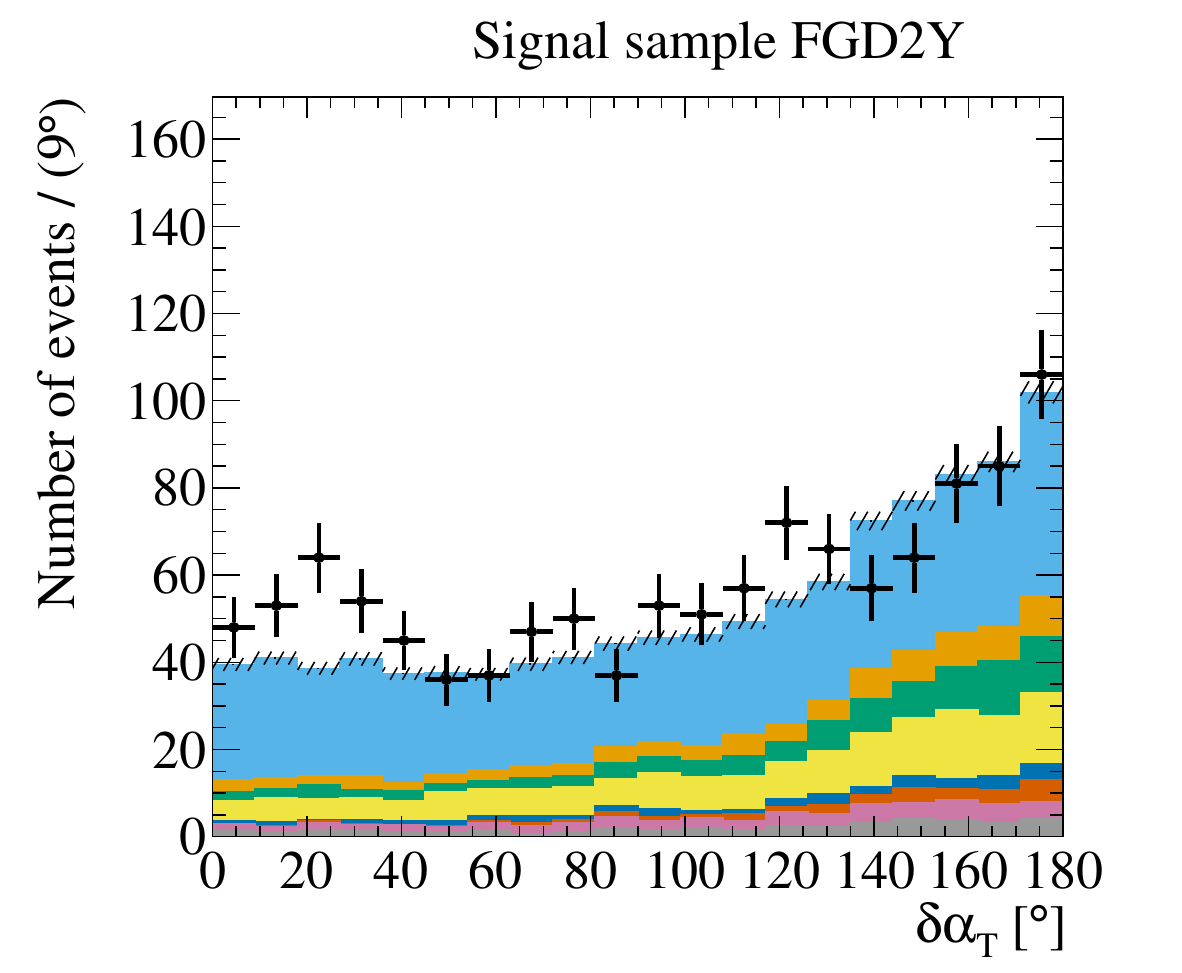}}
\\
\subfloat[$p_{N}$]
{\includegraphics[width=0.33\linewidth]{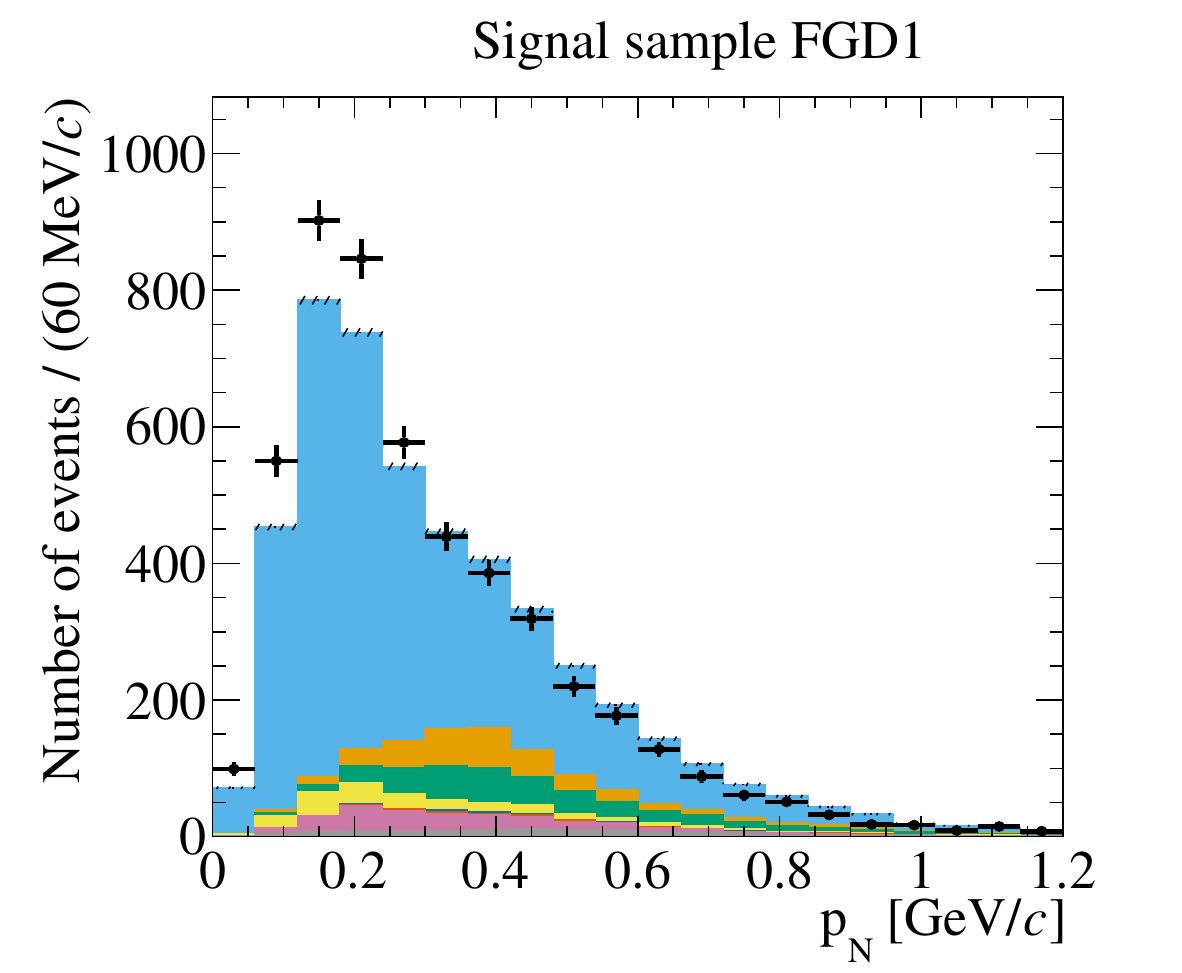}
\includegraphics[width=0.33\linewidth]{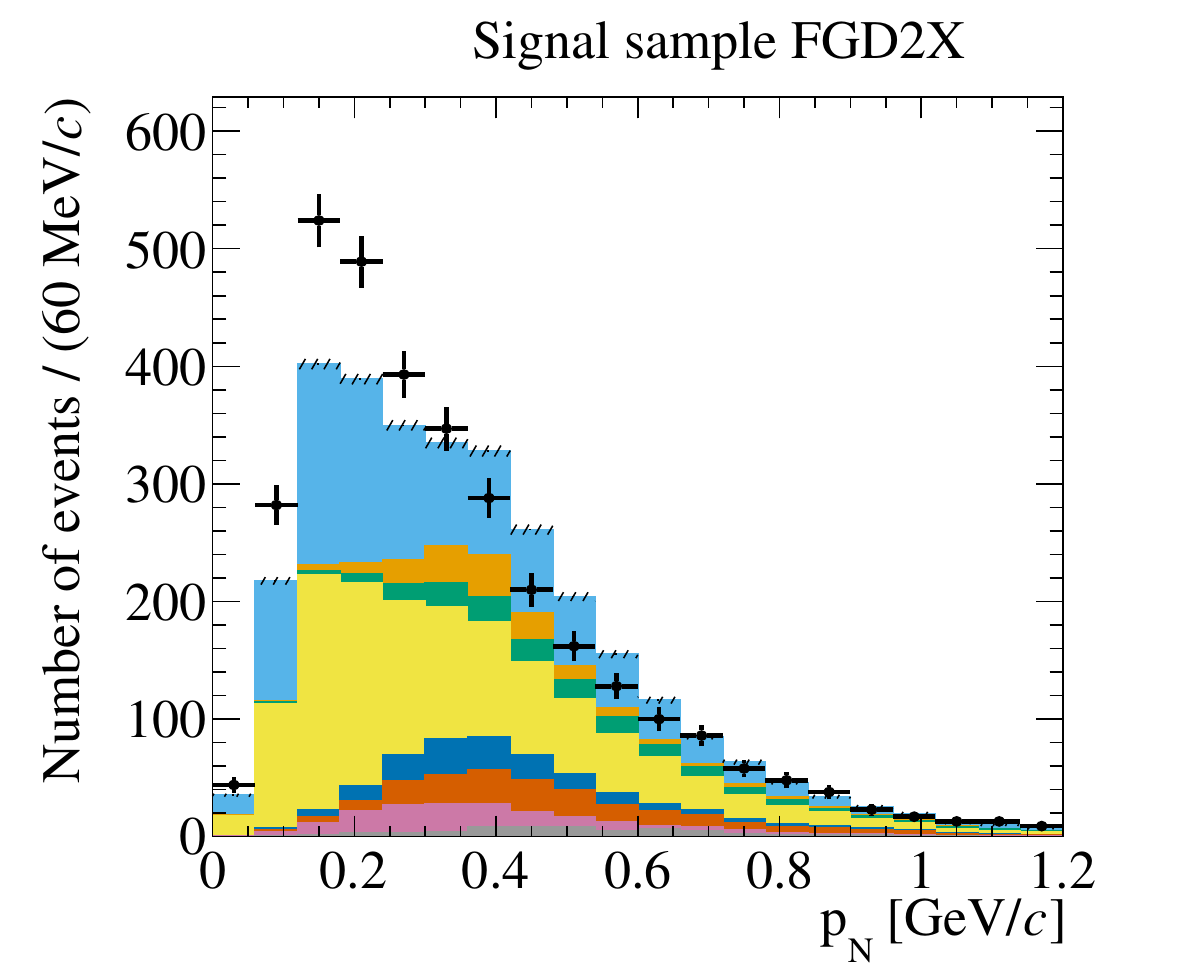}
\includegraphics[width=0.33\linewidth]{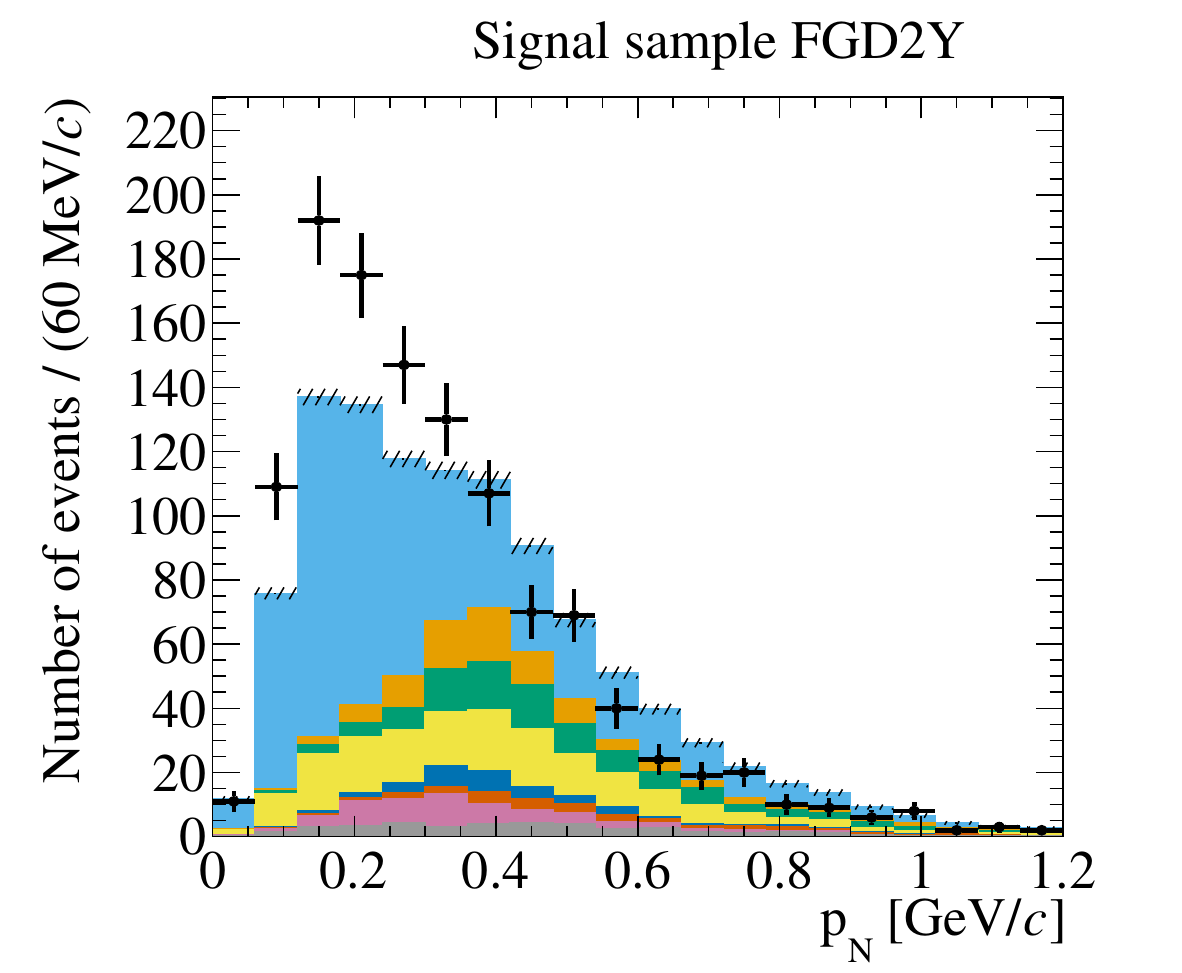}}
\\
\subfloat[$\cos\theta_{\mu}$]
{\includegraphics[width=0.33\linewidth]{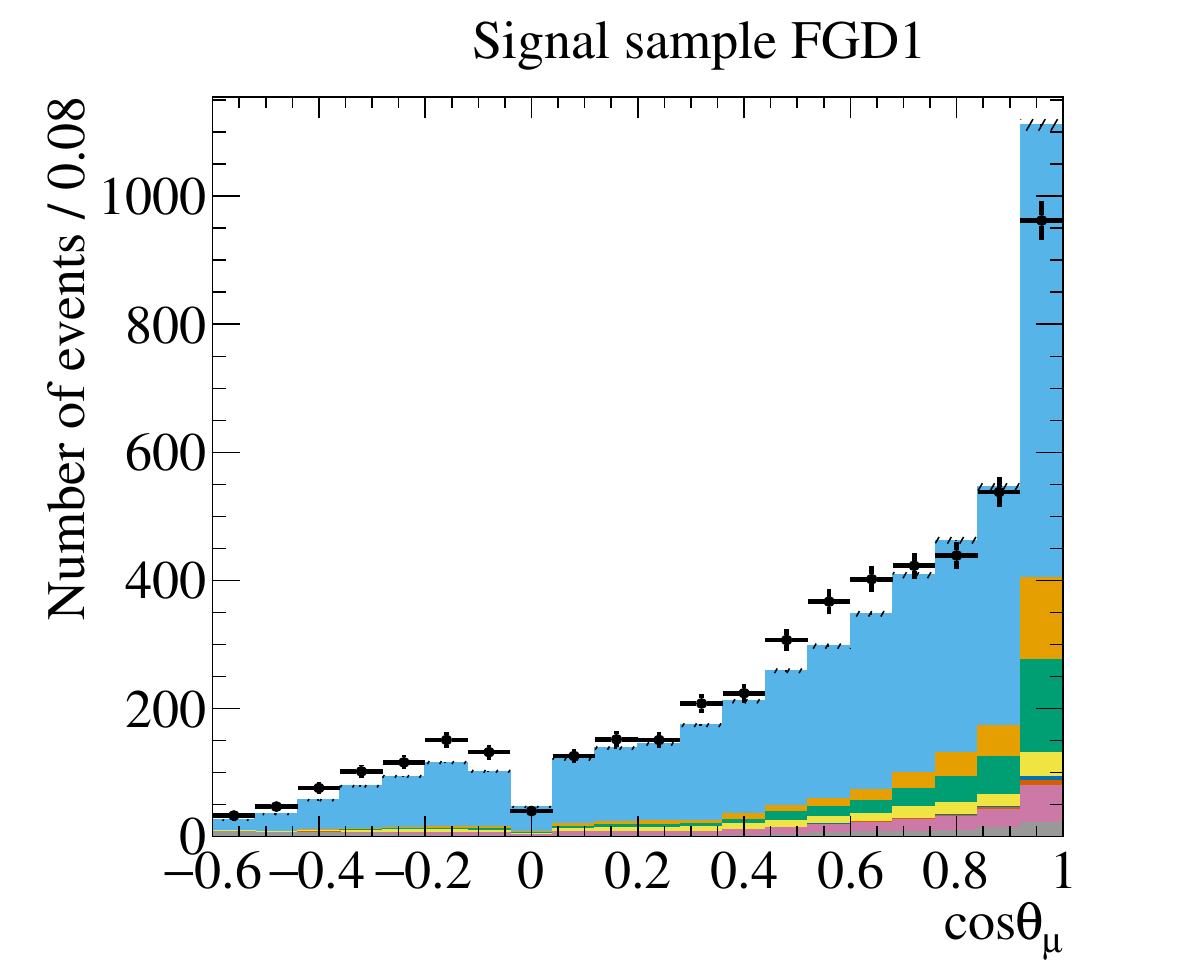}
\includegraphics[width=0.33\linewidth]{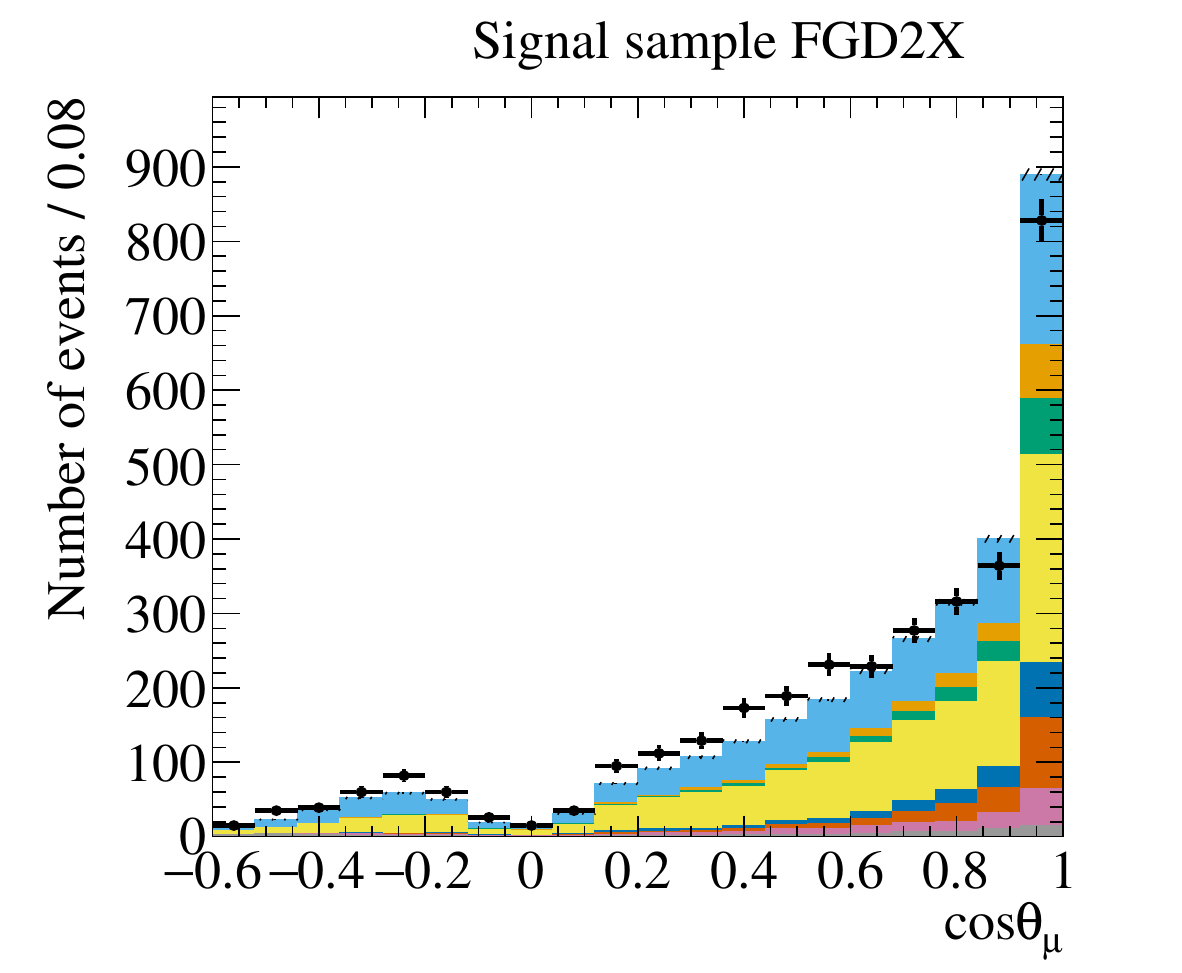}
\includegraphics[width=0.33\linewidth]{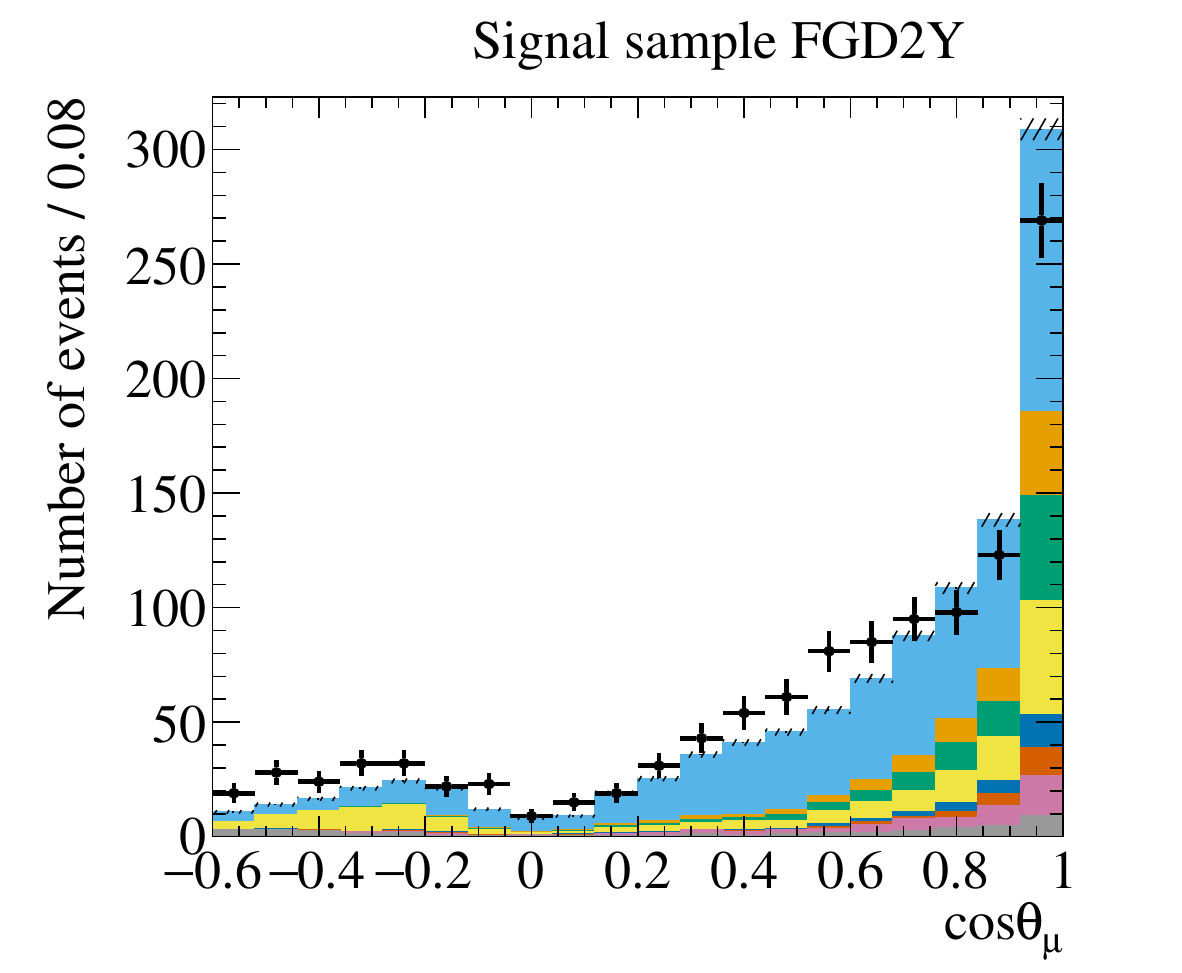}}
\caption{Selected $\nu_{\mu}$ CC$0\pi Np$ signal sample distributions as a function of four observables ($\delta p_{T}$ (a), $\delta \alpha_{T}$ (b), $p_{N}$ (c) and $\cos\theta_{\mu}$ (d)) used in this measurement. All three signal sub-samples (described in Sec.~\ref{subsec:sig_def}) are merged together in each FGD1, FGD2X and FGD2Y case. The MC distributions are normalized to the same POT as data and they are decomposed into true topology categories with corresponding fractions shown in the brackets. Black slash shows MC statistical uncertainties.}
\label{fig:sig_samp_dis}
\end{figure*}

Tab.~\ref{tab:sig_stat_data} summarizes the selected signal sample statistics from the data. In total there are 9469 $\nu_{\mu}$ CC$0\pi Np$ candidates selected in the FGD1 and FGD2 FV with an overall signal purity 74.7\% (78.6\% for FGD1, 70.7\% for FGD2X and 69.0\% for FGD2Y sub-samples) and signal selection efficiency 25.6\% (26.5\% for carbon and 22.8\% for oxygen). Fig.~\ref{fig:sig_samp_dis} shows the selected signal sample distributions as a function of the four observables used in this measurement for the data and MC. In the MC, the distributions are decomposed into true topology categories defined as:
\begin{itemize}
\item \textbf{C - $\nu_{\mu}$ CC$0\pi Np$}: the signal $\nu_{\mu}$ CC$0\pi Np$ interactions on the carbon target in the FGD1 or FGD2 FV.

\item \textbf{C - $\nu_{\mu}$ CC$1\pi^{+}$}: the $\nu_{\mu}$ CC interactions on the carbon target in the FGD1 or FGD2 FV with a positively charged pion produced in the final state. This interaction channel is listed separately because it comprises the largest background in the signal sample (Sec.~\ref{subsec:cont_sample}).

\item \textbf{C - other}: other neutrino-nucleus interactions on the carbon target in the FGD1 or FGD2 FV not belonging to the two carbon categories listed above. The dominant channel comes from $\nu_{\mu}$ CC$1\pi^{0}$ with a neutral pion produced, which comprises the second largest background in the signal sample (Sec.~\ref{subsec:cont_sample}).

\item \textbf{O - $\nu_{\mu}$ CC$0\pi Np$}: the signal $\nu_{\mu}$ CC$0\pi Np$ interactions on the oxygen target in the FGD1 or FGD2 FV.

\item \textbf{O - $\nu_{\mu}$ CC$1\pi^{+}$}: the $\nu_{\mu}$ CC interactions on the oxygen target in the FGD1 or FGD2 FV with a positively charged pion produced in the final state.

\item \textbf{O - other}: other neutrino-nucleus interactions on the oxygen target in the FGD1 or FGD2 FV not belonging to the two oxygen categories listed above.

\item \textbf{Non C/O}: neutrino–nucleus interactions occurring within the FGD1 or FGD2 FV but not on carbon or oxygen targets. This also includes $\nu_{\mu}$ CC$0\pi Np$ interactions taking place on other nuclear targets present in the FGD detector materials. Such $\nu_{\mu}$ CC$0\pi Np$ interactions are not considered part of the signal.

\item \textbf{OOFV}: out-of-fiducial-volume events, defined as the neutrino-nucleus interactions that happened outside the FGD1 or FGD2 FV.
\end{itemize}
The selected signal sample statistics from the MC, decomposed into true topology categories, are summarized in Tab.~\ref{tab:sig_stat_MC}.

\begin{table}[!htbp]
\centering
\caption{Number of selected data events for the $\nu_{\mu}$ CC$0\pi Np$ signal sample.}
\label{tab:sig_stat_data}
\renewcommand{\arraystretch}{1.2}
\begin{tabular}{c|ccc}
\hline
\hline
Signal sample & FGD1 & FGD2X & FGD2Y \\
\hline
TPC $\mu$ + TPC $p$ & 2259 & 1803 & 600 \\
TPC $\mu$ + FGD $p$ (+ $Np$) & 1425 & 768 & 242 \\
FGD $\mu$ + TPC $p$ (+ $Np$) & 1312 & 739 & 321 \\
\hline
Total & 4996 & 3310 & 1163 \\
\hline
\hline
\end{tabular}
\end{table}

\begin{table}[!htbp]
\centering
\caption{Number of selected MC events for the $\nu_{\mu}$ CC$0\pi Np$ signal sample, decomposed into true topology categories. The MC statistics are normalized to the same POT as data. The errors indicate the MC statistical uncertainties.}
\label{tab:sig_stat_MC}
\renewcommand{\arraystretch}{1.2}
\begin{tabular}{c|ccc}
\hline
\hline
True topology & FGD1 & FGD2X & FGD2Y \\
\hline
C-$\nu_{\mu}$CC$0\pi Np$ & 3567.56$\pm$13.84 & 1033.29$\pm$7.47 & 566.82$\pm$5.48  \\
C-$\nu_{\mu}$CC$1\pi^{+}$ & 316.28$\pm$5.16 & 167.20$\pm$3.57 & 86.40$\pm$2.62 \\
C-other & 358.43$\pm$5.28 & 163.65$\pm$3.54 & 98.51$\pm$2.72 \\
O-$\nu_{\mu}$CC$0\pi Np$ & 183.83$\pm$3.20 & 1175.73$\pm$8.02 & 170.22$\pm$3.14  \\
O-$\nu_{\mu}$CC$1\pi^{+}$ & 13.69$\pm$1.02 & 175.12$\pm$3.63 & 31.38$\pm$1.53 \\
O-other & 18.92$\pm$1.22 & 206.40$\pm$4.04 & 26.22$\pm$1.47 \\
Non C / O & 213.57$\pm$3.60 & 145.57$\pm$3.03 & 51.31$\pm$1.79 \\
OOFV & 104.96$\pm$3.06 & 66.89$\pm$2.51 & 40.86$\pm$1.78 \\
\hline
Total & 4777.23$\pm$16.77 & 3133.85$\pm$13.80 & 1071.73$\pm$8.06 \\
\hline
\hline
\end{tabular}
\end{table}

The data exhibit an excess of events near the peaks of the $\delta p_{T}$ and $p_{N}$ distributions compared to the MC predictions. Similar discrepancies have been reported in cross-section measurements using similar interaction channels on an argon target~\cite{MicroBooNE:2023cmw,MicroBooNE:2024yzp}. Additionally, the MC simulation tends to overestimate the data for muons produced in the very forward direction ($\cos\theta_{\mu} \sim 1$). This feature is consistent with observations in previous T2K cross-section measurements (see, for example, Refs.~\cite{T2K:2020jav,T2K:2018rnz,T2K:2020sbd}). A potential origin of this discrepancy is the mis-modeling of neutrino-nucleus interactions in the low energy transfer region, where the impulse approximation is applied despite its known limitations in this regime.

\subsection{Control Sample}
\label{subsec:cont_sample}

Based on MC simulations, the primary backgrounds in the $\nu_{\mu}$ CC$0\pi Np$ signal sample originate from $\nu_{\mu}$ CC$1\pi^{+}$ ($\sim 9.4\%$) and $\nu_{\mu}$ CC$1\pi^{0}$ ($\sim 3.6\%$) pion production topologies. The $\nu_{\mu}$ CC$1\pi^{0}$ background, in which a $\pi^{0}$ is produced in the final state, is misidentified as signal primarily when the $\pi^{0}$ fails to be tagged by the signal selection. The $\nu_{\mu}$ CC$1\pi^{+}$ background is misidentified when the primary $\pi^{+}$ is not reconstructed in the FGDs. In the low-momentum region ($p_{\pi^{+}} < 0.2$~GeV/c), pions may remain undetected if their tracks are too short for reconstruction within the FGDs and the subsequent Michel electrons are not successfully tagged. Conversely, in the high-momentum region ($p_{\pi^{+}} > 0.2$~GeV/c), pion tracks can be fragmented by secondary interactions in the FGD materials, where the resulting segments are sometimes too short to be successfully reconstructed.

In this analysis, the $\nu_{\mu}$ CC$1\pi^{0}$ and other minor background components are subtracted from the signal sample based on MC predictions. For the more significant $\nu_{\mu}$ CC$1\pi^{+}$ background, a data-driven approach is employed to obtain a more precise estimation of its contribution to the selected data. To this end, a dedicated control sample enriched in $\nu_{\mu}$ CC$1\pi^{+}$ interactions was developed. During the cross-section extraction (Sec.~\ref{subsubsec:temp_fit}), the MC prediction for the $\nu_{\mu}$ CC$1\pi^{+}$ background is compared to the data control sample simultaneously with the signal sample. This allows the MC prediction of the $\nu_{\mu}$ CC$1\pi^{+}$ fraction in the signal sample to be adjusted accordingly. To ensure the control sample accurately represents the signal sample backgrounds, two sub-samples are defined:
\begin{itemize}
\item \textbf{CCproton ME}: A negatively charged muon and at least one proton are selected, following a procedure similar to the signal selection, but requiring the detection of at least one Michel electron (ME) in the FGDs. This sub-sample targets the \textit{lower} $\pi^{+}$ momentum region of the $\nu_{\mu}$ CC$1\pi^{+}$ background.

\item \textbf{CCproton $1\pi^{+}$}: A negatively charged muon and at least one proton are selected, alongside an identified $\pi^{+}$ track entering the forward TPC. This sub-sample targets the \textit{higher} $\pi^{+}$ momentum region of the $\nu_{\mu}$ CC$1\pi^{+}$ background.
\end{itemize}
The selection flow for the control samples parallels that of the signal sample (Sec.~\ref{subsec:sig_selection}). The fundamental difference lies in the inversion of the charged pion veto cuts; whereas the signal selection rejects these events, the control samples specifically require the presence of Michel electrons in the FGDs or $\pi^{+}$ tracks in the TPCs. For Michel electron detection, the tagging efficiency is $\sim$ 57.6\% (MC) / 56.0\% (data) in FGD1 and $\sim$ 31.0\% (MC) / 30.0\% (data) in FGD2. The lower tagging efficiency in FGD2 is mainly due to the poorer reconstruction performance for short Michel electron tracks, similar to the case of short protons. In this analysis, the detector systematic uncertainty on the Michel electron selection is evaluated as the difference in tagging efficiency between data and MC, which is determined with dedicated FGD cosmic-muon samples~\cite{T2K:2015sqm}. Since this difference is $<$ 2\% for both FGD1 and FGD2, its contribution to the total detector systematic uncertainty is negligible.

\begin{figure*}
\centering
\subfloat[$\delta p_{T}$]
{\includegraphics[width=0.33\linewidth]{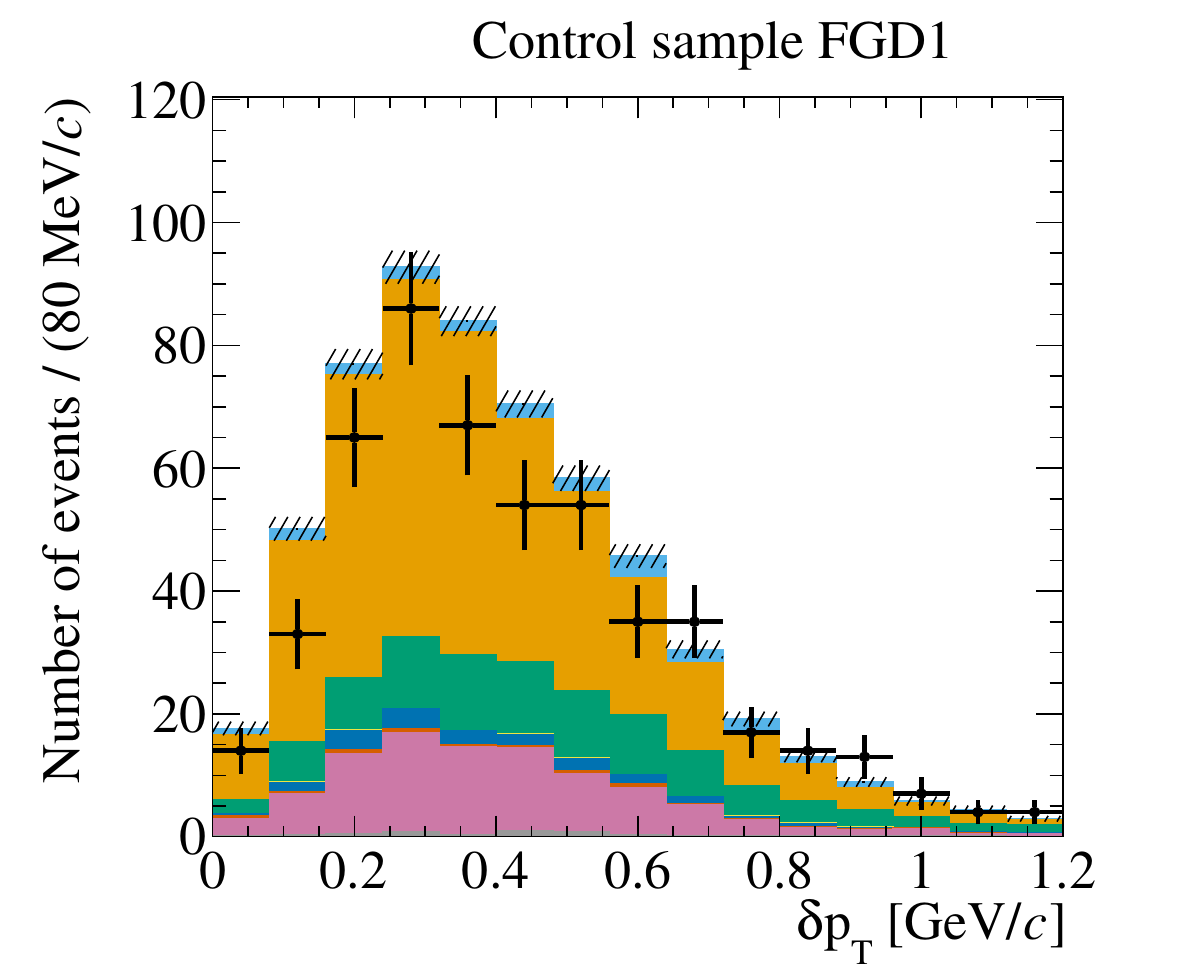}
\includegraphics[width=0.33\linewidth]{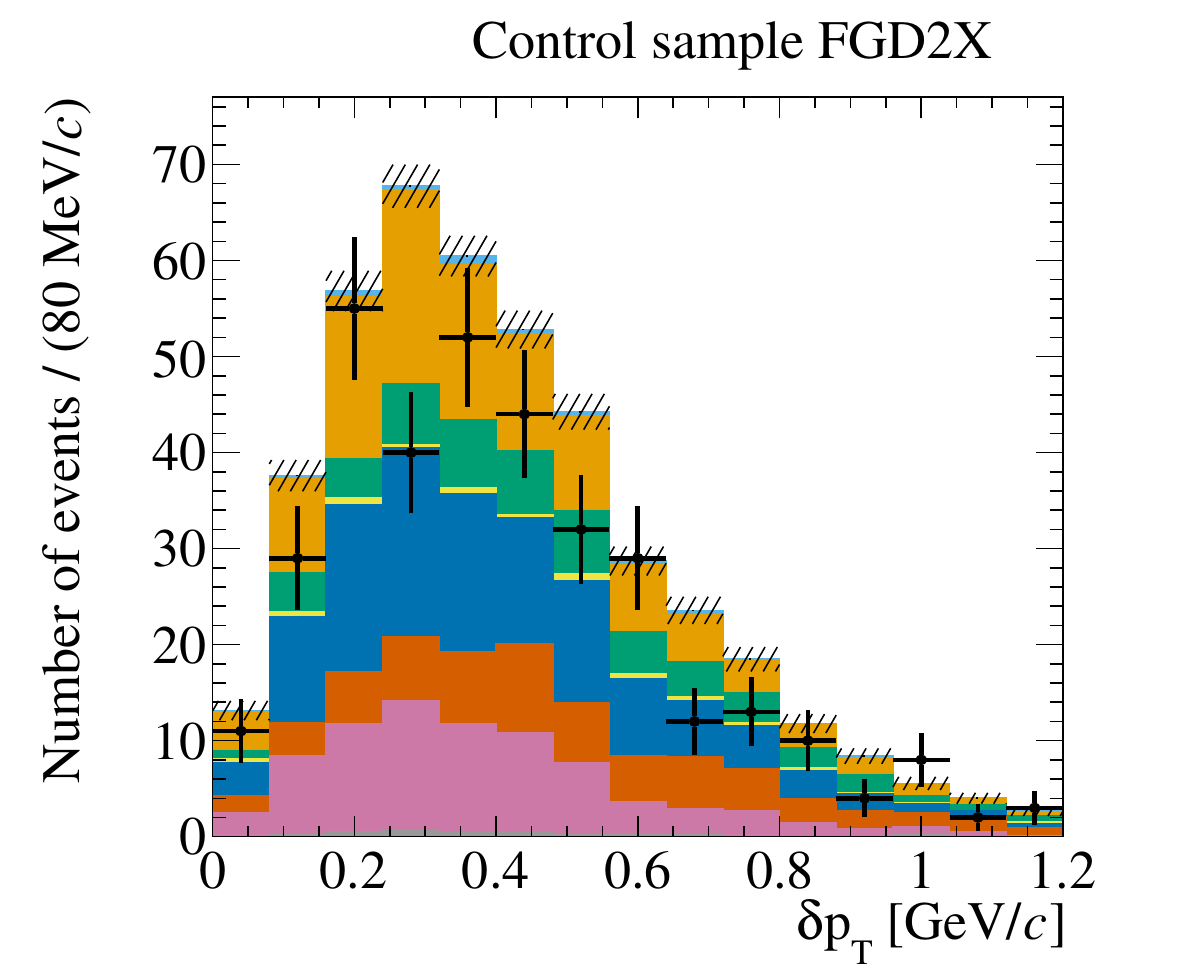}
\includegraphics[width=0.33\linewidth]{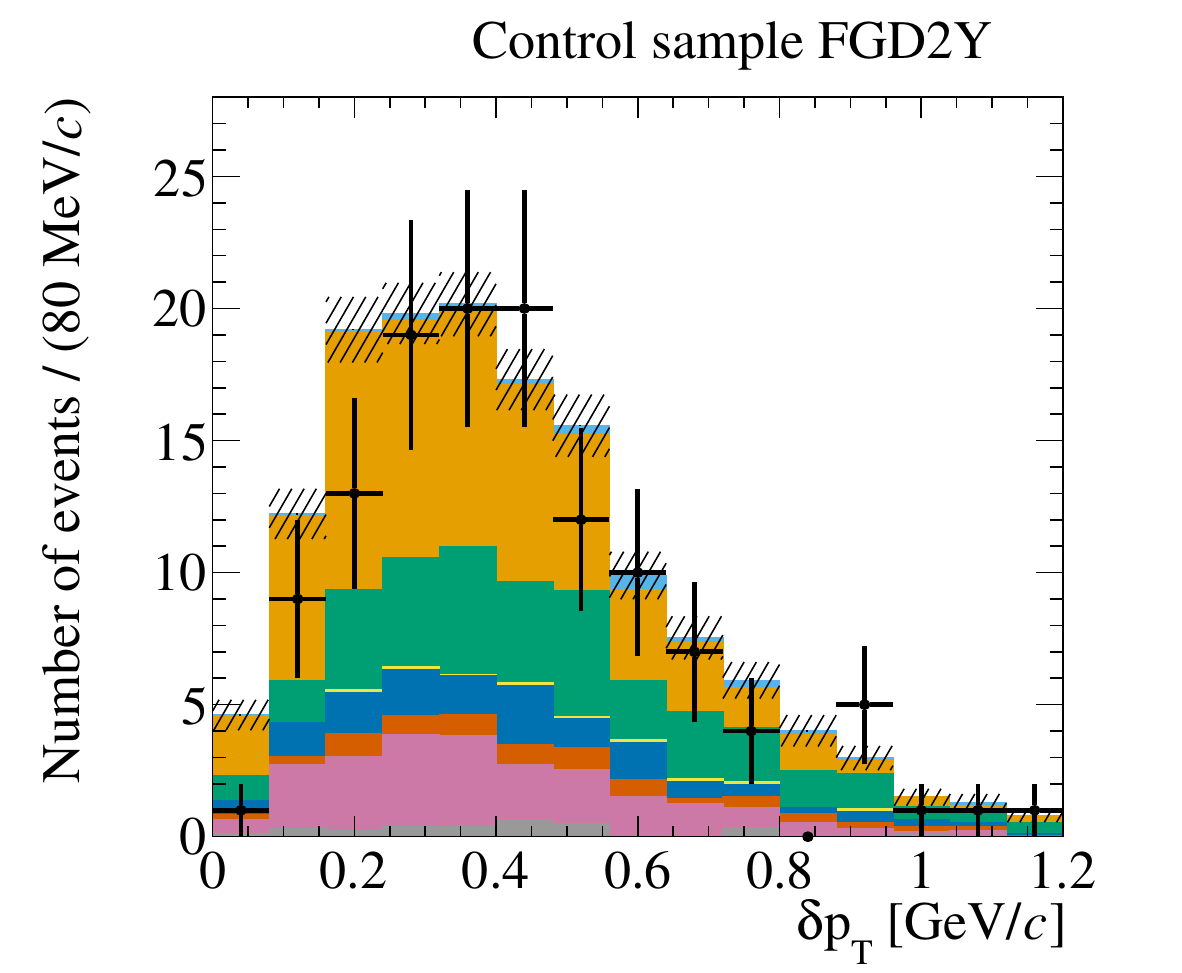}}
\\
\subfloat[$\delta \alpha_{T}$]
{\includegraphics[width=0.33\linewidth]{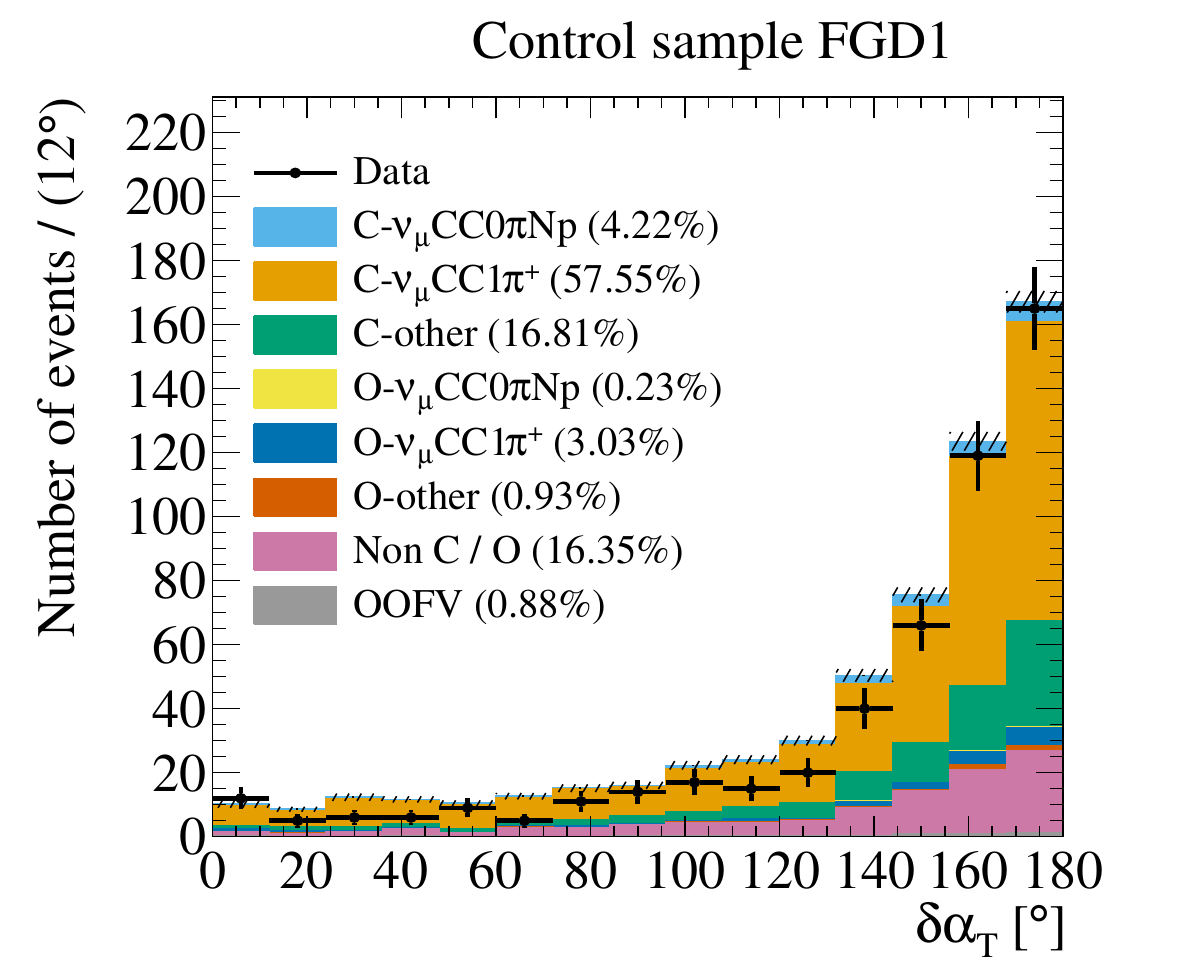}
\includegraphics[width=0.33\linewidth]{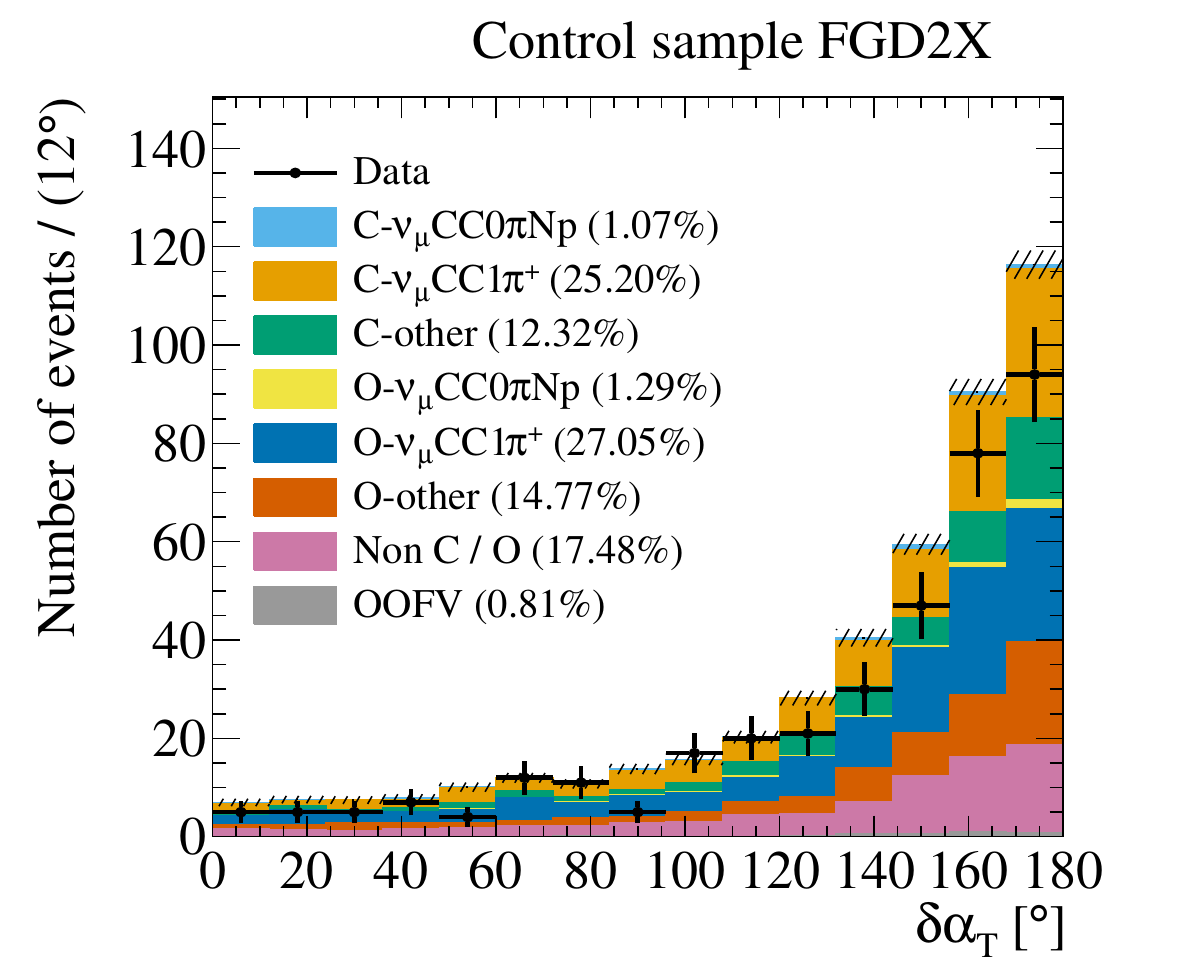}
\includegraphics[width=0.33\linewidth]{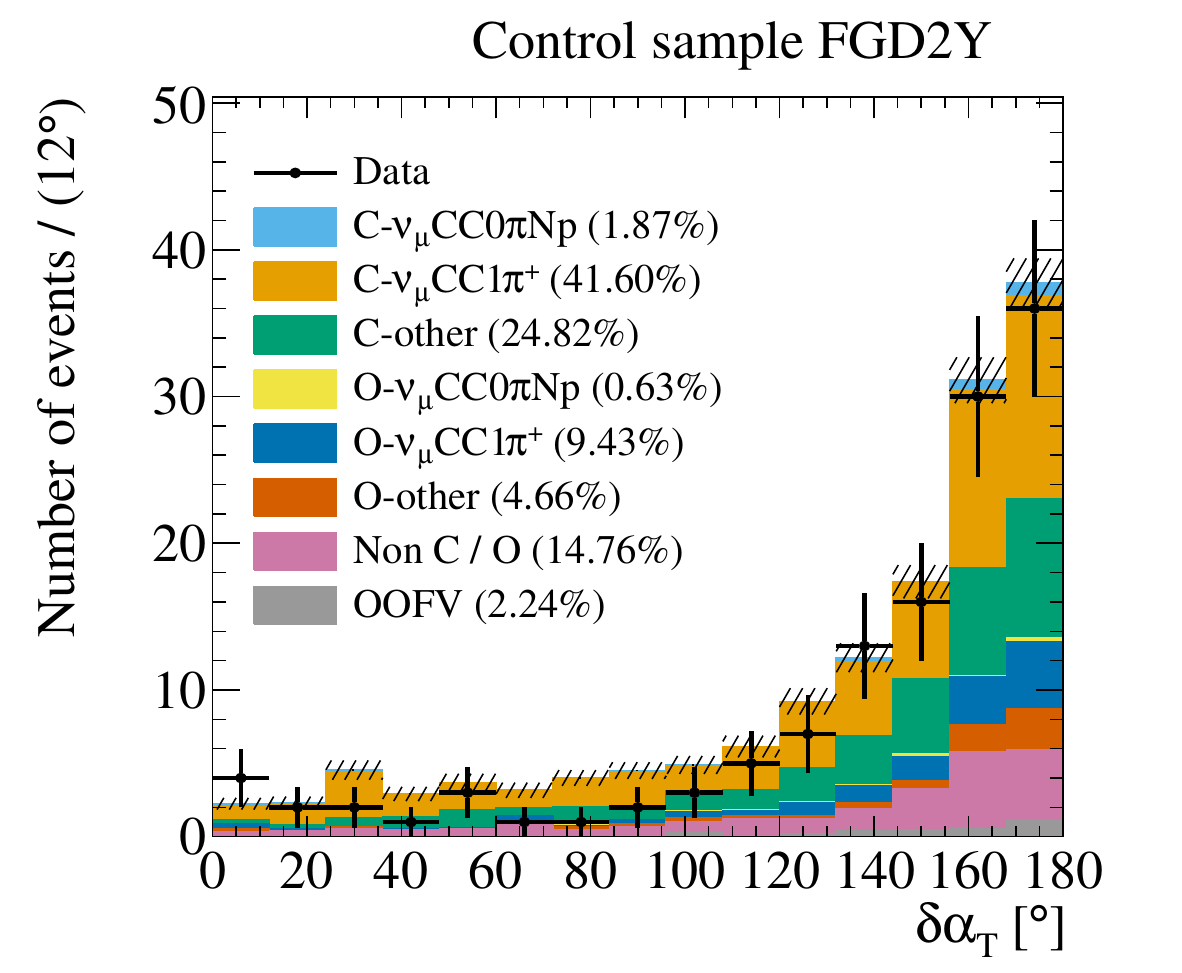}}
\\
\subfloat[$p_{N}$]
{\includegraphics[width=0.33\linewidth]{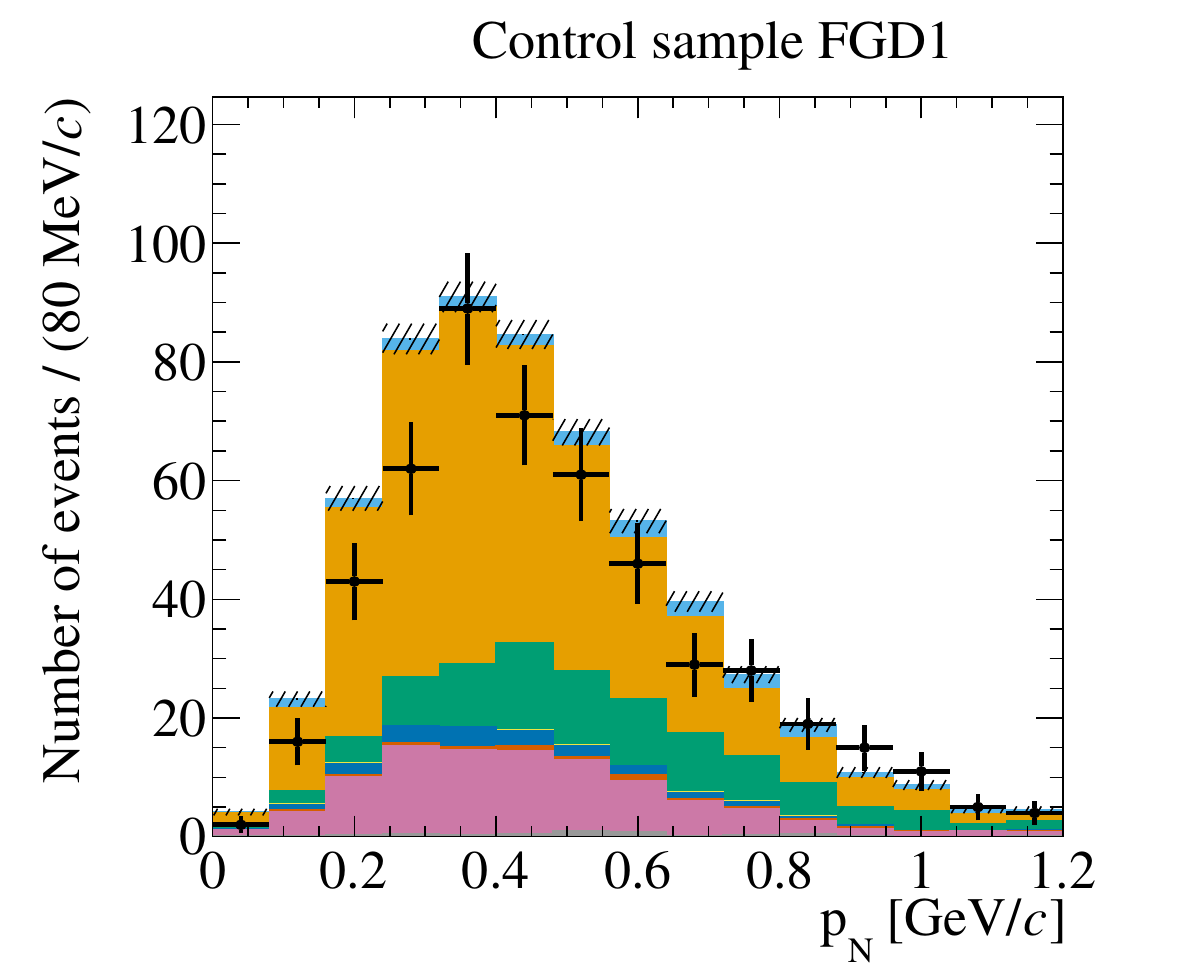}
\includegraphics[width=0.33\linewidth]{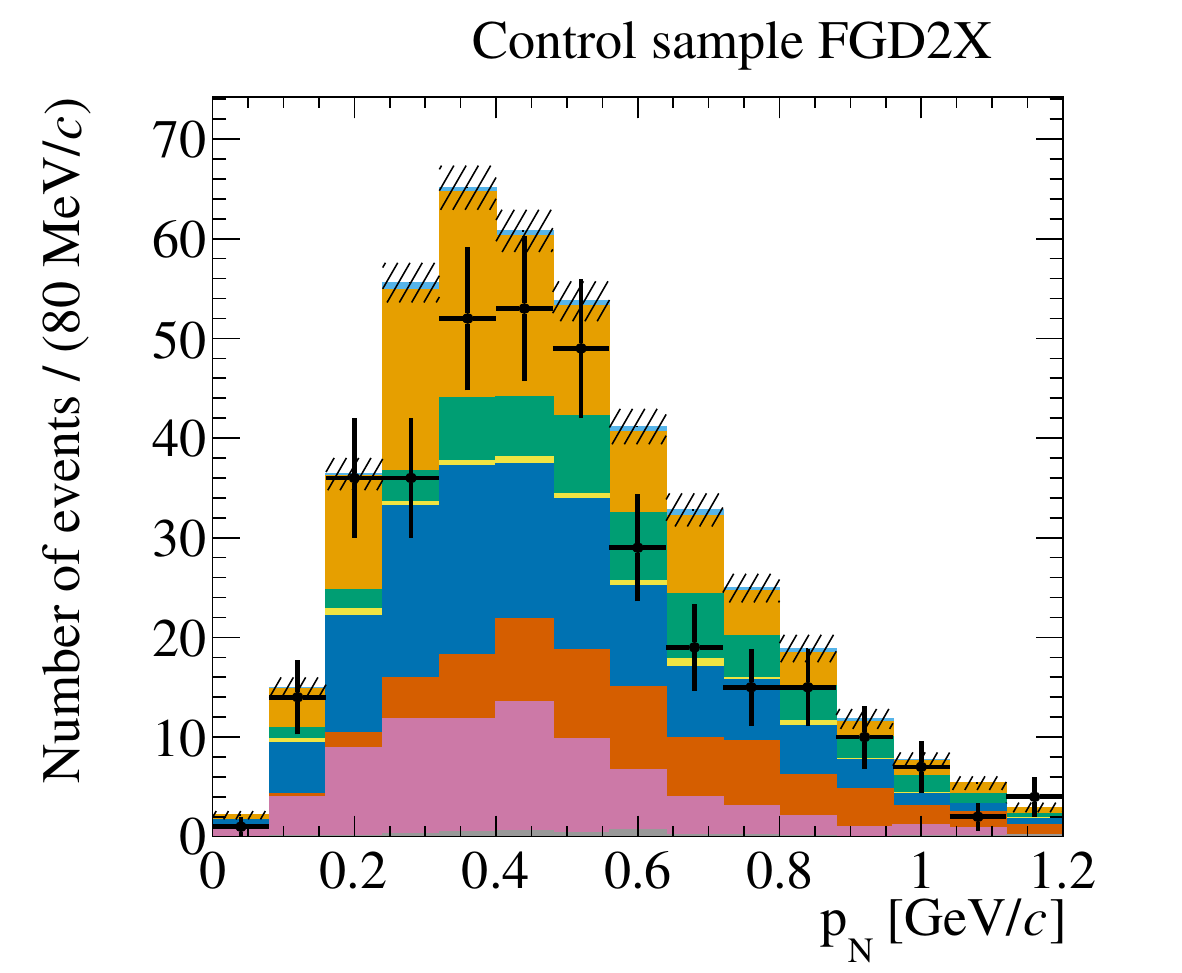}
\includegraphics[width=0.33\linewidth]{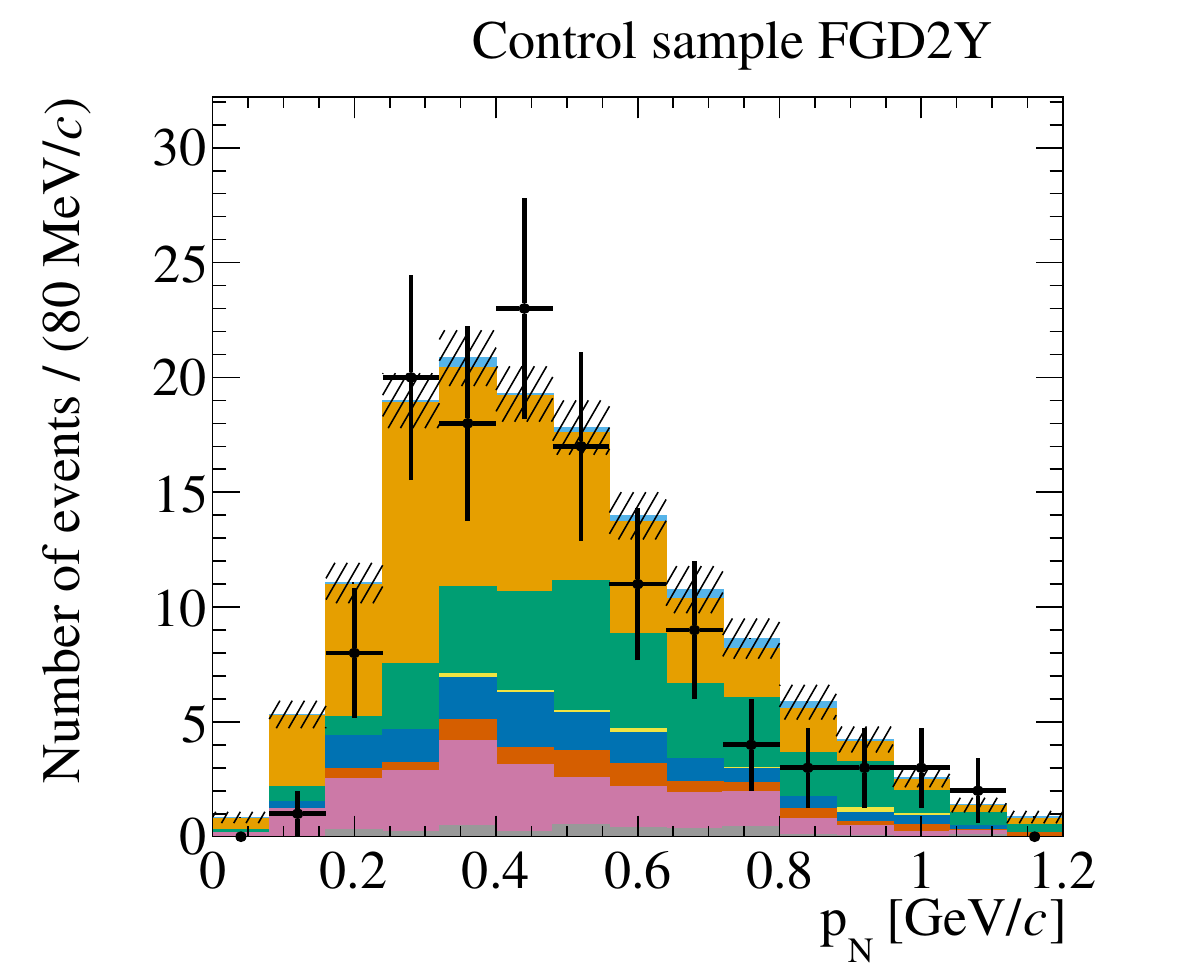}}
\\
\subfloat[$\cos\theta_{\mu}$]
{\includegraphics[width=0.33\linewidth]{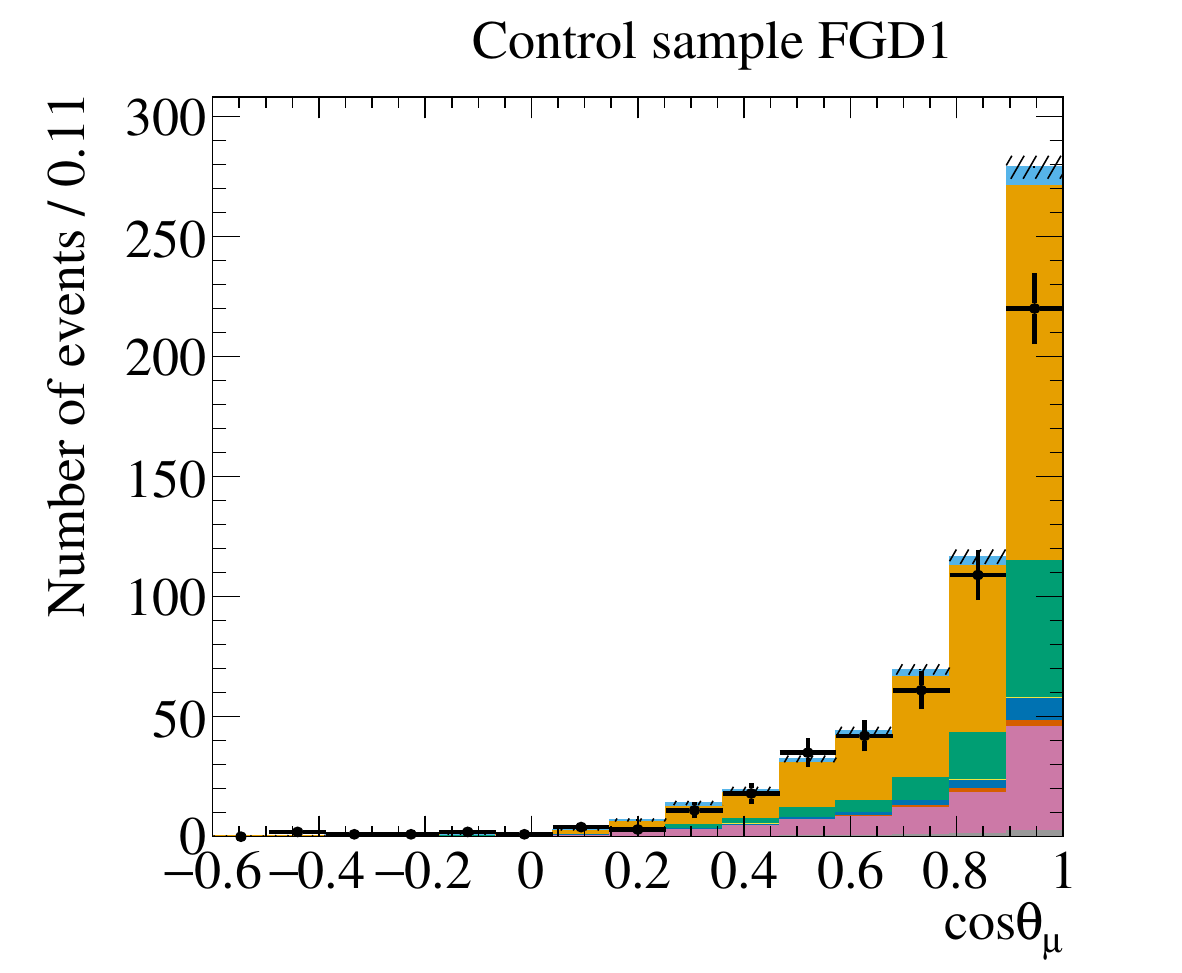}
\includegraphics[width=0.33\linewidth]{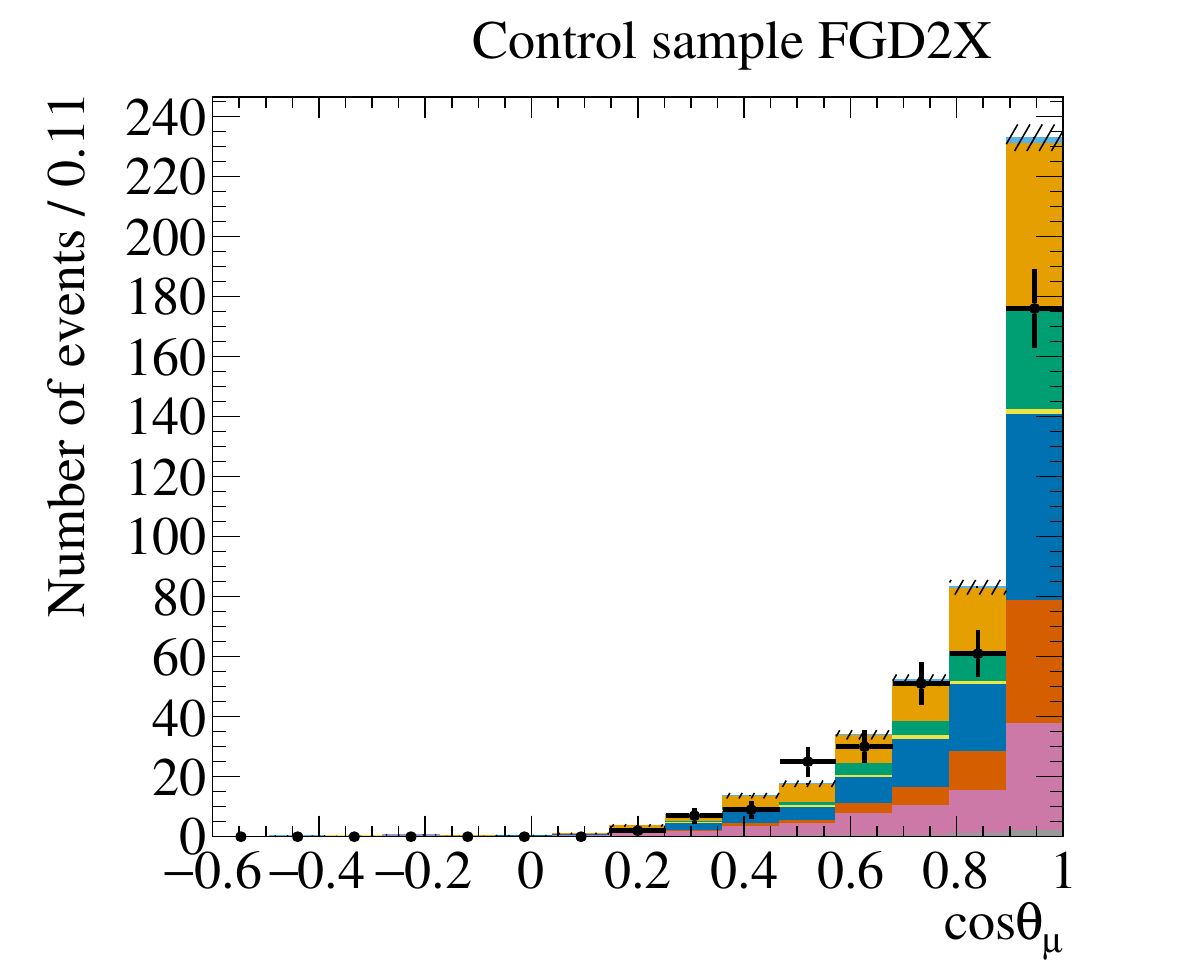}
\includegraphics[width=0.33\linewidth]{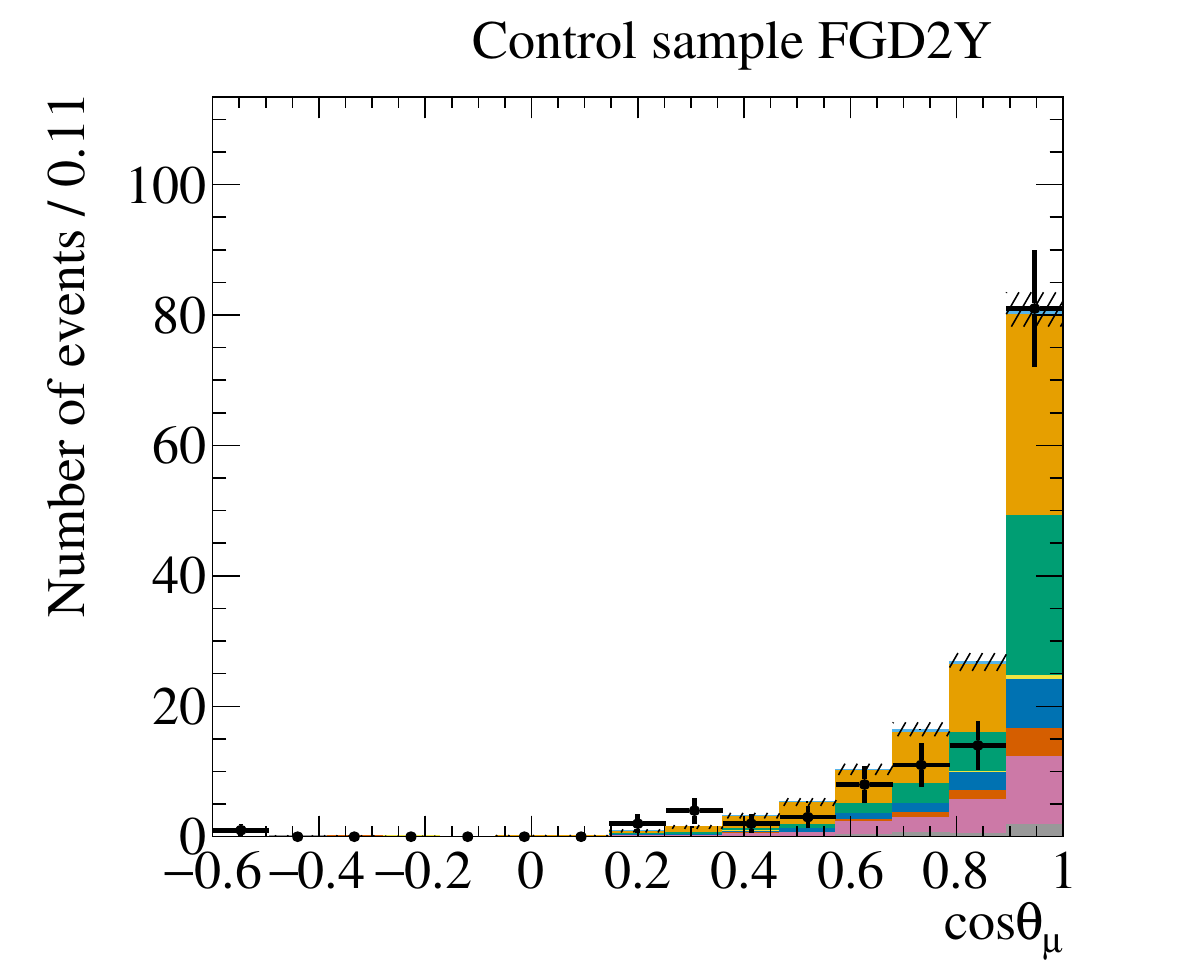}}
\caption{Selected control sample distributions as a function of four observables ($\delta p_{T}$ (a), $\delta \alpha_{T}$ (b), $p_{N}$ (c) and $\cos\theta_{\mu}$ (d)) used in this measurement. The two sub-samples are merged together in each FGD1, FGD2X and FGD2Y case. The MC distributions are normalized to the same POT as the data and they are decomposed into true topology categories with corresponding fractions shown in the brackets. Black slash shows MC statistical uncertainties.}
\label{fig:cont_samp_dis}
\end{figure*}

Tab.~\ref{tab:cont_stat_data} summarizes the selected control sample statistics obtained from the data, while the corresponding event distributions as a function of the analysis observables are presented in Fig.~\ref{fig:cont_samp_dis}. The selected control sample statistics from the MC, decomposed into true topology categories, are summarized in Tab.~\ref{tab:cont_stat_MC}. Based on MC estimates, the selected control samples are dominated by $\nu_{\mu}$ CC$1\pi^{+}$ interactions, with purities of 60.6\%, 52.3\%, and 51.0\% for the FGD1, FGD2X, and FGD2Y sub-samples, respectively. A significant contribution from multi-pion production (categorized under ``other'') is also observed, specifically in cases where only a single primary pion is identified. The ranges of observable values of the control samples provide extensive coverage of the $\nu_{\mu}$ CC$1\pi^{+}$ background present in the signal samples. Furthermore, the $\nu_{\mu}$ CC$1\pi^{+}$ interactions generally exhibit larger values for the KI observables $\delta p_{T}$, $\delta\alpha_{T}$, and $p_{N}$ compared to the signal $\nu_{\mu}$ CC$0\pi Np$ interactions. This shift occurs because the pion kinematics are excluded from the observable calculations, resulting in an enhanced apparent kinematic imbalance.

\begin{table}[!htbp]
\centering
\caption{Number of selected data events in each control sample.}
\label{tab:cont_stat_data}
\renewcommand{\arraystretch}{1.2}
\begin{tabular}{c|ccc}
\hline
\hline
Sample & FGD1 & FGD2X & FGD2Y \\
\hline
CCproton ME & 283 & 162 & 57 \\
CCproton $1\pi^{+}$ & 227 & 199 & 69 \\
\hline
Total & 510 & 361 & 126 \\
\hline
\hline
\end{tabular}
\end{table}

\begin{table}[!htbp]
\centering
\caption{Number of selected MC events for the control sample, decomposed into true topology categories. The MC statistics are normalized to the same POT as data. The errors indicate the MC statistical uncertainties.}
\label{tab:cont_stat_MC}
\renewcommand{\arraystretch}{1.2}
\begin{tabular}{c|ccc}
\hline
\hline
True topology & FGD1 & FGD2X & FGD2Y \\
\hline
C-$\nu_{\mu}$CC$0\pi Np$ & 24.92$\pm$1.14 & 4.80$\pm$0.58 & 2.73$\pm$0.38  \\
C-$\nu_{\mu}$CC$1\pi^{+}$ & 339.51$\pm$5.11 & 112.71$\pm$2.95 & 60.79$\pm$2.24 \\
C-other & 99.17$\pm$2.83 & 55.09$\pm$2.08 & 36.27$\pm$1.70 \\
O-$\nu_{\mu}$CC$0\pi Np$ & 1.36$\pm$0.25 & 5.77$\pm$0.56 & 0.92$\pm$0.26 \\
O-$\nu_{\mu}$CC$1\pi^{+}$ & 17.88$\pm$1.19 & 120.96$\pm$3.13 & 13.78$\pm$1.01 \\
O-other & 5.49$\pm$0.67 & 66.07$\pm$2.34 & 6.81$\pm$0.74 \\
Non C / O & 96.47$\pm$2.55 & 78.18$\pm$2.30 & 21.56$\pm$1.20 \\
OOFV & 5.17$\pm$0.65 & 3.63$\pm$0.53 & 3.27$\pm$0.53 \\
\hline
Total & 589.96$\pm$6.65 & 447.21$\pm$5.87 & 146.13$\pm$3.38 \\
\hline
\hline
\end{tabular}
\end{table}

\section{Cross-section Extraction}
\label{sec:xsec_extra}

\subsection{Basic Method}
\label{subsec:xsec_method}

The cross section is extracted using a template-fit unfolding method~\cite{T2K:2025kdk,T2K:2025kda,T2K:2023qjb} with the GUNDAM software~\cite{GUNDAM}. The fit model is constructed from the selected MC samples and incorporates a set of parameters that allow for variations in both shape and normalization. The extraction process is performed in two stages. First (Sec.~\ref{subsubsec:temp_fit}), the MC templates are fitted to the data in the reconstructed observable space, utilizing the selected signal and control samples. A binned likelihood fit is employed to determine the parameter values that best reproduce the data. Second (Sec.~\ref{subsubsec:xsec_def_calc}), the cross section is calculated in the true observable space by subtracting the fitted background and unfolding detector effects. Specifically, this is achieved by taking the true signal component of the templates, reweighted according to the best-fit parameter values.

\subsubsection{Template Fit}
\label{subsubsec:temp_fit}

To perform the template fit, we define a likelihood function of the form
\begin{equation}
L(\vec{\theta}) = L_{\text{stat}}(\vec{\theta}) \times L_{\text{syst}}(\vec{\theta}),
\end{equation}
where $L_{\text{stat}}$ is the statistical term comparing the MC templates to the data, and $L_{\text{syst}}$ represents the systematic uncertainty (or penalty) term. The vector $\vec{\theta}$ comprises the parameters to be estimated in the fit, including the template and systematic parameters described later. In practice, the fit is performed by minimizing the negative log-likelihood with respect to $\vec{\theta}$. A test statistic $\chi^{2}(\vec{\theta})$ is introduced and defined as:
\begin{equation}
\begin{split}
\chi^{2}(\vec{\theta})
& = \chi^{2}_{\text{stat}}(\vec{\theta}) + \chi^{2}_{\text{syst}}(\vec{\theta}) \\
& \approx -2\log L_{\text{stat}}(\vec{\theta}) - 2\log L_{\text{syst}}(\vec{\theta}).
\end{split}
\label{eq:chi2_total}
\end{equation}

The statistical term $\chi^{2}_{stat}(\vec{\theta})$ is defined assuming a Poisson distribution in each selected sample bin:
\begin{equation}
\begin{split}
\chi^{2}_{stat}(\vec{\theta}) \approx 2 \sum_{j}^{reco} 
& \left[ \beta_{j} N_{j}^{exp}(\vec{\theta}) - N_{j}^{obs} \right. \\ 
& \left. + N_{j}^{obs} \log \left( \frac{N_{j}^{obs}}{\beta_{j} N_{j}^{exp}(\vec{\theta})} \right) + \frac{(\beta_{j} - 1)^{2}}{2\sigma_{\beta_{j}}^{2}} \right].
\end{split}
\label{eq:chi2_stat}
\end{equation}
In Eq.~\ref{eq:chi2_stat}, we incorporate the ``Barlow-Beeston'' correction~\cite{BARLOW1993219,T2K:2023qjb} with the Conway approximation~\cite{Conway:2011in}, where the parameters $\beta_{j}$ are introduced to take into account the statistical fluctuations arising from the finite size of the MC samples. The statistical term is calculated by summing over all bin indices $j$ across the selected signal and control samples; the corresponding binning definitions are summarized in Appendix~\ref{subsec:samp_binning}. Here, $N_{j}^{\text{obs}}$ denotes the number of observed events in the data for bin $j$, while $N_{j}^{\text{exp}}(\vec{\theta})$ represents the expected number of events predicted by the MC template for the same bin. During the fit, the $\beta_{j}$ parameter can be solved analytically with the equation:
\begin{equation}
\beta_{j}^{2} + \left(  N_{j}^{exp} \sigma_{\beta_{j}}^{2} - 1 \right) \beta_{j} - N_{j}^{obs} \sigma_{\beta_{j}}^{2} = 0,
\end{equation}
where the uncertainty of $\beta_{j}$ is evaluated with the formula $\sigma_{\beta_{j}} = 1 / \sqrt{N_{j}^{exp}}$.

$N_{j}^{exp}(\vec{\theta})$ can be expressed as the sum of the signal and background events located in the same bin:
\begin{equation}
N_{j}^{exp}(\vec{\theta}) = \sum_{sig \in j} \omega_{sig}(\vec{\theta}) + \sum_{bkg \in j} \omega_{bkg}(\vec{\theta}).
\label{eq:pred_number}
\end{equation}
For each signal and background event, we introduce a weight factor, $\omega_{sig}(\vec{\theta})$ and $\omega_{bkg}(\vec{\theta})$, respectively, where both depend on the parameters $\vec{\theta}$ implemented in the MC template. The weight factor is expressed as:
\begin{equation}
\begin{cases}
\omega_{sig}(\vec{\theta}) = t_{i} \times d_{j} \times f_{k} \times \prod_{a}^{sig,model} R_{a}(x_{a}) \\
\omega_{bkg}(\vec{\theta}) = d_{j} \times f_{k} \times \prod_{a}^{bkg,model} R_{a}(x_{a}).
\end{cases}
\label{eq:event_weight}
\end{equation}
In these expressions, $t_{i}$ represents the template parameter applied exclusively to the true signal events; it normalizes the cross section in the true observable bin $i$, as defined in Appendix~\ref{subsec:xsec_binning}. These template parameters are left unconstrained in the fit, providing the MC template with maximum flexibility. The remaining quantities represent systematic effects and are incorporated as nuisance parameters, constrained by their respective prior values and uncertainties. $d_{j}$ is the detector systematic parameter defined for each selected sample bin $j$, acting as a normalization factor to account for detector effects (Sec.~\ref{subsubsec:syst_det}). $f_{k}$ is the neutrino flux systematic parameter defined for each true neutrino energy bin $k$, serving as a normalization factor for flux uncertainties (Sec.~\ref{subsubsec:syst_flux}). The product $\prod_{a}^{sig (bkg),model} R_{a}(x_{a})$ represents the total weight associated with all neutrino interaction modeling systematic uncertainties affecting a given signal (background) event. Each factor $R_{a}(x_{a})$ is a response function implementing the effect of a specific cross-section systematic uncertainty, as described in Sec.~\ref{subsubsec:syst_xsec}, where $x_{a}$ denotes the corresponding systematic parameter.

The penalty term of the likelihood, $\chi^{2}_{syst}(\vec{\theta})$ is defined assuming the multi-variate Gaussian distribution as the prior constraint:
\begin{equation}
\chi^{2}_{syst}(\vec{\theta}) \approx (\vec{\theta}_{syst} - \vec{\theta}_{syst}^{prior})^{t} (V_{syst}^{cov})^{-1} (\vec{\theta}_{syst} - \vec{\theta}_{syst}^{prior}).
\end{equation}
where $\vec{\theta}_{syst} = (\vec{d},\vec{f},\vec{x})$ includes the detector, neutrino flux and cross-section modeling systematic parameters.

The $\chi^{2}$ function of Eq.~\ref{eq:chi2_total} is minimized to find the vector of best-fit parameters $\vec{\theta}_{best-fit}$ and their covariance matrix $V_{best-fit}$. The best-fit parameter values and covariance matrix are propagated to the next step to calculate the cross sections.

\subsubsection{Cross-section Definition and Calculation}
\label{subsubsec:xsec_def_calc}

The cross section is extracted as:
\begin{equation}
\frac{d^{2}\sigma_{i}^{T}}{(dXdY)_{i}} = \frac{N_{i,sig}^{T}}{\Phi_{\nu} \cdot n_{T} \cdot \epsilon_{i}^{T}} \cdot \frac{1}{(\Delta X \Delta Y)_{i}}.
\label{eq:xsec}
\end{equation}
In this definition, $(dXdY)_{i}$ refers to the combinations of two analysis observables ($d\delta p_{T}\,d\delta \alpha_{T}$ or $dp_{N}\,d\cos\theta_{\mu}$) and $T$ indicates the target (carbon or oxygen). We measure the differential cross section in each true observable bin $i$. The $\Phi_{\nu}$, $n_{T}$ and $\epsilon_{i}^{T}$ in the denominator are the integrated incoming neutrino flux, the number of target nucleons and the signal selection efficiency in the true observable bin $i$, respectively. The $N_{i,sig}^{T}$ in the numerator is the unfolded signal event number after the background subtraction, which can be expressed as
\begin{equation}
N_{i,sig}^{T} = \sum_{sig \in i} \omega_{sig}^{T}(\vec{\theta}),
\label{eq:sig_number}
\end{equation}
where the weight factor has the same form as the $\omega_{sig}(\vec{\theta})$ in Eq.~\ref{eq:event_weight}.

The signal selection efficiency in each true observable bin $i$ is computed as
\begin{equation}
\epsilon_{i}^{T} = \frac{N_{i, sig}^{T}}{N_{i, tot}^{T}},
\end{equation}
where $N_{i, sig}^{T}$ is the number of selected MC signal events in bin $i$, evaluated with Eq.~\ref{eq:sig_number}, and $N_{i, tot}^{T}$ is the total number of predicted signal events from the MC template in bin $i$. It can be expressed as
\begin{equation}
N_{i, tot}^{T} = \sum_{sig, tot \in i} \omega_{sig, tot}^{T}(\vec{\theta}),
\label{eq:sig_tot_number}
\end{equation}
where the weight factor has the form
\begin{equation}
\omega_{sig, tot}^{T}(\vec{\theta}) = t_{i} \times f_{k} \times \prod_{a}^{sig,model} R_{a}(x_{a}).
\label{eq:sig_weight}
\end{equation}

The cross-section values and their associated uncertainties are calculated using a toy throw approach, where the template and systematic parameters are varied based on their best-fit values and the resulting covariance matrix. For each toy, a vector of randomized parameter values is generated as
\begin{equation}
\vec{\theta}_{\text{toy}} = \vec{\theta}_{\text{best-fit}} + L \vec{r}_{\text{toy}},
\end{equation}
where $L$ is the lower triangular matrix obtained from the Cholesky decomposition~\cite{golub2013matrix} of the best-fit covariance matrix ($V_{\text{best-fit}} = L L^{T}$), and $\vec{r}_{\text{toy}}$ is a vector of random values sampled from a standard Gaussian distribution ($\mu=0, \sigma=1$). Once $\vec{\theta}_{\text{toy}}$ is generated, it is propagated through Eq.~\ref{eq:sig_number} and Eq.~\ref{eq:sig_tot_number} to determine the cross section for each true observable bin. This measurement utilizes 2000 toy throws. The final cross-section result in each bin is defined as the mean of the toy distribution, while the corresponding uncertainty is derived from the covariance matrix $V$ calculated over the full ensemble of toys, defined as
\begin{equation}
\begin{split}
V_{ij} = \frac{1}{N_{toy}} \sum_{t = 1}^{N_{toy}} 
& \left[ \left( \frac{d^{2}\sigma}{dXdY}_{i,t} - \left\langle \frac{d^{2}\sigma}{dXdY} \right\rangle_{i} \right) \times \right. \\
& \left. \left( \frac{d^{2}\sigma}{dXdY}_{j,t} - \left\langle \frac{d^{2}\sigma}{dXdY} \right\rangle_{j} \right) \right],
\end{split}
\label{eq:xsec_cov}
\end{equation}
where $<...>$ averages over all toys in the bin $i$ or $j$.

\subsection{Systematic Uncertainties}

The systematic uncertainty sources in this measurement include detector effects, neutrino flux prediction, neutrino-nucleus cross-section modeling and estimation of target nucleon numbers.

\subsubsection{Detector Systematic Uncertainties}
\label{subsubsec:syst_det}

Detector systematic uncertainties account for potential mismodeling in the simulation of particle propagation within the ND280 detector and the electronics response of the sub-detectors. These also include uncertainties arising from the event reconstruction and selection procedures, stemming from: (1) the modeling of re-interactions of protons and pions leaving the nucleus within detector materials; (2) the uncertainty of the incident neutrino beam direction with respect to the ND280 orientation; (3) the normalization of externally induced backgrounds, such as OOFV and pile-up interactions; (4) track reconstruction, including track charge identification and matching between neighboring sub-detectors; and (5) the reconstruction performance of individual sub-detectors (TPC, FGD, and ECAL).

These uncertainties are evaluated using dedicated control samples to characterize the discrepancies between data and simulation for each contribution. The dominant uncertainty sources are the vertex reconstruction performance in FGD2, which drives confusion between the FGD2X and FGD2Y sub-samples. The evaluation of this uncertainty source is described in detail in Ref.~\cite{T2K:2020jav}. In this analysis, we estimate its contribution to the total detector systematic uncertainty to be $\sim$ 33\% for the FGD2 samples. Besides, another dominant uncertainty source comes from the ECAL track reconstruction efficiency, which significantly impacts $\pi^{0}$ tagging.

In the fit, detector effects are implemented as nuisance parameters that act as normalization factors for each reconstructed analysis bin. These parameters encapsulate the combined impact of all detector-related systematics on the event rate. Their prior uncertainties and correlations are described by a covariance matrix generated from 500 toy experiments, where the underlying detector systematic parameters are varied according to their respective uncertainties. The resulting bin content distributions are used to define the uncertainty for each normalization parameter. These distributions are observed to generally follow a Gaussian profile in all bins. Overall, the prior uncertainties of detector systematic parameters range between approximately 8\% and 10\%.

\subsubsection{Neutrino Flux Systematic Uncertainties}
\label{subsubsec:syst_flux}

The neutrino flux systematic uncertainties represent the uncertainties in the predicted neutrino flux reaching the ND280 detector. Several factors limit the precision of this prediction, including the modeling of hadronic interactions within the graphite target and surrounding materials, the characterization of the proton beam profile, and the uncertainties associated with the magnetic horn current, field, and alignment. Similar to the detector systematics, flux parameters are implemented as normalization factors in bins of true neutrino energy. These parameters account for the collective effect of flux uncertainties on the predicted event rate and are constrained by a covariance matrix that encodes prior knowledge of the flux prediction. The prior uncertainties have been significantly reduced with the help of dedicated hadron production measurements by the NA61/SHINE experiment~\cite{NA61SHINE:2015bad,NA61SHINE:2018rhe}. Appendix~\ref{sec:breakdown_flux_syst} shows the uncertainty on the $\nu_{\mu}$ flux prediction at ND280 used in this analysis, broken down by the uncertainty sources.

In this analysis, the neutrino flux systematic uncertainties enter the cross-section extraction through two mechanisms. First, the neutrino flux parameters introduced in Eq.~\ref{eq:event_weight} modify the weights for both signal and background events, directly impacting the numerator of the cross-section formula in Eq.~\ref{eq:xsec}. Second, these parameters propagate to the calculation of the integrated neutrino flux, which appears in the denominator of Eq.~\ref{eq:xsec}. Overall, the prior uncertainties of flux systematic parameters range between approximately 6\% and 8\%.

\subsubsection{Neutrino Interaction Modeling Systematic Uncertainties}
\label{subsubsec:syst_xsec}

In this measurement, the signal is defined by a $\nu_{\mu}$ CC$0\pi Np$ final-state topology, which receives contributions primarily from CCQE, multi-nucleon (predominantly 2p2h), and CCRES interactions, with a smaller contribution from CCDIS. The CCRES and CCDIS components enter the signal sample only in cases where the produced pion is absorbed in the nucleus. The background is dominated by single-pion production (SPP) topologies, mainly CC$1\pi^{+}$ and CC$1\pi^{0}$, which arise predominantly from CCRES and CCDIS interactions. For each interaction type, a set of systematic parameters has been developed to take into account known model uncertainties~\cite{T2K:2023qjb}:
\begin{itemize}
\item \textbf{CCQE}: (1) One parameter that modifies the axial mass $M_{A}^{QE}$, used to parametrize the nucleon axial form factor within a dipole model. (2) Three parameters that scale the event normalizations in the high-$Q^{2}$ region ($Q^{2} > 0.25$~GeV$^{2}$), where $Q^{2}$ denotes the squared four-momentum transfer. (3) Five parameters that control the nucleon shell occupancies within the SF model of the nucleon initial state: two for carbon (S-shell and P-shell) and three for oxygen (S-shell, P$_{1/2}$-shell, and P$_{3/2}$-shell). (4) Two parameters that independently adjust the short-range correlation (SRC) fractions for carbon and oxygen targets in the SF model.

\item \textbf{2p2h}: (1) One parameter that controls the overall normalization of the 2p2h contribution. (2) Two parameters that modify the relative contributions of meson-exchange currents and nucleon–nucleon correlations for neutron–proton ($np$) and neutron–neutron ($nn$) pairs. These parameters are applied separately to carbon and oxygen targets. (3) One additional parameter that adjusts the relative fraction of $np$ and $nn$ nucleon pairs for both carbon and oxygen targets.

\item \textbf{FSI}: (1) Six parameters that describe uncertainties associated with pion FSI, including elastic scattering, charge exchange, absorption, and inelastic interactions. (2) Six parameters that describe nucleon FSI, accounting for uncertainties in the total interaction cross section and single-pion production processes. The nucleon FSI parameters are defined separately for carbon, oxygen, and all other targets (grouped into a single category).

\item \textbf{SPP}: (1) Three Rein–Sehgal model parameters that describe uncertainties in the axial mass $M_{A}^{RES}$, the normalization $C_{A}^{5}$ of the CCRES axial form factor, and the nonresonant background contribution $I_{1/2}$. (2) One parameter that modifies the $\Delta$-resonance decay kinematics. (3) Two parameters that adjust the binding energies for CCRES interactions on carbon and oxygen targets separately. (4) One parameter that scales the overall CC$1\pi^{0}$ normalization.

\item \textbf{Other}: Two parameters that account for uncertainties in the overall CCDIS normalization and in the normalization of interactions on non-carbon and non-oxygen targets.
\end{itemize}
In the template fit, these systematic parameters are implemented as response functions that return event weights for a given parameter value ($R_{a}(x_{a})$ in Eq.~\ref{eq:event_weight}) and are built for each simulated event separately. The prior uncertainties for neutrino interaction systematic parameters span a wide range: from under 10\% for the CCQE axial mass and CCRES Rein-Sehgal model parameters, to approximately 30\%-40\% for the SF and FSI model parameters, and reaching as high as 100\% for the 2p2h model parameters. Consequently, the modeling of neutrino interactions represents the largest source of prior uncertainties among the various systematic sources.

\subsubsection{Target Nucleon Number Systematic Uncertainties}

The systematic uncertainties associated with the number of target nucleons arise from the mass estimation of the target nuclei, specifically carbon and oxygen, within the FGD1 and FGD2 FVs. The dominant contributions to these uncertainties originate from the measured densities of the constituent materials in the FGD detectors~\cite{T2KND280FGD:2012umz}. The numbers of target nucleons in carbon and oxygen nuclei and their uncertainties used in this measurement are
\begin{equation}
\begin{cases}
n_{T}^{C} = (7.403 \pm 0.035) \times 10^{29} \\
n_{T}^{O} = (2.396 \pm 0.018) \times 10^{29}.
\end{cases}
\end{equation}
These uncertainties are incorporated into the cross-section calculation (Eq.~\ref{eq:xsec}) by sampling the number of target nucleons for each toy from a Gaussian distribution.

\subsection{Validation of Cross-section Extraction}

The primary objectives of the validation are to ensure the stability of the template fit (Sec.~\ref{subsubsec:valid_covfit}) and to verify that the cross-section extraction is sufficiently robust and maintains minimal dependence on the neutrino-nucleus interaction models implemented in the MC template (Sec.~\ref{subsubsec:valid_FDS}).

\subsubsection{Coverage Fits}
\label{subsubsec:valid_covfit}

\begin{figure}
\centering
\subfloat[$\delta p_{T}$--$\delta \alpha_{T}$]
{\includegraphics[width=0.45\textwidth]{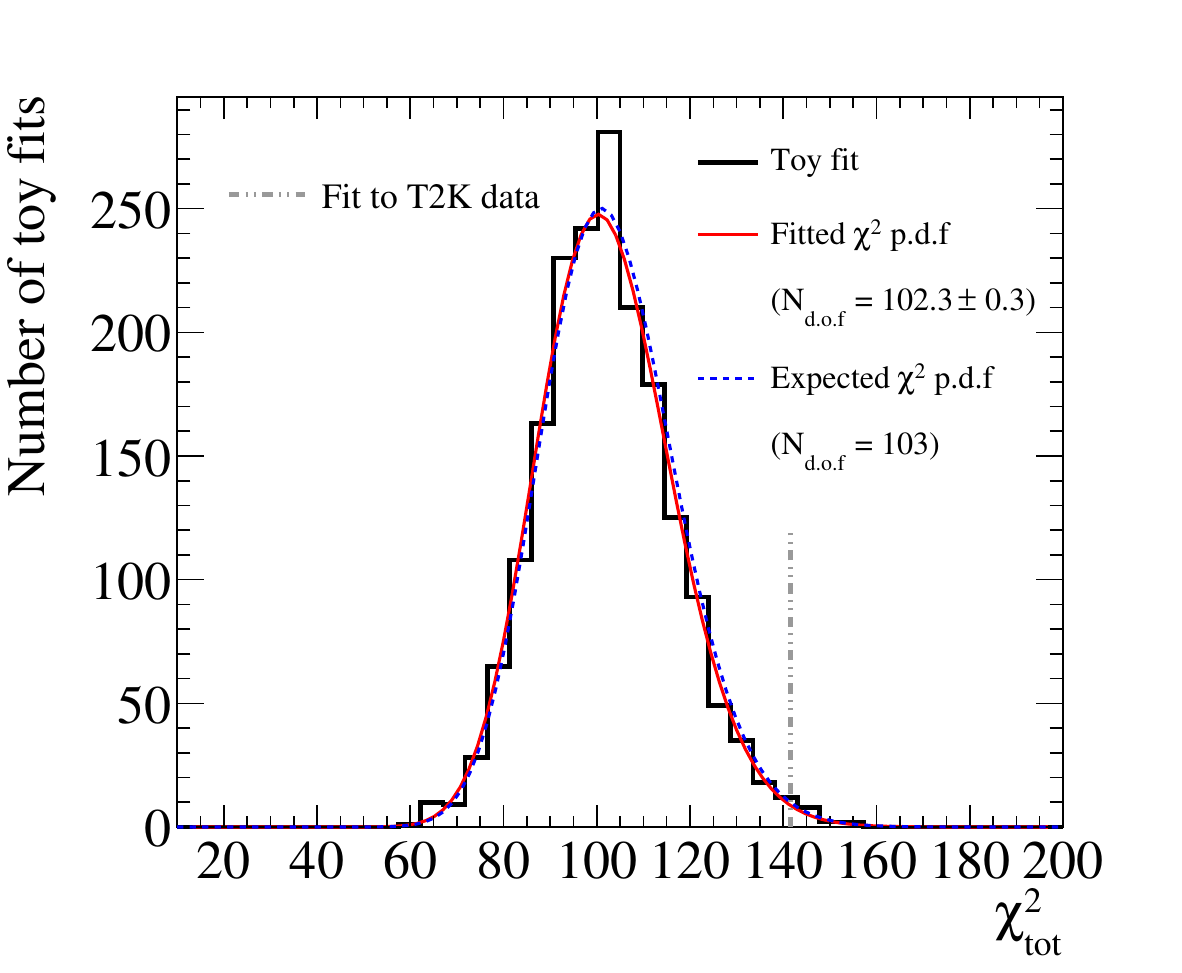}}
\\
\subfloat[$p_{N}$--$\cos\theta_{\mu}$]
{\includegraphics[width=0.45\textwidth]{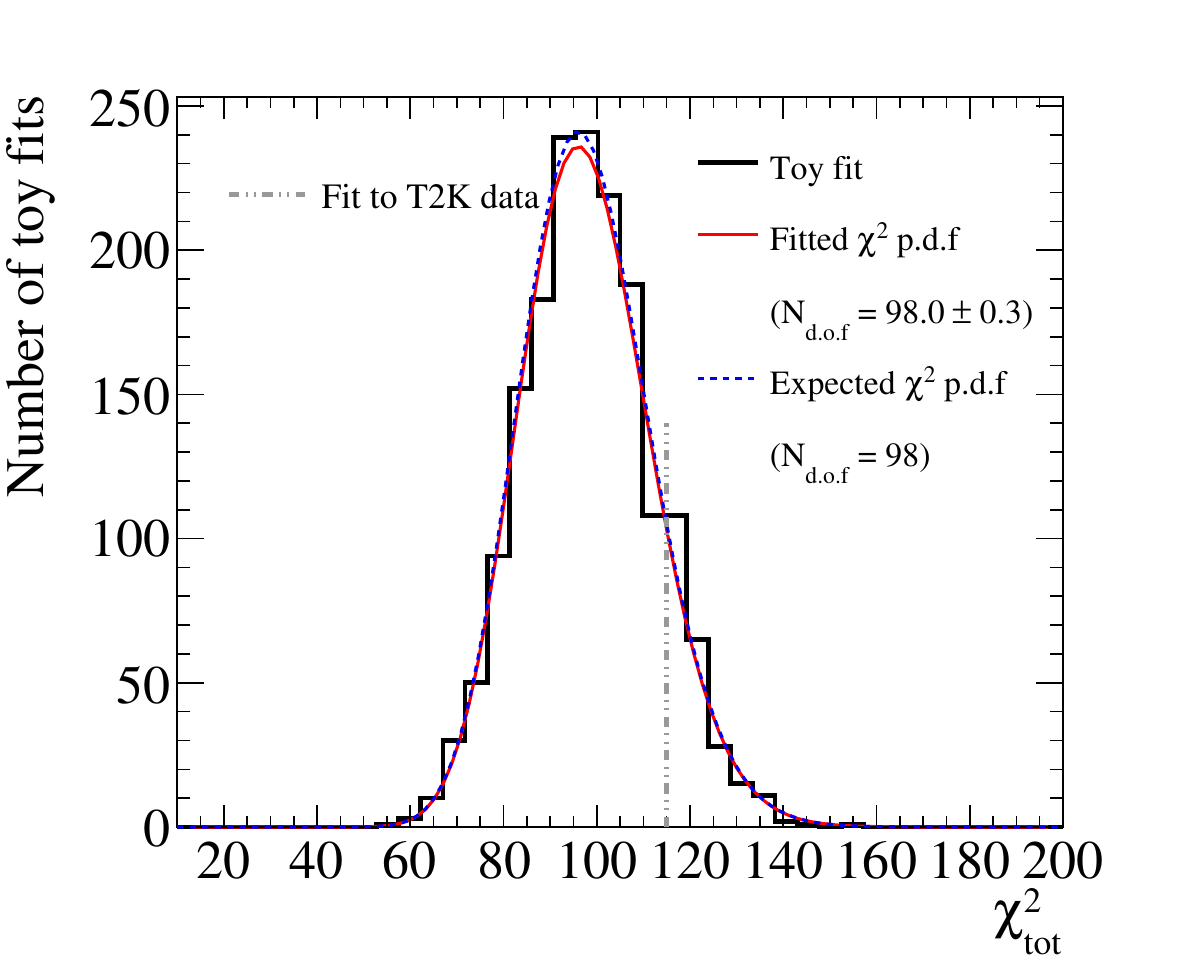}}
\caption{Post-fit $\chi^{2}$ distributions obtained from coverage fits using the $\delta p_{T}$--$\delta \alpha_{T}$ (a) and $p_{N}$--$\cos\theta_{\mu}$ (b) binning schemes. The black histograms represent the post-fit $\chi^{2}$ results from the template fit. The solid red curves denote the fitted $\chi^{2}$ probability density functions (p.d.f.) for these results, while the dashed blue curves show the expected $\chi^{2}$ p.d.f. corresponding to the number of degrees of freedom (d.o.f.). The vertical dashed gray line indicates the $\chi^{2}$ result from T2K data fit (Sec.~\ref{sec:result}).}
\label{fig:valid_cov_chi2}
\end{figure}

The coverage validation incorporates random statistical and systematic fluctuations into the input MC template to generate an ensemble of pseudo-data sets. A fit is performed for each pseudo-data set to assess the stability and performance of the extraction. Specifically, the post-fit $\chi^{2}$ distribution is analyzed. Ideally, this should follow a $\chi^{2}$ probability density function corresponding to the expected degrees of freedom. In this measurement, the number of degrees of freedom is defined as the total number of reconstructed sample bins minus the number of signal template parameters. Fig.~\ref{fig:valid_cov_chi2} displays the post-fit $\chi^{2}$ results from 1000 pseudo-data sets. The distributions for both binning schemes are in good agreement with the theoretical expectations.

\subsubsection{Physics Pseudo-data Fits}
\label{subsubsec:valid_FDS}

The objective of the physics pseudo-data fits is to validate the model independence of the cross-section extraction with respect to the neutrino-nucleus interaction models employed by the MC template. Pseudo-data sets are generated by reweighting the simulation to reproduce various interaction models, accounting for variations in both the signal and background. Furthermore, this analysis also utilizes alternate MC samples produced with different neutrino interaction generator. The cross sections extracted from each pseudo-data fit are compared with the true signal to verify the agreement within uncertainties. A failure in these tests would suggest a potential bias in the template fit model, necessitating the introduction of additional freedom to the template. The model variations evaluated in this analysis are classified into three broad categories:
\begin{itemize}
\item \textbf{Signal model variation}: (1) Variation of the nucleon initial-state model used for CCQE interactions from the SF to alternative nuclear models, including the local Fermi gas (LFG)~\cite{Nieves:2011pp, Bourguille:2020bvw} and the continuum random phase approximation (CRPA)~\cite{Kolbe:1999au,Jachowicz:2002rr}. (2) Variation of the non-QE normalizations in CC$0\pi Np$ signal events, primarily affecting the 2p2h and SPP channels where pions are absorbed via FSI. (3) Data-driven modification of proton kinematics, in which the proton kinematic distributions ($p$–$\cos\theta$) from the MC template are reweighted based on the observed data-to-MC discrepancies. This is motivated by the strong sensitivity of final-state hadron kinematics to nuclear effects. (4) Variation of FSI strength, where the probabilities of pion absorption and nucleon interactions are increased by three times their prior uncertainties.

\item \textbf{Background model variation}: (1) Variation of the final-state hadron kinematic shapes in the SPP channel, motivated by uncertainties in the $\Delta$-resonance decay. (2) Low-$Q^{2}$ suppression, implemented as a normalization of SPP events in the low-$Q^{2}$ region based on observations from the MINERvA experiment~\cite{MINERvA:2019kfr}. (3) Variation of the CC$1\pi^{0}$ normalization, where the overall normalization is increased by 90\% (corresponding to three times the prior uncertainty).

\item \textbf{Neutrino generator alteration}: In this pseudo-data set, the NEUT MC prediction is reweighted to an alternative generated by the GENIE v2.8.0 generator~\cite{Andreopoulos:2009rq}, which incorporates a distinct set of nuclear effect models. The primary discrepancies in this GENIE prediction lie in the treatment of the nuclear ground state and FSI, which are modeled using the LFG model and a GENIE-specific INC approach, respectively. The reweighting is performed in the $p_{p}$ (leading proton momentum) versus $q_{0}$ (energy transfer) space, which is expected to capture the dominant differences in neutrino interaction modeling between NEUT and GENIE.
\end{itemize}

To check the consistency between the extracted cross sections and the pseudo-data ones, we define and compute the cross-section $\chi^{2}$, $\chi^{2}_{xsec}$, as:
\begin{equation}
\begin{split}
\chi^{2}_{xsec} = \sum_{i,j}
& \left[ \left( \frac{d^{2}\sigma}{dXdY}_{i,pred} - \frac{d^{2}\sigma}{dXdY}_{i,meas} \right) (V^{-1})_{ij} \right. \\
& \left. \left( \frac{d^{2}\sigma}{dXdY}_{j,pred} - \frac{d^{2}\sigma}{dXdY}_{j,meas} \right) \right],
\end{split}
\label{eq:xsec_chi2}
\end{equation}
where $\frac{d^{2}\sigma}{dXdY}_{i,meas}$ is the extracted cross section, $\frac{d^{2}\sigma}{dXdY}_{i,pred}$ is the pseudo-data cross section and $V$ is the covariance matrix of the extracted cross section defined by Eq.~\ref{eq:xsec_cov}. The sum loops over all cross-section bins defined in Appendix~\ref{subsec:xsec_binning}. Furthermore, to better validate the results, the $p$-value of each $\chi^{2}_{xsec}$ is computed based on the $\chi^{2}_{xsec}$ distribution generated with the same method as Sec.~\ref{subsubsec:valid_covfit}, but where only systematic fluctuations are enabled. The statistical fluctuations are not included because the pseudo-data sets are reweighted directly from the MC template so there is 100\% statistical correlation between the nominal MC and pseudo-data distributions. Tab.~\ref{tab:fds_xsec_chi2} summarizes the $\chi^{2}_{xsec}$ and corresponding $p$-value results from each physics pseudo-data set. High $p$-values (close to one) are obtained for all pseudo-data sets in both binning schemes, indicating the robustness of the cross-section extraction framework.

\begin{table*}
\centering
\caption{$\chi^{2}_{xsec}$ values and corresponding $p$-values from the physics pseudo-data fits.}
\label{tab:fds_xsec_chi2}
\renewcommand{\arraystretch}{1.2}
\begin{tabular}{c|c|c}
\hline
\hline
\multirow{2}{*}{Pseudo-data} & \multicolumn{2}{c}{$\chi^{2}_{xsec}$ ($p$-value)} \\
\cline{2-3}
 & $\delta p_{T}$--$\delta \alpha_{T}$ ($N_{bin} = 25$) & $p_{N}$--$\cos\theta_{\mu}$ ($N_{bin} = 30$) \\
\hline
SF $\rightarrow$ LFG reweight & 0.89 (1.00) & 0.62 (1.00) \\
SF $\rightarrow$ CRPA reweight & 0.97 (1.00) & 1.35 (1.00) \\
Non-QE normalization alteration & 1.88 (1.00) & 0.95 (1.00) \\
Proton kinematics alteration & 0.77 (1.00) & 0.90 (1.00) \\
FSI strength alteration & 2.27 (0.99) & 3.12 (0.99) \\
\hline
SPP hadron kinematics alteration & 1.53 (1.00) & 0.90 (1.00) \\
SPP low $Q^{2}$ suppression & 0.63 (1.00) & 1.22 (1.00) \\
CC$1\pi^{0}$ normalization alteration & 1.28 (1.00) & 1.82 (1.00) \\
\hline
NEUT $\rightarrow$ GENIE reweight & 3.19 (0.98) & 3.71 (0.99) \\
\hline
\hline
\end{tabular}
\end{table*}

\section{Results}
\label{sec:result}

This section presents the cross-section results measured from the fit to T2K data.

\subsection{$\delta p_{T}$--$\delta \alpha_{T}$ Combination}
\label{subsec:result_1}

The template fit has post-fit $\chi^{2} = 141.54$ corresponding to a $p$-value around 0.008 ($N_{dof} = 103$). This $p$-value is lower than the often employed significance threshold of 0.05. Dedicated studies were performed to investigate the origin of the low $p$-value, indicating that it is primarily driven by the limited flexibility of the background interaction modeling in the $\delta p_{T}$--$\delta \alpha_{T}$ space. Additional test that introduce more freedom into the background modeling yields higher $p$-value ($\sim 0.09$). Despite these adjustments, the extracted signal cross sections remain consistent with the original results, exhibiting both similar central values and comparable uncertainties. This supports the robustness of the original signal cross-section results which are presented in this article. More details are summarized in Appendix~\ref{sec:disc_small_pval}.

\begin{figure*}
\centering
\subfloat[Carbon]
{\includegraphics[width=0.33\linewidth]{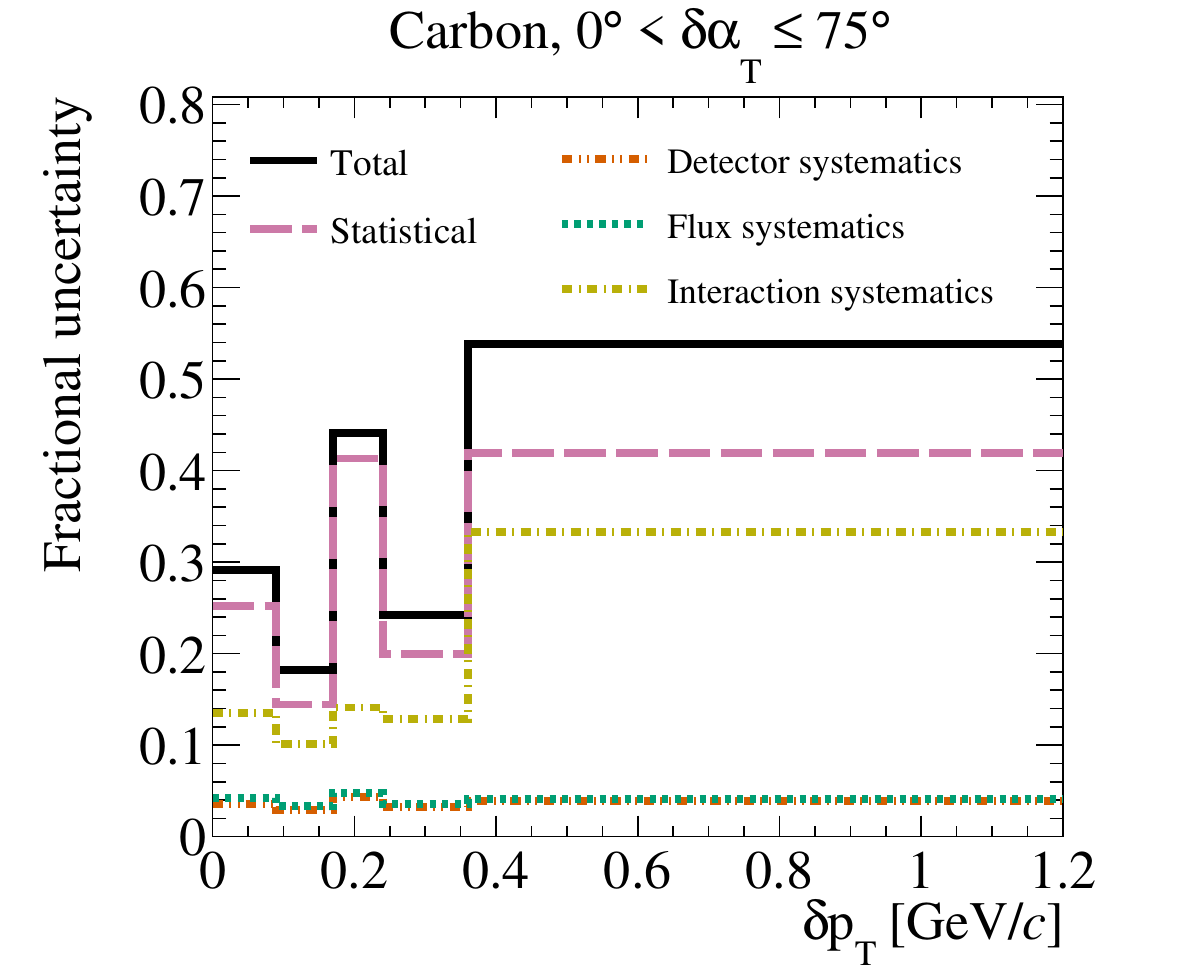}
\includegraphics[width=0.33\linewidth]{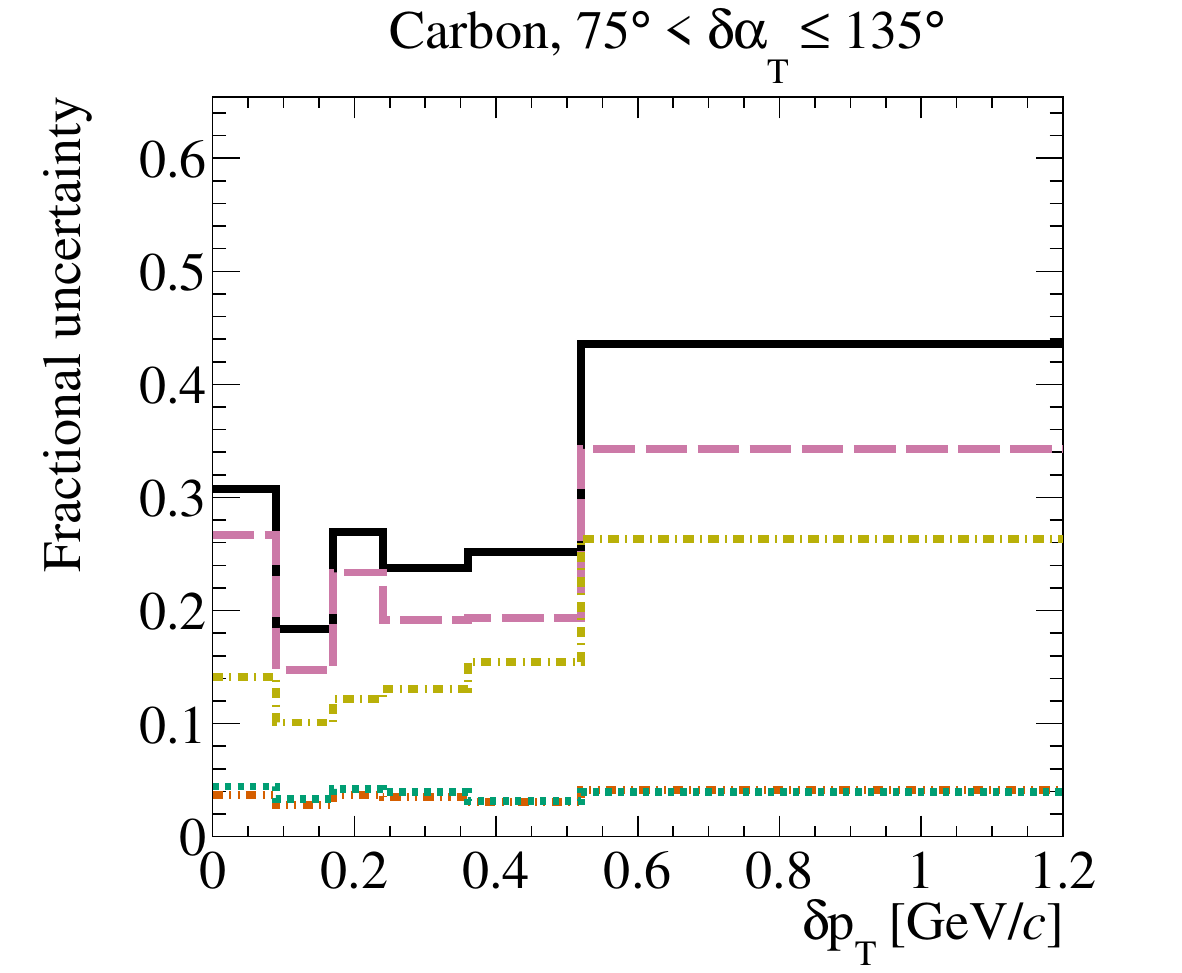}
\includegraphics[width=0.33\linewidth]{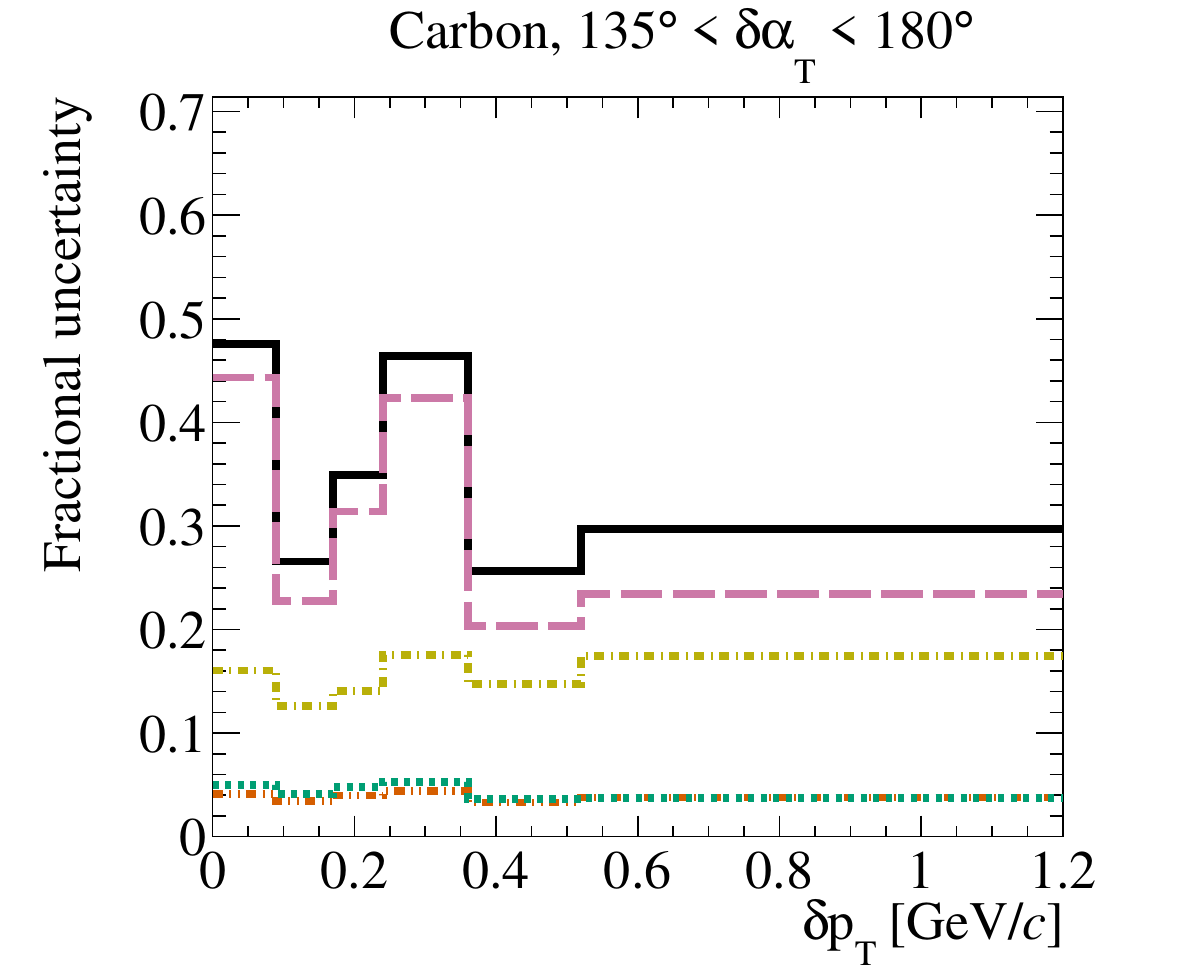}}
\\
\subfloat[Oxygen]
{\includegraphics[width=0.33\linewidth]{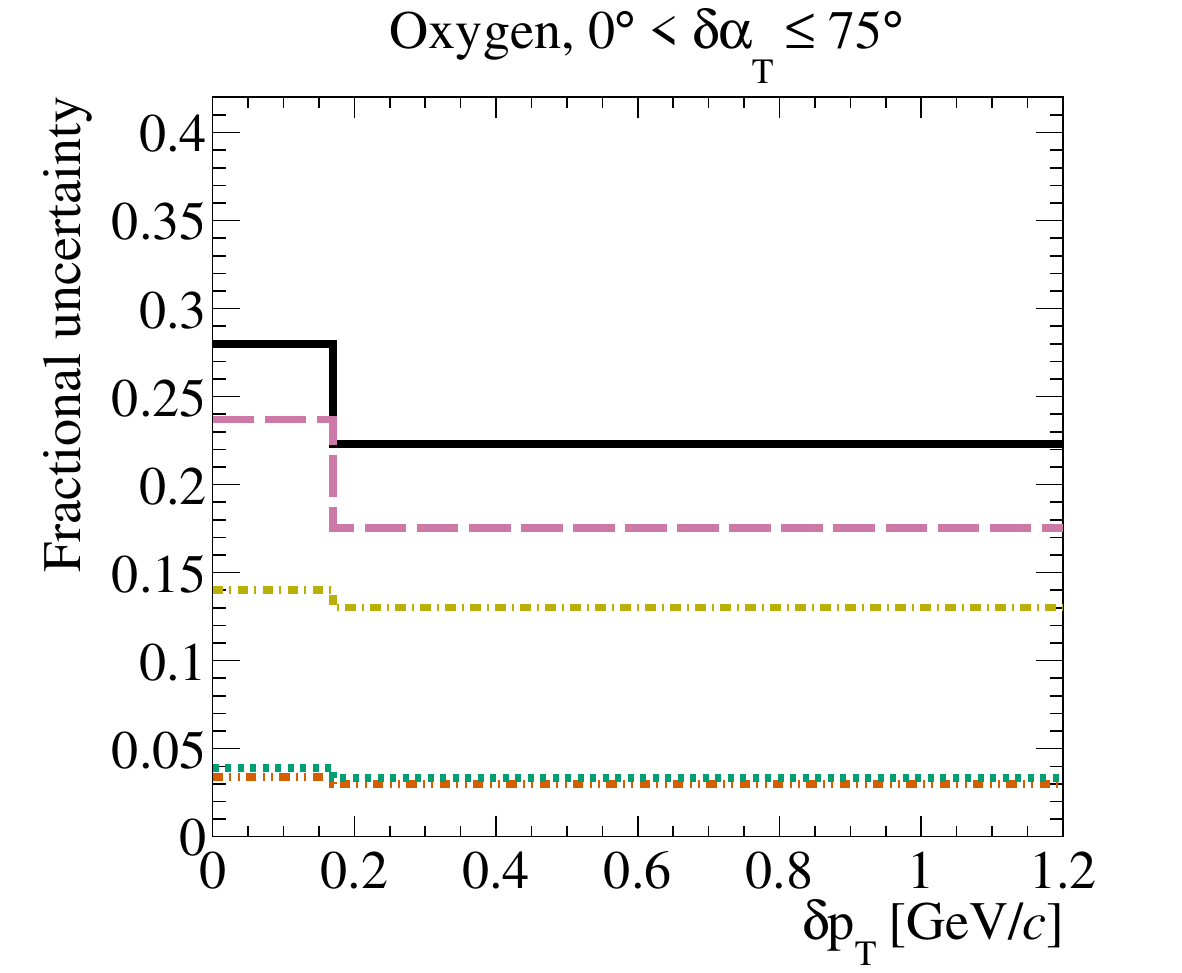}
\includegraphics[width=0.33\linewidth]{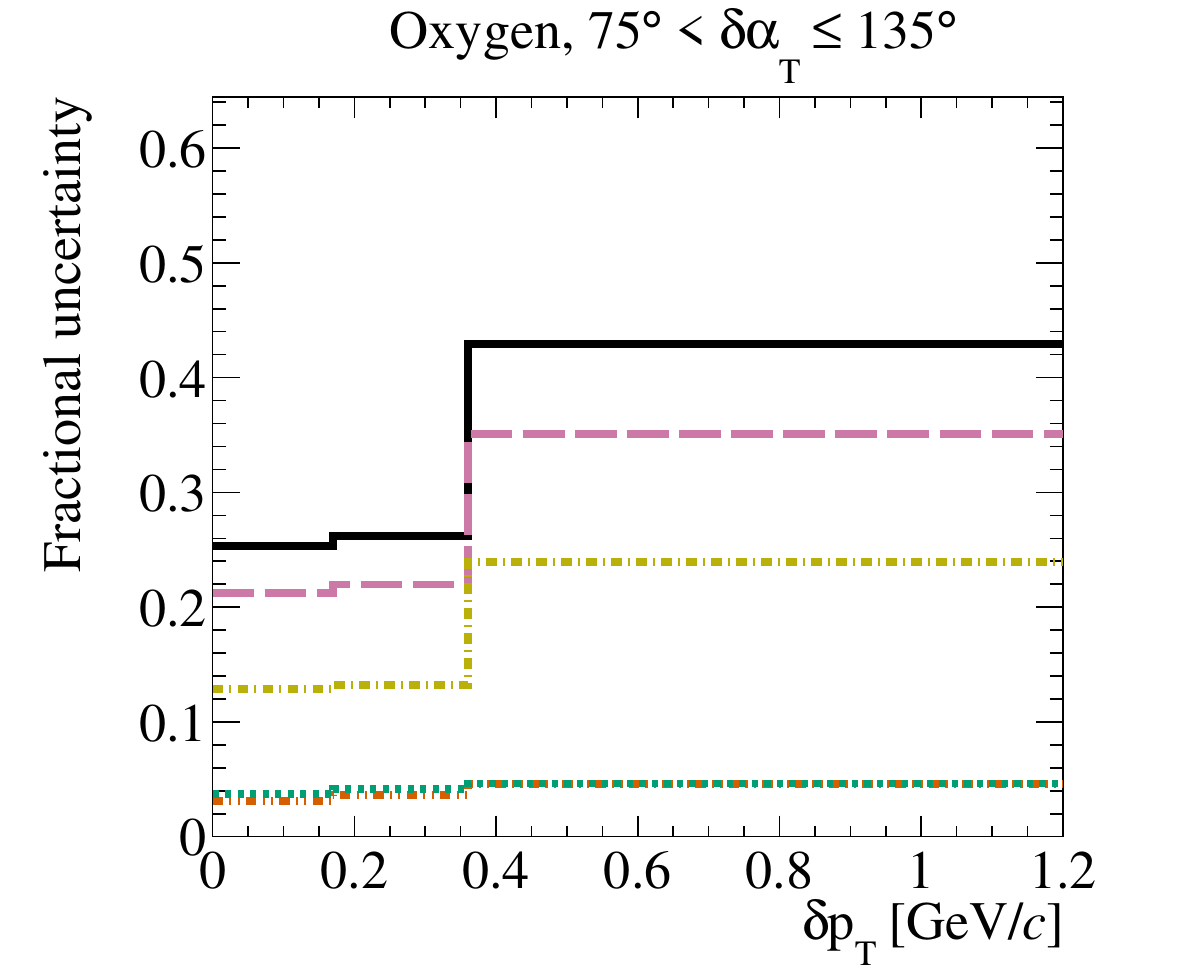}
\includegraphics[width=0.33\linewidth]{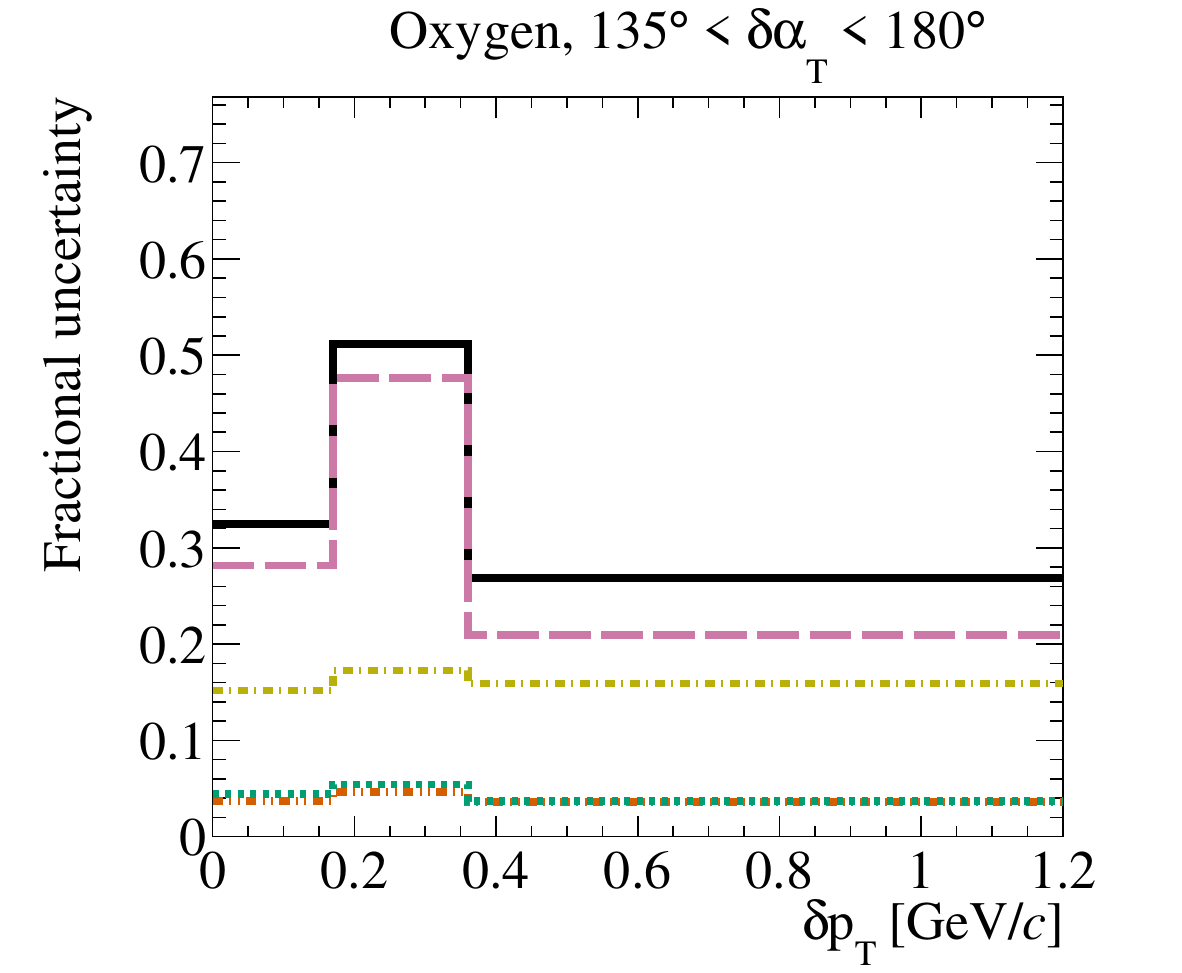}}
\caption{Measurement uncertainties for the $\nu_{\mu}$ CC$0\pi Np$ cross sections on carbon (a) and oxygen (b) targets, derived from the fit to T2K data in $\delta p_{T}$--$\delta \alpha_{T}$ space. In addition to the total uncertainty (black), contributions from individual sources are shown: statistics (pink) and various systematics, including detector (orange), neutrino flux (bluish green), and neutrino interaction (yellow). The systematic uncertainty from number of target nucleons is omitted as it contributes less than 1\% for both targets. The fractional uncertainty in each bin is defined as the ratio of the absolute uncertainty to the corresponding cross-section value.}
\label{fig:realdata_xsec_err_1}
\end{figure*}

\begin{figure}
\centering
\subfloat[$\delta p_{T}$--$\delta \alpha_{T}$]
{\includegraphics[width=0.45\textwidth]{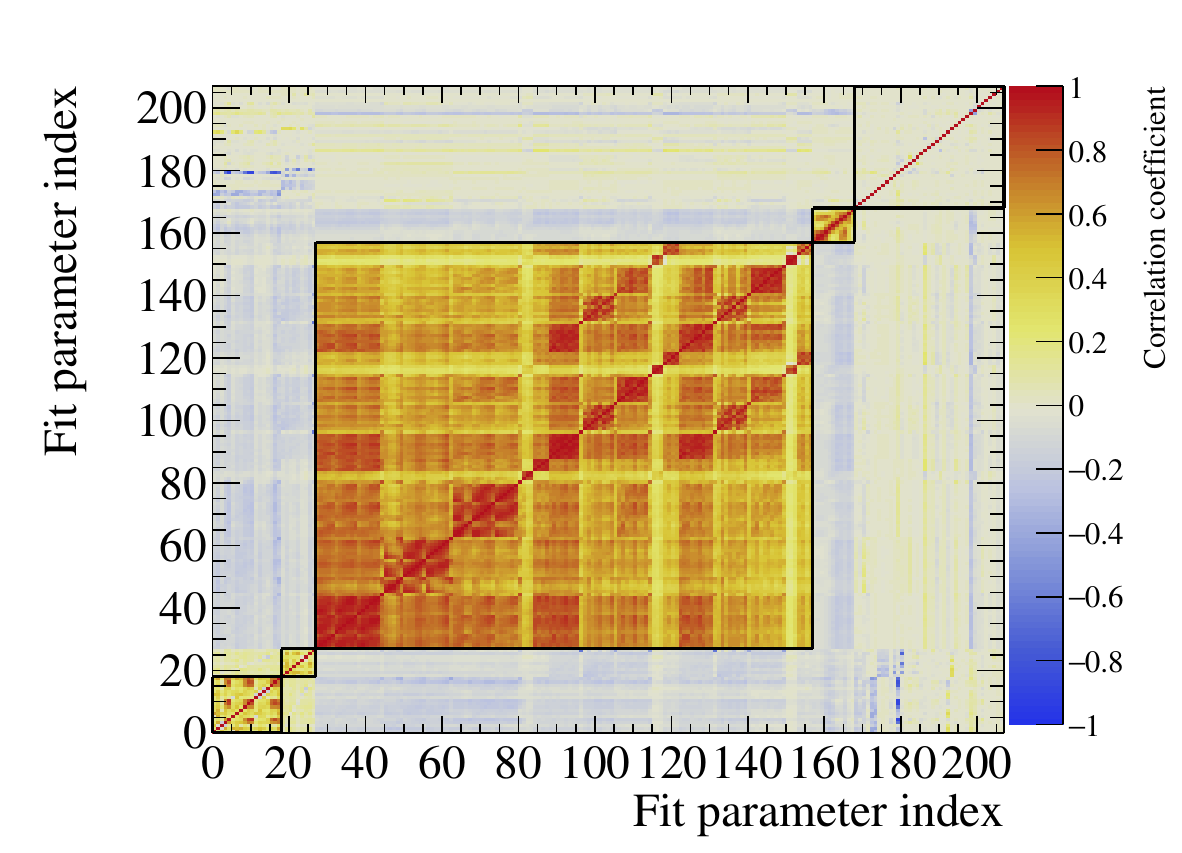}}
\\
\subfloat[$p_{N}$--$\cos\theta_{\mu}$]
{\includegraphics[width=0.45\textwidth]{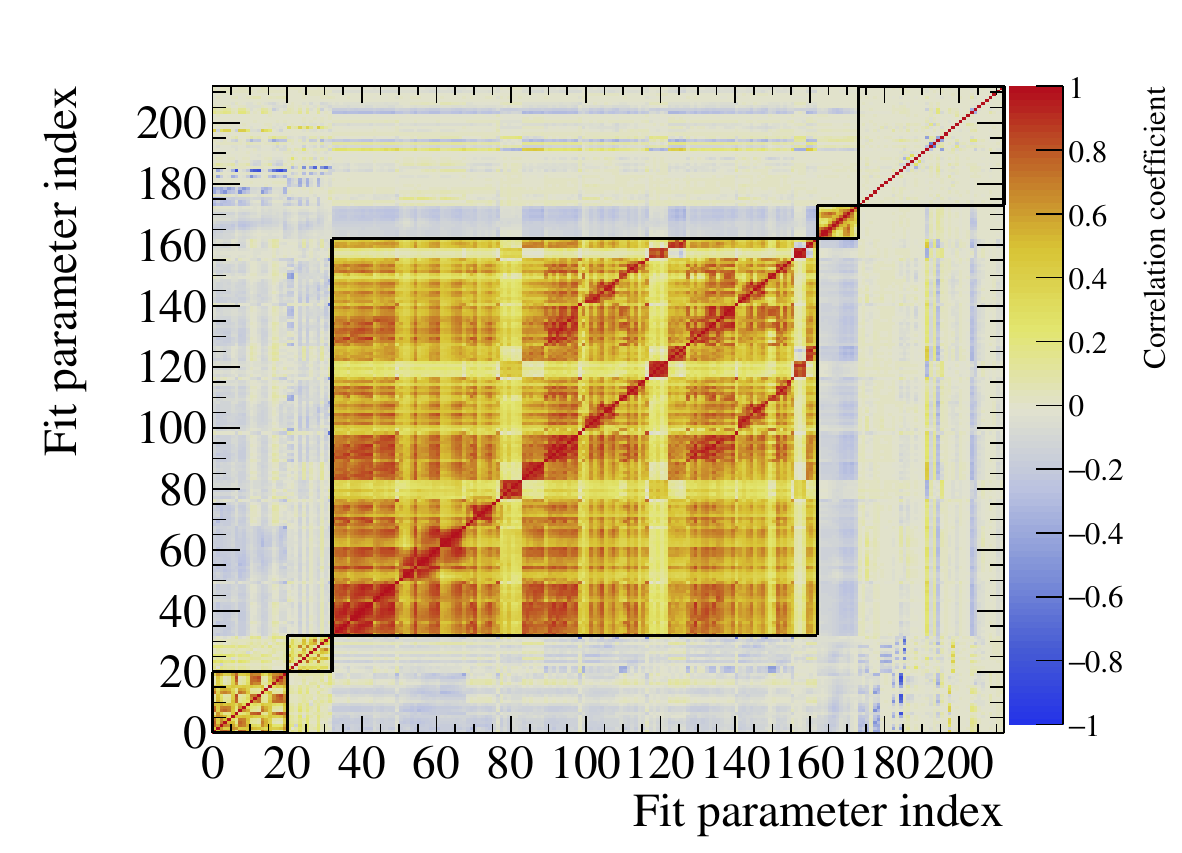}}
\caption{Post-fit correlation matrices for the fit parameters obtained from the T2K data fit in the (a) $\delta p_{T}$-$\delta \alpha_{T}$ and (b) $p_{N}$-$\cos\theta_{\mu}$ spaces. The diagonal blocks (delineated by black lines) represent distinct parameter sets, which include (from bottom-left to top-right): carbon template parameters, oxygen template parameters, detector systematic parameters, neutrino flux systematic parameters, and neutrino interaction systematic parameters.}
\label{fig:postfit_corr_matrix}
\end{figure}

Fig.~\ref{fig:realdata_xsec_1} presents the extracted $\nu_{\mu}$ CC$0\pi Np$ signal cross sections for carbon and oxygen targets. The associated measurement uncertainties are detailed in Fig.~\ref{fig:realdata_xsec_err_1}, which includes a breakdown of contributions from various sources. To estimate the impact of each source, only the relevant parameters are varied during the cross-section calculation stage (Sec.~\ref{subsubsec:xsec_def_calc}), while all other parameters remain fixed at their best-fit values. The uncertainty evaluation follows the method described in Sec.~\ref{subsubsec:xsec_def_calc}, except that only the subset of the post-fit covariance matrix corresponding to the relevant parameters is considered. Statistical uncertainties are evaluated via the template parameters ($t_{i}$ in Eq.~\ref{eq:event_weight}), as these govern the signal normalization in each analysis bin. Systematic uncertainties from each source are similarly assessed using their corresponding parameter sets. For example, to evaluate the contribution from the detector systematic uncertainties, only the parameters $d_{j}$ (Eq.~\ref{eq:event_weight}) are varied. It should be noted that because these parameter sets become correlated after the fit, as illustrated by the post-fit correlation matrix in Fig.~\ref{fig:postfit_corr_matrix}, a precise decoupling of individual contributions is non-trivial. Nonetheless, this procedure provides a reliable qualitative estimate of each uncertainty source. The total measurement uncertainties range from 20\% to 70\% across the cross-section bins. The statistical error remains the dominant contribution in all bins, indicating that data statistics are the primary limiting factor for the precision of this measurement. Among the systematic sources, the modeling of neutrino-nucleus interactions represents the most significant contribution, exceeding those from the detector and neutrino flux uncertainties.

\begin{figure*}
\centering
\subfloat[Carbon]
{\includegraphics[width=0.33\linewidth]{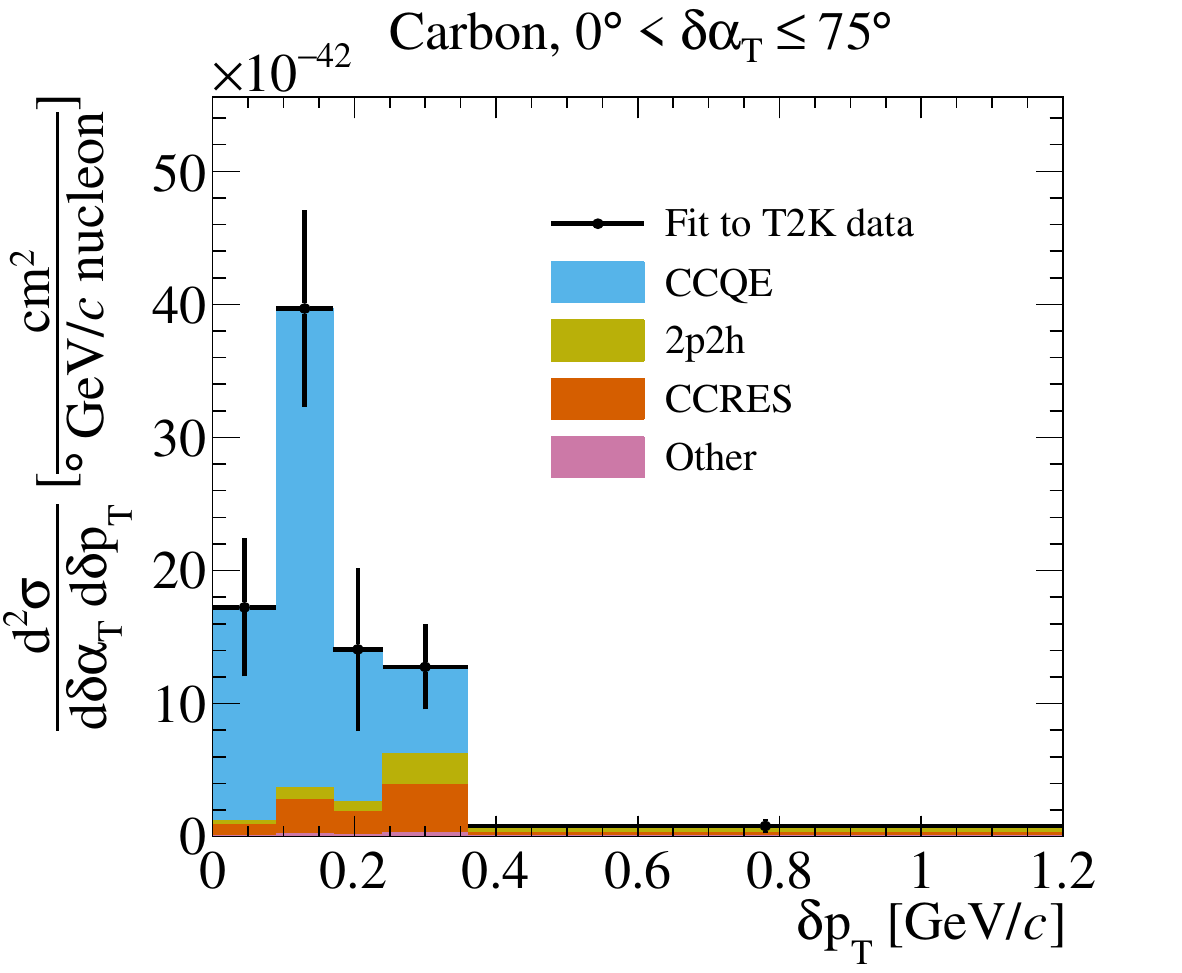}
\includegraphics[width=0.33\linewidth]{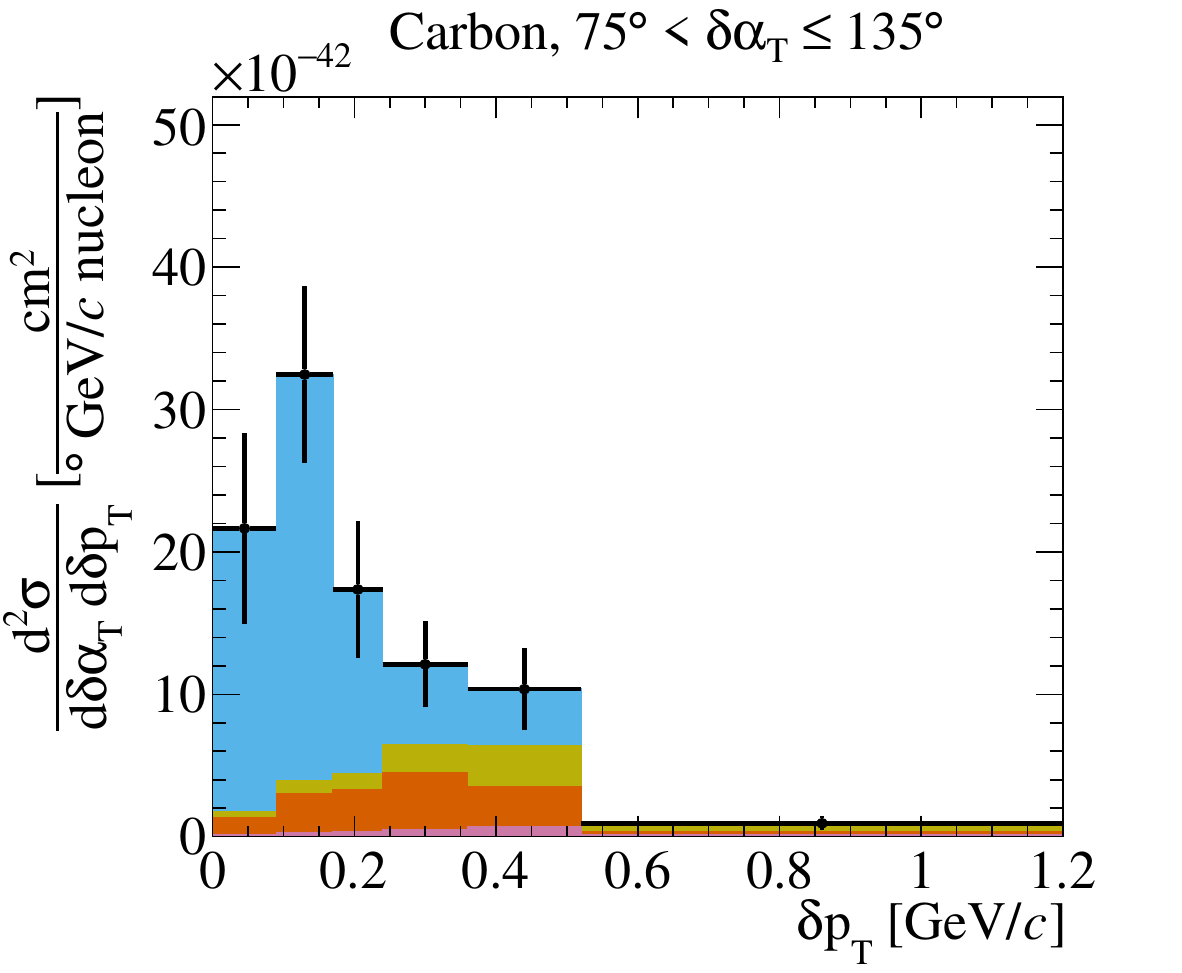}
\includegraphics[width=0.33\linewidth]{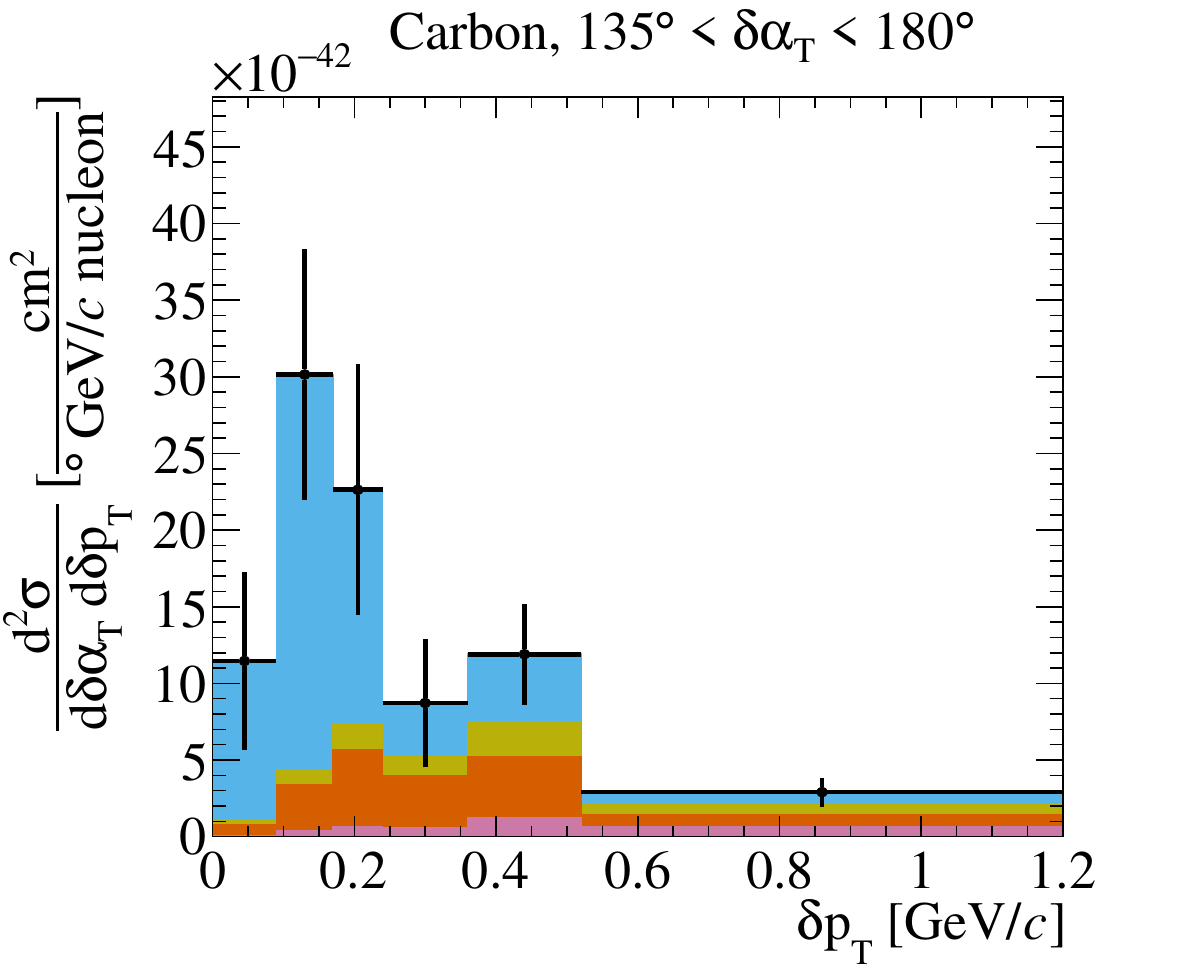}}
\\
\subfloat[Oxygen]
{\includegraphics[width=0.33\linewidth]{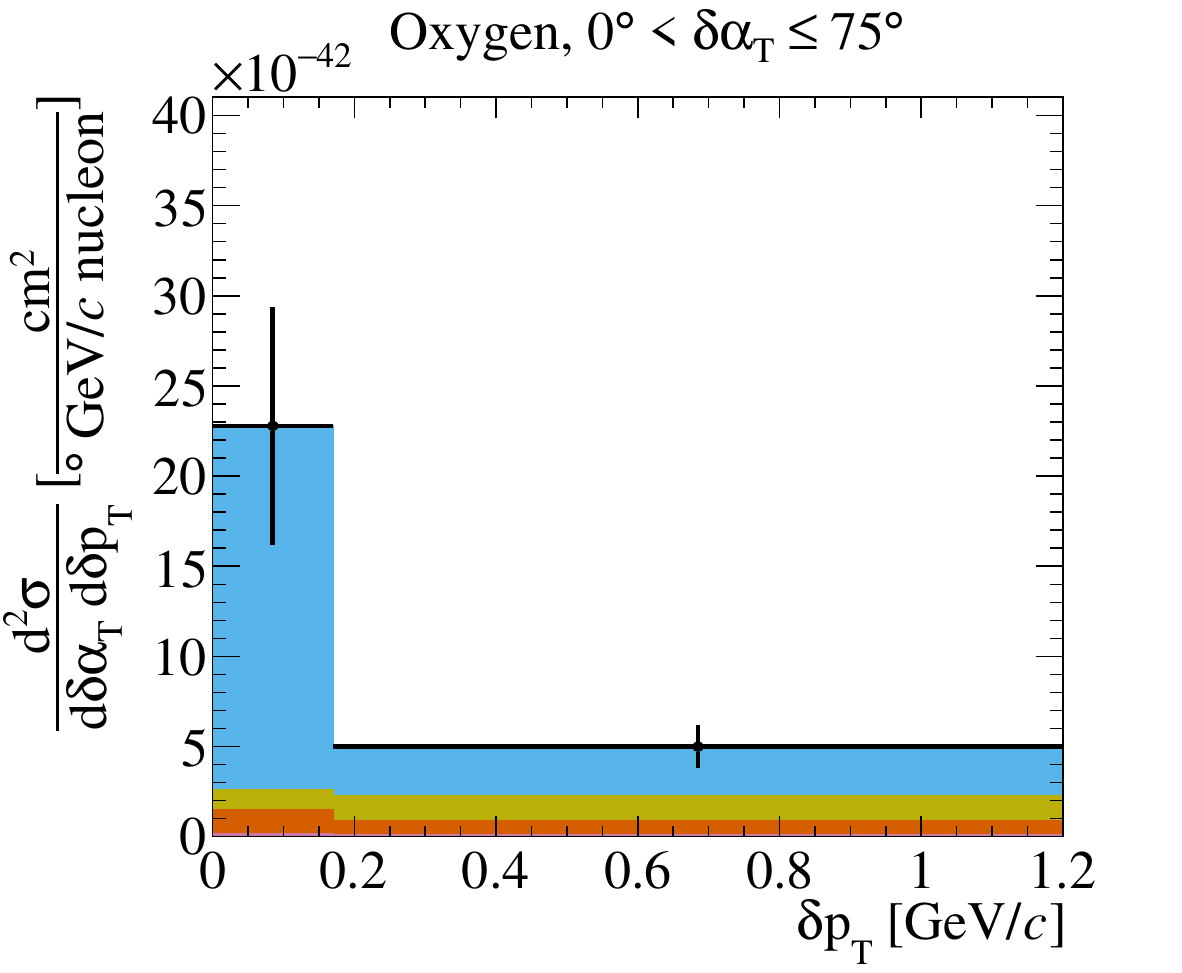}
\includegraphics[width=0.33\linewidth]{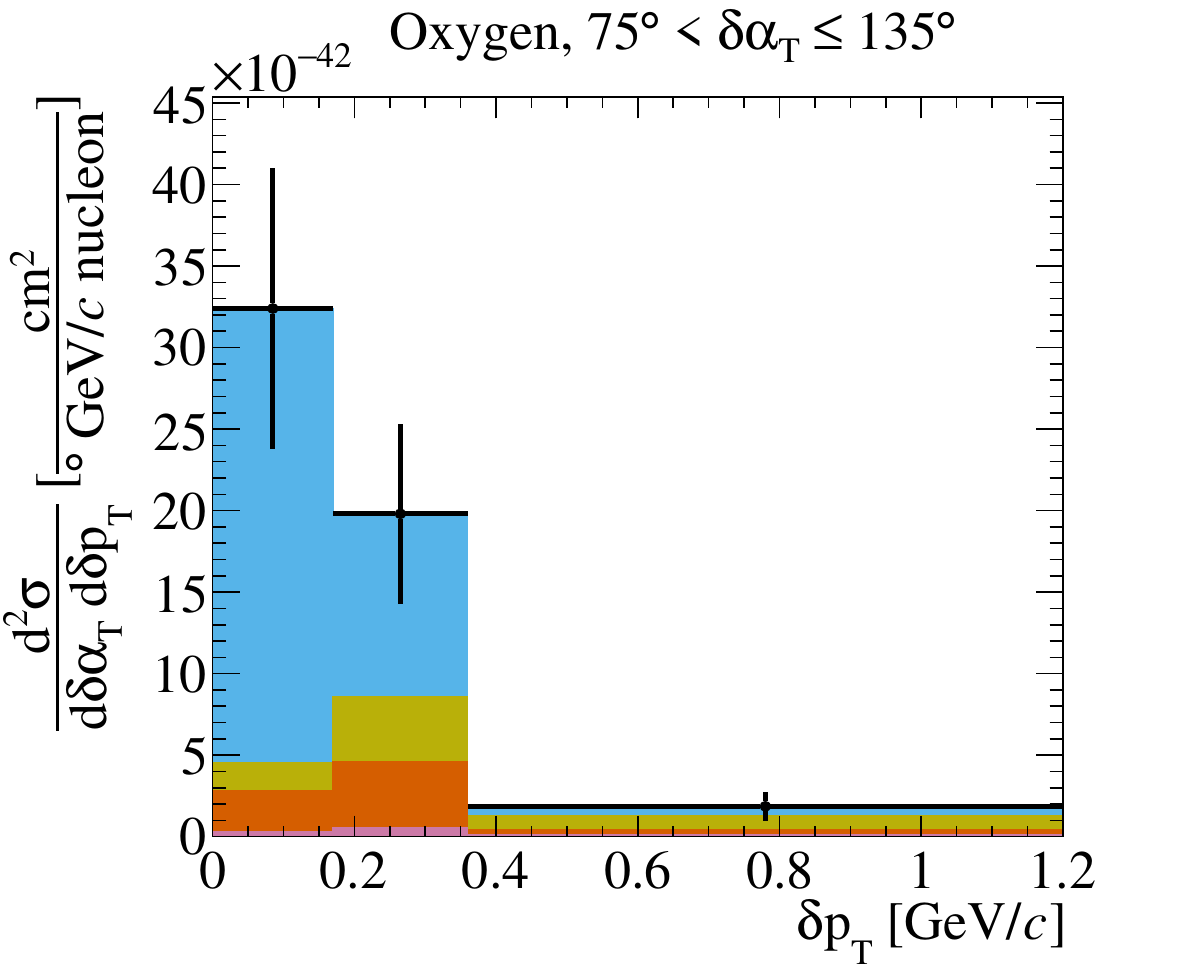}
\includegraphics[width=0.33\linewidth]{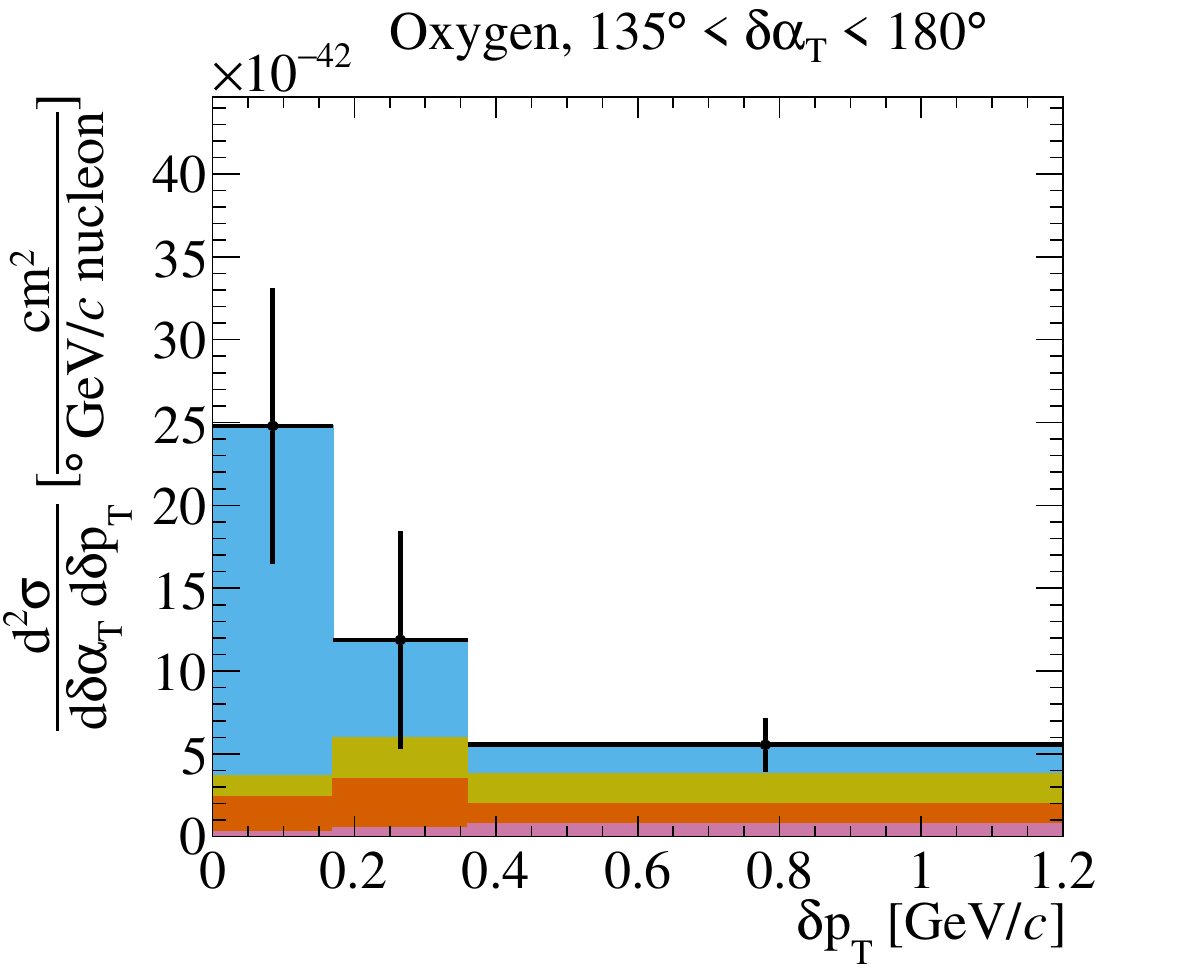}}
\caption{Measurements of the $\nu_{\mu}$ CC$0\pi Np$ signal cross section on carbon (a) and oxygen (b) targets from the T2K data fit in the $\delta p_{T}$--$\delta \alpha_{T}$ space. The post-fit MC template is decomposed into different interaction types, including CCQE, 2p2h, CCRES and other (all remaining interaction types). The error bar represents the combined statistical and systematic uncertainty.}
\label{fig:realdata_xsec_1}
\end{figure*}

\subsection{$p_{N}$--$\cos\theta_{\mu}$ Combination}

\begin{figure*}
\centering
\subfloat[Carbon]{
\begin{tabular}{c}
\includegraphics[width=0.32\linewidth]{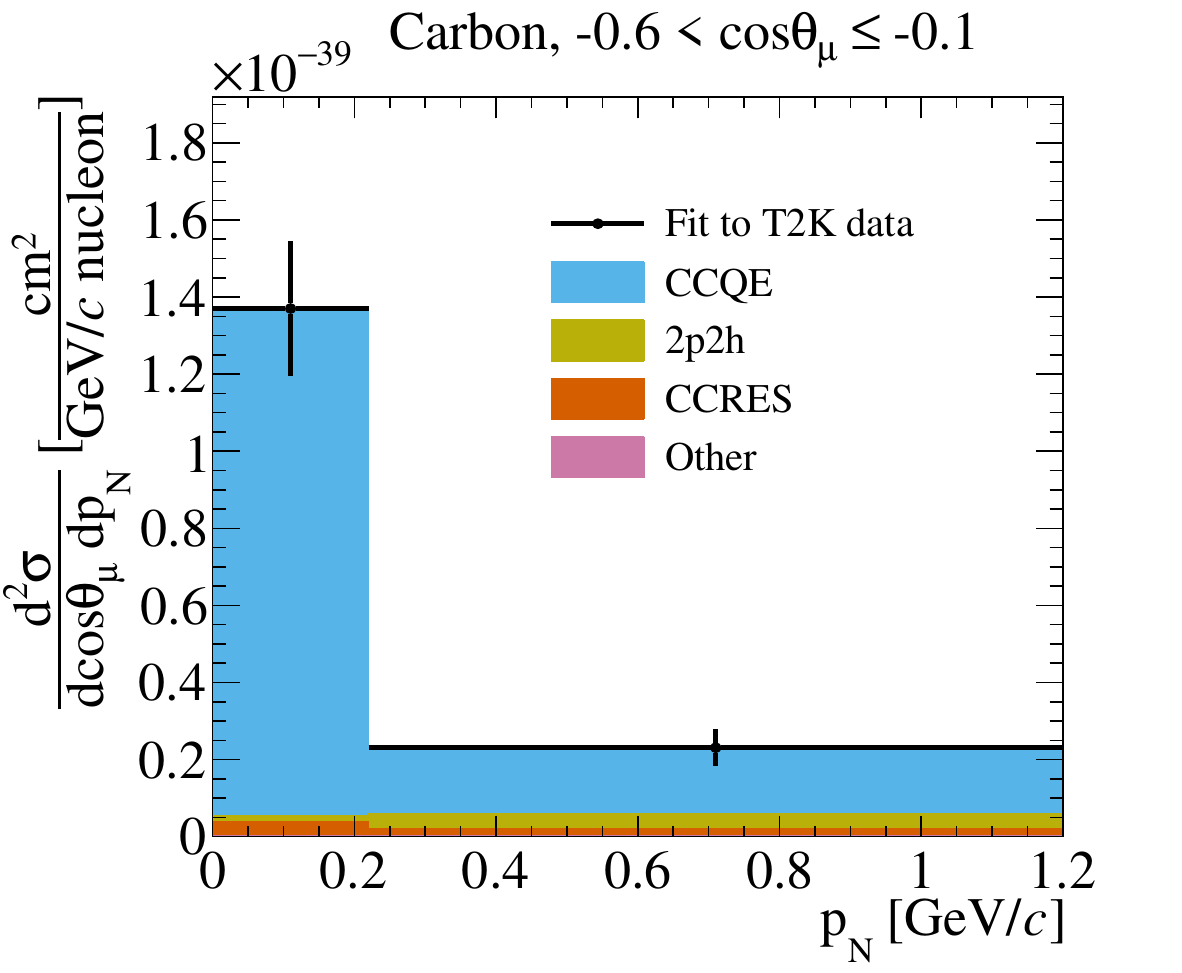}
\includegraphics[width=0.32\linewidth]{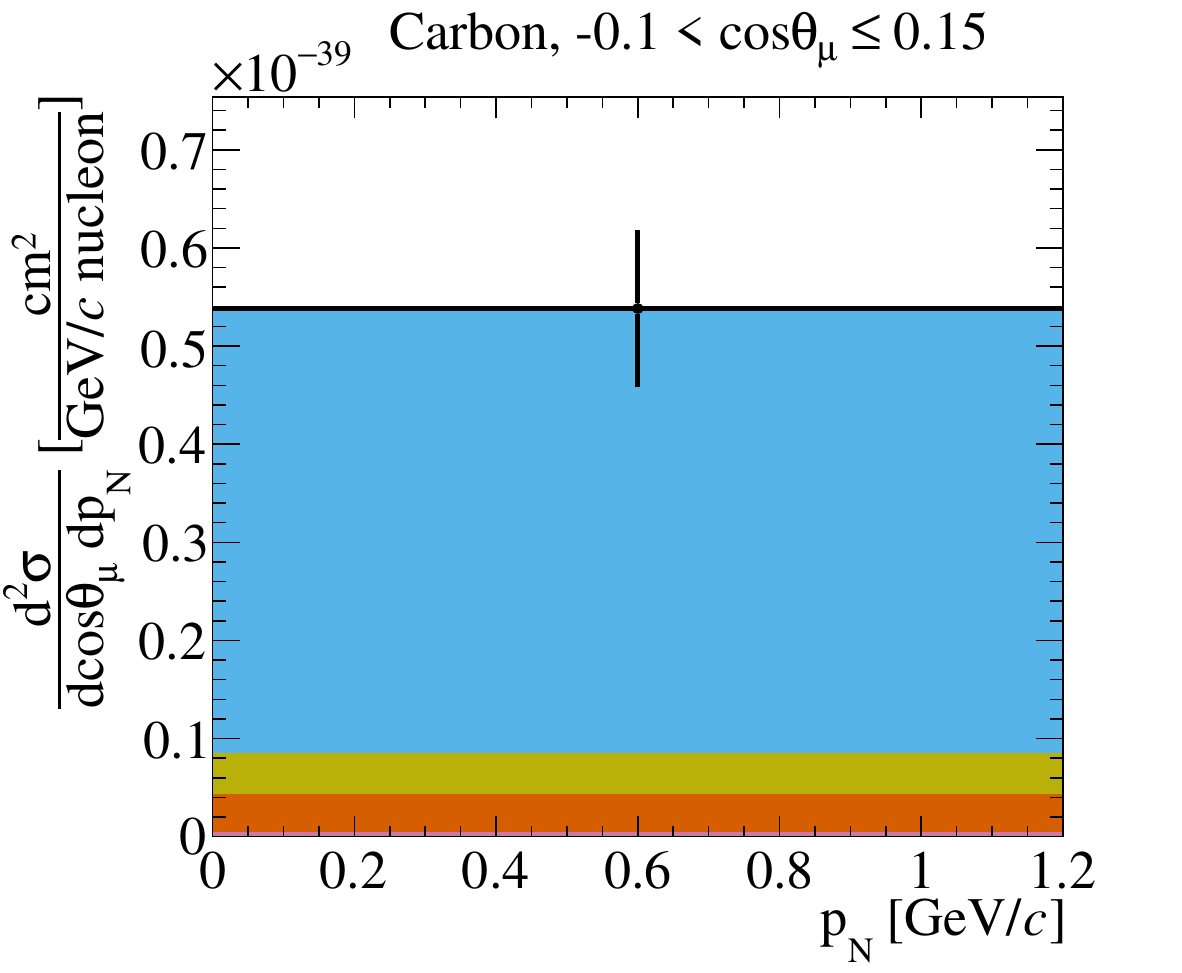}
\includegraphics[width=0.32\linewidth]{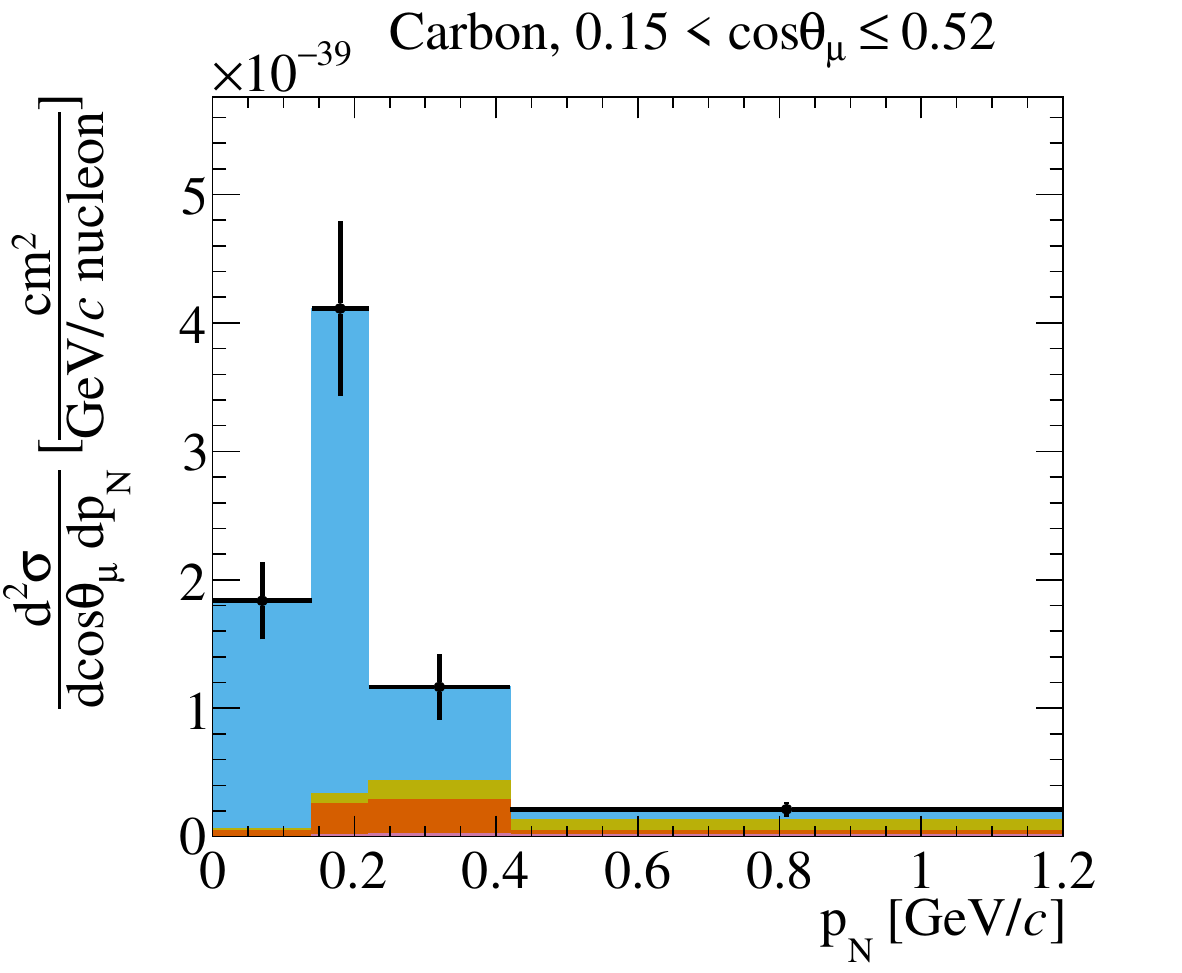} \\
\includegraphics[width=0.32\linewidth]{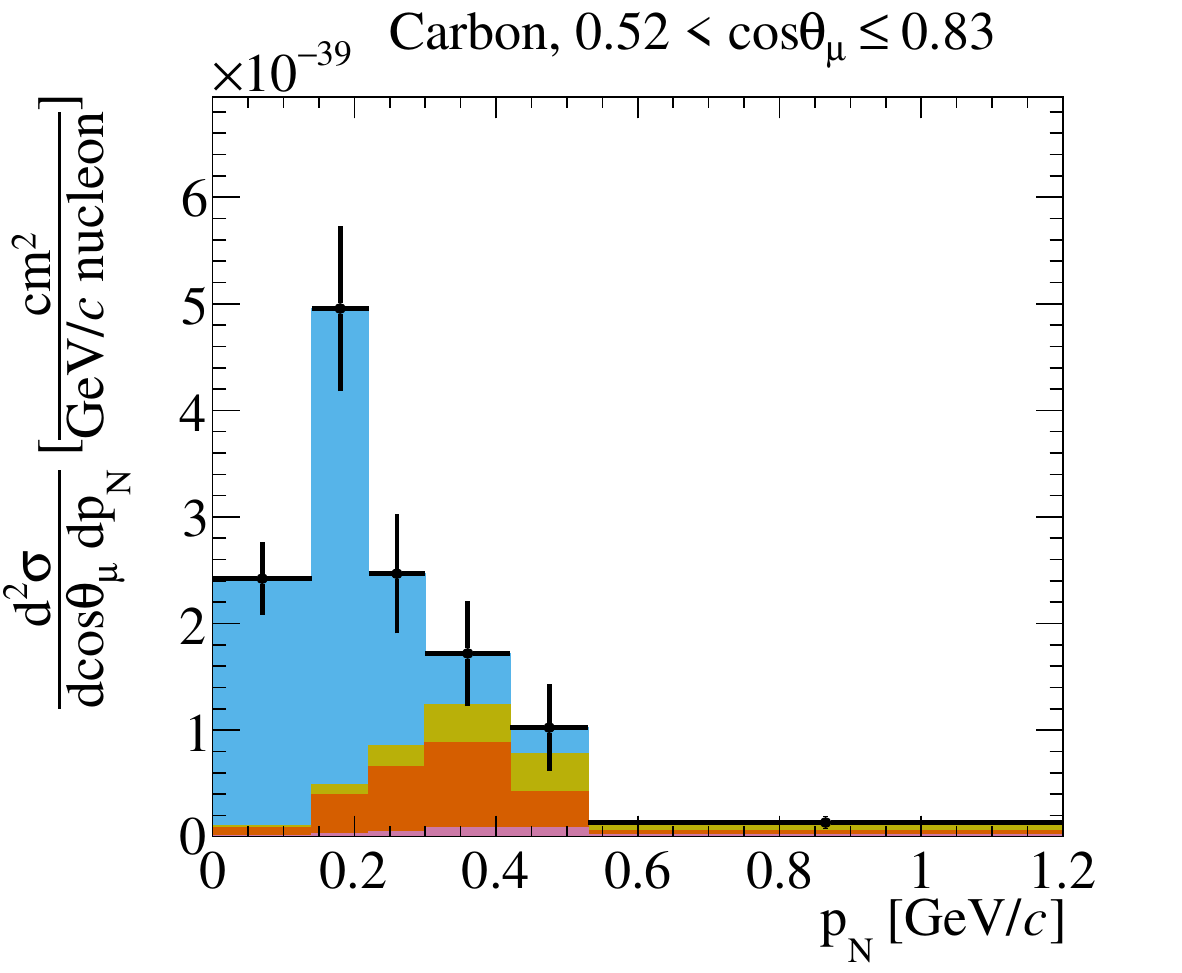}
\includegraphics[width=0.32\linewidth]{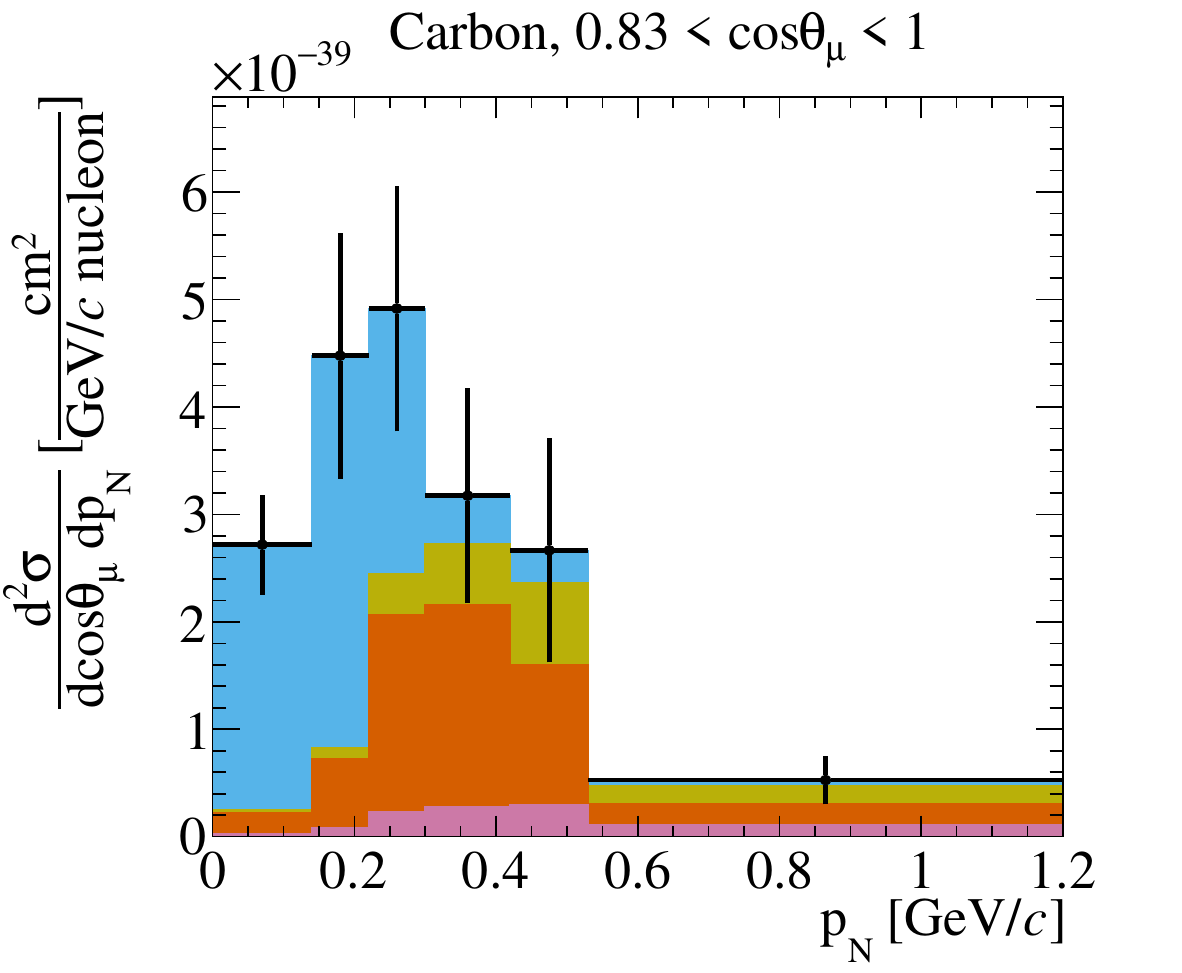}
\end{tabular}
}
\\
\subfloat[Oxygen]{
\begin{tabular}{c}
\includegraphics[width=0.32\linewidth]{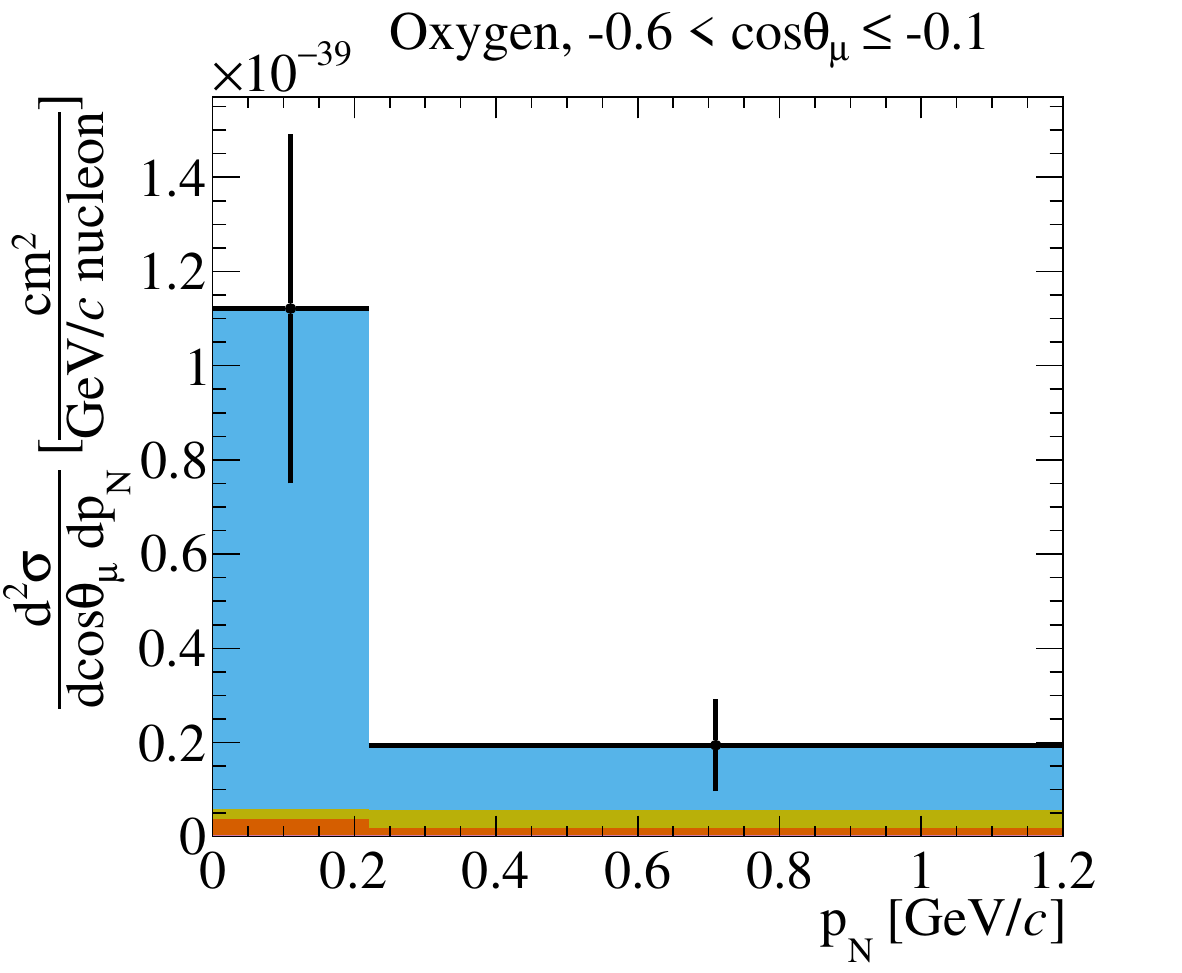}
\includegraphics[width=0.32\linewidth]{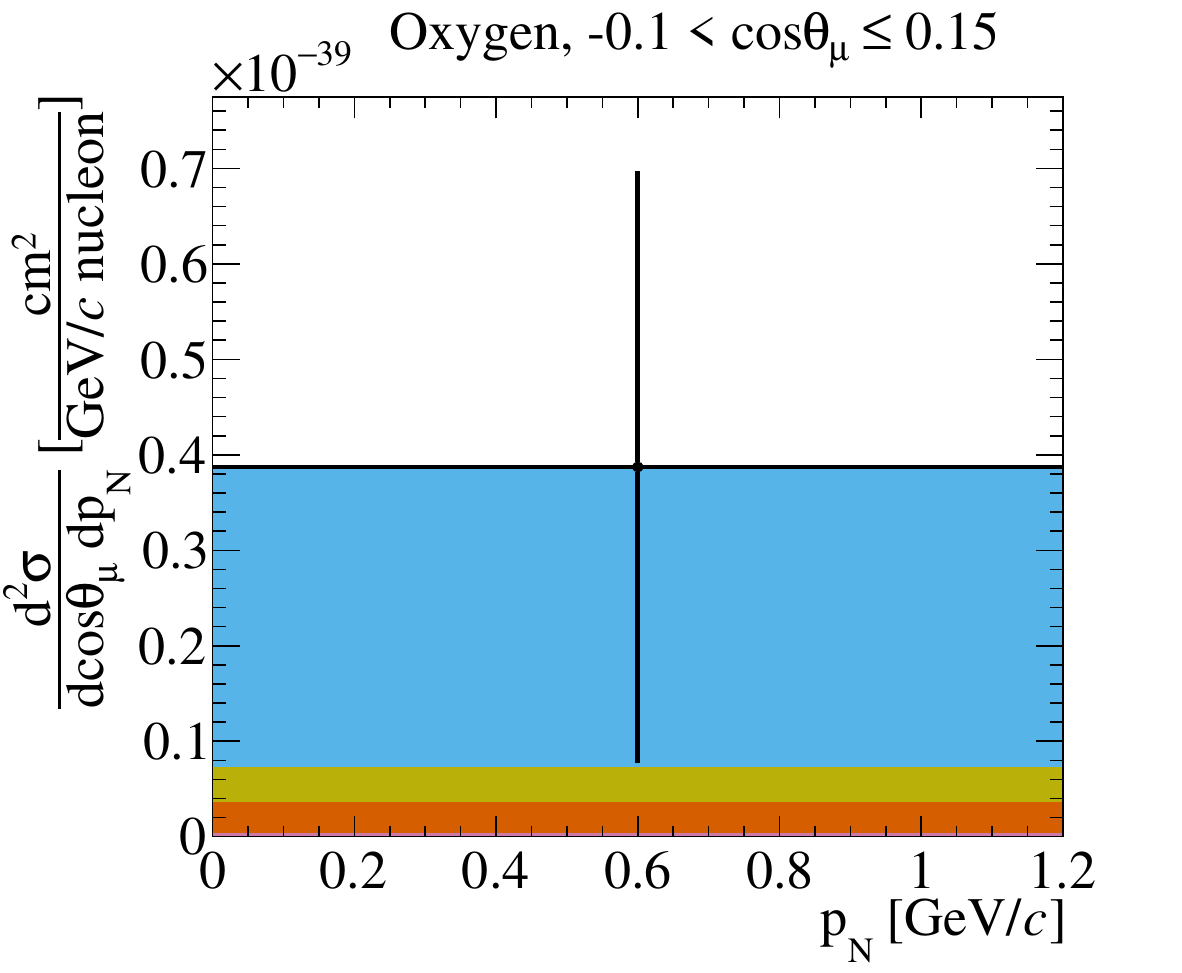}
\includegraphics[width=0.32\linewidth]{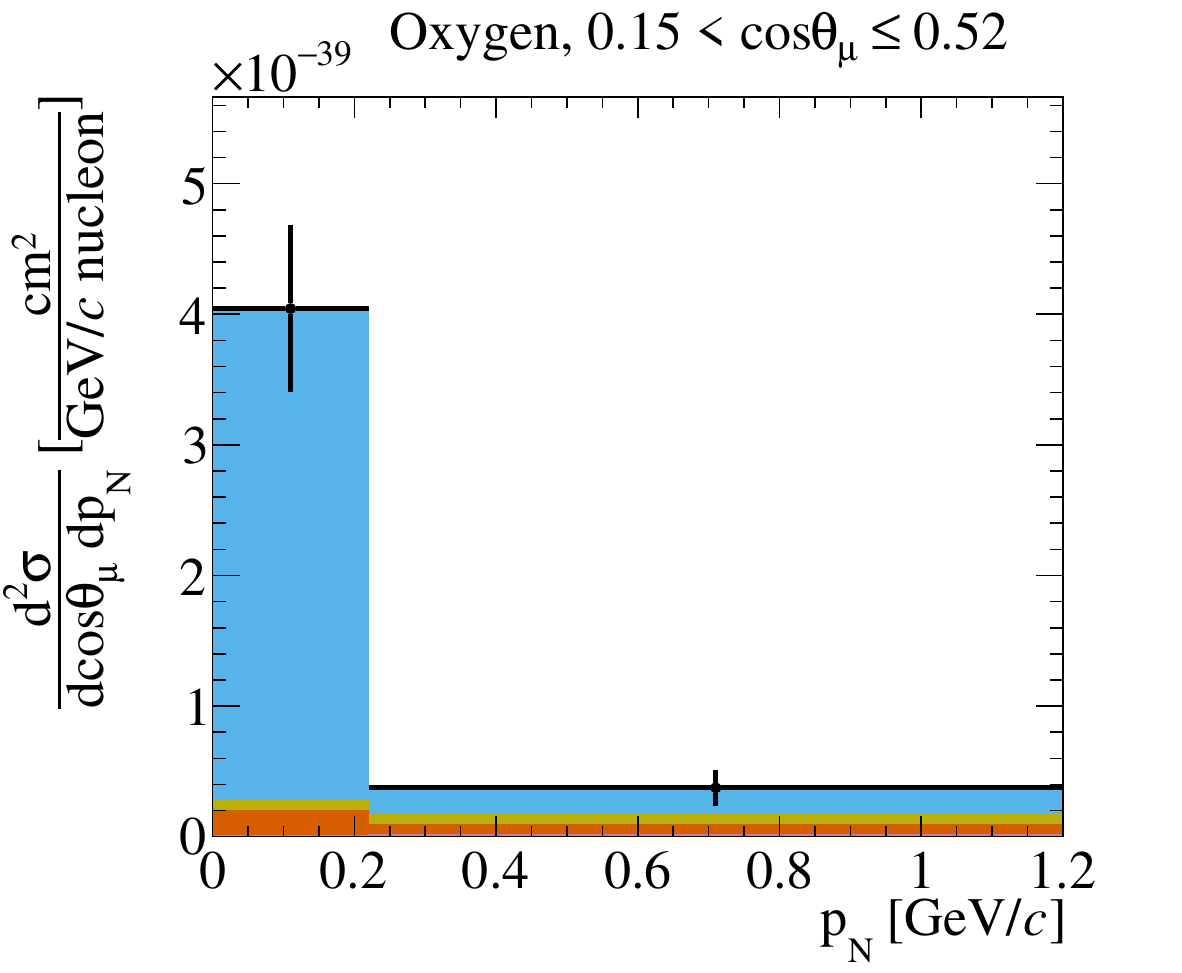} \\
\includegraphics[width=0.32\linewidth]{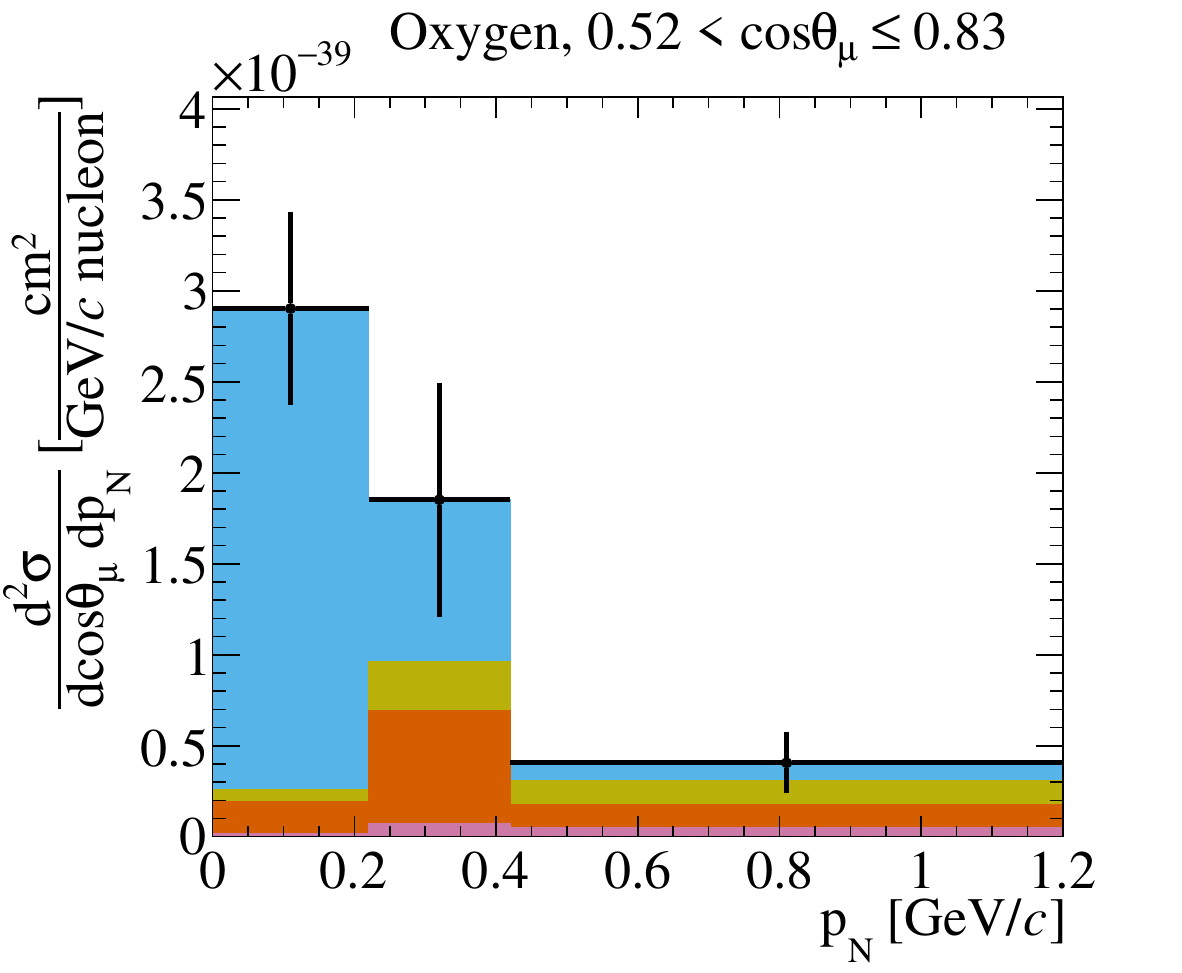}
\includegraphics[width=0.32\linewidth]{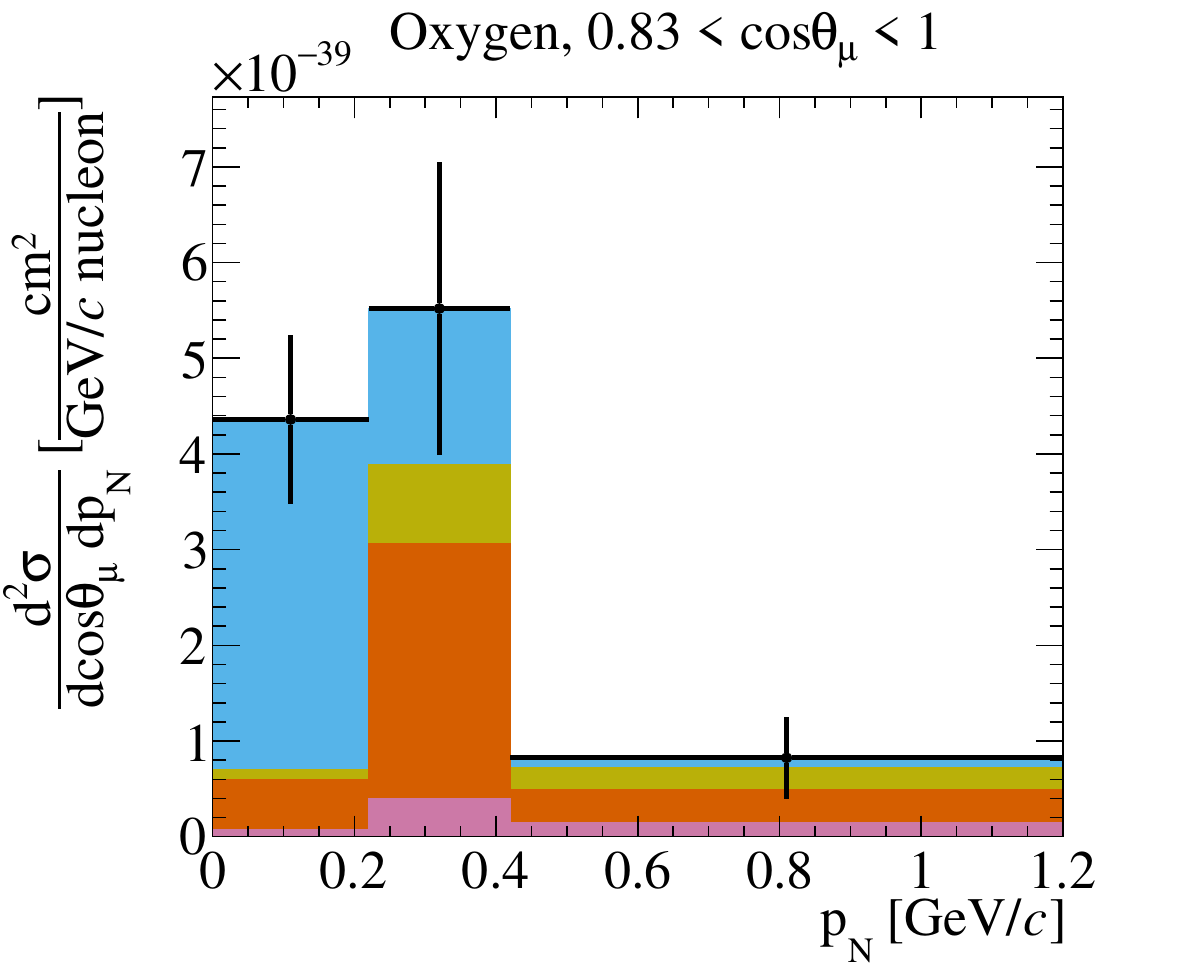}
\end{tabular}
}
\caption{Measurements of the $\nu_{\mu}$ CC$0\pi Np$ signal cross section on carbon (a) and oxygen (b) targets from the T2K data fit in the $p_{N}$--$\cos\theta_{\mu}$ space. The post-fit MC template is decomposed into different interaction types, including CCQE, 2p2h, CCRES and other (all remaining interaction types). The error bar represents the combined statistical and systematic uncertainty.}
\label{fig:realdata_xsec_2}
\end{figure*}

\begin{figure*}
\centering
\subfloat[Carbon]{
\begin{tabular}{c}
\includegraphics[width=0.32\linewidth]{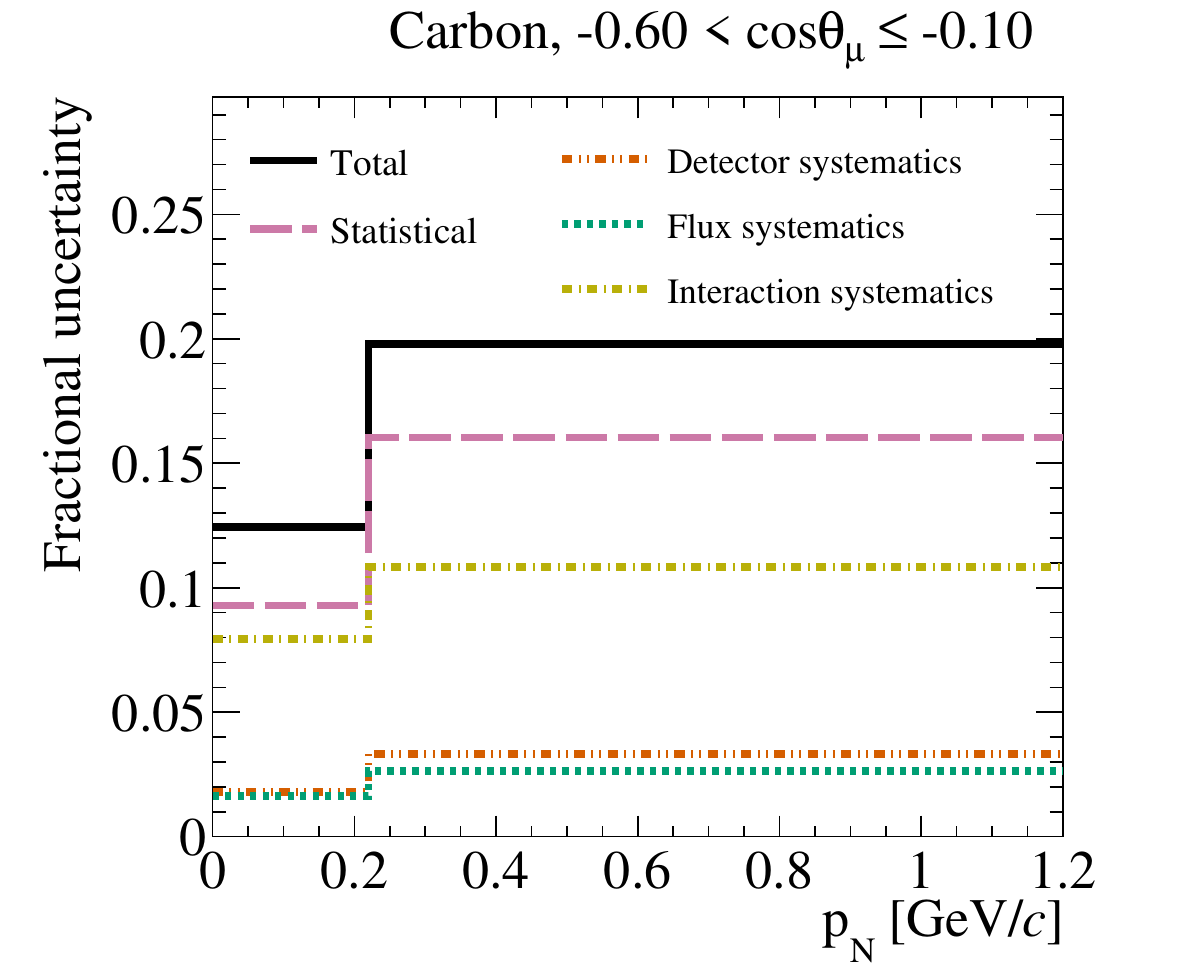}
\includegraphics[width=0.32\linewidth]{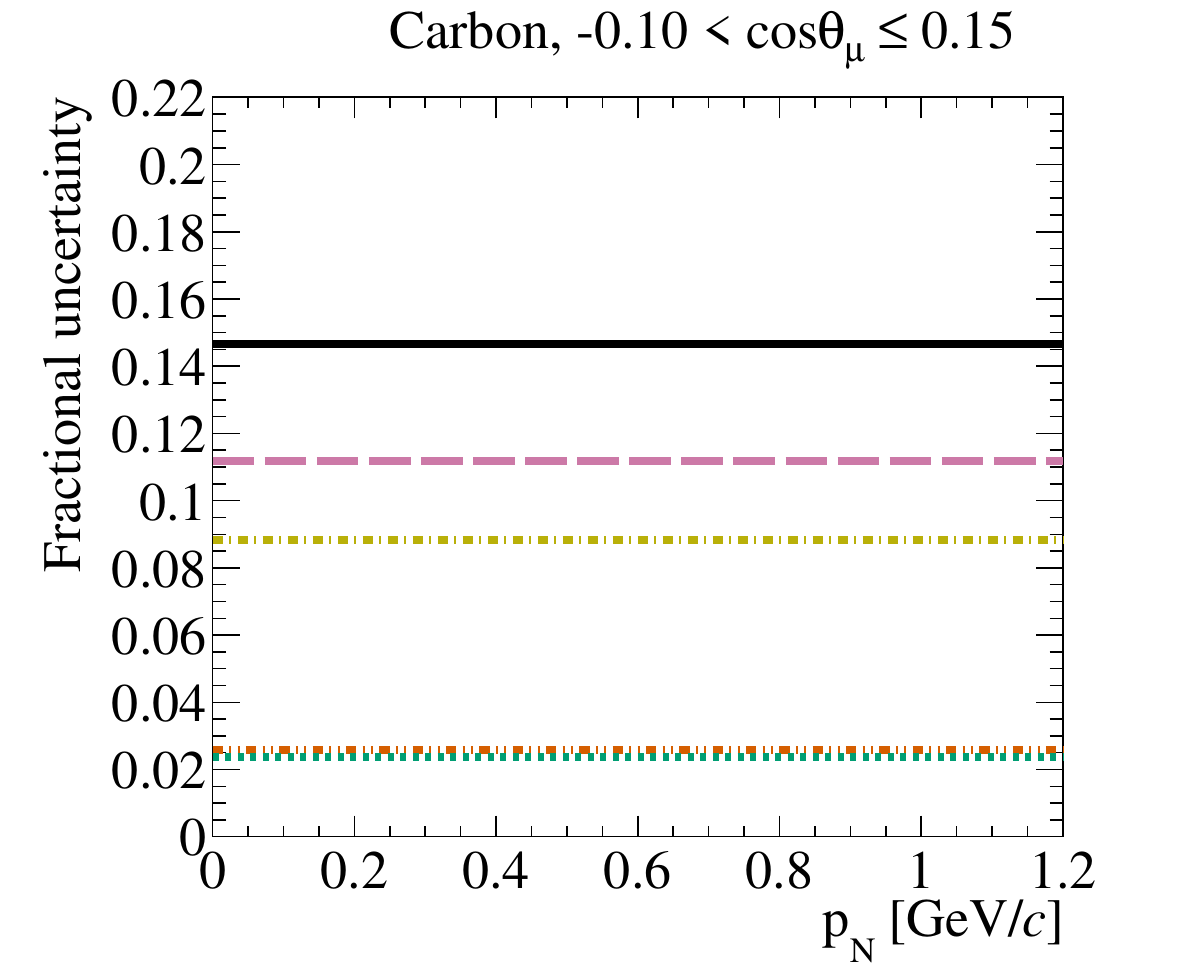}
\includegraphics[width=0.32\linewidth]{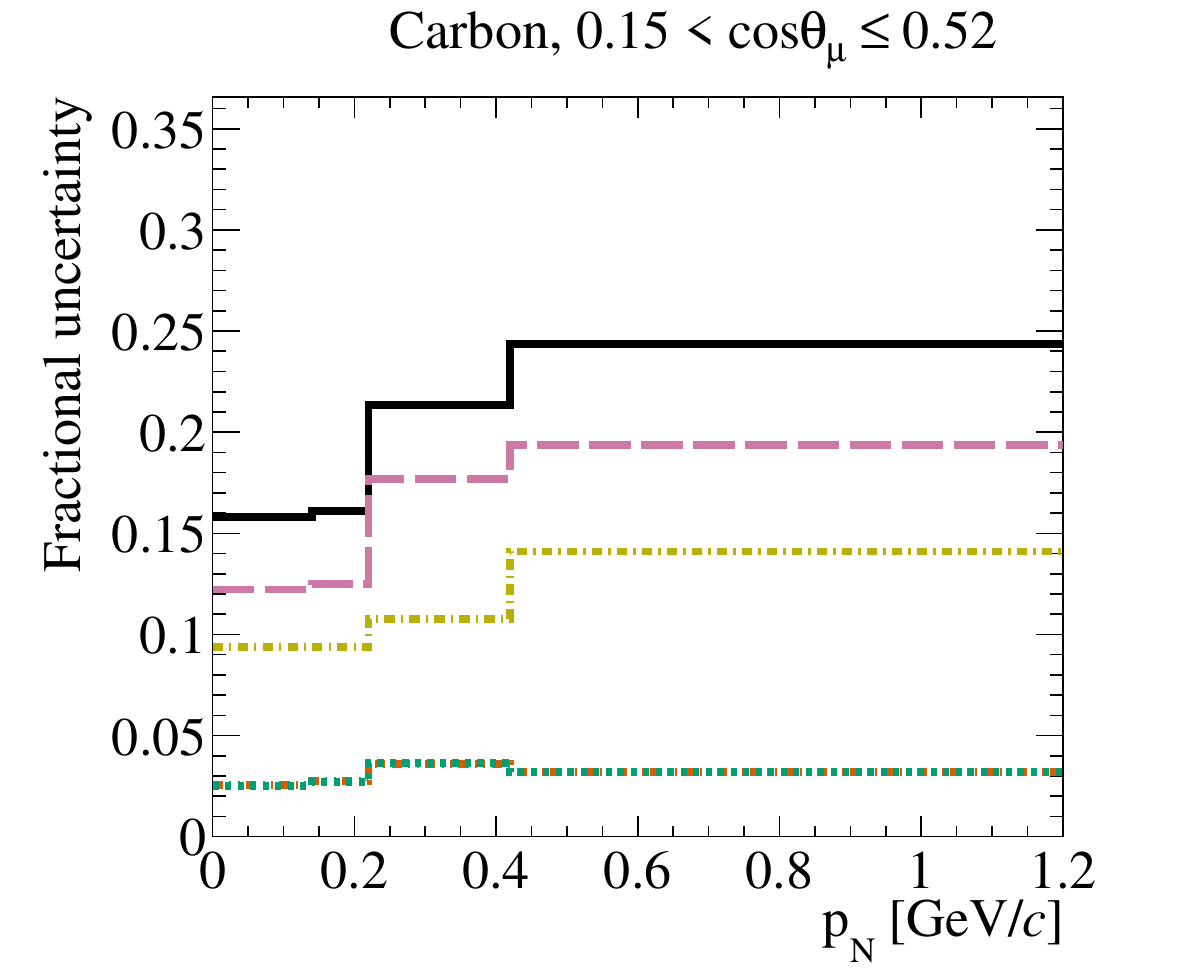} \\
\includegraphics[width=0.32\linewidth]{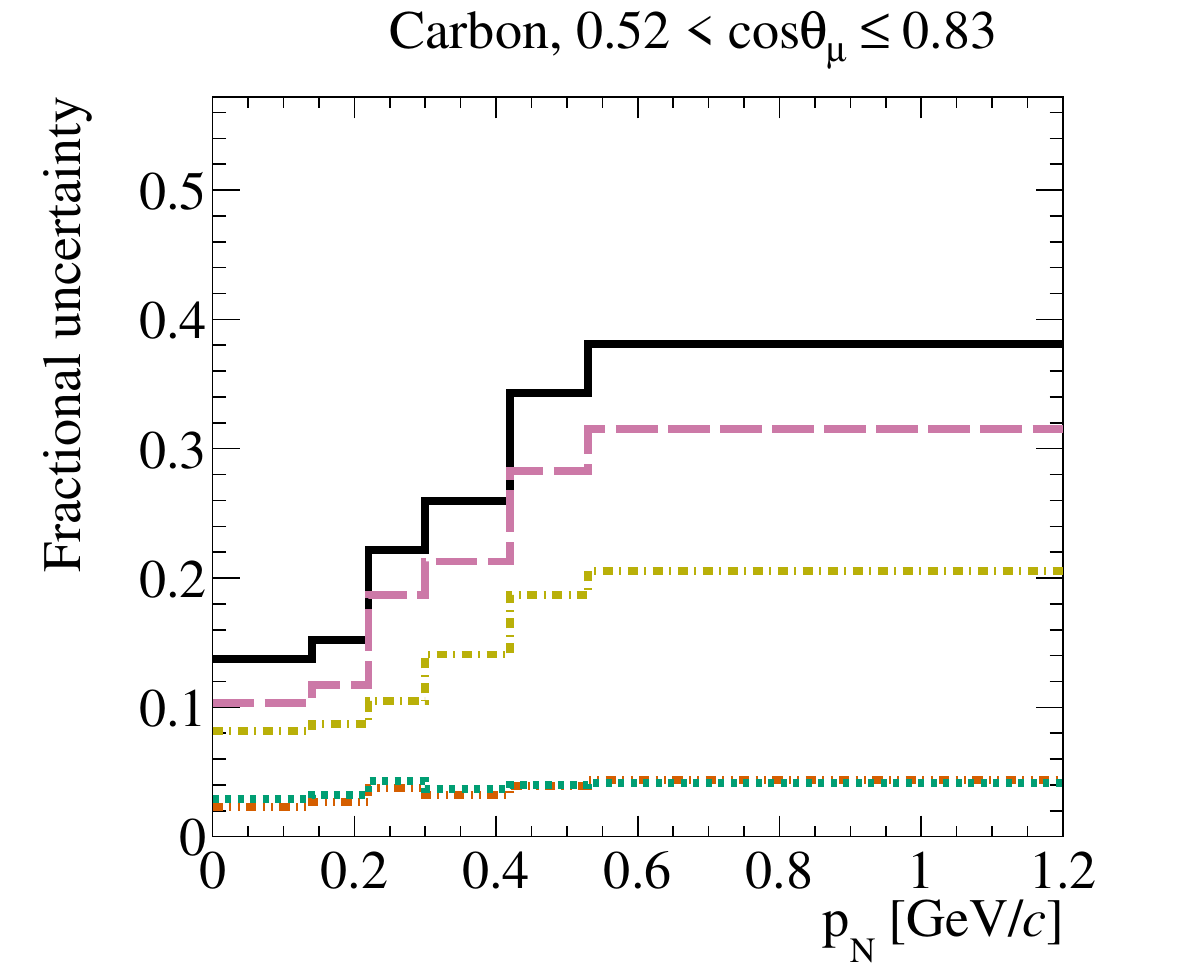}
\includegraphics[width=0.32\linewidth]{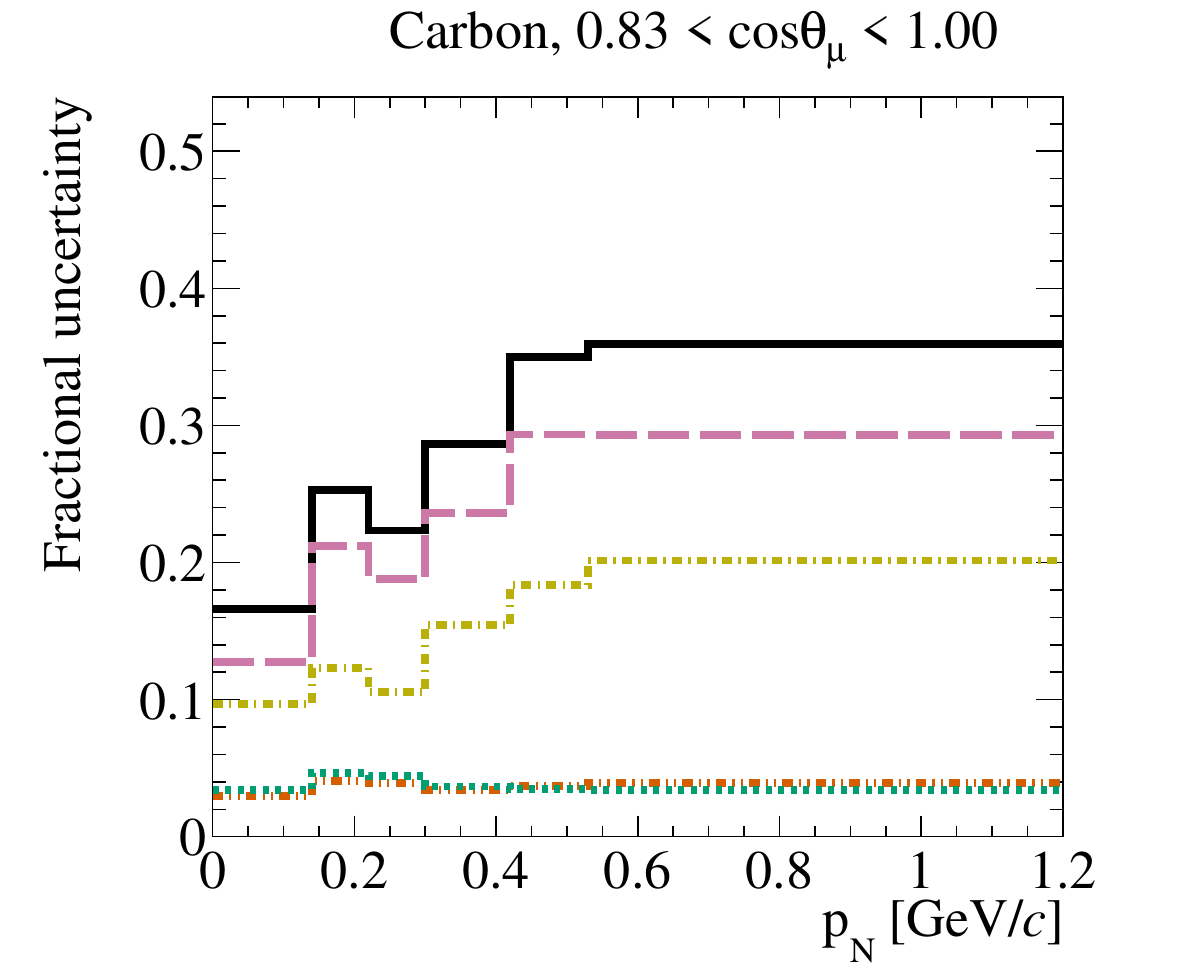}
\end{tabular}
}
\\
\subfloat[Oxygen]{
\begin{tabular}{c}
\includegraphics[width=0.32\linewidth]{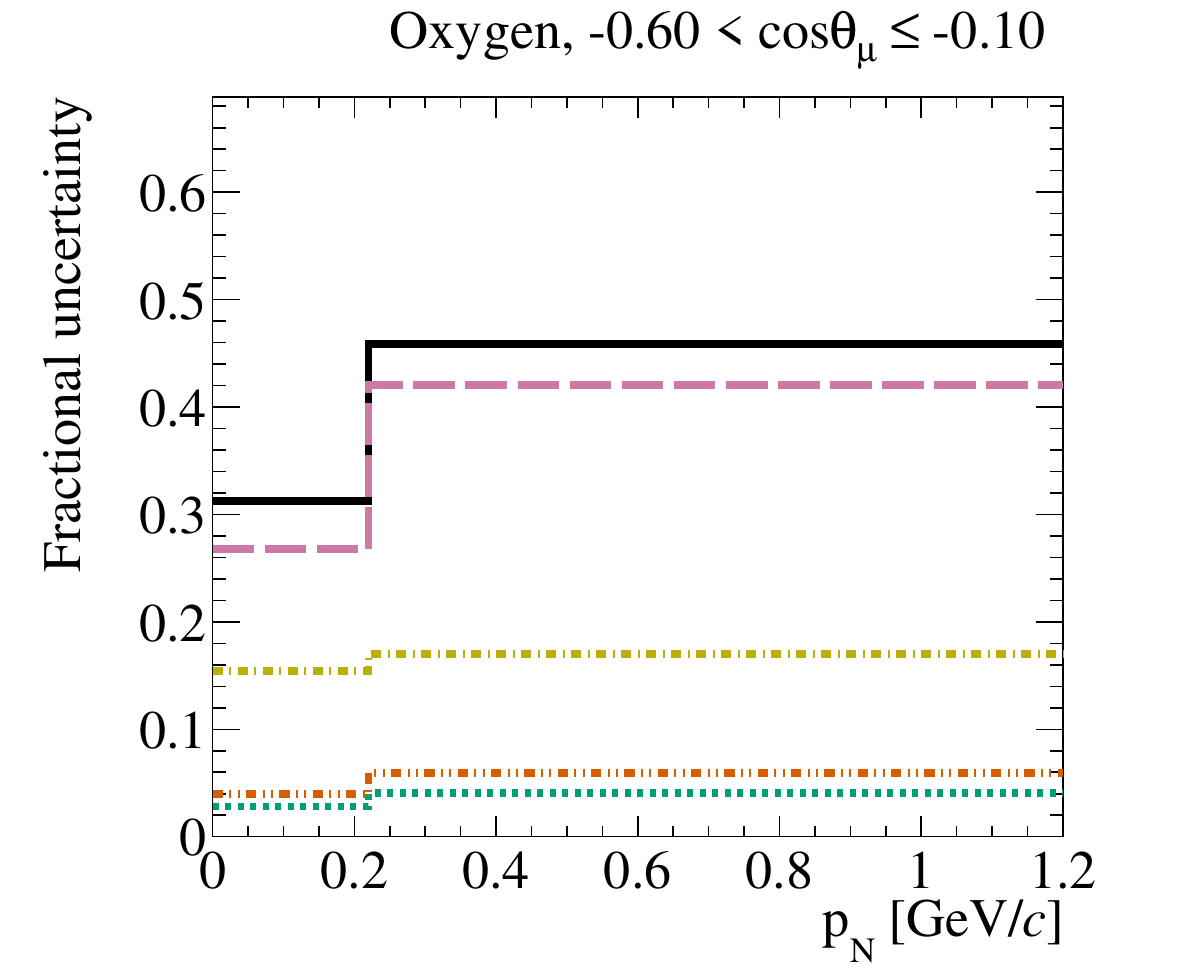}
\includegraphics[width=0.32\linewidth]{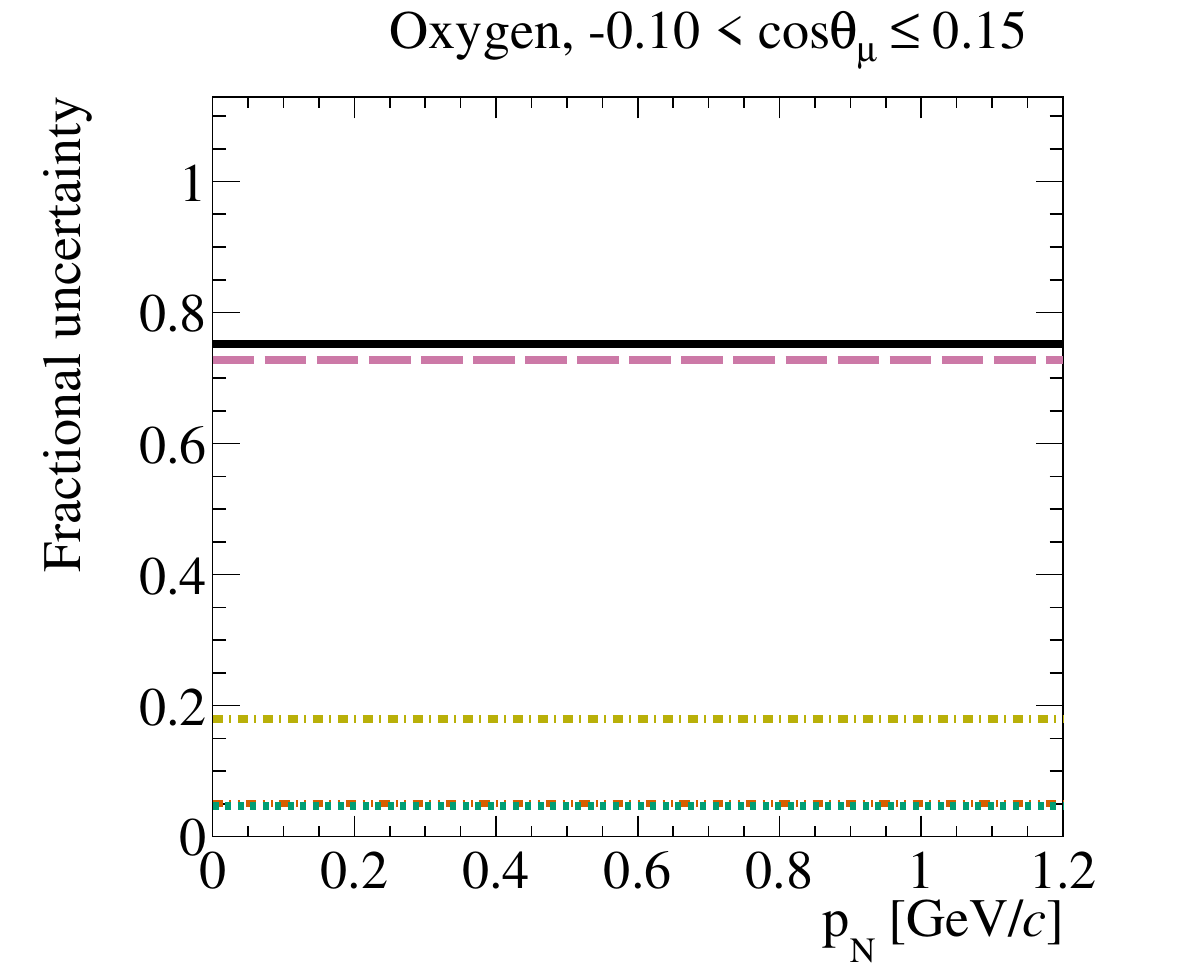}
\includegraphics[width=0.32\linewidth]{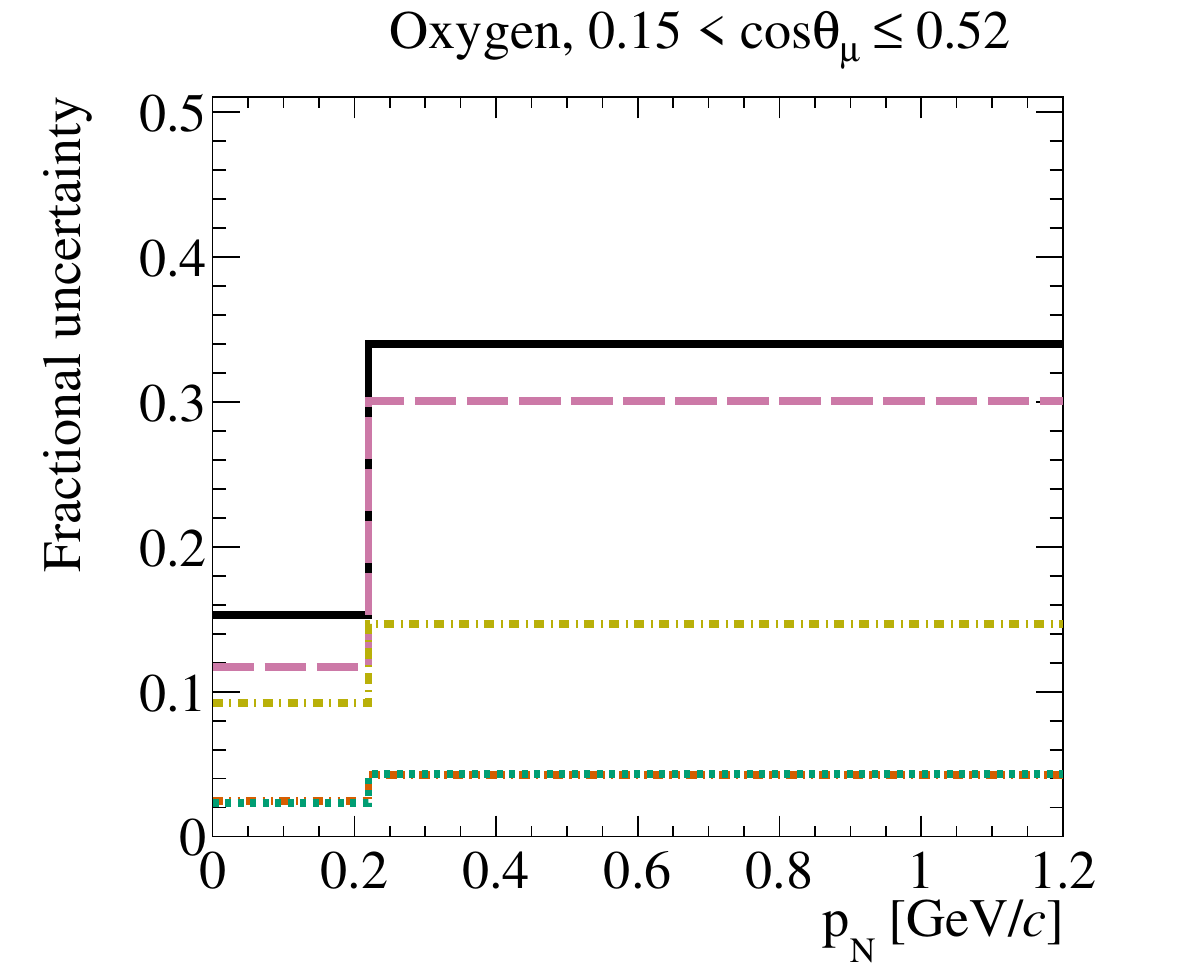} \\
\includegraphics[width=0.32\linewidth]{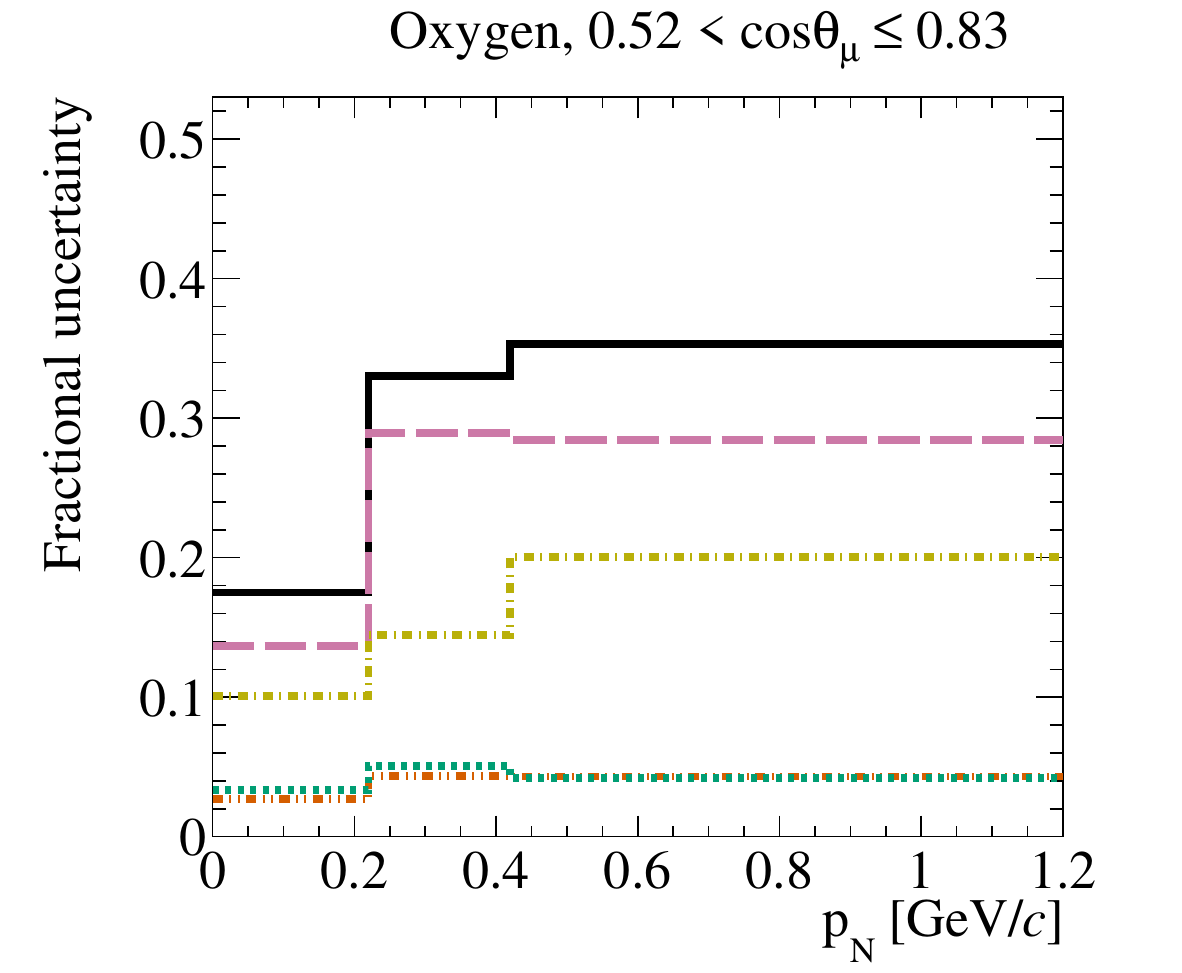}
\includegraphics[width=0.32\linewidth]{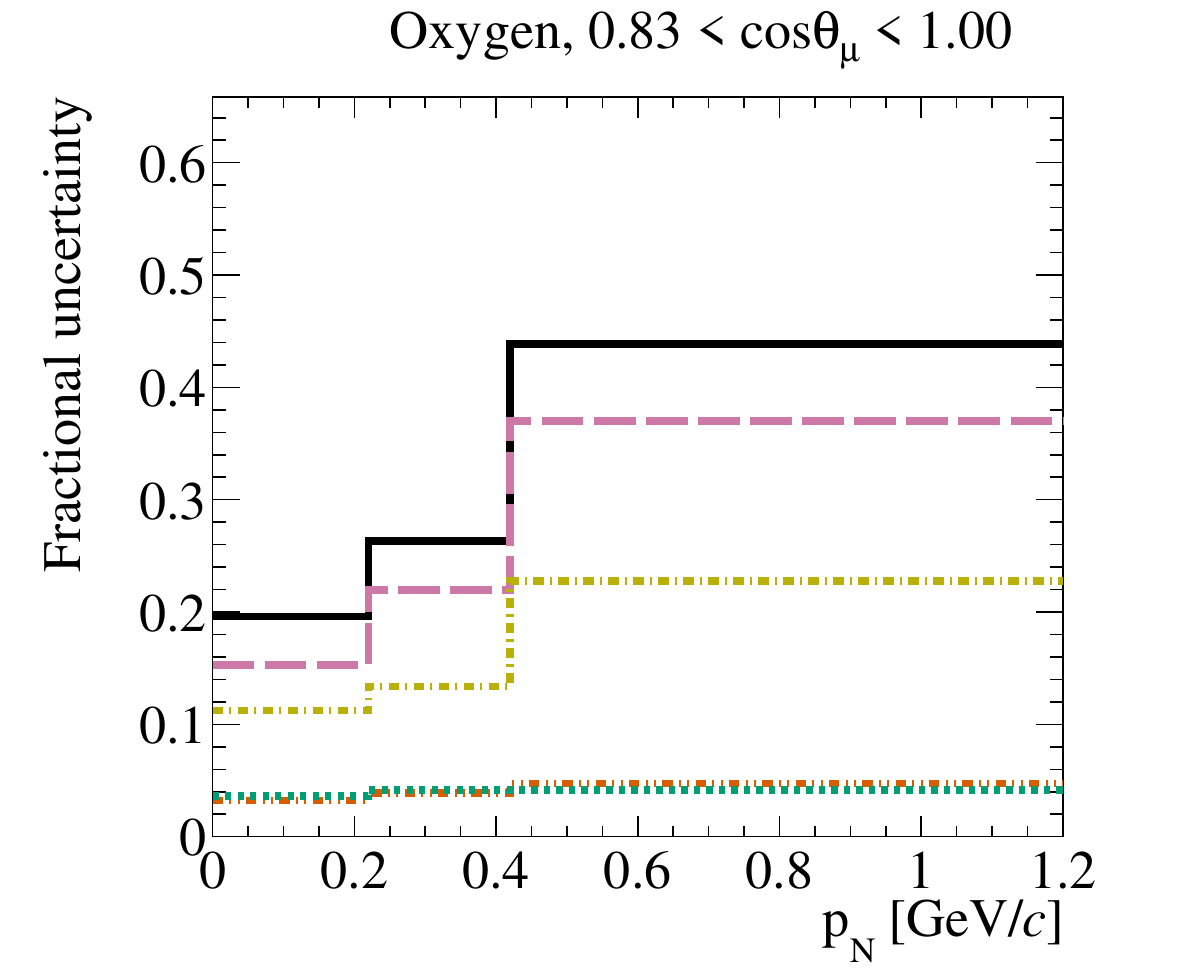}
\end{tabular}
}
\caption{Measurement uncertainties for the $\nu_{\mu}$ CC$0\pi Np$ cross sections on carbon (a) and oxygen (b) targets, derived from the fit to T2K data in $p_{N}$--$\cos\theta_{\mu}$ space. In addition to the total uncertainty (black), contributions from individual sources are shown: statistics (pink) and various systematics, including detector (orange), neutrino flux (bluish green), and neutrino interaction (yellow). The systematic uncertainty from number of target nucleons is omitted as it contributes less than 1\% for both targets. The fractional uncertainty in each bin is defined as the ratio of the absolute uncertainty to the corresponding cross-section value.}
\label{fig:realdata_xsec_err_2}
\end{figure*}

The template fit has post-fit $\chi^{2} = 114.95$ corresponding to a $p$-value around 0.122 ($N_{dof} = 98$), which is above the threshold of 0.05. Fig.~\ref{fig:realdata_xsec_2} presents the extracted $\nu_{\mu}$ CC$0\pi Np$ signal cross-section results on the carbon and oxygen targets. The measurement uncertainties are shown in Fig.~\ref{fig:realdata_xsec_err_2}. The overall measurement uncertainties range from 15\% to 80\% across different cross-section bins, where the statistical uncertainty is dominant in all bins. Concerning the systematic uncertainties, similar to the $\delta p_{T}$--$\delta \alpha_{T}$ results, the modeling of neutrino-nucleus interactions dominates the contribution.

\section{Comparisons with Models}
\label{sec:model_comp}

In this section, we compare the extracted cross sections with a range of predictions based on different nuclear effect models. We focus on two primary categories of nuclear modeling: Sec.~\ref{subsec:modelcomp_ngs_mni} presents the comparison across various models describing CCQE interactions from different neutrino event generators and Sec.~\ref{subsec:modelcomp_FSI} focuses on the comparison of different FSI models. The models presented here are complementary to those in Ref.~\cite{T2K:2026rru}, which focuses more on general model predictions, whereas this paper presents models focusing on specific interaction modes.

\subsection{Comparisons with Generator Models}
\label{subsec:modelcomp_ngs_mni}

We consider five model predictions generated using NEUT v6.1.0~\cite{Hayato:2021heg} and GENIE v3.6.2~\cite{GENIE:2021npt}. Within NEUT, three configurations are examined: the relativistic mean-field model with an energy-dependent potential (ED-RMF)~\cite{McKean:2025khb}, the SF model~\cite{Benhar:1994hw} featuring an updated proton SF table for carbon~\cite{Ankowski:2024ntv}, and the LFG model~\cite{Nieves:2011pp,Bourguille:2020bvw}. The NEUT SF prediction generated here uses the same configuration as that presented in Ref.~\cite{T2K:2026rru}. While the SF and LFG models primarily describe the nucleon ground state, the ED-RMF model applies relativistic mean field (RMF) theory~\cite{Serot:1984ey} to the ground state and utilizes the relativistic distorted wave impulse approximation (RDWIA) to account for elastic FSI; inelastic FSI remains described by the INC model~\cite{McKean:2025khb}. This treatment of FSI is different from the traditional approach used in the SF and LFG configurations, which apply an INC model based on the plane-wave impulse approximation (PWIA). For multi-nucleon interactions, 2p2h processes are described using the Valencia approach~\cite{Nieves:2011pp,Gran:2013kda}, with three-particle three-hole (3p3h) contributions included as well~\cite{Prasad:2024gnv}. 

In GENIE, we consider two models based on the G21\_11b comprehensive model configuration (CMC). Both models employ the same LFG description for the ground state but differ in their treatment of the inclusive CCQE cross section: one utilizes the continuum random phase approximation (CRPA)~\cite{Kolbe:1999au,Jachowicz:2002rr}, while the other adopts the SuSAv2 model~\cite{Gonzalez-Jimenez:2014eqa,Dolan:2019bxf}. Additionally, both configurations use the SuSAv2 framework for 2p2h interactions and the hN2018 model~\cite{Dytman:2021ohr} for FSI. Hereafter, these models are denoted as ``GENIE CRPA'' and ``GENIE SuSAv2,'' respectively. We note that both models were implemented in GENIE following a similar approach (described in \cite{Dolan:2019bxf}) which uses pre-computed hadron tensors to predict the inclusive cross section, but sample the nucleon kinematics from an LFG ground state. As a result, the SuSAv2 and CRPA model reproduce the input theory calculations for the CCQE cross section as a function of lepton kinematics, but the correlations between lepton and hadron kinematics will be indicative of a LFG model.

\begin{figure*}
\centering
\subfloat[Carbon]
{\includegraphics[width=0.33\linewidth]{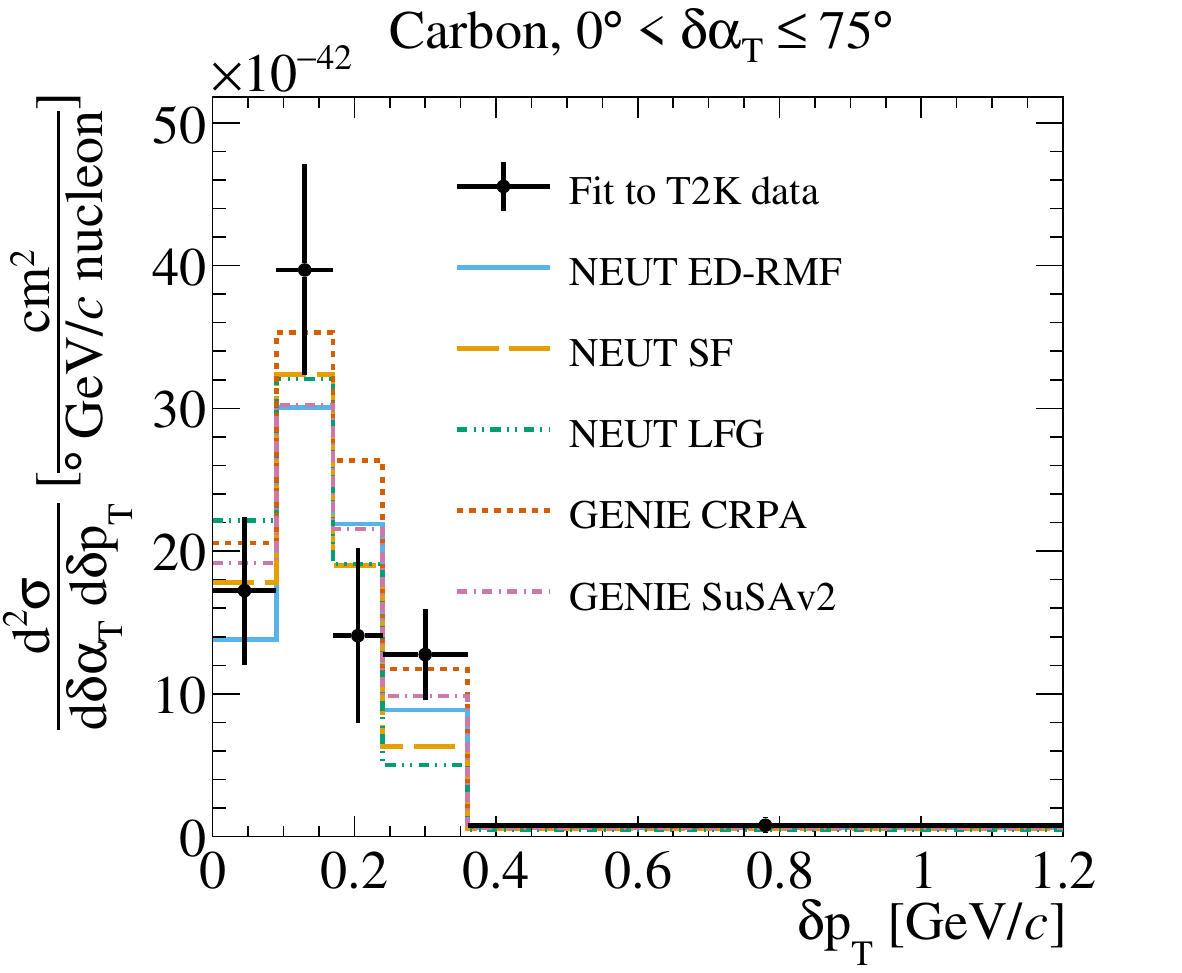}
\includegraphics[width=0.33\linewidth]{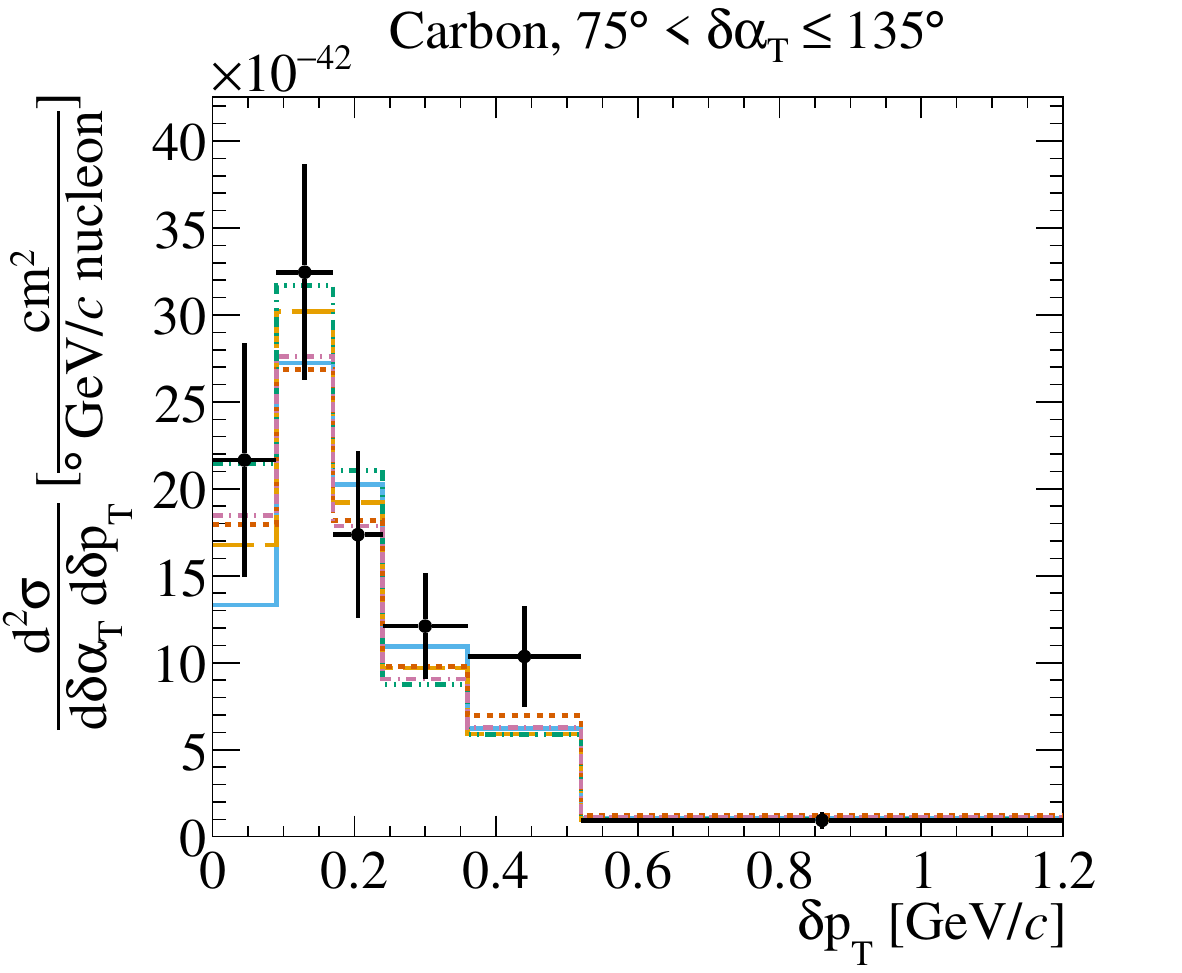}
\includegraphics[width=0.33\linewidth]{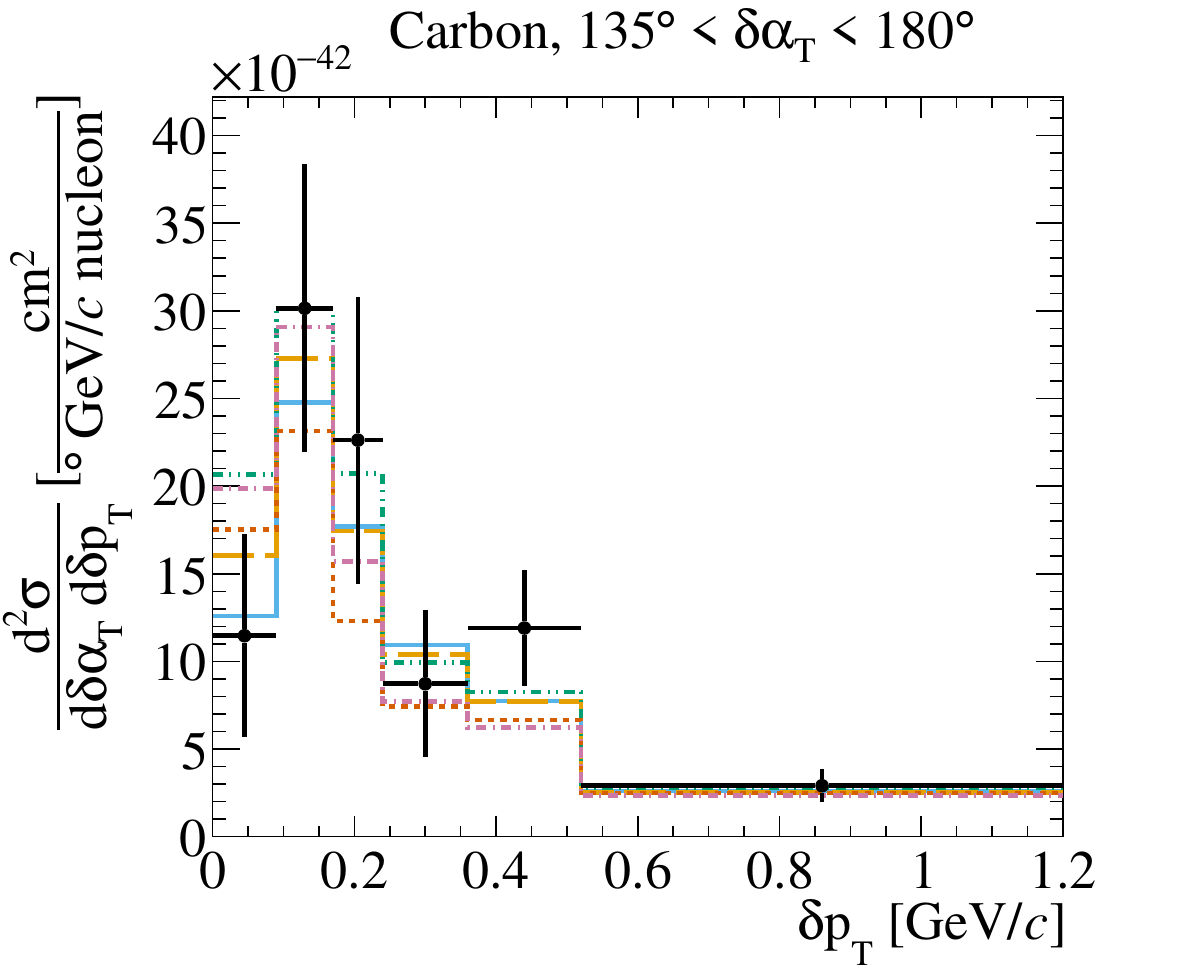}}
\\
\subfloat[Oxygen]
{\includegraphics[width=0.33\linewidth]{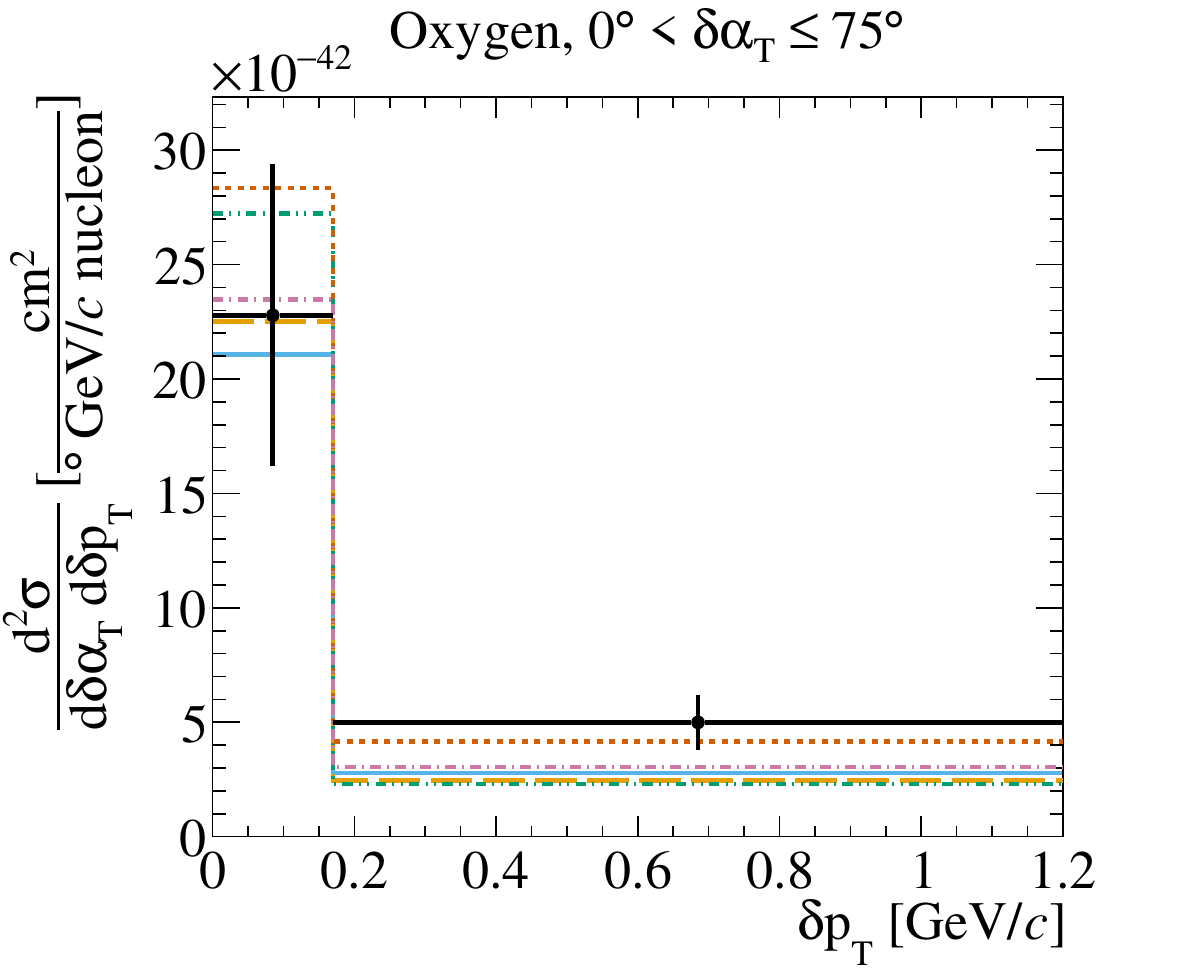}
\includegraphics[width=0.33\linewidth]{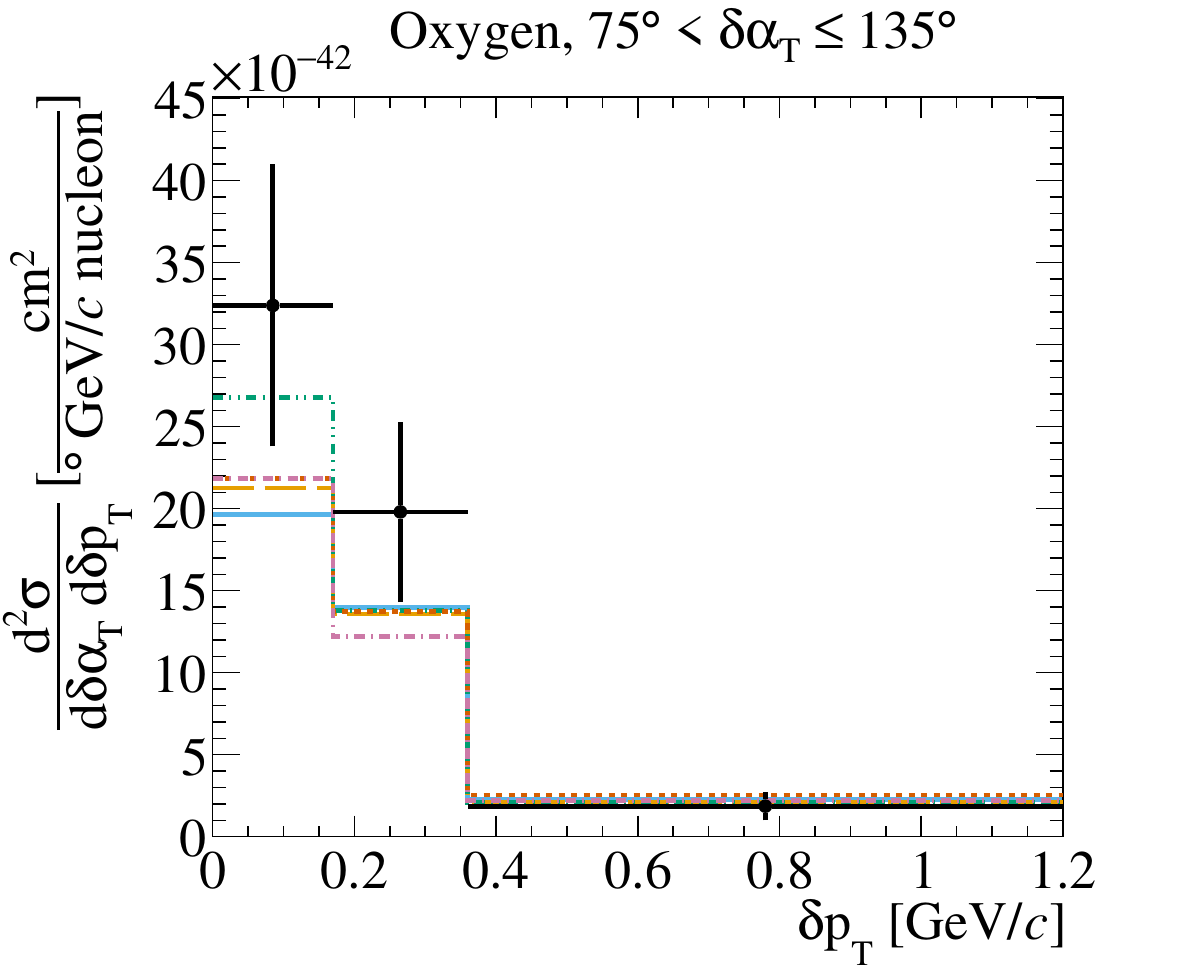}
\includegraphics[width=0.33\linewidth]{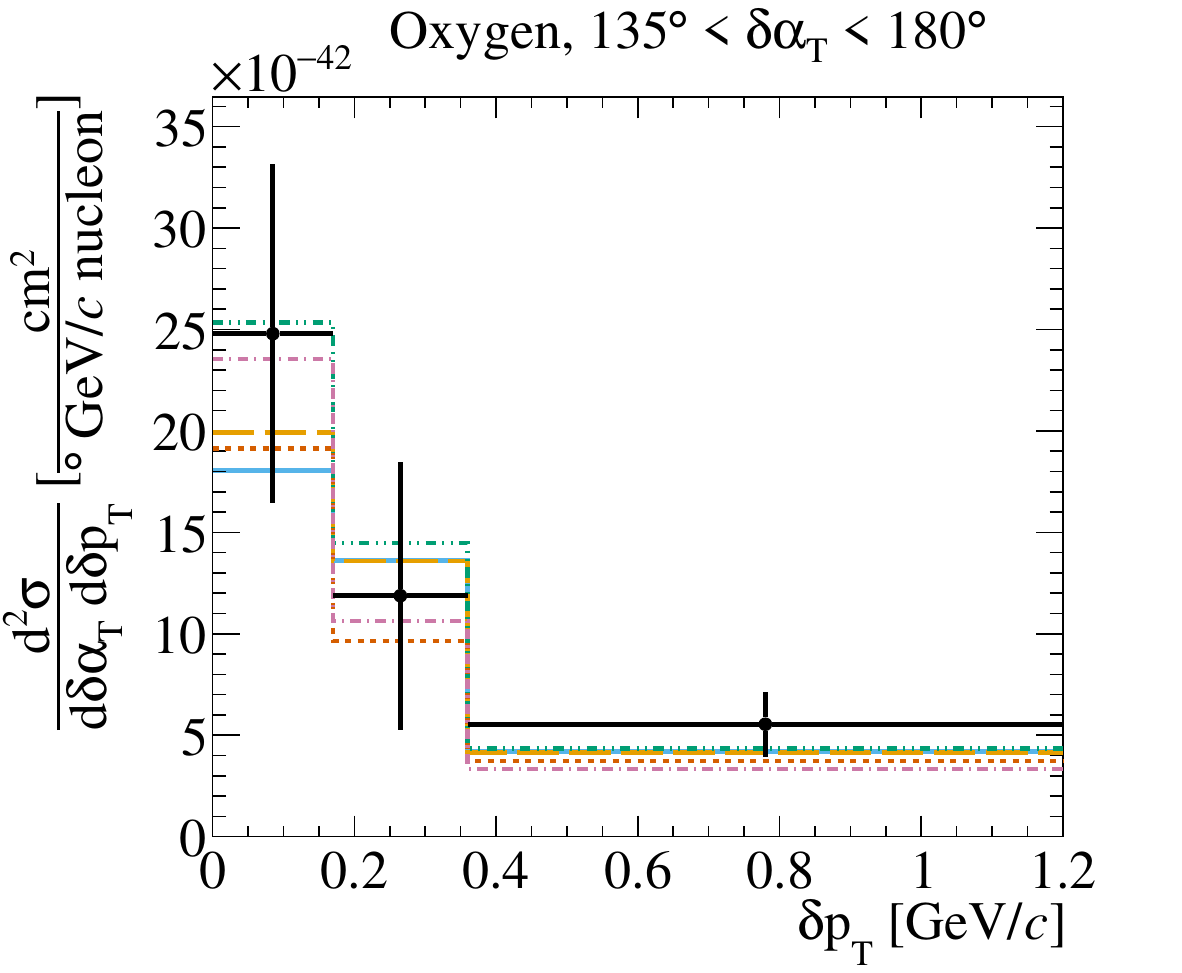}}
\caption{Measurements of the $\nu_{\mu}$ CC$0\pi Np$ double-differential cross sections on carbon (a) and oxygen (b) targets, obtained from the fit to T2K data in $\delta p_{T}$--$\delta \alpha_{T}$ space. Error bars represent the combined statistical and systematic uncertainties. The results are compared with various neutrino-nucleus interaction models, including NEUT ED-RMF (sky blue), NEUT SF (orange), NEUT LFG (bluish green), GENIE CRPA (vermillion) and GENIE SuSAv2 (reddish purple).}
\label{fig:modelcomp_plot_inistate_1}
\end{figure*}

\begin{figure*}
\centering
\subfloat[Carbon]{
\begin{tabular}{c}
\includegraphics[width=0.32\linewidth]{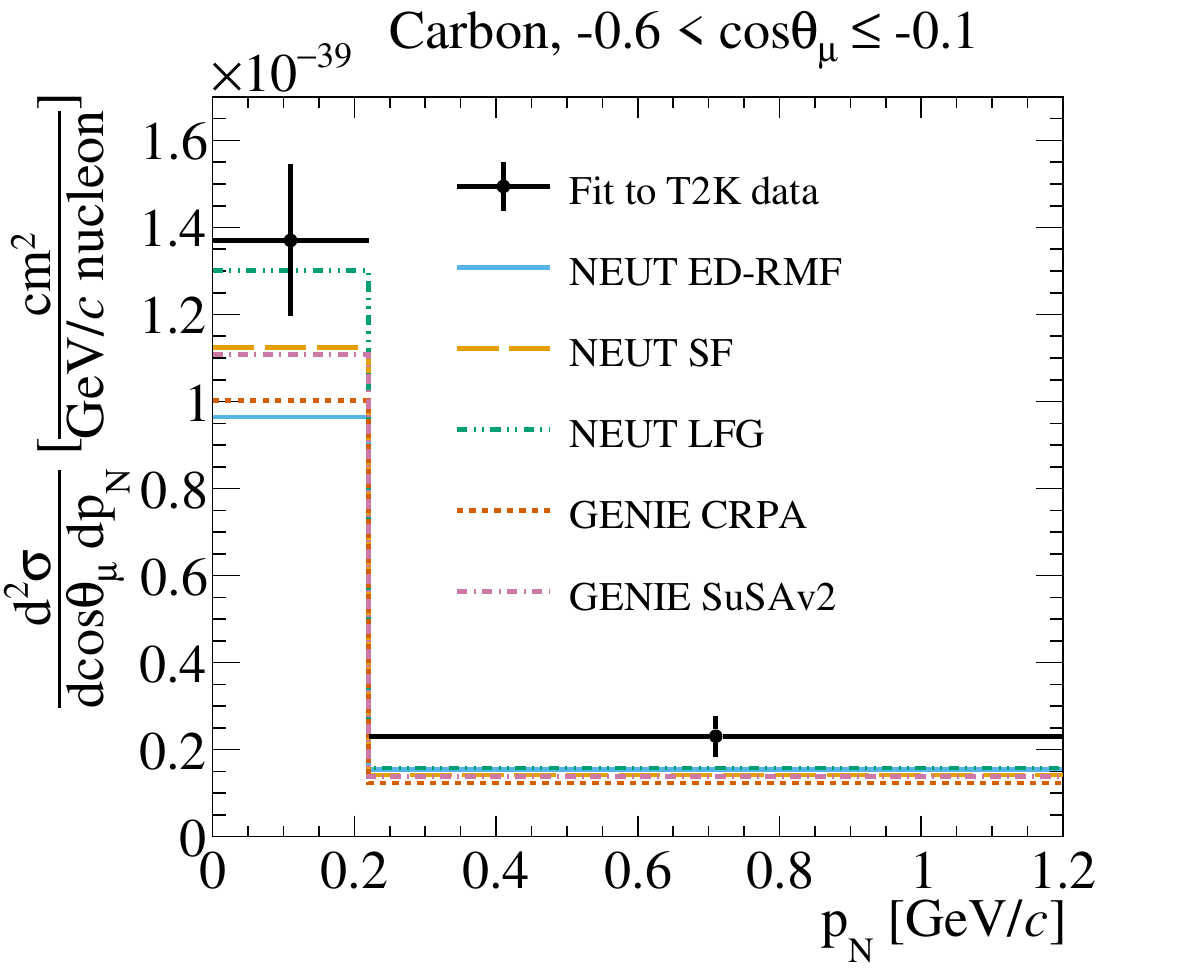}
\includegraphics[width=0.32\linewidth]{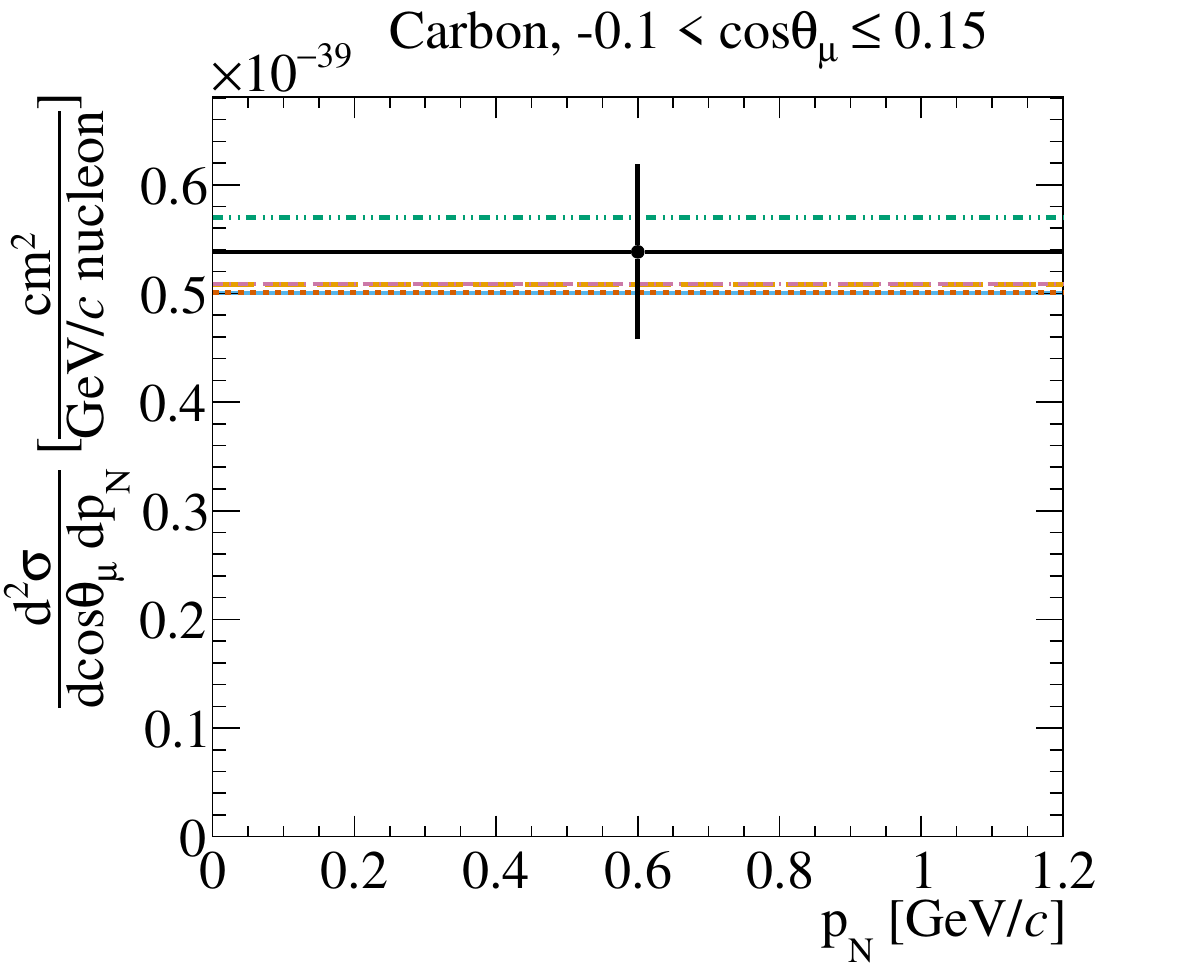}
\includegraphics[width=0.32\linewidth]{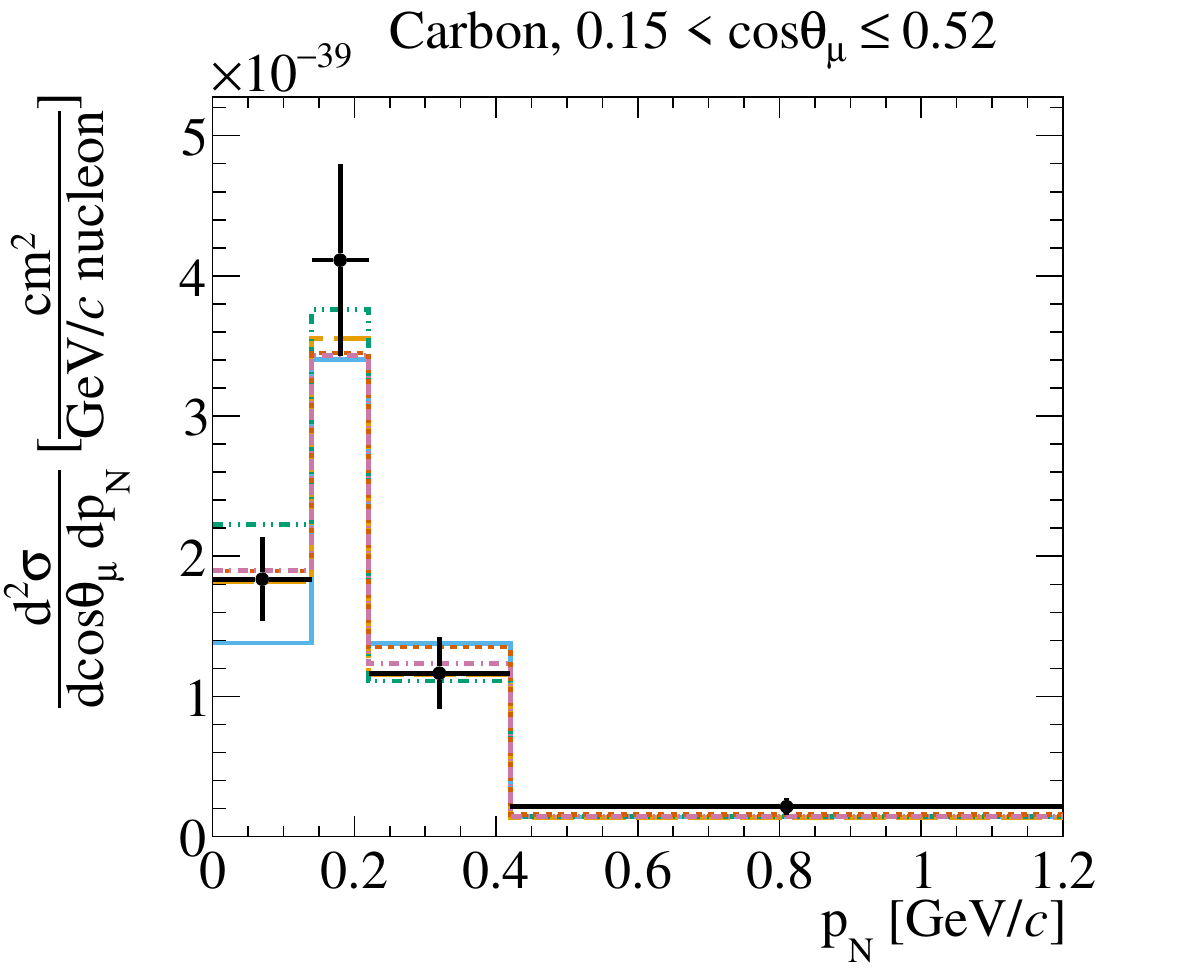} \\
\includegraphics[width=0.32\linewidth]{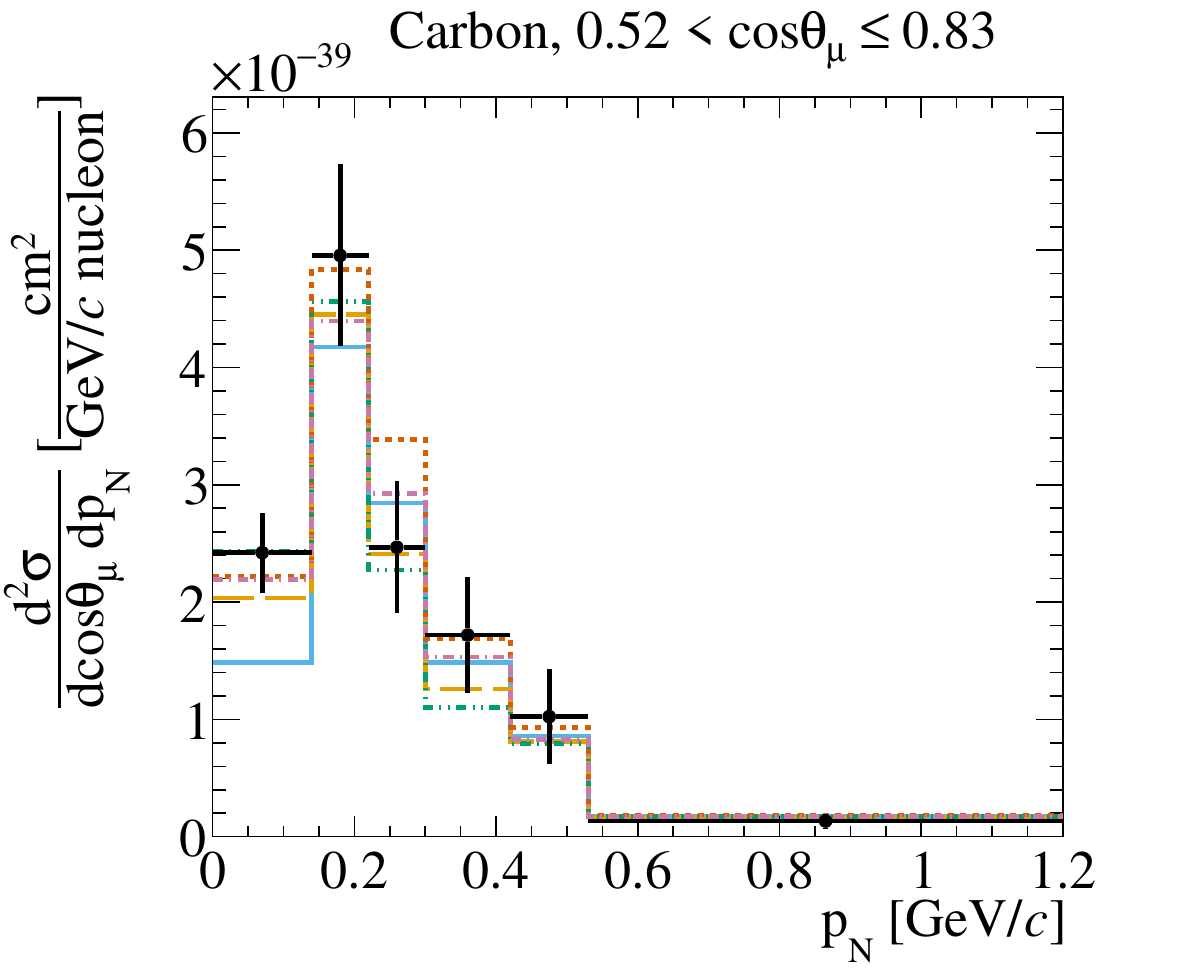}
\includegraphics[width=0.32\linewidth]{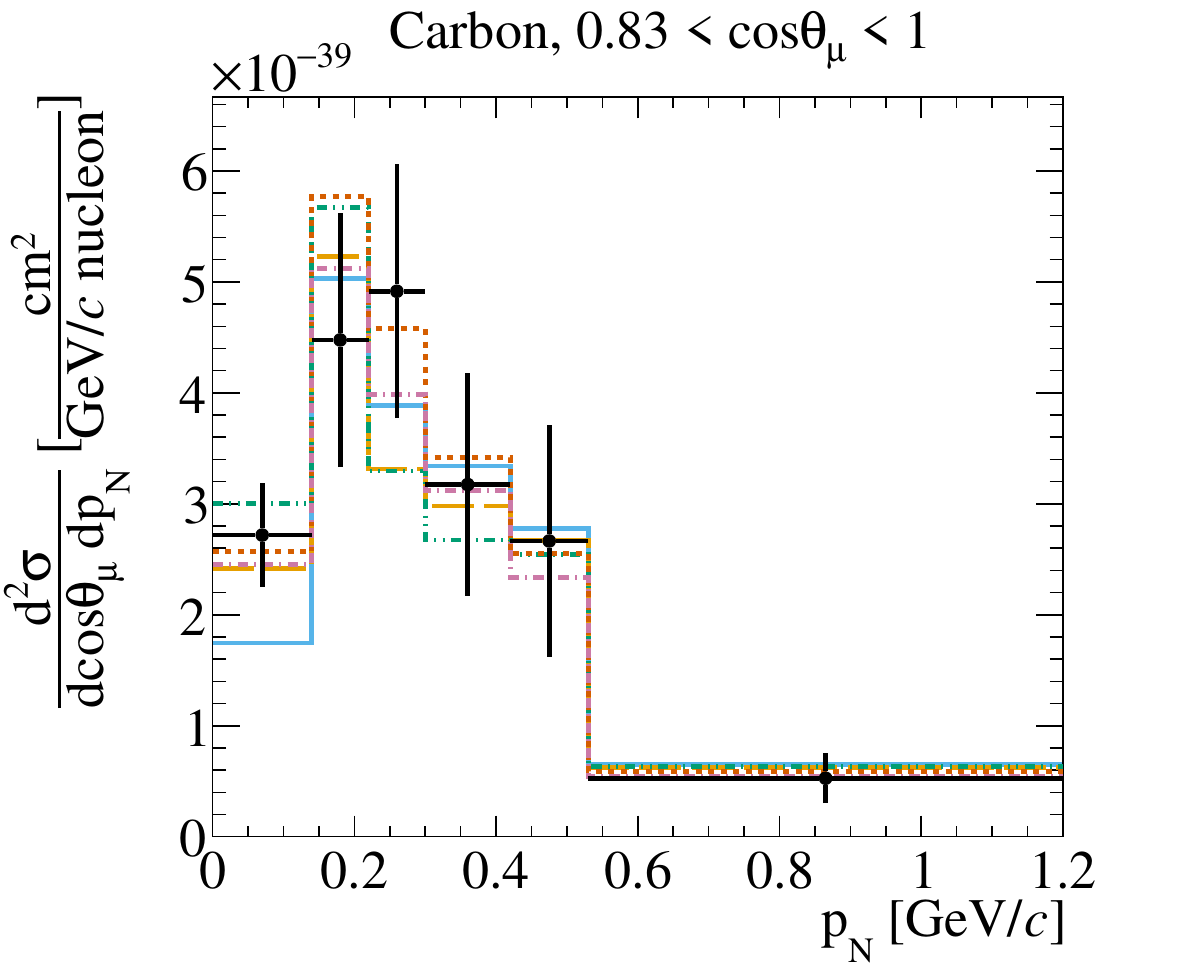}
\end{tabular}
}
\\
\subfloat[Oxygen]{
\begin{tabular}{c}
\includegraphics[width=0.32\linewidth]{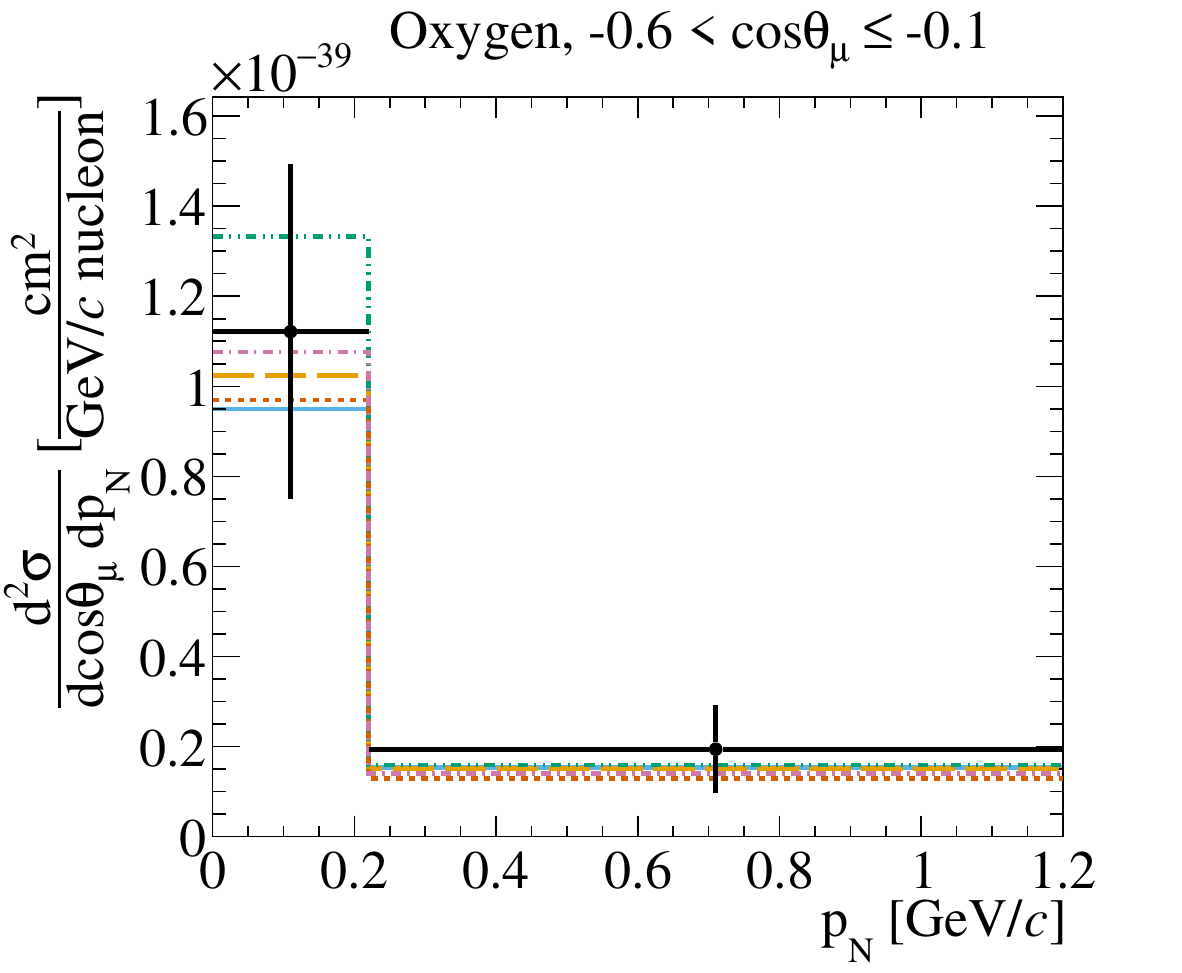}
\includegraphics[width=0.32\linewidth]{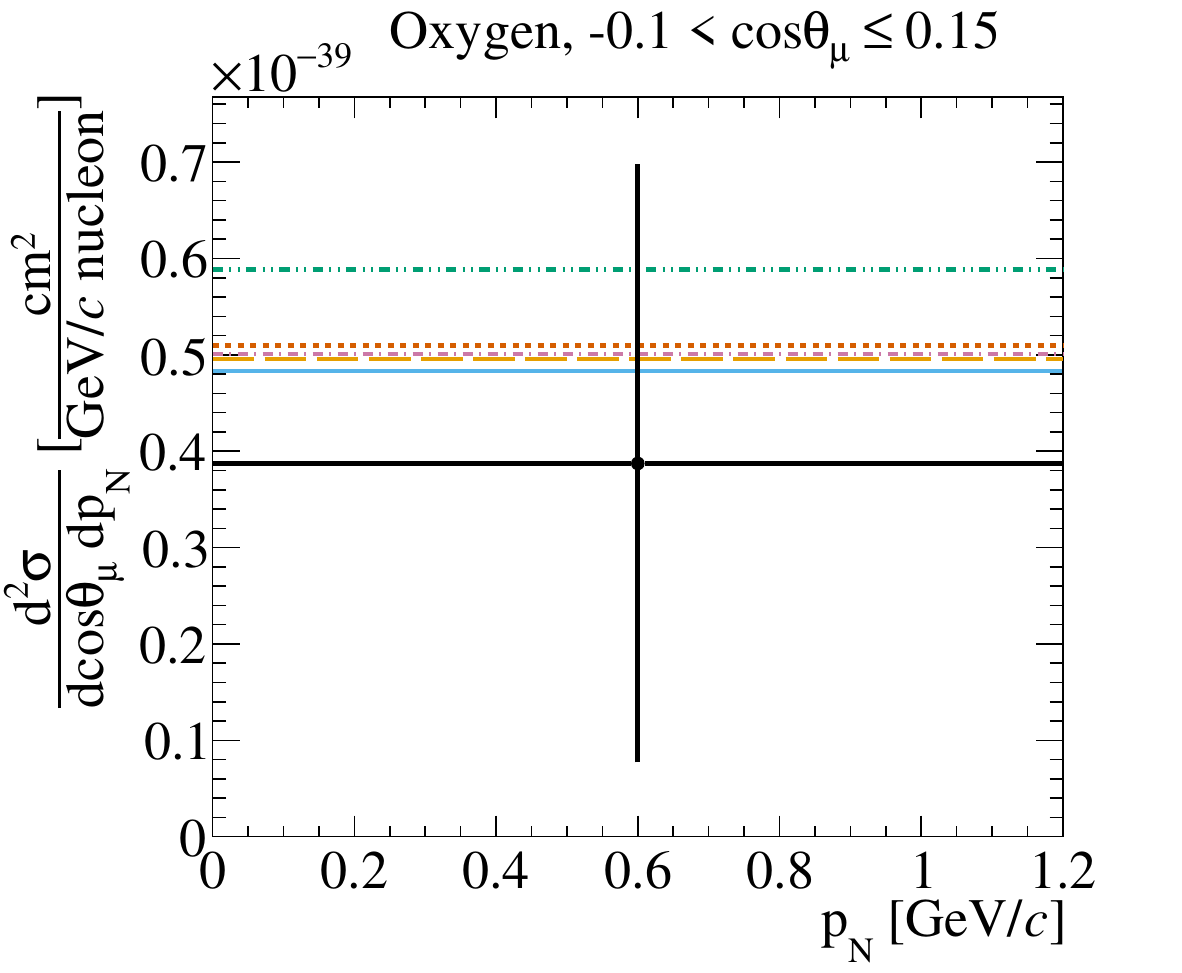}
\includegraphics[width=0.32\linewidth]{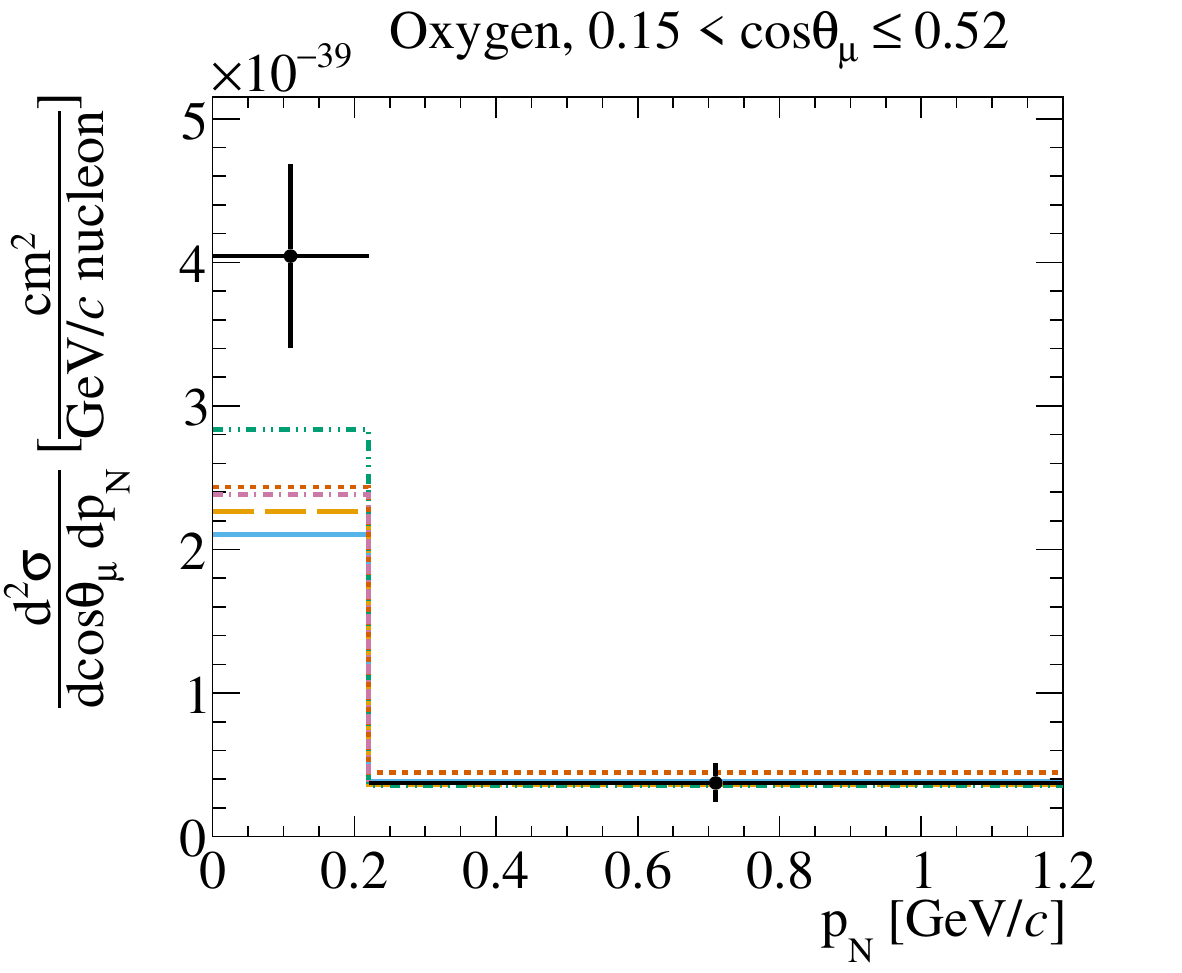} \\
\includegraphics[width=0.32\linewidth]{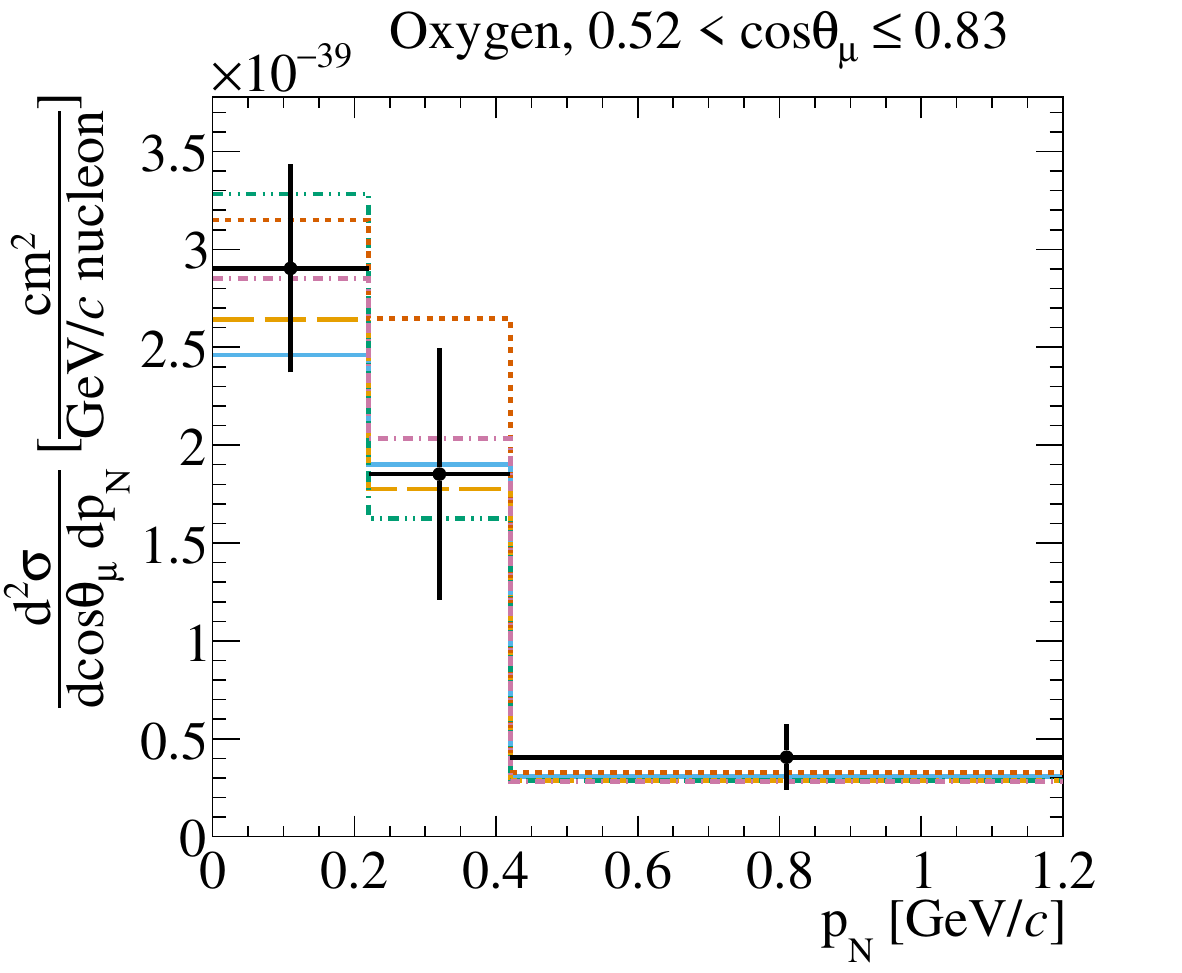}
\includegraphics[width=0.32\linewidth]{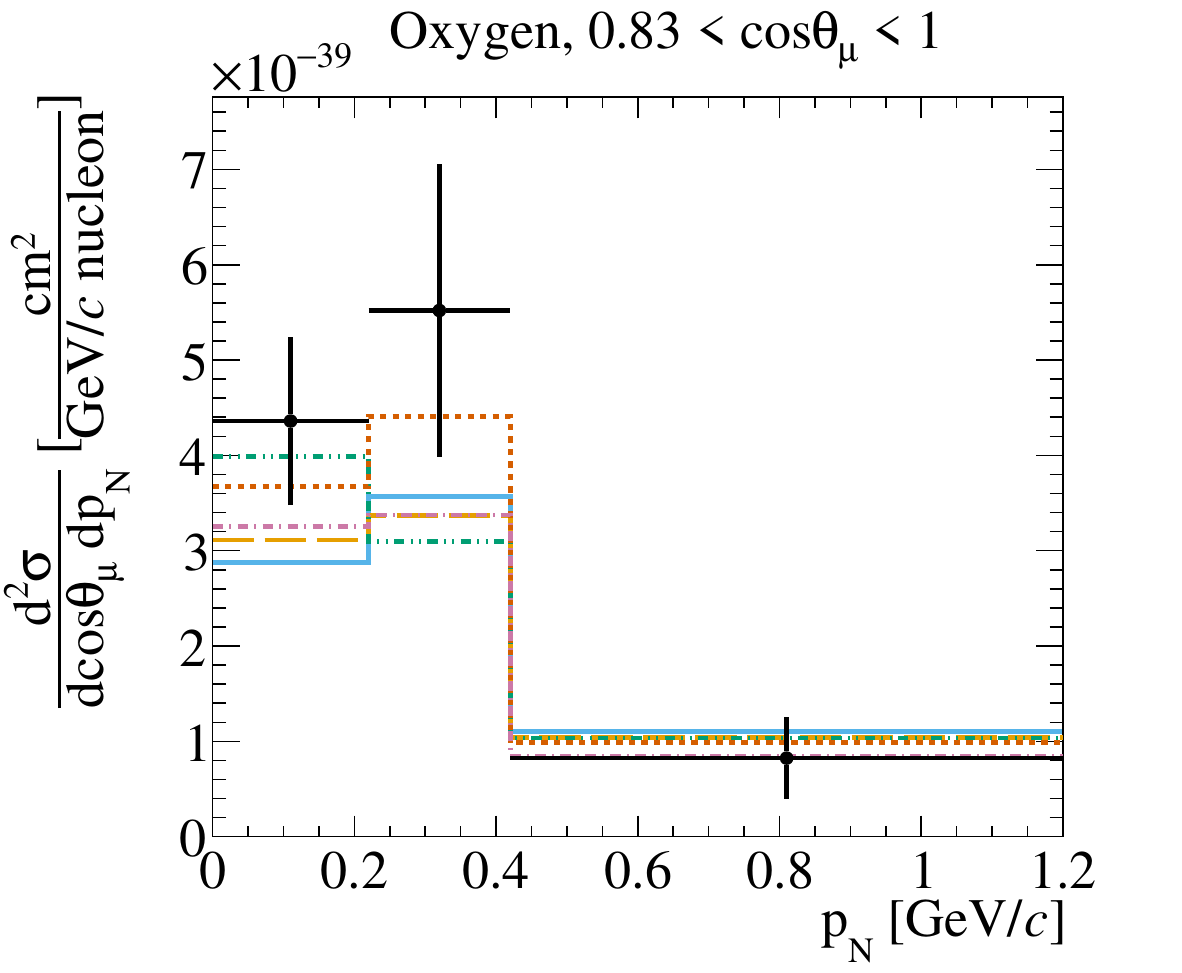}
\end{tabular}
}
\caption{Measurements of the $\nu_{\mu}$ CC$0\pi Np$ double-differential cross sections on carbon (a) and oxygen (b) targets, obtained from the fit to T2K data in $p_{N}$--$\cos\theta_{\mu}$ space. Error bars represent the combined statistical and systematic uncertainties. The results are compared with various neutrino-nucleus interaction models, including NEUT ED-RMF (sky blue), NEUT SF (orange), NEUT LFG (bluish green), GENIE CRPA (vermillion) and GENIE SuSAv2 (reddish purple).}
\label{fig:modelcomp_plot_inistate_2}
\end{figure*}

Fig.~\ref{fig:modelcomp_plot_inistate_1} and Fig.~\ref{fig:modelcomp_plot_inistate_2} present comparisons of the results in the $\delta p_{T}$--$\delta \alpha_{T}$ and $p_{N}$--$\cos\theta_{\mu}$ spaces, respectively, for the five models described above. Tab.~\ref{tab:modelcomp_tab_inistate_1} and Tab.~\ref{tab:modelcomp_tab_inistate_2} summarize the corresponding $\chi^{2}$ values and $p$-values. The $\chi^{2}$ is computed using Eq.~\ref{eq:xsec_chi2}, and the associated $p$-value is evaluated assuming a $\chi^{2}$ distribution with the number of degrees of freedom equals to the number of bins in which the cross section is extracted. A small $p$-value (below or equal to the 0.05 threshold adopted in this measurement) indicates that the model prediction is disfavored by the data. In the $\delta p_{T}$--$\delta \alpha_{T}$ phase space, the NEUT SF model provides the best description of the joint carbon and oxygen dataset, yielding an overall $p$-value of 0.70, followed by the NEUT ED-RMF model with a $p$-value of 0.51. The GENIE SuSAv2 model also demonstrates a relatively high level of agreement with the combined data, resulting in a $p$-value of 0.33. Conversely, both the NEUT LFG and GENIE CRPA models are disfavored by the joint measurement, as indicated by their lower $p$-values.

\begin{table*}[!htbp]
\caption{$\chi^{2}$ values and corresponding $p$-values (in parentheses) for the $\nu_{\mu}$ CC$0\pi Np$ cross-section comparisons with various neutrino-nucleus interaction models from Sec.~\ref{subsec:modelcomp_ngs_mni} in $\delta p_{T}$--$\delta \alpha_{T}$ space. The $\chi^{2}$ and $p$-values are computed separately for the joint carbon and oxygen data, as well as for carbon-only and oxygen-only data. Additionally, the $\chi^{2}$ and $p$-values are provided for different $\delta\alpha_{T}$ regions. Cells are highlighted with a light gray background where $p \le 0.05$.}
\label{tab:modelcomp_tab_inistate_1}
\centering
\renewcommand{\arraystretch}{1.2}
\begin{tabular}{c|c|c|c|c|c|c|c}
\hline 
\hline 
\multicolumn{2}{c|}{Target / $\delta\alpha_{T}$ region} & $N_{bin}$ & NEUT ED-RMF & NEUT SF & NEUT LFG & GENIE CRPA & GENIE SuSAv2 \\ 
\hline 
\multicolumn{2}{c|}{C + O} & 25 & 24.1 (0.51) & 20.8 (0.70) & \cellcolor{lightgray} 45.1 (0.01) & \cellcolor{lightgray} 39.7 (0.03) & 27.5 (0.33) \\ 
\hline 
\multicolumn{2}{c|}{C-only} & 17 & 17.5 (0.42) & 11.0 (0.86) & 26.0 (0.07) & \cellcolor{lightgray} 27.6 (0.05) & 18.4 (0.36) \\ 
\hline 
\multicolumn{2}{c|}{O-only} & 8 & 7.6 (0.48) & 7.5 (0.49) & 8.2 (0.42) & 6.4 (0.60) & 6.3 (0.61) \\ 
\hline 
\multirow{3}{*}{C + O} & $0^{\circ} < \delta\alpha_{T} \leqslant 75^{\circ}$ & 7 & 6.9 (0.43) & 9.2 (0.24) & \cellcolor{lightgray} 15.8 (0.03) & 5.4 (0.62) & 4.6 (0.71) \\ 
 & $75^{\circ} < \delta\alpha_{T} \leqslant 135^{\circ}$ & 9 & 11.0 (0.28) & 8.2 (0.52) & 8.6 (0.48) & 9.4 (0.40) & 9.4 (0.40) \\ 
 & $135^{\circ} < \delta\alpha_{T} < 180^{\circ}$ & 9 & 4.4 (0.88) & 4.1 (0.90) & 7.1 (0.63) & 9.2 (0.42) & 10.1 (0.34) \\ 
\hline 
\multirow{3}{*}{C-only} & $0^{\circ} < \delta\alpha_{T} \leqslant 75^{\circ}$ & 5 & 4.5 (0.49) & 4.1 (0.53) & 7.2 (0.20) & 4.5 (0.48) & 2.3 (0.80) \\ 
 & $75^{\circ} < \delta\alpha_{T} \leqslant 135^{\circ}$ & 6 & 6.9 (0.33) & 5.3 (0.50) & 7.1 (0.31) & 5.4 (0.49) & 5.8 (0.45) \\ 
 & $135^{\circ} < \delta\alpha_{T} < 180^{\circ}$ & 6 & 2.8 (0.83) & 3.0 (0.81) & 5.0 (0.54) & 8.1 (0.23) & 7.4 (0.28) \\ 
\hline 
\multirow{3}{*}{O-only} & $0^{\circ} < \delta\alpha_{T} \leqslant 75^{\circ}$ & 2 & 3.6 (0.16) & 4.6 (0.10) & 5.4 (0.07) & 1.1 (0.57) & 2.7 (0.26) \\ 
 & $75^{\circ} < \delta\alpha_{T} \leqslant 135^{\circ}$ & 3 & 4.0 (0.26) & 3.4 (0.34) & 1.9 (0.60) & 3.9 (0.27) & 4.0 (0.26) \\ 
 & $135^{\circ} < \delta\alpha_{T} < 180^{\circ}$ & 3 & 1.2 (0.76) & 1.0 (0.80) & 0.8 (0.85) & 1.7 (0.64) & 1.9 (0.59) \\ 
\hline 
\hline 
\end{tabular}
\end{table*}

\begin{table*}[!htbp]
\caption{$\chi^{2}$ values and corresponding $p$-values (in parentheses) for the $\nu_{\mu}$ CC$0\pi Np$ cross-section comparisons with various neutrino-nucleus interaction models from Sec.~\ref{subsec:modelcomp_ngs_mni} in $p_{N}$--$\cos\theta_{\mu}$ space. The $\chi^{2}$ and $p$-values are computed separately for the joint carbon and oxygen data, as well as for carbon-only and oxygen-only data. Additionally, the $\chi^{2}$ and $p$-values are provided for different $\cos\theta_{\mu}$ regions. Cells are highlighted with a light gray background where $p \le 0.05$.}
\label{tab:modelcomp_tab_inistate_2}
\centering
\renewcommand{\arraystretch}{1.2}
\begin{tabular}{c|c|c|c|c|c|c|c}
\hline 
\hline 
\multicolumn{2}{c|}{Target / $\cos\theta_{\mu}$ region} & $N_{bin}$ & NEUT ED-RMF & NEUT SF & NEUT LFG & GENIE CRPA & GENIE SuSAv2 \\ 
\hline 
\multicolumn{2}{c|}{C + O} & 30 & 40.5 (0.10) & 27.9 (0.57) & 36.4 (0.19) & \cellcolor{lightgray} 48.4 (0.02) & 27.8 (0.58) \\ 
\hline 
\multicolumn{2}{c|}{C-only} & 19 & 26.0 (0.13) & 14.5 (0.76) & 21.3 (0.32) & 27.2 (0.10) & 13.9 (0.79) \\ 
\hline 
\multicolumn{2}{c|}{O-only} & 11 & 13.4 (0.27) & 11.9 (0.37) & 10.9 (0.45) & 13.8 (0.25) & 11.6 (0.39) \\ 
\hline 
\multirow{5}{*}{C + O} & $-0.6 < \cos\theta_{\mu} \leqslant -0.1$ & 4 & 7.1 (0.13) & 4.9 (0.30) & 4.1 (0.40) & 8.5 (0.08) & 5.6 (0.23) \\ 
 & $-0.1 < \cos\theta_{\mu} \leqslant 0.15$ & 2 & 0.3 (0.88) & 0.2 (0.90) & 0.8 (0.69) & 0.3 (0.86) & 0.2 (0.90) \\ 
 & $0.15 < \cos\theta_{\mu} \leqslant 0.52$ & 6 & \cellcolor{lightgray} 16.9 (0.01) & 9.1 (0.17) & 7.1 (0.31) & 9.8 (0.13) & 8.7 (0.19) \\ 
 & $0.52 < \cos\theta_{\mu} \leqslant 0.83$ & 9 & 13.8 (0.13) & 4.4 (0.89) & 6.8 (0.66) & 9.5 (0.39) & 4.4 (0.89) \\ 
 & $0.83 < \cos\theta_{\mu} < 1$ & 9 & 11.5 (0.24) & 10.1 (0.34) & 13.7 (0.14) & 3.3 (0.95) & 5.5 (0.79) \\ 
\hline 
\multirow{5}{*}{C-only} & $-0.6 < \cos\theta_{\mu} \leqslant -0.1$ & 2 & \cellcolor{lightgray} 6.5 (0.04) & 4.5 (0.11) & 2.4 (0.31) & \cellcolor{lightgray} 7.7 (0.02) & 5.0 (0.08) \\ 
 & $-0.1 < \cos\theta_{\mu} \leqslant 0.15$ & 1 & 0.2 (0.64) & 0.1 (0.71) & 0.2 (0.69) & 0.2 (0.64) & 0.1 (0.71) \\ 
 & $0.15 < \cos\theta_{\mu} \leqslant 0.52$ & 4 & 7.0 (0.14) & 2.0 (0.74) & 3.8 (0.43) & 2.2 (0.69) & 2.1 (0.71) \\ 
 & $0.52 < \cos\theta_{\mu} \leqslant 0.83$ & 6 & \cellcolor{lightgray} 12.7 (0.05) & 4.2 (0.65) & 4.9 (0.56) & 5.8 (0.44) & 3.5 (0.75) \\ 
 & $0.83 < \cos\theta_{\mu} < 1$ & 6 & 5.3 (0.50) & 3.2 (0.79) & 6.2 (0.41) & 1.4 (0.96) & 1.1 (0.98) \\ 
\hline 
\multirow{5}{*}{O-only} & $-0.6 < \cos\theta_{\mu} \leqslant -0.1$ & 2 & 0.3 (0.85) & 0.2 (0.89) & 0.5 (0.76) & 0.5 (0.76) & 0.3 (0.86) \\ 
 & $-0.1 < \cos\theta_{\mu} \leqslant 0.15$ & 1 & 0.1 (0.76) & 0.1 (0.73) & 0.4 (0.52) & 0.2 (0.69) & 0.1 (0.71) \\ 
 & $0.15 < \cos\theta_{\mu} \leqslant 0.52$ & 2 & \cellcolor{lightgray} 9.3 (0.01) & \cellcolor{lightgray} 7.8 (0.02) & 3.6 (0.17) & \cellcolor{lightgray} 6.7 (0.03) & \cellcolor{lightgray} 6.8 (0.03) \\ 
 & $0.52 < \cos\theta_{\mu} \leqslant 0.83$ & 3 & 0.9 (0.82) & 0.7 (0.88) & 1.3 (0.73) & 2.2 (0.54) & 0.7 (0.88) \\ 
 & $0.83 < \cos\theta_{\mu} < 1$ & 3 & 6.0 (0.11) & 5.2 (0.16) & 3.6 (0.31) & 1.6 (0.66) & 3.9 (0.27) \\ 
\hline 
\hline 
\end{tabular}
\end{table*}

When evaluating the targets independently, the NEUT SF model yields the highest level of agreement for the carbon-only measurement, followed by the ED-RMF and GENIE SuSAv2 models, which exhibit similar performance. The NEUT LFG and GENIE CRPA models show slightly improved $p$-values compared to the joint results, although GENIE CRPA remains disfavored. For the oxygen-only measurement, the preference shifts toward GENIE SuSAv2. In this case, both NEUT LFG and GENIE CRPA exhibit significantly higher $p$-values and are no longer excluded by the data. However, it is worth noting that another reason for the lack of model exclusion in the oxygen-only measurement is the limited oxygen statistics in the selected samples, which restrict the discriminating power among models.

The NEUT SF and ED-RMF models provide a better description of the high-$\delta \alpha_{T}$ region, where FSI becomes dominant. Because the ED-RMF implementation in NEUT describes the nucleon initial state using characteristics similar to the SF model, the primary distinction lies in the FSI modeling: ED-RMF employs the RDWIA, whereas SF utilizes a PWIA-based INC approach. The data exhibit a comparable preference for both modeling strategies in this regime. Conversely, in the low-$\delta \alpha_{T}$ region, specifically for $\delta \alpha_{T} \leqslant 75^{\circ}$, where the CCQE nucleon ground-state contribution dominates, the GENIE SuSAv2 model performs best. While the GENIE CRPA model shows reasonable agreement with the data ($p$-value $> 0.23$) across different $\delta \alpha_{T}$ intervals, its overall performance is poorer, particularly for the carbon target. This may be attributed to the limited capability of the CRPA model to reproduce the cross-section evolution across the $\delta \alpha_{T}$ range. Furthermore, in the GENIE implementation~\cite{Dolan:2021rdd}, the CRPA implementation only modifies the inclusive CC prediction, while the exclusive CC$0\pi Np$ channels continue to rely on the underlying LFG model.

However, there are regions where none of the selected models provides a good description of the data. These are mainly the intermediate $\delta p_{T}$ regions, where all models under-predict the data. For carbon, they correspond to the 0.36--0.52~GeV/c bins with $\delta \alpha_{T} > 75^{\circ}$; for oxygen, they correspond to the 0.17--0.36~GeV/c bin in the $75^{\circ}$--$135^{\circ}$ slice of $\delta \alpha_{T}$. This excess could indicate insufficient contributions from SRCs or multi-nucleon interactions. To check whether this excess could be an artifact, we examined the reconstruction performance by inspecting the migration matrix between the true and reconstructed analysis observables in the cross-section binning (Tab.~\ref{tab:binedge_xsec_1}). The migration is expected to be largest across the falling edge of the $\delta p_{T}$ peak, where events in the peak could migrate to the intermediate $\delta p_{T}$ region because of the finite resolution of the muon and proton kinematic measurements. We found the migration across the falling edge into the observed excess bins to be negligible. In addition, we performed another pseudo-data fit using the MC template, in which the cross sections in these intermediate $\delta p_{T}$ regions were enhanced by hand to match the excess observed in the data. The extracted cross sections agree well with the enhanced ones, and in the other regions they are unchanged with respect to the MC template. These additional studies confirm the robustness of the excess observed in the data fit.

In the $p_{N}$--$\cos\theta_{\mu}$ space, the GENIE SuSAv2 model provides the best overall agreement with the data, with the NEUT SF model yielding a similarly high level of consistency. While the GENIE CRPA model provides the most accurate predictions in the forward-scattering region ($\cos\theta_{\mu} > 0.83$), it struggles to describe the backward-going muons ($\cos\theta_{\mu} \leqslant -0.1$), particularly for the carbon target. Its small overall $p$-value suggests a limited capacity to reproduce the correlations across different muon scattering angles. The NEUT LFG model shows improved agreement relative to its performance in the $\delta p_{T}$--$\delta \alpha_{T}$ space. Conversely, the NEUT ED-RMF model performs less effectively in the $p_{N}$--$\cos\theta_{\mu}$ space than in the $\delta p_{T}$--$\delta \alpha_{T}$ space. This discrepancy is most evident at larger scattering angles ($\cos\theta_{\mu} \leqslant 0.83$), specifically for the carbon measurement, where the model underestimates the data in the $p_{N} < 0.15$\,GeV/c region. It is worth noting that the implementation of the ED-RMF model in NEUT does not contain contributions from two-body currents \cite{Franco-Munoz:2022jcl,Franco-Munoz:2023zoa,Franco-Munoz:2026hxi}, which cause an increase in the total cross section and may help resolve this discrepancy.

\subsection{Comparisons with FSI Models}
\label{subsec:modelcomp_FSI}

We consider three FSI models implemented in GENIE based on the G21\_11b CMC: hA2018, hN2018~\cite{Dytman:2021ohr}, and the Bertini intranuclear cascade (INC) model~\cite{Wright:2015xia}, as implemented in Geant4 v11.3.2. The hN2018 model employs a cascade approach that accounts for multiple interactions within the nucleus, whereas hA2018 is a more empirical model that characterizes a single interaction using hadronic fate probabilities tuned to hadron-nucleus scattering data~\cite{Dytman:2021ohr}. For all FSI configurations, the CCQE interaction is described by the SuSAv2 model utilizing the LFG for the nucleon initial state, and the 2p2h contribution is likewise modeled within the SuSAv2 framework. It is worth noting that the hN2018 prediction here uses the same model configuration as the GENIE SuSAv2 prediction in Sec.~\ref{subsec:modelcomp_ngs_mni}.

\begin{figure*}
\centering
\subfloat[Carbon]
{\includegraphics[width=0.33\linewidth]{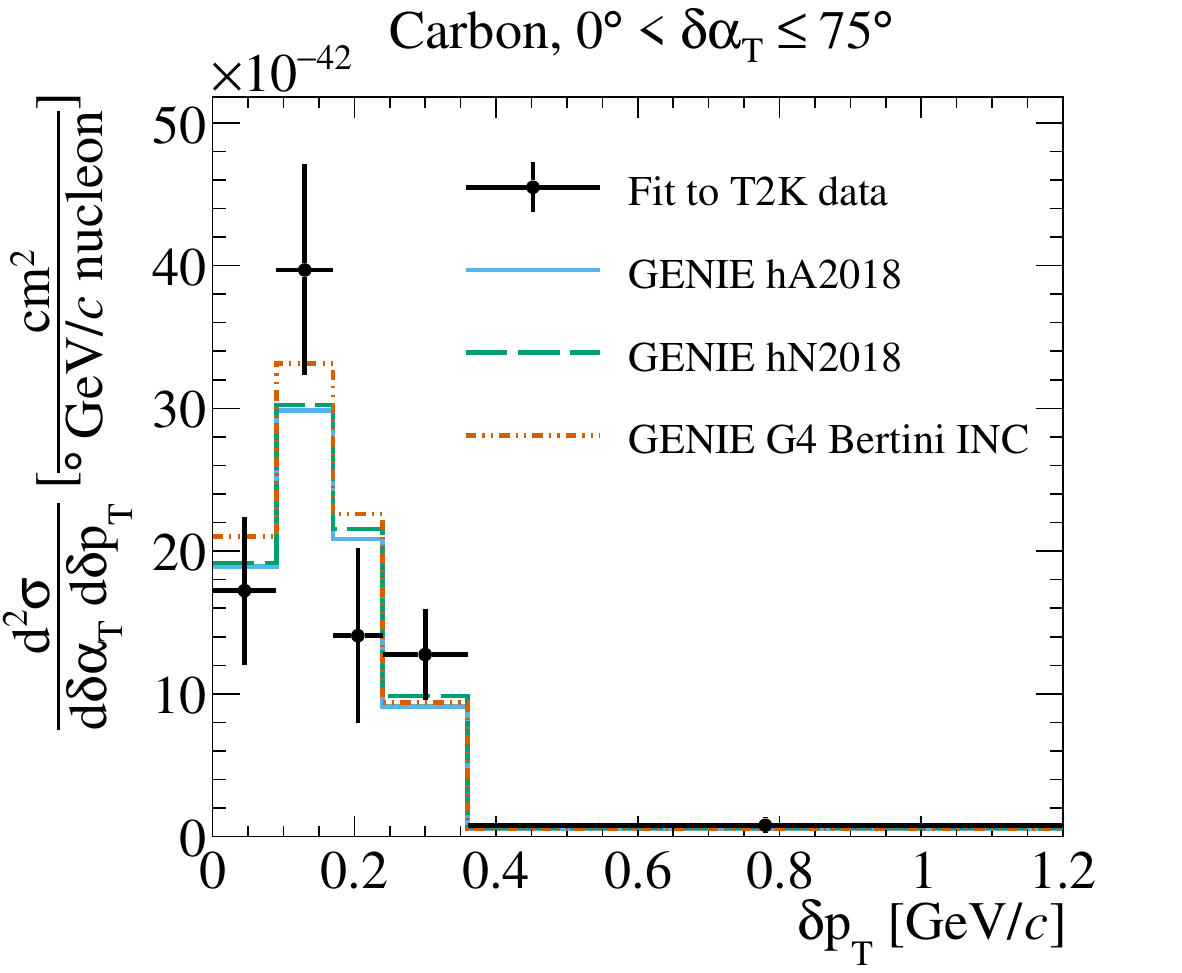}
\includegraphics[width=0.33\linewidth]{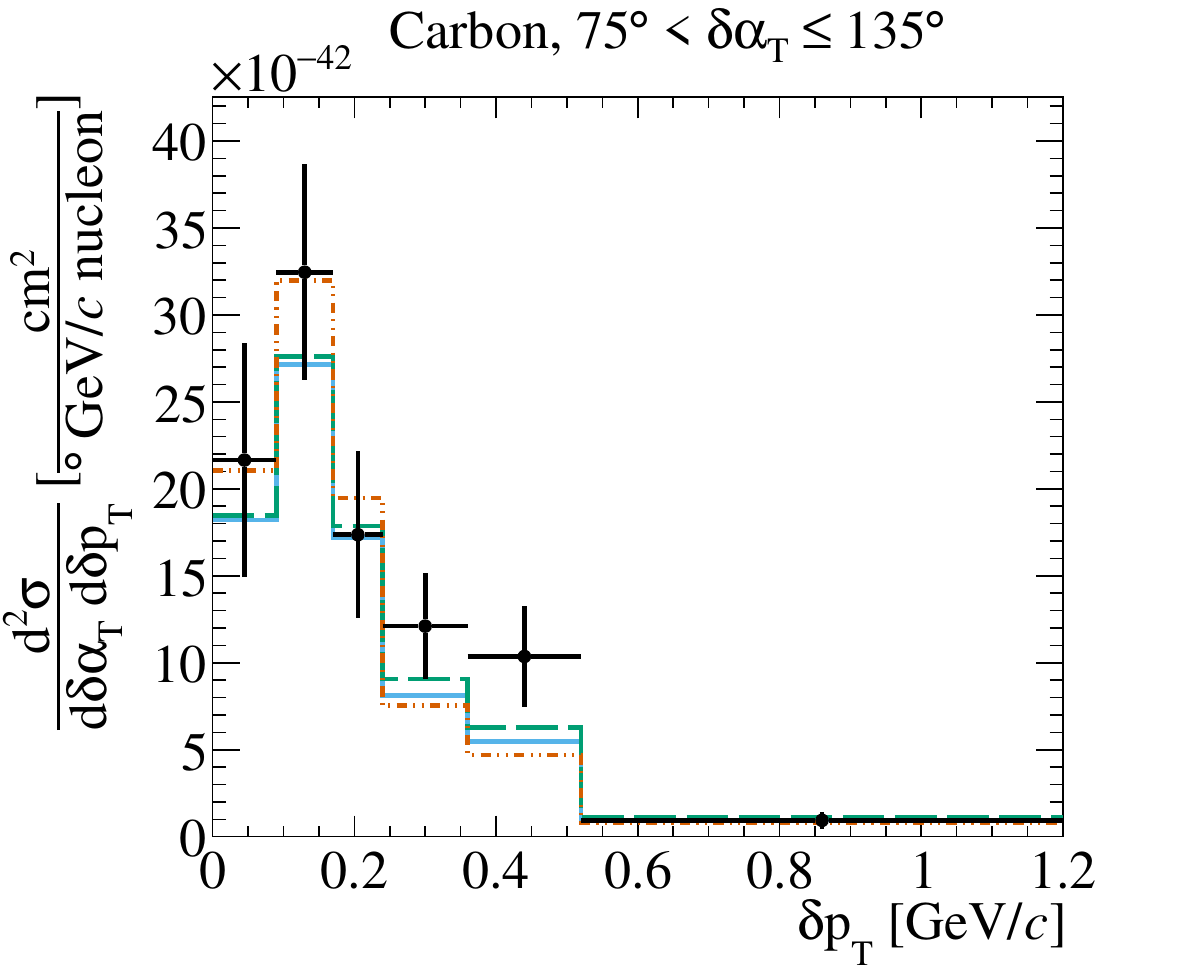}
\includegraphics[width=0.33\linewidth]{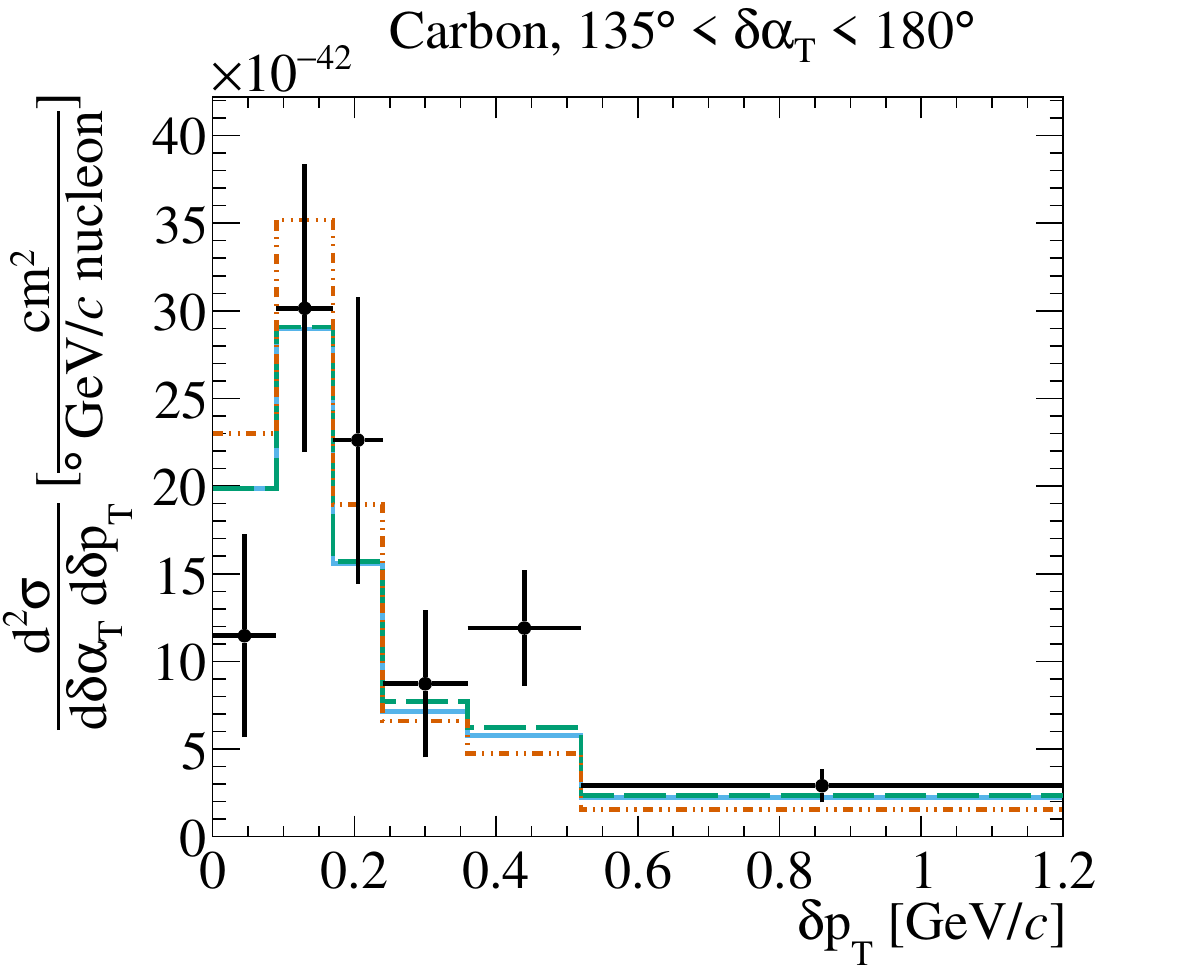}}
\\
\subfloat[Oxygen]
{\includegraphics[width=0.33\linewidth]{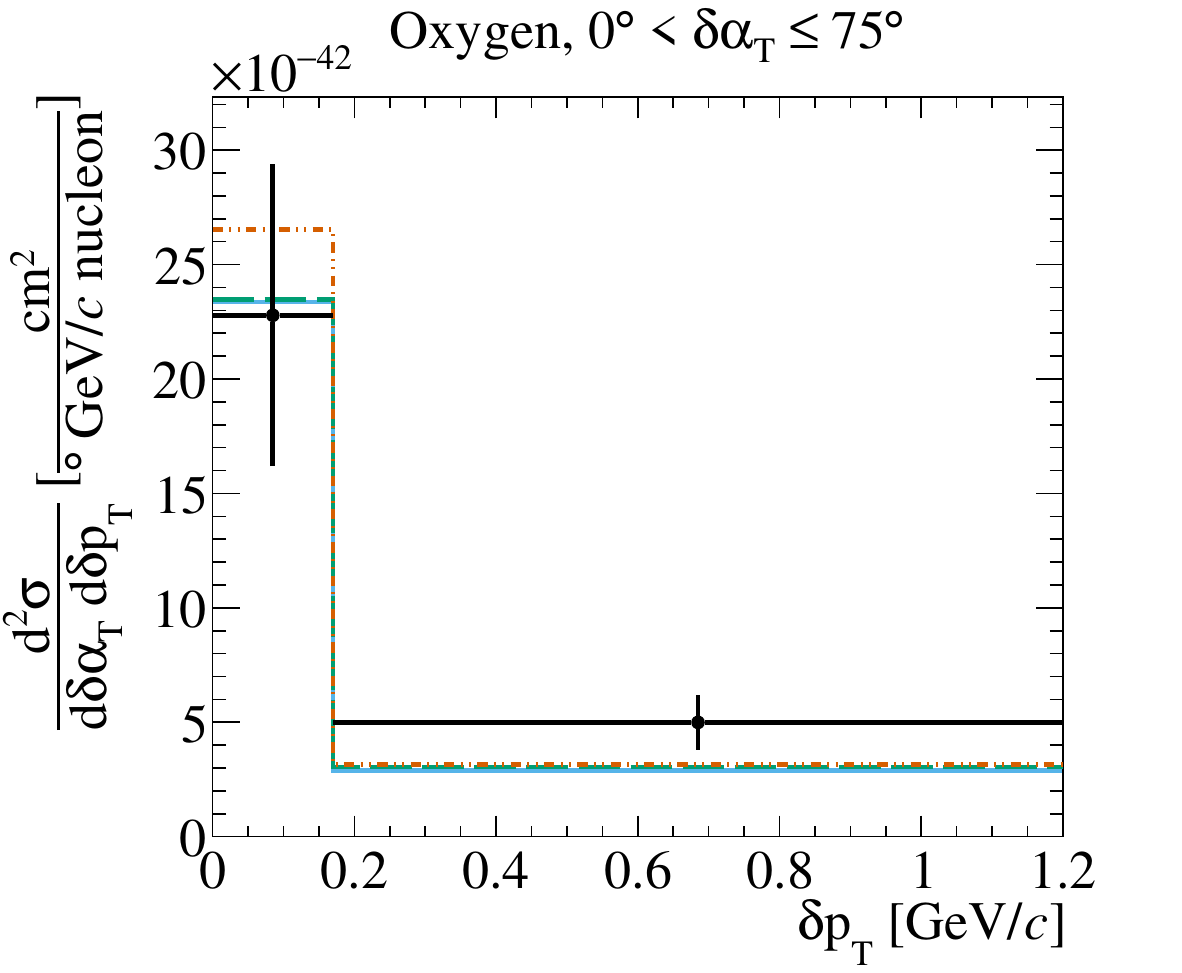}
\includegraphics[width=0.33\linewidth]{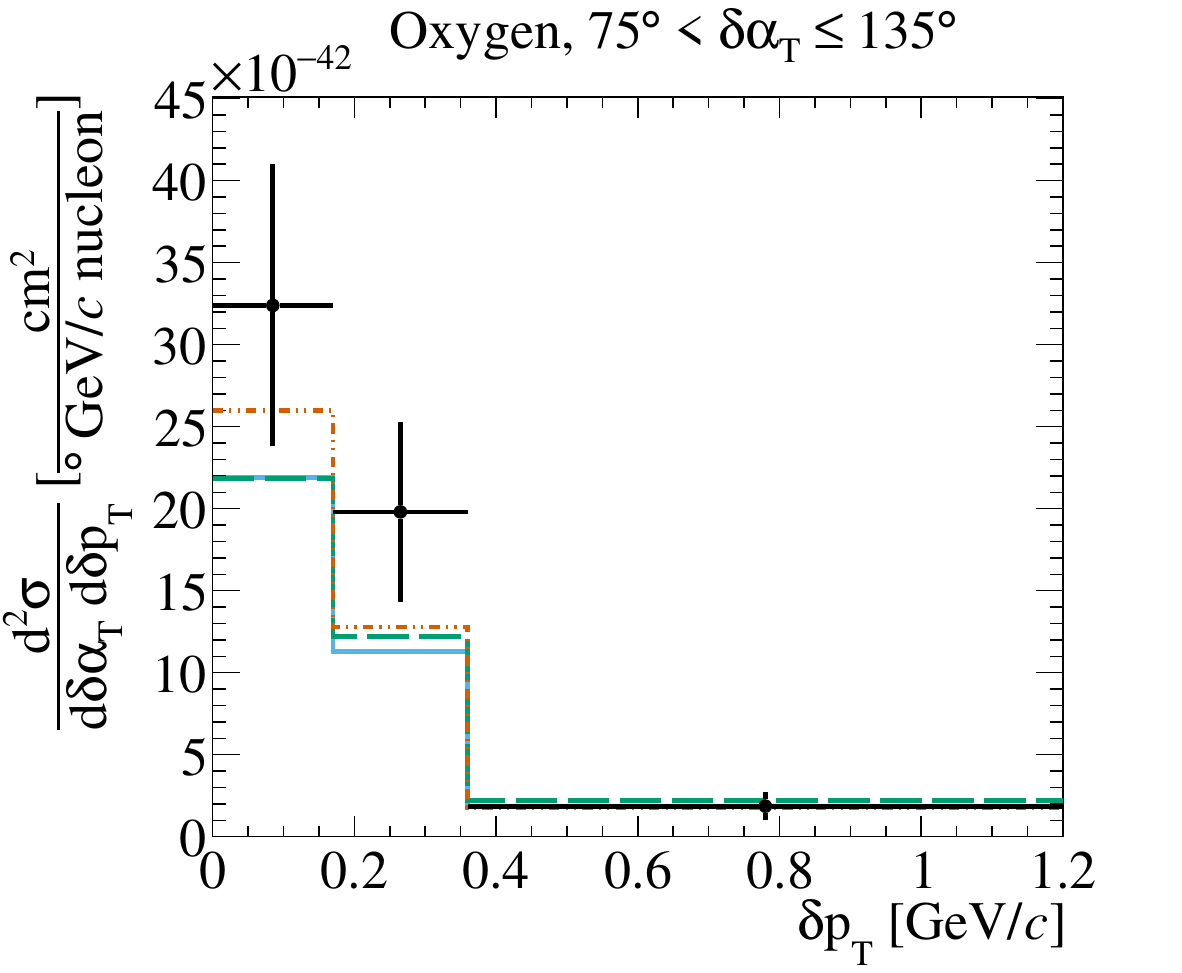}
\includegraphics[width=0.33\linewidth]{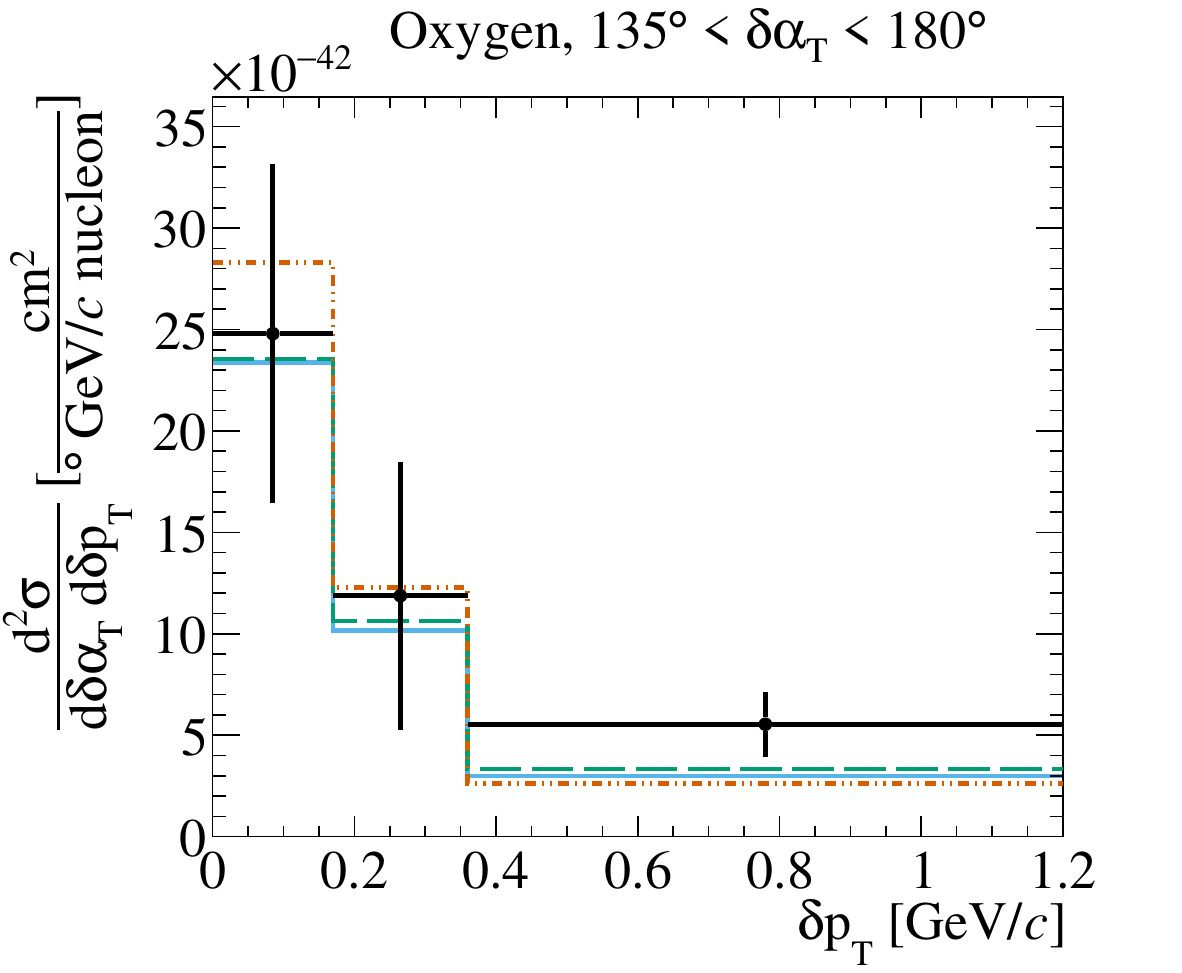}}
\caption{Measurements of the $\nu_{\mu}$ CC$0\pi Np$ double-differential cross sections on carbon (a) and oxygen (b) targets, obtained from the fit to T2K data in $\delta p_{T}$--$\delta \alpha_{T}$ space. Error bars represent the combined statistical and systematic uncertainties. The results are compared with various neutrino-nucleus interaction models, including GENIE hA2018 (sky blue), hN2018 (bluish green) and Geant4-based Bertini INC (vermillion).}
\label{fig:modelcomp_plot_FSI_1}
\end{figure*}

\begin{figure*}
\centering
\subfloat[Carbon]{
\begin{tabular}{c}
\includegraphics[width=0.32\linewidth]{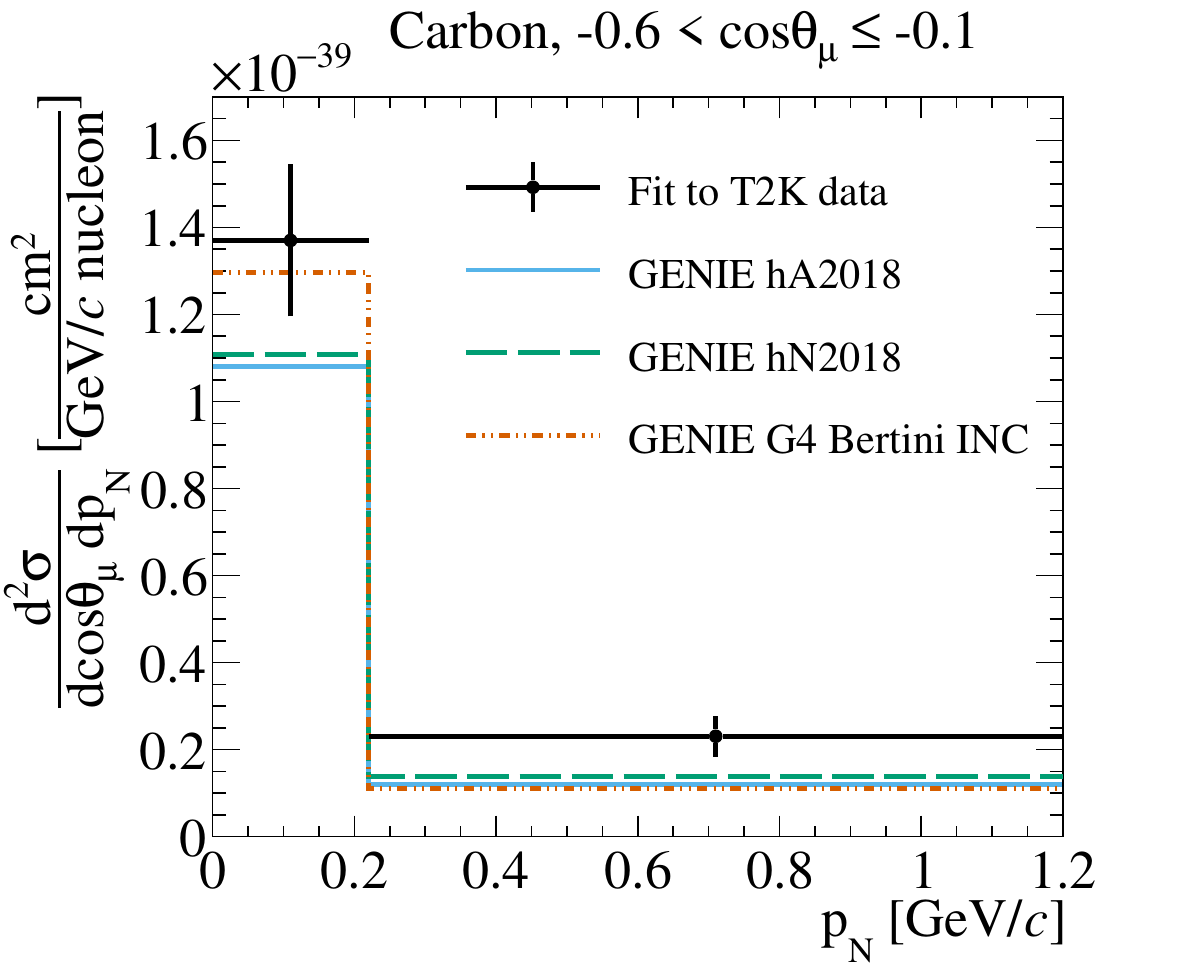}
\includegraphics[width=0.32\linewidth]{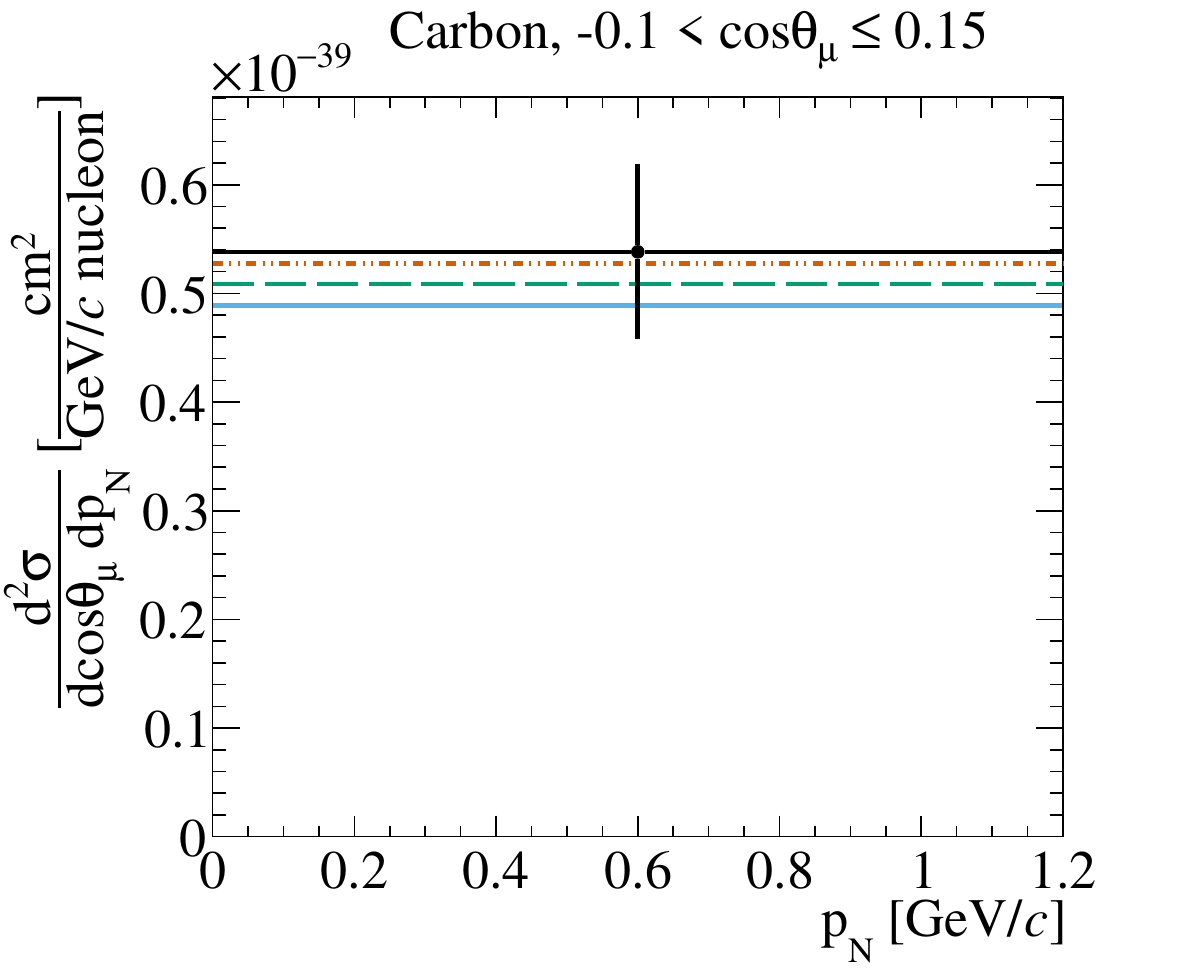}
\includegraphics[width=0.32\linewidth]{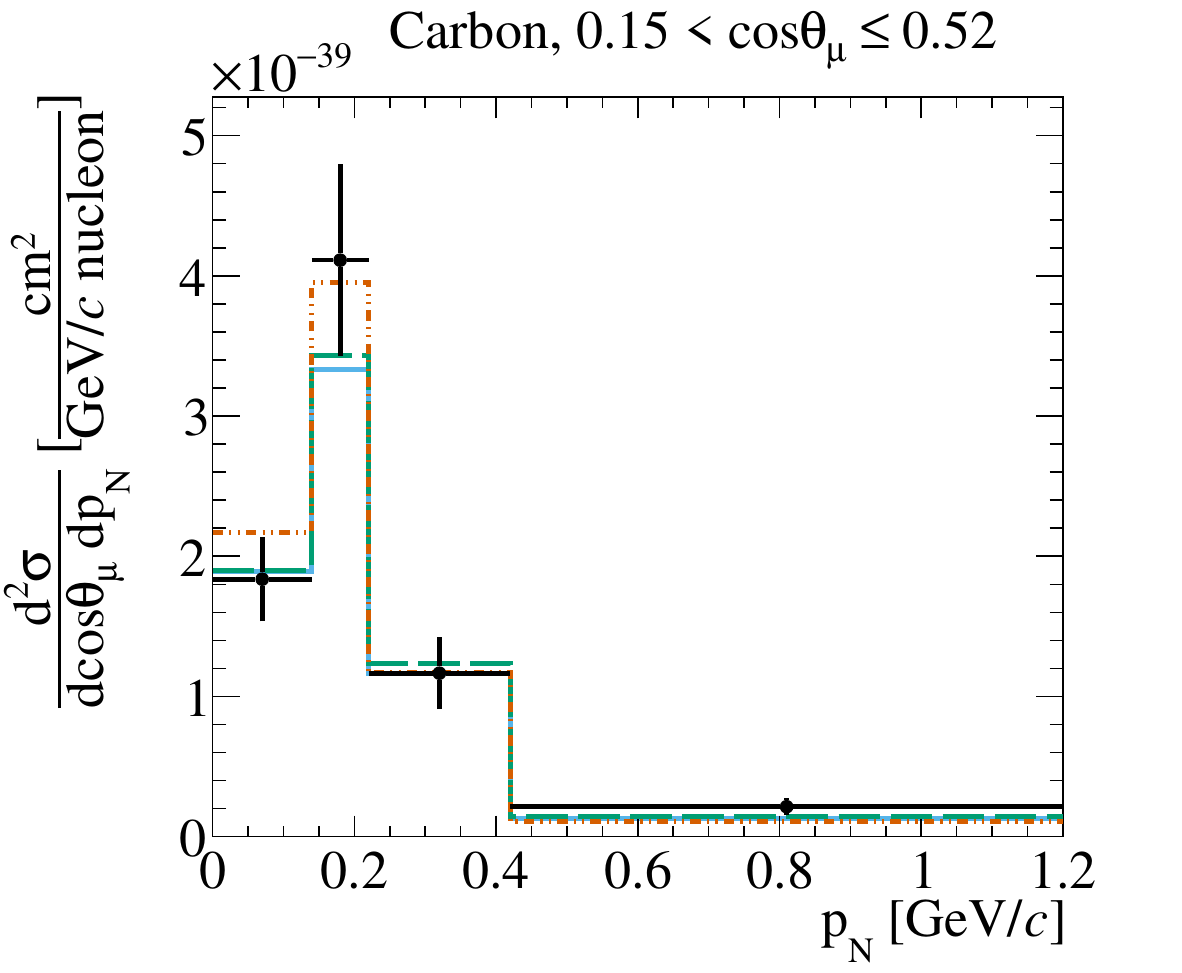} \\
\includegraphics[width=0.32\linewidth]{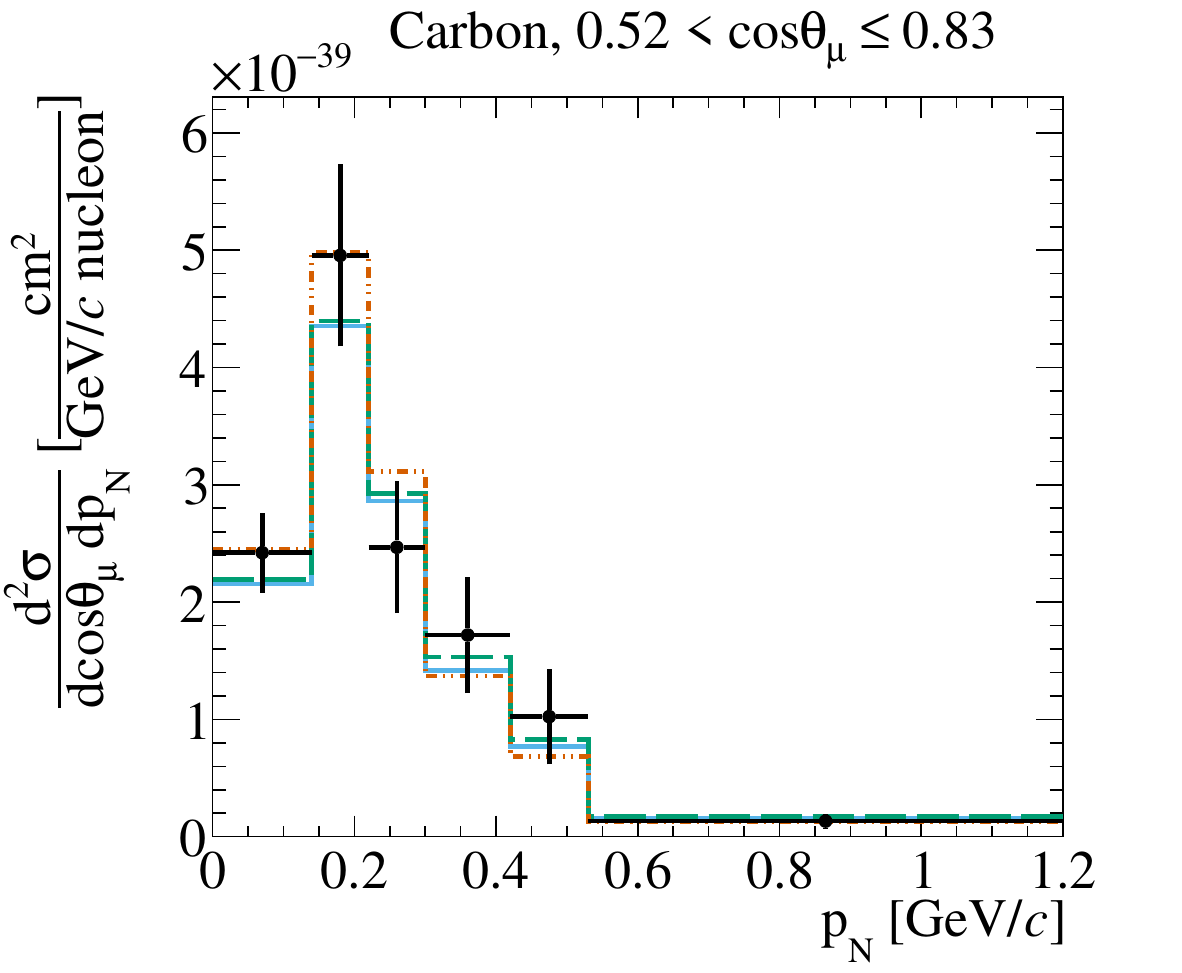}
\includegraphics[width=0.32\linewidth]{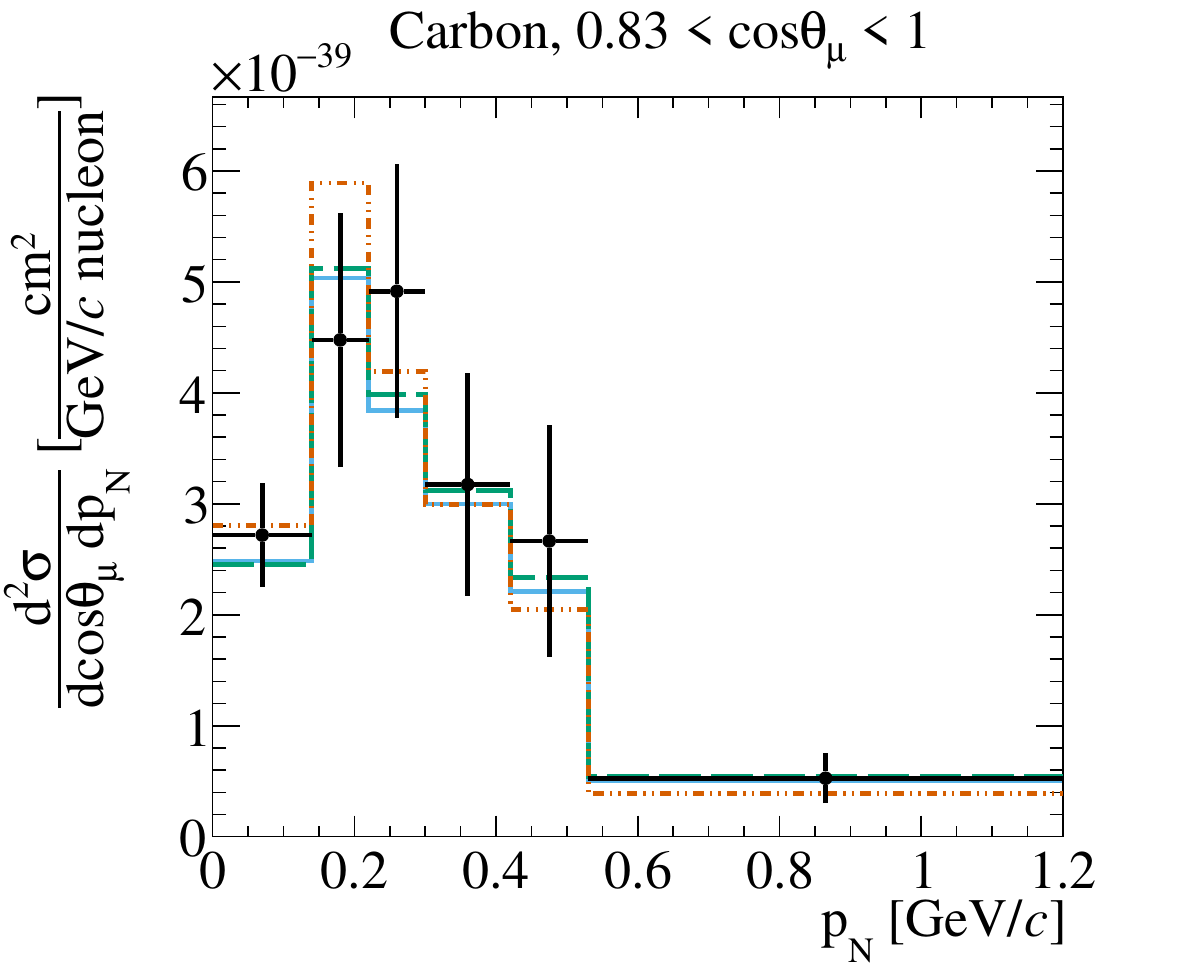}
\end{tabular}
}
\\
\subfloat[Oxygen]{
\begin{tabular}{c}
\includegraphics[width=0.32\linewidth]{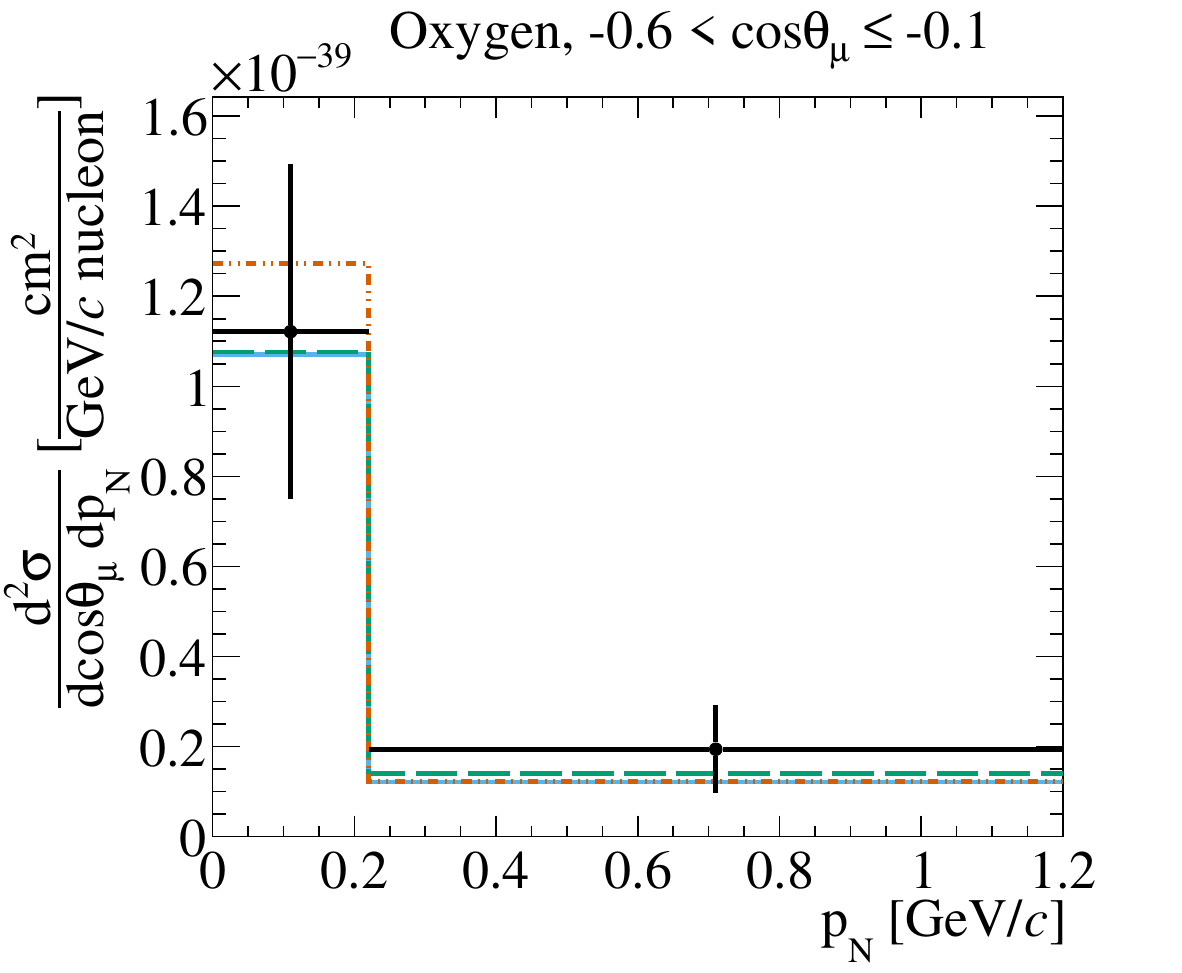}
\includegraphics[width=0.32\linewidth]{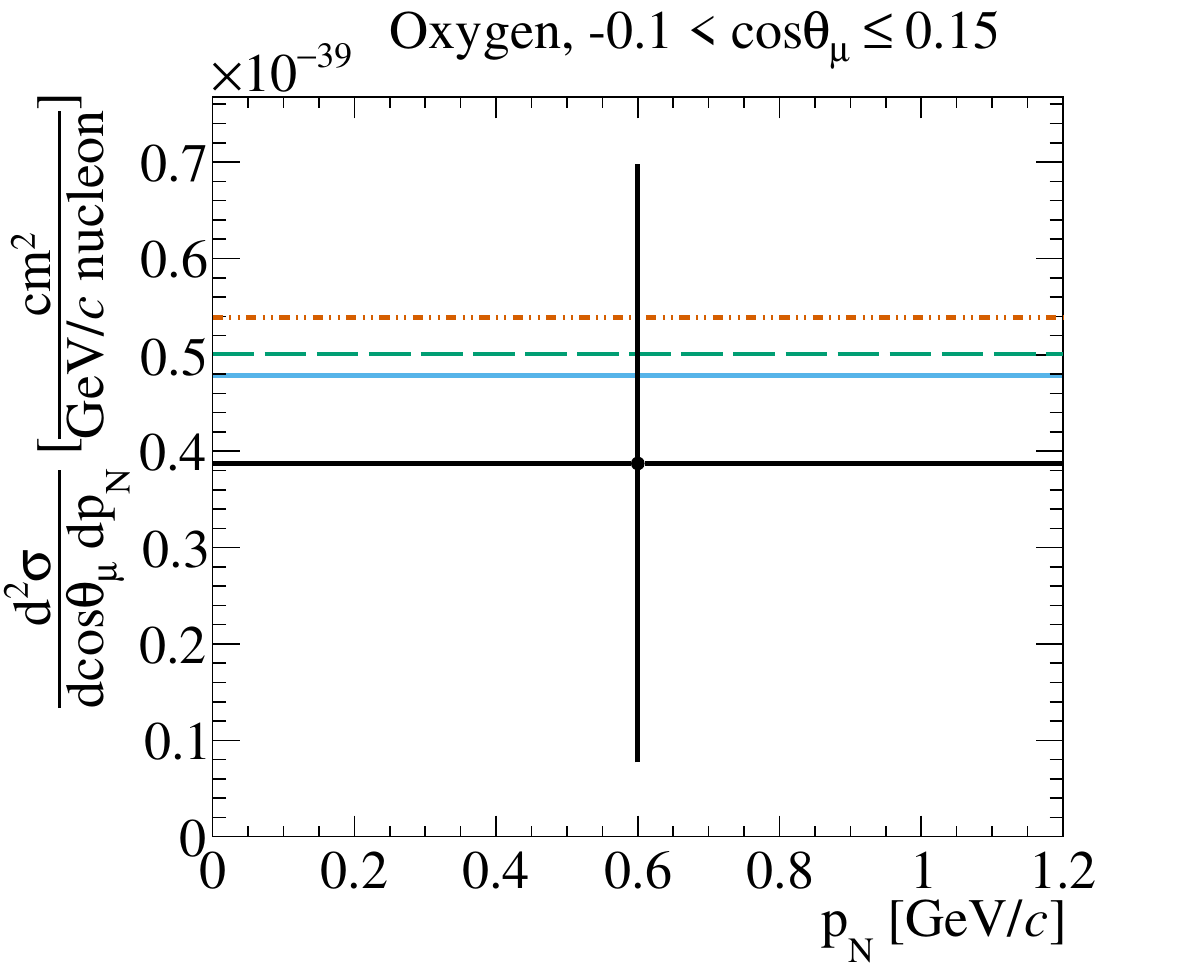}
\includegraphics[width=0.32\linewidth]{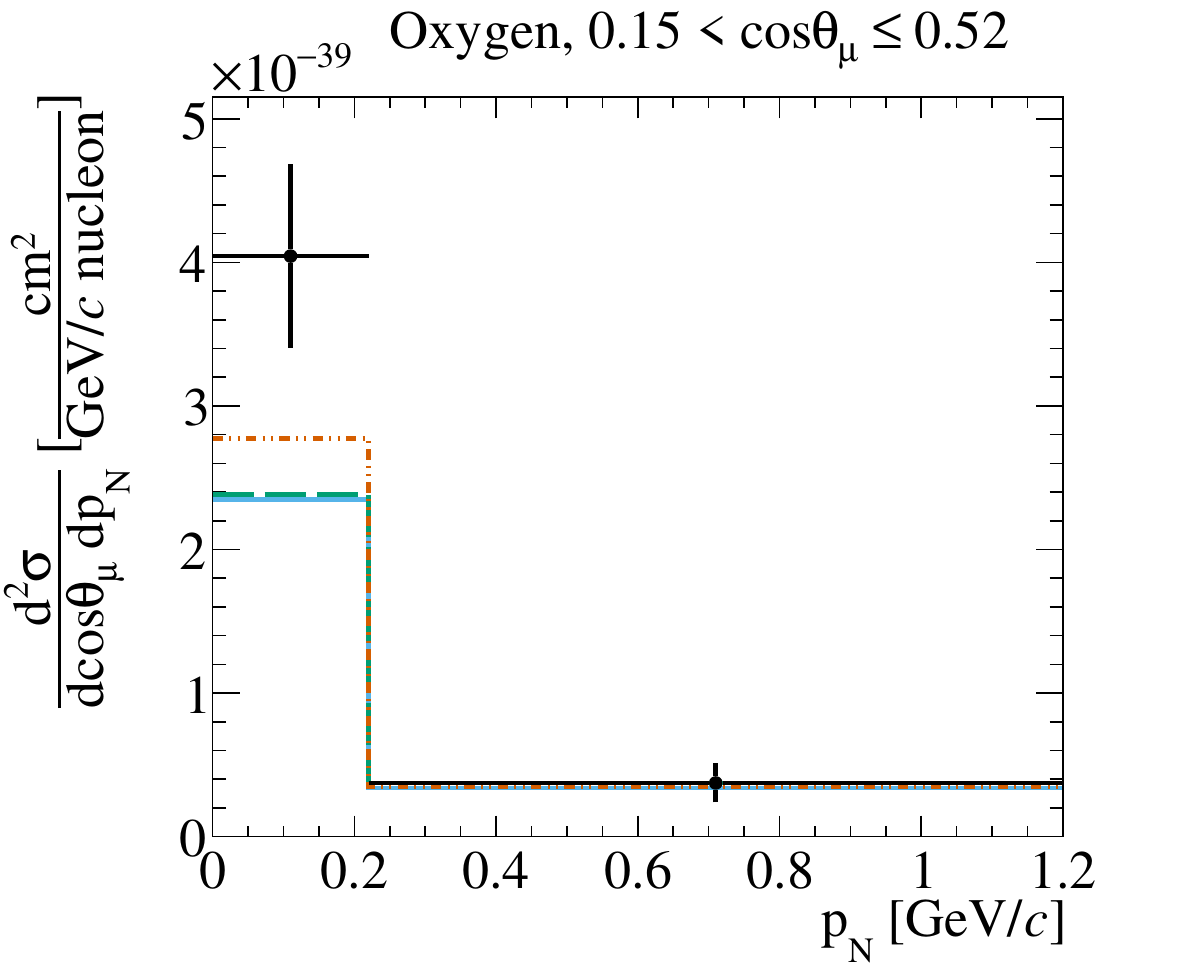} \\
\includegraphics[width=0.32\linewidth]{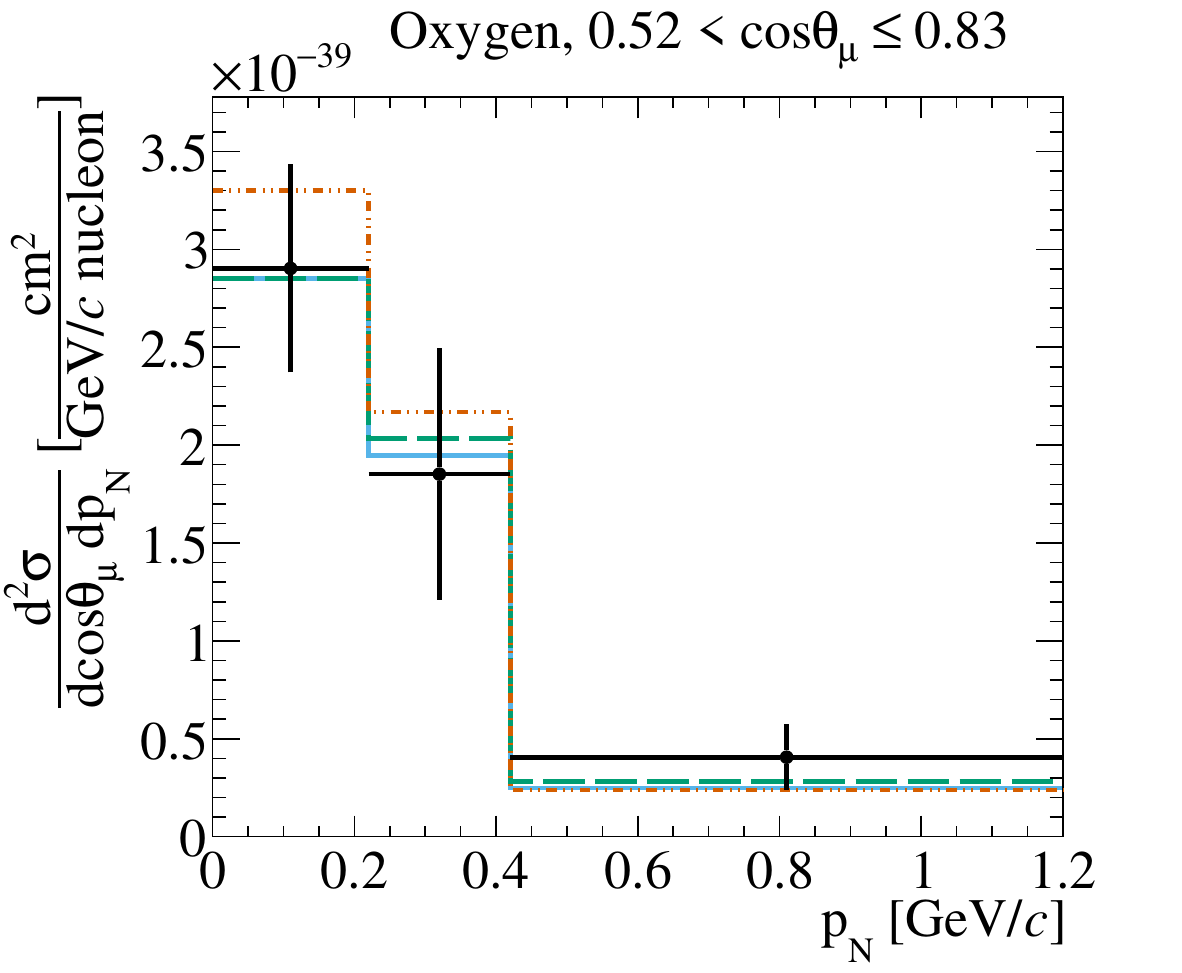}
\includegraphics[width=0.32\linewidth]{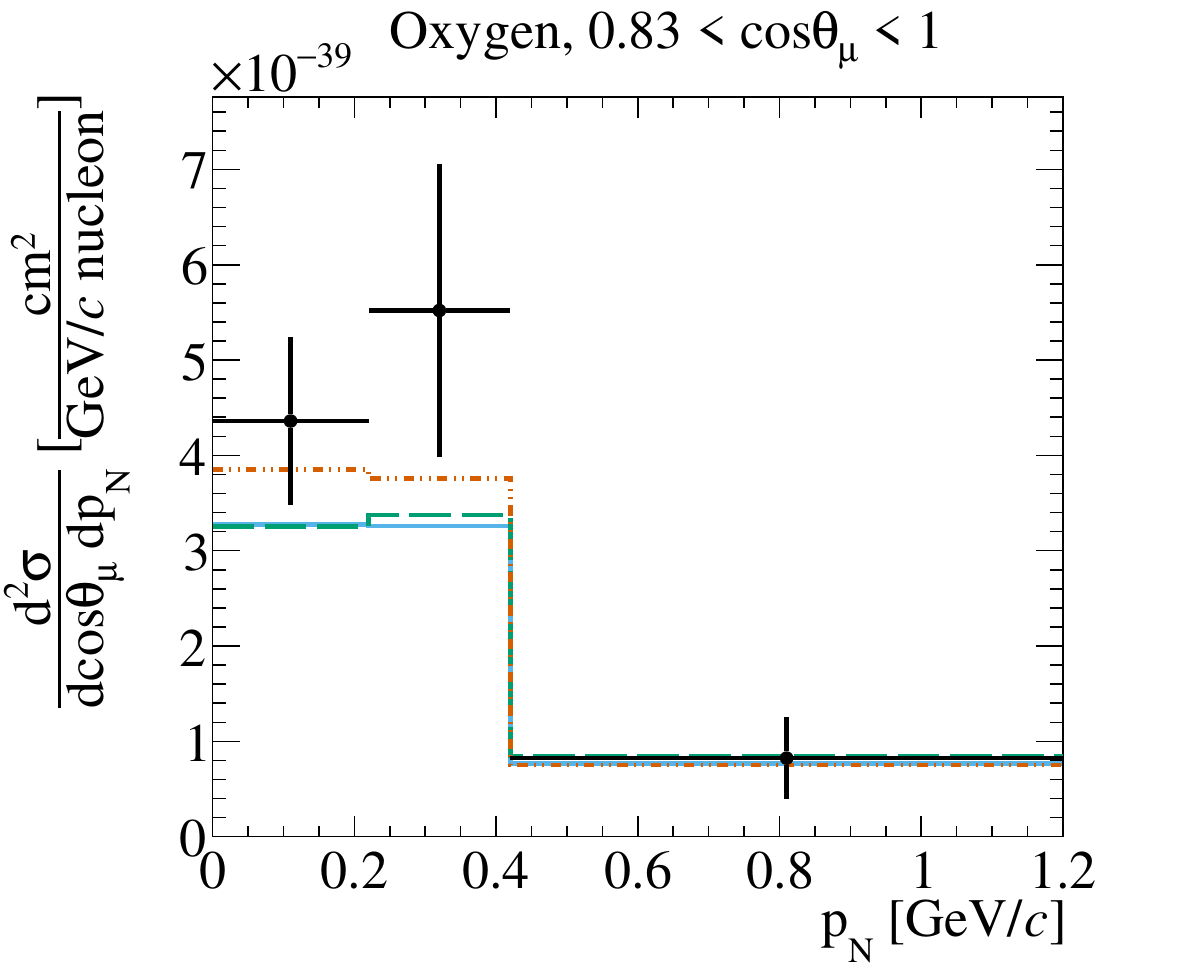}
\end{tabular}
}
\caption{Measurements of the $\nu_{\mu}$ CC$0\pi Np$ double-differential cross sections on carbon (a) and oxygen (b) targets, obtained from the fit to T2K data in $p_{N}$--$\cos\theta_{\mu}$ space. Error bars represent the combined statistical and systematic uncertainties. The results are compared with various neutrino-nucleus interaction models, including GENIE hA2018 (sky blue), hN2018 (bluish green) and Geant4-based Bertini INC (vermillion).}
\label{fig:modelcomp_plot_FSI_2}
\end{figure*}

Fig.~\ref{fig:modelcomp_plot_FSI_1} and Fig.~\ref{fig:modelcomp_plot_FSI_2} present the comparisons in the $\delta p_{T}$--$\delta \alpha_{T}$ and $p_{N}$--$\cos\theta_{\mu}$ spaces, respectively. Tab.~\ref{tab:modelcomp_tab_FSI_1} and Tab.~\ref{tab:modelcomp_tab_FSI_2} summarize the corresponding $\chi^{2}$ values and $p$-values. The measurement favors both the GENIE hA2018 and hN2018 models in the $\delta p_{T}$--$\delta \alpha_{T}$ and $p_{N}$--$\cos\theta_{\mu}$ spaces, with a slight preference for hN2018 observed in the overall comparison as well as in different $\delta \alpha_{T}$ and $\cos\theta_{\mu}$ regions. The GENIE Geant4 Bertini model does not describe the measurement in the $\delta p_{T}$--$\delta \alpha_{T}$ space well, particularly for the carbon sample and in the high-$\delta \alpha_{T}$ region ($\delta \alpha_{T} > 135^{\circ}$), where it over-predicts the $\delta p_{T}$ peak and under-predicts the high-$\delta p_{T}$ tail. This behavior may indicate insufficient FSI strength to reproduce the observed distributions. However, in the $p_{N}$--$\cos\theta_{\mu}$ space, the Bertini model provides a better description of the data, as reflected by a significantly larger $p$-value.

\begin{table*}[!htbp]
\caption{$\chi^{2}$ values and corresponding $p$-values (in parentheses) for the $\nu_{\mu}$ CC$0\pi Np$ cross-section comparisons with various neutrino-nucleus interaction models from Sec.~\ref{subsec:modelcomp_FSI} in $\delta p_{T}$--$\delta \alpha_{T}$ space. The $\chi^{2}$ and $p$-values are computed separately for the joint carbon and oxygen data, as well as for carbon-only and oxygen-only data. Additionally, the $\chi^{2}$ and $p$-values are provided for different $\delta\alpha_{T}$ regions. Cells are highlighted with a light gray background where $p \le 0.05$.}
\label{tab:modelcomp_tab_FSI_1}
\centering
\renewcommand{\arraystretch}{1.2}
\begin{tabular}{c|c|c|c|c|c}
\hline
\hline
\multicolumn{2}{c|}{Target / $\delta\alpha_{T}$ region} & $N_{bin}$ & GENIE hA2018 & GENIE hN2018 & GENIE G4 Bertini INC \\ 
\hline 
\multicolumn{2}{c|}{C + O} & 25 & 28.7 (0.28) & 27.5 (0.33) & \cellcolor{lightgray} 49.6 (0.00) \\ 
\hline 
\multicolumn{2}{c|}{C-only} & 17 & 19.3 (0.31) & 18.4 (0.36) & \cellcolor{lightgray} 32.6 (0.01) \\ 
\hline 
\multicolumn{2}{c|}{O-only} & 8 & 6.7 (0.57) & 6.3 (0.61) & 7.0 (0.53) \\ 
\hline 
\multirow{3}{*}{C + O} & $0^{\circ} < \delta\alpha_{T} \leqslant 75^{\circ}$ & 7 & 5.1 (0.64) & 4.6 (0.71) & 5.6 (0.58) \\ 
 & $75^{\circ} < \delta\alpha_{T} \leqslant 135^{\circ}$ & 9 & 10.0 (0.35) & 9.4 (0.40) & 10.6 (0.30) \\ 
 & $135^{\circ} < \delta\alpha_{T} < 180^{\circ}$ & 9 & 11.7 (0.23) & 10.1 (0.34) & \cellcolor{lightgray} 25.0 (0.00) \\ 
\hline 
\multirow{3}{*}{C-only} & $0^{\circ} < \delta\alpha_{T} \leqslant 75^{\circ}$ & 5 & 2.5 (0.78) & 2.3 (0.80) & 2.5 (0.78) \\ 
 & $75^{\circ} < \delta\alpha_{T} \leqslant 135^{\circ}$ & 6 & 6.6 (0.36) & 5.8 (0.45) & 8.7 (0.19) \\ 
 & $135^{\circ} < \delta\alpha_{T} < 180^{\circ}$ & 6 & 8.2 (0.23) & 7.4 (0.28) & \cellcolor{lightgray} 15.1 (0.02) \\ 
\hline 
\multirow{3}{*}{O-only} & $0^{\circ} < \delta\alpha_{T} \leqslant 75^{\circ}$ & 2 & 3.2 (0.21) & 2.7 (0.26) & 2.6 (0.27) \\ 
 & $75^{\circ} < \delta\alpha_{T} \leqslant 135^{\circ}$ & 3 & 4.1 (0.25) & 4.0 (0.26) & 2.3 (0.52) \\ 
 & $135^{\circ} < \delta\alpha_{T} < 180^{\circ}$ & 3 & 2.6 (0.46) & 1.9 (0.59) & 3.9 (0.28) \\ 
\hline 
\hline
\end{tabular}
\end{table*}

\begin{table*}[!htbp]
\caption{$\chi^{2}$ values and corresponding $p$-values (in parentheses) for the $\nu_{\mu}$ CC$0\pi Np$ cross-section comparisons with various neutrino-nucleus interaction models from Sec.~\ref{subsec:modelcomp_FSI} in $p_{N}$--$\cos\theta_{\mu}$ space. The $\chi^{2}$ and $p$-values are computed separately for the joint carbon and oxygen data, as well as for carbon-only and oxygen-only data. Additionally, the $\chi^{2}$ and $p$-values are provided for different $\cos\theta_{\mu}$ regions. Cells are highlighted with a light gray background where $p \le 0.05$.}
\label{tab:modelcomp_tab_FSI_2}
\centering
\renewcommand{\arraystretch}{1.2}
\begin{tabular}{c|c|c|c|c|c}
\hline
\hline
\multicolumn{2}{c|}{Target / $\cos\theta_{\mu}$ region} & $N_{bin}$ & GENIE hA2018 & GENIE hN2018 & GENIE G4 Bertini INC \\ 
\hline 
\multicolumn{2}{c|}{C + O} & 30 & 29.6 (0.48) & 27.8 (0.58) & 35.1 (0.24) \\ 
\hline 
\multicolumn{2}{c|}{C-only} & 19 & 15.9 (0.66) & 13.9 (0.79) & 20.1 (0.39) \\ 
\hline 
\multicolumn{2}{c|}{O-only} & 11 & 11.9 (0.37) & 11.6 (0.39) & 11.1 (0.44) \\ 
\hline 
\multirow{5}{*}{C + O} & $-0.6 < \cos\theta_{\mu} \leqslant -0.1$ & 4 & 7.9 (0.10) & 5.6 (0.23) & \cellcolor{lightgray} 10.0 (0.04) \\ 
 & $-0.1 < \cos\theta_{\mu} \leqslant 0.15$ & 2 & 0.4 (0.82) & 0.2 (0.90) & 0.2 (0.89) \\ 
 & $0.15 < \cos\theta_{\mu} \leqslant 0.52$ & 6 & 9.1 (0.17) & 8.7 (0.19) & 9.0 (0.17) \\ 
 & $0.52 < \cos\theta_{\mu} \leqslant 0.83$ & 9 & 4.5 (0.88) & 4.4 (0.89) & 6.0 (0.74) \\ 
 & $0.83 < \cos\theta_{\mu} < 1$ & 9 & 5.5 (0.79) & 5.5 (0.79) & 4.7 (0.86) \\ 
\hline 
\multirow{5}{*}{C-only} & $-0.6 < \cos\theta_{\mu} \leqslant -0.1$ & 2 & \cellcolor{lightgray} 6.7 (0.04) & 5.0 (0.08) & \cellcolor{lightgray} 6.4 (0.04) \\ 
 & $-0.1 < \cos\theta_{\mu} \leqslant 0.15$ & 1 & 0.4 (0.54) & 0.1 (0.71) & 0.0 (0.90) \\ 
 & $0.15 < \cos\theta_{\mu} \leqslant 0.52$ & 4 & 2.9 (0.57) & 2.1 (0.71) & 5.3 (0.26) \\ 
 & $0.52 < \cos\theta_{\mu} \leqslant 0.83$ & 6 & 3.5 (0.75) & 3.5 (0.75) & 3.1 (0.80) \\ 
 & $0.83 < \cos\theta_{\mu} < 1$ & 6 & 1.2 (0.98) & 1.1 (0.98) & 2.5 (0.86) \\ 
\hline 
\multirow{5}{*}{O-only} & $-0.6 < \cos\theta_{\mu} \leqslant -0.1$ & 2 & 0.6 (0.75) & 0.3 (0.86) & 0.8 (0.66) \\ 
 & $-0.1 < \cos\theta_{\mu} \leqslant 0.15$ & 1 & 0.1 (0.77) & 0.1 (0.71) & 0.2 (0.62) \\ 
 & $0.15 < \cos\theta_{\mu} \leqslant 0.52$ & 2 & \cellcolor{lightgray} 7.1 (0.03) & \cellcolor{lightgray} 6.8 (0.03) & 4.0 (0.14) \\ 
 & $0.52 < \cos\theta_{\mu} \leqslant 0.83$ & 3 & 0.9 (0.82) & 0.7 (0.88) & 2.2 (0.53) \\ 
 & $0.83 < \cos\theta_{\mu} < 1$ & 3 & 4.0 (0.27) & 3.9 (0.27) & 1.7 (0.63) \\ 
\hline 
\hline
\end{tabular}
\end{table*}

\section{Conclusions}
\label{sec:conclu}

We present the first measurement of the $\nu_{\mu}$ CC$0\pi Np$ differential cross section jointly on carbon and oxygen in the two-dimensional $\delta p_{T}$--$\delta \alpha_{T}$ and $p_{N}$--$\cos\theta_{\mu}$ phase spaces. This measurement efficiently isolates different nuclear-effect contributions in different phase space regions, making it a valuable input for the development and validation of neutrino–nucleus interaction models on carbon and oxygen, with direct relevance to the T2K and Hyper-Kamiokande physics programs.

The measured cross sections have relative uncertainties of 20–70\% in the $\delta p_{T}$--$\delta \alpha_{T}$ space and 15–80\% in the $p_{N}$--$\cos\theta_{\mu}$ space, with statistical uncertainties dominating in all bins. The post-fit $p$-value for the data fit is approximately 0.008 for the $\delta p_{T}$--$\delta \alpha_{T}$ measurement and 0.122 for the $p_{N}$--$\cos\theta_{\mu}$ measurement. The small $p$-value observed in the former has been investigated and is primarily attributed to the limited range of modeling variations for the background predictions allowed by the current systematic parameterization. Notably, this background model, consistent with that used in Ref.~\cite{T2K:2025yoy}, represents the most sophisticated available parameterization for these channels in the relevant energy range. Its inability to fully describe the data, as indicated by the low $p$-value, highlights the discriminating power of the KI observables and indicates the need for improved modeling of the relevant background processes. Additional studies were performed to assess the impact of this limitation on the extracted signal cross sections, and the effect was found to be small, demonstrating the robustness of the reported results.

The measured cross sections were compared with a broad range of models covering the nucleon ground state, multi-nucleon interactions, and FSIs. The comparisons identify two distinct modeling frameworks favored by the data. The first incorporates a SF ground state or ED-RMF model for CCQE interactions, the Valencia model for multi-nucleon correlations and the NEUT-style INC for FSI. An alternative configuration combines a LFG description of the nuclear ground state with the SuSAv2 model for both the inclusive CCQE and 2p2h contributions, together with a GENIE-style INC (hN) for FSI. Despite these successes, discrepancies persist in the intermediate $\delta p_{T}$ and $p_{N}$ regions for both carbon and oxygen targets where the models exhibit deficiencies in describing the data. This potentially indicates that enhanced contributions from SRCs or multi-nucleon interactions are necessary. However, tensions in the $p_{N}$ distributions can be partially alleviated by incorporating the CRPA model for CCQE interactions, suggesting an additional direction for improving neutrino–nucleus interaction modeling in this phase-space region.

The results of this analysis remain limited by statistical uncertainties. The next phase of the T2K experiment, featuring the upgraded ND280 detector~\cite{T2K:2019bbb}, will provide increased statistics through a larger active target mass of carbon and a more intense neutrino beam. The upgraded ND280 includes a new scintillator-based target detector (Super-FGD)~\cite{Abe:2026elv} with enhanced capabilities for reconstructing low-momentum protons and improved angular acceptance for both muons and hadrons. These improvements will significantly enhance future measurements of KI observables and increase sensitivity to nuclear effects. In addition, the improved statistics will allow the correlation between neutrino-nucleus interactions on these two targets to be fully explored through a direct measurement of the oxygen-to-carbon cross-section ratio, which is statistically limited in the present analysis. Collectively, they will enable a more precise understanding of neutrino–nucleus interactions and contribute to reducing systematic uncertainties in neutrino oscillation measurements.

\section*{Acknowledgments}

The T2K collaboration would like to thank the J-PARC staff for superb accelerator performance. We thank the CERN NA61/SHINE Collaboration for providing valuable particle production data. We acknowledge the support of MEXT, JSPS KAKENHI (JP16H06288, JP18K03682, JP18H03701, JP18H05537, JP19J01119, JP19J22440, JP19J22258, JP20H00162, JP20H00149, JP20J20304, JP24K17065) and bilateral programs (JPJSBP120204806, JPJSBP120209601),  Japan; UGent-BOF and FWO-Flanders, Belgium; NSERC, the NRC, and CFI, Canada; the CEA and CNRS/IN2P3, France; the Deutsche Forschungsgemeinschaft (DFG, German Research Foundation) 397763730, 517206441, Germany; the NKFIH (NKFIH 137812 and TKP2021-NKTA-64), Hungary; the INFN, Italy; the Ministry of Science and Higher Education (2023/WK/04) and the National Science Centre (UMO-2018/30/E/ST2/00441, UMO-2022/46/E/ST2/00336 and UMO-2021/43/D/ST2/01504), Poland;  the RSF (RSF 26-12-00495) and the Ministry of Science and Higher Education, Russia;  MICINN  (PID2022-136297NB-I00 /AEI/10.13039/501100011033/ FEDER, UE, PID2024-157541NB-I00 (UAM) and PID2023-146401NB-I00 (US), Severo Ochoa Centres of Excellence Programme 2025-2029 (CEX2024001441-S),  Government of Andalucia (FQM160) and the University of Tokyo ICRR's Inter-University Research Program FY2025 Ref. J1, and ERDF and European Union (UAM: H2020-MSCA-RISE-GA872549- SK2HK) and NextGenerationEU funds (PRTR-C17.I1) and  Generalitat de Catalunya (AGAUR 2021-SGR-01506, CERCA program) University of Seville grant (RYC2022-035203-I funded by MICIU/AEI/10.13039/501100011033, ``ERDF a way of making Europe'' and FSE+, Ayudas ``Atracción de Investigadores con Alto Potencial''. Ref. VIIPPIT-2025, and Secretariat for Universities and Research of the Ministry of Business and Knowledge of the Government of Catalonia and the European Social Fund (2022FI\_B 00336), Spain; the SNSF and SERI (200021\_185012, 200020\_188533, 20FL21\_186178I), Switzerland; the STFC and UKRI, UK; the DOE, USA; and NAFOSTED (103.99-2023.144,IZVSZ2.203433), Vietnam. We also thank CERN for the UA1/NOMAD magnet, DESY for the HERA-B magnet mover system, the BC DRI Group, Prairie DRI Group, ACENET, SciNet, and CalculQuebec consortia in the Digital Research Alliance of Canada, and GridPP in the United Kingdom, and the CNRS/IN2P3 Computing Center in France and NERSC (HEP-ERCAP0028625). In addition, the participation of individual researchers and institutions has been further supported by funds from the ERC (FP7), “la Caixa” Foundation  (ID 100010434, fellowship code LCF/BQ/IN17/11620050), the European Union’s Horizon 2020 Research and Innovation Programme under the Marie Sklodowska-Curie grant agreement numbers 713673 and 754496, and H2020 grant numbers  RISE-GA822070-JENNIFER2 2020 and RISE-GA872549-SK2HK, the Horizon Europe Marie Sklodowska-Curie Staff Exchange project JENNIFER3 grant 101183137; the JSPS, Japan; the Royal Society, UK; French ANR grant number ANR-19-CE31-0001 and ANR-21-CE31-0008; and  Sorbonne Université Emergences programmes; the SNF Eccellenza grant number PCEFP2\_203261;  the VAST-JSPS (No. QTJP01.02/20-22);  and the DOE Early Career programme, USA. For the purposes of open access, the authors have applied a Creative Commons Attribution license to any Author Accepted Manuscript version arising.

\section*{Data availability}

The measurement results have an associated data release that can be found in Ref.~\cite{data_release}.

\bibliographystyle{apsrev4-2}
\bibliography{biblio}

\clearpage

\appendix

\section{Discussion of Small $p$-value in $\delta p_{T}$--$\delta \alpha_{T}$ Space Fit}
\label{sec:disc_small_pval}

As shown in Sec.~\ref{subsec:result_1}, the post-fit $\chi^{2}$ for the $\delta p_{T}$--$\delta \alpha_{T}$ measurement corresponds to a $p$-value of $\sim 0.008$, below the conventional 0.05 threshold. Detailed studies were performed to identify the sources contributing to this low $p$-value. A dominant contribution was found to arise from the limited flexibility of the background interaction systematic model, which induces a tension between the MC template and data. To investigate this further, an alternative fit was performed in which additional freedom was introduced into the background modeling through the inclusion of free background template parameters, analogous to the signal parameters $t_i$ in Eq.~\ref{eq:event_weight}. Due to limited statistics of background interactions in the selected samples, these parameters were defined using a coarser binning than that reported in Tab.~\ref{tab:binedge_xsec_1}. With this modified fit, a substantially improved $p$-value of $\sim 0.09$ (compared to $\sim 0.008$) was obtained. The extracted signal cross sections from this alternative fit were compared to the original results presented in this work, and no significant deviations were observed. Both fits yielded similar measurement uncertainties, with the maximum cross-section discrepancy in any bin remaining below 60\% of the corresponding uncertainty. Moreover, this modified fit has no significant impact on the measured cross sections in the intermediate $\delta p_{T}$ regions where the data excess is observed. The excess persists. This demonstrates the robustness of the differential cross-section measurement in the $\delta p_{T}$--$\delta \alpha_{T}$ space and indicates that the low $p$-value has no significant impact on the reported results.

\section{Analysis Binning}

\subsection{Cross-section Binning}
\label{subsec:xsec_binning}
\vspace{-0.4cm}
\begin{table}[H]
\caption{Cross-section binning used in the $\delta p_{T}$--$\delta \alpha_{T}$ space measurement. The binnings are defined in the true analysis observable space.}
\label{tab:binedge_xsec_1}
\centering
\renewcommand{\arraystretch}{1.2}
\begin{tabular}{c|c|c|c}
\hline
\hline
Target & Bin Index & $\delta \alpha_{T}$ [$^{\circ}$] & $\delta p_{T}$ [GeV/c] \\
\hline
\multirow{18}{*}{Carbon} & 0 & 0.00 - 75.00 & 0.00 - 0.09 \\
 & 1 & 0.00 - 75.00 & 0.09 - 0.17 \\
 & 2 & 0.00 - 75.00 & 0.17 - 0.24 \\
 & 3 & 0.00 - 75.00 & 0.24 - 0.36 \\
 & 4 & 0.00 - 75.00 & 0.36 - 1.20 \\
 & 5 & 75.00 - 135.00 & 0.00 - 0.09 \\
 & 6 & 75.00 - 135.00 & 0.09 - 0.17 \\
 & 7 & 75.00 - 135.00 & 0.17 - 0.24 \\
 & 8 & 75.00 - 135.00 & 0.24 - 0.36 \\
 & 9 & 75.00 - 135.00 & 0.36 - 0.52 \\
 & 10 & 75.00 - 135.00 & 0.52 - 1.20 \\
 & 11 & 135.00 - 180.00 & 0.00 - 0.09 \\
 & 12 & 135.00 - 180.00 & 0.09 - 0.17 \\
 & 13 & 135.00 - 180.00 & 0.17 - 0.24 \\
 & 14 & 135.00 - 180.00 & 0.24 - 0.36 \\
 & 15 & 135.00 - 180.00 & 0.36 - 0.52 \\
 & 16 & 135.00 - 180.00 & 0.52 - 1.20 \\
\hline
\multirow{9}{*}{Oxygen} & 0 & 0.00 - 75.00 & 0.00 - 0.17 \\
 & 1 & 0.00 - 75.00 & 0.17 - 1.20 \\
 & 2 & 75.00 - 135.00 & 0.00 - 0.17 \\
 & 3 & 75.00 - 135.00 & 0.17 - 0.36 \\
 & 4 & 75.00 - 135.00 & 0.36 - 1.20 \\
 & 5 & 135.00 - 180.00 & 0.00 - 0.17 \\
 & 6 & 135.00 - 180.00 & 0.17 - 0.36 \\
 & 7 & 135.00 - 180.00 & 0.36 - 1.20 \\
\hline
\hline
\end{tabular}
\end{table}

Tab.~\ref{tab:binedge_xsec_1} and Tab.~\ref{tab:binedge_xsec_2} summarize the cross-section binning used in the $\delta p_{T}$--$\delta \alpha_{T}$ and $p_{N}$--$\cos\theta_{\mu}$ space measurements.

\begin{table}[H]
\caption{Cross-section binning used in the $p_{N}$--$\cos\theta_{\mu}$ space measurement. The binnings are defined in the true analysis observable space.}
\label{tab:binedge_xsec_2}
\centering
\renewcommand{\arraystretch}{1.2}
\begin{tabular}{c|c|c|c}
\hline
\hline
Target & Bin Index & $\cos\theta_{\mu}$ & $p_{N}$ [GeV/c] \\
\hline
\multirow{20}{*}{Carbon} & 0 & -0.60 - -0.10 & 0.00 - 0.22 \\
 & 1 & -0.60 - -0.10 & 0.22 - 1.20 \\
 & 2 & -0.10 - 0.15 & 0.00 - 1.20 \\
 & 3 & 0.15 - 0.52 & 0.00 - 0.14 \\
 & 4 & 0.15 - 0.52 & 0.14 - 0.22 \\
 & 5 & 0.15 - 0.52 & 0.22 - 0.42 \\
 & 6 & 0.15 - 0.52 & 0.42 - 1.20 \\
 & 7 & 0.52 - 0.83 & 0.00 - 0.14 \\
 & 8 & 0.52 - 0.83 & 0.14 - 0.22 \\
 & 9 & 0.52 - 0.83 & 0.22 - 0.30 \\
 & 10 & 0.52 - 0.83 & 0.30 - 0.42 \\
 & 11 & 0.52 - 0.83 & 0.42 - 0.53 \\
 & 12 & 0.52 - 0.83 & 0.53 - 1.20 \\
 & 13 & 0.83 - 1.00 & 0.00 - 0.14 \\
 & 14 & 0.83 - 1.00 & 0.14 - 0.22 \\
 & 15 & 0.83 - 1.00 & 0.22 - 0.30 \\
 & 16 & 0.83 - 1.00 & 0.30 - 0.42 \\
 & 17 & 0.83 - 1.00 & 0.42 - 0.53 \\
 & 18 & 0.83 - 1.00 & 0.53 - 1.20 \\
\hline
\multirow{12}{*}{Oxygen} & 0 & -0.60 - -0.10 & 0.00 - 0.22 \\
 & 1 & -0.60 - -0.10 & 0.22 - 1.20 \\
 & 2 & -0.10 - 0.15 & 0.00 - 1.20 \\
 & 3 & 0.15 - 0.52 & 0.00 - 0.22 \\
 & 4 & 0.15 - 0.52 & 0.22 - 1.20 \\
 & 5 & 0.52 - 0.83 & 0.00 - 0.22 \\
 & 6 & 0.52 - 0.83 & 0.22 - 0.42 \\
 & 7 & 0.52 - 0.83 & 0.42 - 1.20 \\
 & 8 & 0.83 - 1.00 & 0.00 - 0.22 \\
 & 9 & 0.83 - 1.00 & 0.22 - 0.42 \\
 & 10 & 0.83 - 1.00 & 0.42 - 1.20 \\
\hline
\hline
\end{tabular}
\end{table}

\subsection{Sample Binning}
\label{subsec:samp_binning}

The sample binning is summarized in the following tables:
\begin{itemize}
\item \textbf{$\delta p_{T}$--$\delta \alpha_{T}$ combination}: FGD1 samples (Tab.~\ref{tab:sample_bin_1_FGD1_sig} and Tab.~\ref{tab:sample_bin_1_FGD1_cont}). FGD2X samples (Tab.~\ref{tab:sample_bin_1_FGD2X_sig} and Tab.~\ref{tab:sample_bin_1_FGD2X_cont}). FGD2Y samples (Tab.~\ref{tab:sample_bin_1_FGD2Y_sig} and Tab.~\ref{tab:sample_bin_1_FGD2Y_cont}).

\item \textbf{$p_{N}$--$\cos\theta_{\mu}$ combination}: FGD1 samples (Tab.~\ref{tab:sample_bin_2_FGD1_sig} and Tab.~\ref{tab:sample_bin_2_FGD1_cont}). FGD2X samples (Tab.~\ref{tab:sample_bin_2_FGD2X_sig} and Tab.~\ref{tab:sample_bin_2_FGD2X_cont}). FGD2Y samples (Tab.~\ref{tab:sample_bin_2_FGD2Y_sig} and Tab.~\ref{tab:sample_bin_2_FGD2Y_cont}). 
\end{itemize}

\begin{table}[!htbp]
\caption{FGD1 signal sample binning used in the $\delta p_{T}$--$\delta \alpha_{T}$ space measurement. The bins are defined in the reconstructed observable space. The final bin, labeled ``OOPS'' (out-of-phase-space), contains selected events that do not satisfy the phase space constraints defined in Tab.~\ref{tab:ps_cons}. This bin is included to estimate the migration of signal events across phase space boundaries due to finite detector resolution, utilizing a data-driven method.}
\label{tab:sample_bin_1_FGD1_sig}
\centering
\resizebox{0.45\textwidth}{!}{
\renewcommand{\arraystretch}{1.2}
\begin{tabular}{c|c|c|c}
\hline
\hline
Sample & Bin Index & $\delta \alpha_{T}$ [$^{\circ}$] & $\delta p_{T}$ [GeV/c] \\
\hline
\multirow{18}{*}{TPC $\mu$ + TPC $p$} & 0 & 0.00 - 75.00 & 0.00 - 0.09 \\
 & 1 & 0.00 - 75.00 & 0.09 - 0.17 \\
 & 2 & 0.00 - 75.00 & 0.17 - 0.24 \\
 & 3 & 0.00 - 75.00 & 0.24 - 0.36 \\
 & 4 & 0.00 - 75.00 & 0.36 - 1.20 \\
 & 5 & 75.00 - 135.00 & 0.00 - 0.09 \\
 & 6 & 75.00 - 135.00 & 0.09 - 0.17 \\
 & 7 & 75.00 - 135.00 & 0.17 - 0.24 \\
 & 8 & 75.00 - 135.00 & 0.24 - 0.36 \\
 & 9 & 75.00 - 135.00 & 0.36 - 0.52 \\
 & 10 & 75.00 - 135.00 & 0.52 - 1.20 \\
 & 11 & 135.00 - 180.00 & 0.00 - 0.09 \\
 & 12 & 135.00 - 180.00 & 0.09 - 0.17 \\
 & 13 & 135.00 - 180.00 & 0.17 - 0.24 \\
 & 14 & 135.00 - 180.00 & 0.24 - 0.36 \\
 & 15 & 135.00 - 180.00 & 0.36 - 0.52 \\
 & 16 & 135.00 - 180.00 & 0.52 - 1.20 \\
 & 17 & OOPS & OOPS \\
\hline
\multirow{18}{*}{TPC $\mu$ + FGD $p$ (+ $Np$)} & 0 & 0.00 - 75.00 & 0.00 - 0.09 \\
 & 1 & 0.00 - 75.00 & 0.09 - 0.17 \\
 & 2 & 0.00 - 75.00 & 0.17 - 0.24 \\
 & 3 & 0.00 - 75.00 & 0.24 - 0.36 \\
 & 4 & 0.00 - 75.00 & 0.36 - 1.20 \\
 & 5 & 75.00 - 135.00 & 0.00 - 0.09 \\
 & 6 & 75.00 - 135.00 & 0.09 - 0.17 \\
 & 7 & 75.00 - 135.00 & 0.17 - 0.24 \\
 & 8 & 75.00 - 135.00 & 0.24 - 0.36 \\
 & 9 & 75.00 - 135.00 & 0.36 - 0.52 \\
 & 10 & 75.00 - 135.00 & 0.52 - 1.20 \\
 & 11 & 135.00 - 180.00 & 0.00 - 0.09 \\
 & 12 & 135.00 - 180.00 & 0.09 - 0.17 \\
 & 13 & 135.00 - 180.00 & 0.17 - 0.24 \\
 & 14 & 135.00 - 180.00 & 0.24 - 0.36 \\
 & 15 & 135.00 - 180.00 & 0.36 - 0.52 \\
 & 16 & 135.00 - 180.00 & 0.52 - 1.20 \\
 & 17 & OOPS & OOPS \\
\hline
\multirow{18}{*}{FGD $\mu$ + TPC $p$ (+ $Np$)} & 0 & 0.00 - 75.00 & 0.00 - 0.09 \\
 & 1 & 0.00 - 75.00 & 0.09 - 0.17 \\
 & 2 & 0.00 - 75.00 & 0.17 - 0.24 \\
 & 3 & 0.00 - 75.00 & 0.24 - 0.36 \\
 & 4 & 0.00 - 75.00 & 0.36 - 1.20 \\
 & 5 & 75.00 - 135.00 & 0.00 - 0.09 \\
 & 6 & 75.00 - 135.00 & 0.09 - 0.17 \\
 & 7 & 75.00 - 135.00 & 0.17 - 0.24 \\
 & 8 & 75.00 - 135.00 & 0.24 - 0.36 \\
 & 9 & 75.00 - 135.00 & 0.36 - 0.52 \\
 & 10 & 75.00 - 135.00 & 0.52 - 1.20 \\
 & 11 & 135.00 - 180.00 & 0.00 - 0.09 \\
 & 12 & 135.00 - 180.00 & 0.09 - 0.17 \\
 & 13 & 135.00 - 180.00 & 0.17 - 0.24 \\
 & 14 & 135.00 - 180.00 & 0.24 - 0.36 \\
 & 15 & 135.00 - 180.00 & 0.36 - 0.52 \\
 & 16 & 135.00 - 180.00 & 0.52 - 1.20 \\
 & 17 & OOPS & OOPS \\
\hline
\hline
\end{tabular}
}
\end{table}

\begin{table}[!htbp]
\caption{FGD1 control sample binning used in the $\delta p_{T}$--$\delta \alpha_{T}$ space measurement. The binnings are defined in the reconstructed analysis observable space.}
\label{tab:sample_bin_1_FGD1_cont}
\centering
\renewcommand{\arraystretch}{1.2}
\begin{tabular}{c|c|c|c}
\hline
\hline
Sample & Bin Index & $\delta \alpha_{T}$ [$^{\circ}$] & $\delta p_{T}$ [GeV/c] \\
\hline
\multirow{3}{*}{CCproton ME} & 0 & 0.00 - 135.00 & 0.00 - 1.20 \\
 & 1 & 135.00 - 180.00 & 0.00 - 0.36 \\
 & 2 & 135.00 - 180.00 & 0.36 - 1.20 \\
\hline
\multirow{4}{*}{CCproton 1$\pi^{+}$} & 0 & 0.00 - 135.00 & 0.00 - 0.36 \\
 & 1 & 0.00 - 135.00 & 0.36 - 1.20 \\
 & 2 & 135.00 - 180.00 & 0.00 - 0.36 \\
 & 3 & 135.00 - 180.00 & 0.36 - 1.20 \\
\hline
\hline
\end{tabular}
\end{table}

\begin{table}[H]
\caption{FGD2X signal sample binning used in the $\delta p_{T}$--$\delta \alpha_{T}$ space measurement. The bins are defined in the reconstructed observable space. The final bin, labeled ``OOPS'' (out-of-phase-space), contains selected events that do not satisfy the phase space constraints defined in Tab.~\ref{tab:ps_cons}. This bin is included to estimate the migration of signal events across phase space boundaries due to finite detector resolution, utilizing a data-driven method.}
\label{tab:sample_bin_1_FGD2X_sig}
\centering
\resizebox{0.45\textwidth}{!}{
\renewcommand{\arraystretch}{1.2}
\begin{tabular}{c|c|c|c}
\hline
\hline
Sample & Bin Index & $\delta \alpha_{T}$ [$^{\circ}$] & $\delta p_{T}$ [GeV/c] \\
\hline
\multirow{9}{*}{TPC $\mu$ + TPC $p$} & 0 & 0.00 - 75.00 & 0.00 - 0.17 \\
 & 1 & 0.00 - 75.00 & 0.17 - 1.20 \\
 & 2 & 75.00 - 135.00 & 0.00 - 0.17 \\
 & 3 & 75.00 - 135.00 & 0.17 - 0.36 \\
 & 4 & 75.00 - 135.00 & 0.36 - 1.20 \\
 & 5 & 135.00 - 180.00 & 0.00 - 0.17 \\
 & 6 & 135.00 - 180.00 & 0.17 - 0.36 \\
 & 7 & 135.00 - 180.00 & 0.36 - 1.20 \\
 & 8 & OOPS & OOPS \\
\hline
\multirow{9}{*}{TPC $\mu$ + FGD $p$ (+ $Np$)} & 0 & 0.00 - 75.00 & 0.00 - 0.17 \\
 & 1 & 0.00 - 75.00 & 0.17 - 1.20 \\
 & 2 & 75.00 - 135.00 & 0.00 - 0.17 \\
 & 3 & 75.00 - 135.00 & 0.17 - 0.36 \\
 & 4 & 75.00 - 135.00 & 0.36 - 1.20 \\
 & 5 & 135.00 - 180.00 & 0.00 - 0.17 \\
 & 6 & 135.00 - 180.00 & 0.17 - 0.36 \\
 & 7 & 135.00 - 180.00 & 0.36 - 1.20 \\
 & 8 & OOPS & OOPS \\
\hline
\multirow{9}{*}{FGD $\mu$ + TPC $p$ (+ $Np$)} & 0 & 0.00 - 75.00 & 0.00 - 0.17 \\
 & 1 & 0.00 - 75.00 & 0.17 - 1.20 \\
 & 2 & 75.00 - 135.00 & 0.00 - 0.17 \\
 & 3 & 75.00 - 135.00 & 0.17 - 0.36 \\
 & 4 & 75.00 - 135.00 & 0.36 - 1.20 \\
 & 5 & 135.00 - 180.00 & 0.00 - 0.17 \\
 & 6 & 135.00 - 180.00 & 0.17 - 0.36 \\
 & 7 & 135.00 - 180.00 & 0.36 - 1.20 \\
 & 8 & OOPS & OOPS \\
\hline
\hline
\end{tabular}
}
\end{table}

\begin{table}[H]
\caption{FGD2X control sample binning used in the $\delta p_{T}$--$\delta \alpha_{T}$ space measurement. The binnings are defined in the reconstructed analysis observable space.}
\label{tab:sample_bin_1_FGD2X_cont}
\centering
\renewcommand{\arraystretch}{1.2}
\begin{tabular}{c|c|c|c}
\hline
\hline
Sample & Bin Index & $\delta \alpha_{T}$ [$^{\circ}$] & $\delta p_{T}$ [GeV/c] \\
\hline
\multirow{3}{*}{CCproton ME} & 0 & 0.00 - 135.00 & 0.00 - 1.20 \\
 & 1 & 135.00 - 180.00 & 0.00 - 0.36 \\
 & 2 & 135.00 - 180.00 & 0.36 - 1.20 \\
\hline
\multirow{4}{*}{CCproton 1$\pi^{+}$} & 0 & 0.00 - 135.00 & 0.00 - 0.36 \\
 & 1 & 0.00 - 135.00 & 0.36 - 1.20 \\
 & 2 & 135.00 - 180.00 & 0.00 - 0.36 \\
 & 3 & 135.00 - 180.00 & 0.36 - 1.20 \\
\hline
\hline
\end{tabular}
\end{table}

\begin{table}[H]
\caption{FGD2Y signal sample binning used in the $\delta p_{T}$--$\delta \alpha_{T}$ space measurement. The bins are defined in the reconstructed observable space. The final bin, labeled ``OOPS'' (out-of-phase-space), contains selected events that do not satisfy the phase space constraints defined in Tab.~\ref{tab:ps_cons}. This bin is included to estimate the migration of signal events across phase space boundaries due to finite detector resolution, utilizing a data-driven method.}
\label{tab:sample_bin_1_FGD2Y_sig}
\centering
\resizebox{0.45\textwidth}{!}{
\renewcommand{\arraystretch}{1.2}
\begin{tabular}{c|c|c|c}
\hline
\hline
Sample & Bin Index & $\delta \alpha_{T}$ [$^{\circ}$] & $\delta p_{T}$ [GeV/c] \\
\hline
\multirow{10}{*}{TPC $\mu$ + TPC $p$} & 0 & 0.00 - 75.00 & 0.00 - 0.17 \\
 & 1 & 0.00 - 75.00 & 0.17 - 0.36 \\
 & 2 & 0.00 - 75.00 & 0.36 - 1.20 \\
 & 3 & 75.00 - 135.00 & 0.00 - 0.17 \\
 & 4 & 75.00 - 135.00 & 0.17 - 0.36 \\
 & 5 & 75.00 - 135.00 & 0.36 - 1.20 \\
 & 6 & 135.00 - 180.00 & 0.00 - 0.17 \\
 & 7 & 135.00 - 180.00 & 0.17 - 0.36 \\
 & 8 & 135.00 - 180.00 & 0.36 - 1.20 \\
 & 9 & OOPS & OOPS \\
\hline
\multirow{9}{*}{TPC $\mu$ + FGD $p$ (+ $Np$)} & 0 & 0.00 - 75.00 & 0.00 - 0.17 \\
 & 1 & 0.00 - 75.00 & 0.17 - 1.20 \\
 & 2 & 75.00 - 135.00 & 0.00 - 0.17 \\
 & 3 & 75.00 - 135.00 & 0.17 - 0.36 \\
 & 4 & 75.00 - 135.00 & 0.36 - 1.20 \\
 & 5 & 135.00 - 180.00 & 0.00 - 0.17 \\
 & 6 & 135.00 - 180.00 & 0.17 - 0.36 \\
 & 7 & 135.00 - 180.00 & 0.36 - 1.20 \\
 & 8 & OOPS & OOPS \\
\hline
\multirow{9}{*}{FGD $\mu$ + TPC $p$ (+ $Np$)} & 0 & 0.00 - 75.00 & 0.00 - 0.17 \\
 & 1 & 0.00 - 75.00 & 0.17 - 1.20 \\
 & 2 & 75.00 - 135.00 & 0.00 - 0.17 \\
 & 3 & 75.00 - 135.00 & 0.17 - 0.36 \\
 & 4 & 75.00 - 135.00 & 0.36 - 1.20 \\
 & 5 & 135.00 - 180.00 & 0.00 - 0.17 \\
 & 6 & 135.00 - 180.00 & 0.17 - 0.36 \\
 & 7 & 135.00 - 180.00 & 0.36 - 1.20 \\
 & 8 & OOPS & OOPS \\
\hline
\hline
\end{tabular}
}
\end{table}

\begin{table}[H]
\caption{FGD2Y control sample binning used in the $\delta p_{T}$--$\delta \alpha_{T}$ space measurement. The binnings are defined in the reconstructed analysis observable space.}
\label{tab:sample_bin_1_FGD2Y_cont}
\centering
\renewcommand{\arraystretch}{1.2}
\begin{tabular}{c|c|c|c}
\hline
\hline
Sample & Bin Index & $\delta \alpha_{T}$ [$^{\circ}$] & $\delta p_{T}$ [GeV/c] \\
\hline
\multirow{3}{*}{CCproton ME} & 0 & 0.00 - 135.00 & 0.00 - 1.20 \\
 & 1 & 135.00 - 180.00 & 0.00 - 0.36 \\
 & 2 & 135.00 - 180.00 & 0.36 - 1.20 \\
\hline
\multirow{4}{*}{CCproton 1$\pi^{+}$} & 0 & 0.00 - 135.00 & 0.00 - 0.36 \\
 & 1 & 0.00 - 135.00 & 0.36 - 1.20 \\
 & 2 & 135.00 - 180.00 & 0.00 - 0.36 \\
 & 3 & 135.00 - 180.00 & 0.36 - 1.20 \\
\hline
\hline
\end{tabular}
\end{table}

\begin{table}[H]
\caption{FGD1 signal sample binning used in the $p_{N}$--$\cos\theta_{\mu}$ space measurement. The bins are defined in the reconstructed observable space. The final bin, labeled ``OOPS'' (out-of-phase-space), contains selected events that do not satisfy the phase space constraints defined in Tab.~\ref{tab:ps_cons}. This bin is included to estimate the migration of signal events across phase space boundaries due to finite detector resolution, utilizing a data-driven method.}
\label{tab:sample_bin_2_FGD1_sig}
\centering
\resizebox{0.45\textwidth}{!}{
\renewcommand{\arraystretch}{1.2}
\begin{tabular}{c|c|c|c}
\hline
\hline
Sample & Bin Index & $\cos\theta_{\mu}$ & $p_{N}$ [GeV/c] \\
\hline
\multirow{18}{*}{TPC $\mu$ + TPC $p$} & 0 & -0.60 - 0.15 & 0.00 - 1.20 \\
 & 1 & 0.15 - 0.52 & 0.00 - 0.14 \\
 & 2 & 0.15 - 0.52 & 0.14 - 0.22 \\
 & 3 & 0.15 - 0.52 & 0.22 - 0.42 \\
 & 4 & 0.15 - 0.52 & 0.42 - 1.20 \\
 & 5 & 0.52 - 0.83 & 0.00 - 0.14 \\
 & 6 & 0.52 - 0.83 & 0.14 - 0.22 \\
 & 7 & 0.52 - 0.83 & 0.22 - 0.30 \\
 & 8 & 0.52 - 0.83 & 0.30 - 0.42 \\
 & 9 & 0.52 - 0.83 & 0.42 - 0.53 \\
 & 10 & 0.52 - 0.83 & 0.53 - 1.20 \\
 & 11 & 0.83 - 1.00 & 0.00 - 0.14 \\
 & 12 & 0.83 - 1.00 & 0.14 - 0.22 \\
 & 13 & 0.83 - 1.00 & 0.22 - 0.30 \\
 & 14 & 0.83 - 1.00 & 0.30 - 0.42 \\
 & 15 & 0.83 - 1.00 & 0.42 - 0.53 \\
 & 16 & 0.83 - 1.00 & 0.53 - 1.20 \\
 & 17 & OOPS & OOPS \\
\hline
\multirow{18}{*}{TPC $\mu$ + FGD $p$ (+ $Np$)} & 0 & -0.60 - 0.15 & 0.00 - 1.20 \\
 & 1 & 0.15 - 0.52 & 0.00 - 0.14 \\
 & 2 & 0.15 - 0.52 & 0.14 - 0.22 \\
 & 3 & 0.15 - 0.52 & 0.22 - 0.42 \\
 & 4 & 0.15 - 0.52 & 0.42 - 1.20 \\
 & 5 & 0.52 - 0.83 & 0.00 - 0.14 \\
 & 6 & 0.52 - 0.83 & 0.14 - 0.22 \\
 & 7 & 0.52 - 0.83 & 0.22 - 0.30 \\
 & 8 & 0.52 - 0.83 & 0.30 - 0.42 \\
 & 9 & 0.52 - 0.83 & 0.42 - 0.53 \\
 & 10 & 0.52 - 0.83 & 0.53 - 1.20 \\
 & 11 & 0.83 - 1.00 & 0.00 - 0.14 \\
 & 12 & 0.83 - 1.00 & 0.14 - 0.22 \\
 & 13 & 0.83 - 1.00 & 0.22 - 0.30 \\
 & 14 & 0.83 - 1.00 & 0.30 - 0.42 \\
 & 15 & 0.83 - 1.00 & 0.42 - 0.53 \\
 & 16 & 0.83 - 1.00 & 0.53 - 1.20 \\
 & 17 & OOPS & OOPS \\
\hline
\multirow{9}{*}{FGD $\mu$ + TPC $p$ (+ $Np$)} & 0 & -0.60 - -0.10 & 0.00 - 0.22 \\
 & 1 & -0.60 - -0.10 & 0.22 - 1.20 \\
 & 2 & -0.10 - 0.15 & 0.00 - 1.20 \\
 & 3 & 0.15 - 0.52 & 0.00 - 0.14 \\
 & 4 & 0.15 - 0.52 & 0.14 - 0.22 \\
 & 5 & 0.15 - 0.52 & 0.22 - 0.42 \\
 & 6 & 0.15 - 0.52 & 0.42 - 1.20 \\
 & 7 & 0.52 - 1.00 & 0.00 - 1.20 \\
 & 8 & OOPS & OOPS \\
\hline
\hline
\end{tabular}
}
\end{table}

\begin{table}[H]
\caption{FGD1 control sample binning used in the $p_{N}$--$\cos\theta_{\mu}$ space measurement. The binnings are defined in the reconstructed analysis observable space.}
\label{tab:sample_bin_2_FGD1_cont}
\centering
\renewcommand{\arraystretch}{1.2}
\begin{tabular}{c|c|c|c}
\hline
\hline
Sample & Bin Index & $\cos\theta_{\mu}$ & $p_{N}$ [GeV/c] \\
\hline
\multirow{6}{*}{CCproton ME} & 0 & -0.60 - 0.52 & 0.00 - 1.20 \\
 & 1 & 0.52 - 0.83 & 0.00 - 0.42 \\
 & 2 & 0.52 - 0.83 & 0.42 - 1.20 \\
 & 3 & 0.83 - 1.00 & 0.00 - 0.42 \\
 & 4 & 0.83 - 1.00 & 0.42 - 0.53 \\
 & 5 & 0.83 - 1.00 & 0.53 - 1.20 \\
\hline
\multirow{6}{*}{CCproton 1$\pi^{+}$} & 0 & -0.60 - 0.52 & 0.00 - 1.20 \\
 & 1 & 0.52 - 0.83 & 0.00 - 0.42 \\
 & 2 & 0.52 - 0.83 & 0.42 - 1.20 \\
 & 3 & 0.83 - 1.00 & 0.00 - 0.42 \\
 & 4 & 0.83 - 1.00 & 0.42 - 0.53 \\
 & 5 & 0.83 - 1.00 & 0.53 - 1.20 \\
\hline
\hline
\end{tabular}
\end{table}

\begin{table}[H]
\caption{FGD2X signal sample binning used in the $p_{N}$--$\cos\theta_{\mu}$ space measurement. The bins are defined in the reconstructed observable space. The final bin, labeled ``OOPS'' (out-of-phase-space), contains selected events that do not satisfy the phase space constraints defined in Tab.~\ref{tab:ps_cons}. This bin is included to estimate the migration of signal events across phase space boundaries due to finite detector resolution, utilizing a data-driven method.}
\label{tab:sample_bin_2_FGD2X_sig}
\centering
\resizebox{0.45\textwidth}{!}{
\renewcommand{\arraystretch}{1.2}
\begin{tabular}{c|c|c|c}
\hline
\hline
Sample & Bin Index & $\cos\theta_{\mu}$ & $p_{N}$ [GeV/c] \\
\hline
\multirow{10}{*}{TPC $\mu$ + TPC $p$} & 0 & -0.60 - 0.15 & 0.00 - 1.20 \\
 & 1 & 0.15 - 0.52 & 0.00 - 0.22 \\
 & 2 & 0.15 - 0.52 & 0.22 - 1.20 \\
 & 3 & 0.52 - 0.83 & 0.00 - 0.22 \\
 & 4 & 0.52 - 0.83 & 0.22 - 0.42 \\
 & 5 & 0.52 - 0.83 & 0.42 - 1.20 \\
 & 6 & 0.83 - 1.00 & 0.00 - 0.22 \\
 & 7 & 0.83 - 1.00 & 0.22 - 0.42 \\
 & 8 & 0.83 - 1.00 & 0.42 - 1.20 \\
 & 9 & OOPS & OOPS \\
\hline
\multirow{10}{*}{TPC $\mu$ + FGD $p$ (+ $Np$)} & 0 & -0.60 - 0.15 & 0.00 - 1.20 \\
 & 1 & 0.15 - 0.52 & 0.00 - 0.22 \\
 & 2 & 0.15 - 0.52 & 0.22 - 1.20 \\
 & 3 & 0.52 - 0.83 & 0.00 - 0.22 \\
 & 4 & 0.52 - 0.83 & 0.22 - 0.42 \\
 & 5 & 0.52 - 0.83 & 0.42 - 1.20 \\
 & 6 & 0.83 - 1.00 & 0.00 - 0.22 \\
 & 7 & 0.83 - 1.00 & 0.22 - 0.42 \\
 & 8 & 0.83 - 1.00 & 0.42 - 1.20 \\
 & 9 & OOPS & OOPS \\
\hline
\multirow{8}{*}{FGD $\mu$ + TPC $p$ (+ $Np$)} & 0 & -0.60 - -0.10 & 0.00 - 0.22 \\
 & 1 & -0.60 - -0.10 & 0.22 - 1.20 \\
 & 2 & -0.10 - 0.15 & 0.00 - 1.20 \\
 & 3 & 0.15 - 0.52 & 0.00 - 0.22 \\
 & 4 & 0.15 - 0.52 & 0.22 - 1.20 \\
 & 5 & 0.52 - 1.00 & 0.00 - 0.22 \\
 & 6 & 0.52 - 1.00 & 0.22 - 1.20 \\
 & 7 & OOPS & OOPS \\
\hline
\hline
\end{tabular}
}
\end{table}

\begin{table}[H]
\caption{FGD2X control sample binning used in the $p_{N}$--$\cos\theta_{\mu}$ space measurement. The binnings are defined in the reconstructed analysis observable space.}
\label{tab:sample_bin_2_FGD2X_cont}
\centering
\renewcommand{\arraystretch}{1.2}
\begin{tabular}{c|c|c|c}
\hline
\hline
Sample & Bin Index & $\cos\theta_{\mu}$ & $p_{N}$ [GeV/c] \\
\hline
\multirow{5}{*}{CCproton ME} & 0 & -0.60 - 0.52 & 0.00 - 1.20 \\
 & 1 & 0.52 - 0.83 & 0.00 - 0.42 \\
 & 2 & 0.52 - 0.83 & 0.42 - 1.20 \\
 & 3 & 0.83 - 1.00 & 0.00 - 0.42 \\
 & 4 & 0.83 - 1.00 & 0.42 - 1.20 \\
\hline
\multirow{5}{*}{CCproton 1$\pi^{+}$} & 0 & -0.60 - 0.52 & 0.00 - 1.20 \\
 & 1 & 0.52 - 0.83 & 0.00 - 0.42 \\
 & 2 & 0.52 - 0.83 & 0.42 - 1.20 \\
 & 3 & 0.83 - 1.00 & 0.00 - 0.42 \\
 & 4 & 0.83 - 1.00 & 0.42 - 1.20 \\
\hline
\hline
\end{tabular}
\end{table}

\begin{table}[H]
\caption{FGD2Y signal sample binning used in the $p_{N}$--$\cos\theta_{\mu}$ space measurement. The bins are defined in the reconstructed observable space. The final bin, labeled ``OOPS'' (out-of-phase-space), contains selected events that do not satisfy the phase space constraints defined in Tab.~\ref{tab:ps_cons}. This bin is included to estimate the migration of signal events across phase space boundaries due to finite detector resolution, utilizing a data-driven method.}
\label{tab:sample_bin_2_FGD2Y_sig}
\centering
\resizebox{0.45\textwidth}{!}{
\renewcommand{\arraystretch}{1.2}
\begin{tabular}{c|c|c|c}
\hline
\hline
Sample & Bin Index & $\cos\theta_{\mu}$ & $p_{N}$ [GeV/c] \\
\hline
\multirow{14}{*}{TPC $\mu$ + TPC $p$} & 0 & -0.60 - 0.15 & 0.00 - 1.20 \\
 & 1 & 0.15 - 0.52 & 0.00 - 0.22 \\
 & 2 & 0.15 - 0.52 & 0.22 - 1.20 \\
 & 3 & 0.52 - 0.83 & 0.00 - 0.14 \\
 & 4 & 0.52 - 0.83 & 0.14 - 0.22 \\
 & 5 & 0.52 - 0.83 & 0.22 - 0.30 \\
 & 6 & 0.52 - 0.83 & 0.30 - 0.42 \\
 & 7 & 0.52 - 0.83 & 0.42 - 1.20 \\
 & 8 & 0.83 - 1.00 & 0.00 - 0.22 \\
 & 9 & 0.83 - 1.00 & 0.22 - 0.30 \\
 & 10 & 0.83 - 1.00 & 0.30 - 0.42 \\
 & 11 & 0.83 - 1.00 & 0.42 - 0.53 \\
 & 12 & 0.83 - 1.00 & 0.53 - 1.20 \\
 & 13 & OOPS & OOPS \\
\hline
\multirow{8}{*}{TPC $\mu$ + FGD $p$ (+ $Np$)} & 0 & -0.60 - 0.52 & 0.00 - 1.20 \\
 & 1 & 0.52 - 0.83 & 0.00 - 0.22 \\
 & 2 & 0.52 - 0.83 & 0.22 - 0.42 \\
 & 3 & 0.52 - 0.83 & 0.42 - 1.20 \\
 & 4 & 0.83 - 1.00 & 0.00 - 0.22 \\
 & 5 & 0.83 - 1.00 & 0.22 - 0.42 \\
 & 6 & 0.83 - 1.00 & 0.42 - 1.20 \\
 & 7 & OOPS & OOPS \\
\hline
\multirow{7}{*}{FGD $\mu$ + TPC $p$ (+ $Np$)} & 0 & -0.60 - -0.10 & 0.00 - 0.22 \\
 & 1 & -0.60 - -0.10 & 0.22 - 1.20 \\
 & 2 & -0.10 - 0.15 & 0.00 - 1.20 \\
 & 3 & 0.15 - 0.52 & 0.00 - 0.22 \\
 & 4 & 0.15 - 0.52 & 0.22 - 1.20 \\
 & 5 & 0.52 - 1.00 & 0.00 - 1.20 \\
 & 6 & OOPS & OOPS \\
\hline
\hline
\end{tabular}
}
\end{table}

\begin{table}[H]
\caption{FGD2Y control sample binning used in the $p_{N}$--$\cos\theta_{\mu}$ space measurement. The binnings are defined in the reconstructed analysis observable space.}
\label{tab:sample_bin_2_FGD2Y_cont}
\centering
\renewcommand{\arraystretch}{1.2}
\begin{tabular}{c|c|c|c}
\hline
\hline
Sample & Bin Index & $\cos\theta_{\mu}$ & $p_{N}$ [GeV/c] \\
\hline
\multirow{3}{*}{CCproton ME} & 0 & -0.60 - 0.83 & 0.00 - 1.20 \\
 & 1 & 0.83 - 1.00 & 0.00 - 0.53 \\
 & 2 & 0.83 - 1.00 & 0.53 - 1.20 \\
\hline
\multirow{3}{*}{CCproton 1$\pi^{+}$} & 0 & -0.60 - 0.83 & 0.00 - 1.20 \\
 & 1 & 0.83 - 1.00 & 0.00 - 0.53 \\
 & 2 & 0.83 - 1.00 & 0.53 - 1.20 \\
\hline
\hline
\end{tabular}
\end{table}

\section{Breakdown of Neutrino Flux Systematic Uncertainty Sources}
\label{sec:breakdown_flux_syst}

\begin{figure}[htpb!]
\centering
\includegraphics[width=0.45\textwidth]{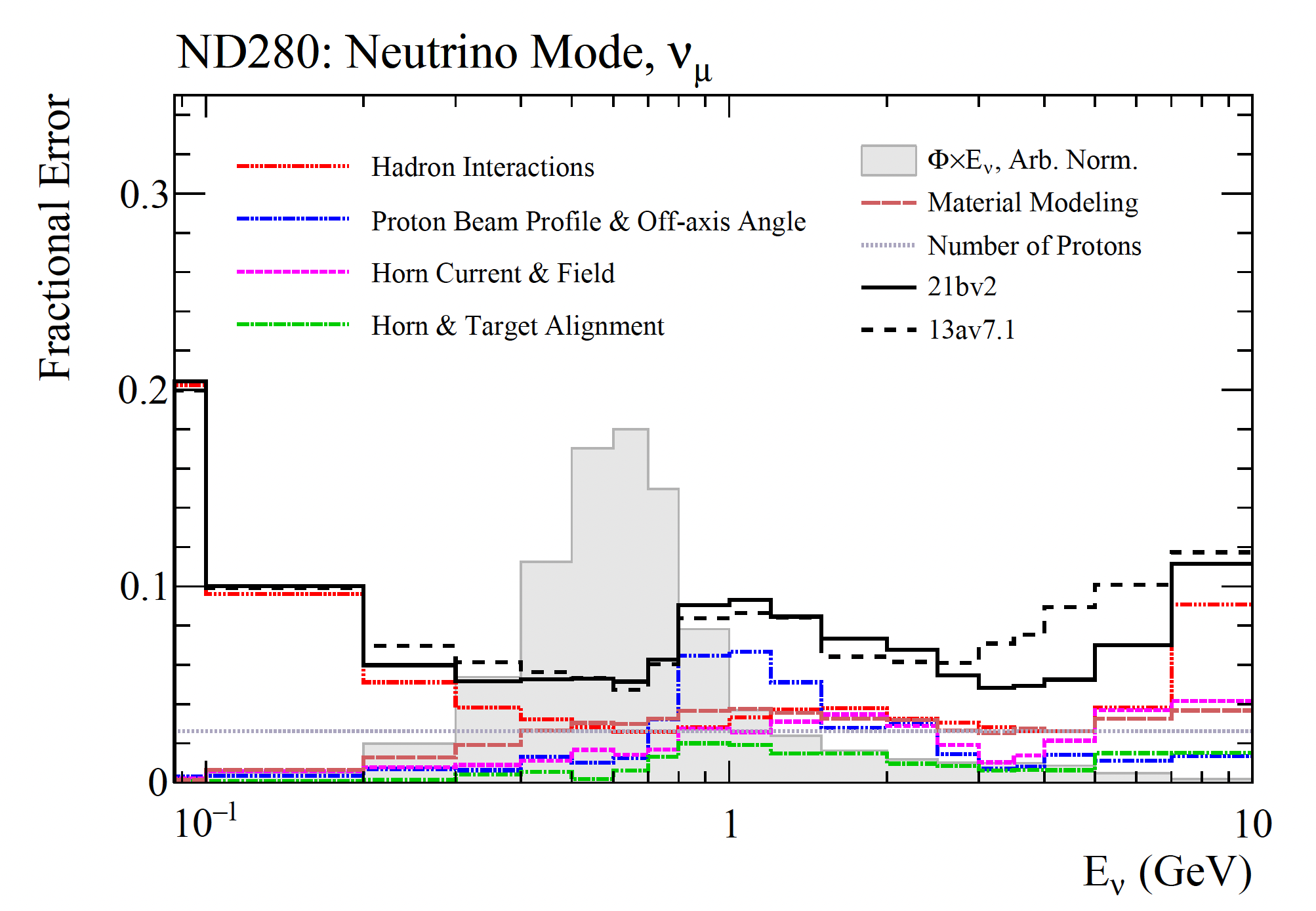}
\caption{Uncertainty on the $\nu_{\mu}$ flux prediction at ND280 as a function of incident neutrino energy, broken down by sources. The solid black line (labeled ``21bv2'') shows the total flux uncertainty used in this analysis. The gray shaded region indicates the shape of the $\nu_{\mu}$ flux.}
\label{fig:nuflux_syst_decomp}
\end{figure}

Figure~\ref{fig:nuflux_syst_decomp} shows the uncertainty on the $\nu_{\mu}$ flux prediction at ND280 as a function of incident neutrino energy, broken down by sources. The prediction is tuned using hadron production measurements from the NA61/SHINE experiment~\cite{NA61SHINE:2015bad,NA61SHINE:2018rhe}. Near the flux peak (${\sim}0.6$~GeV), where most CC$0\pi$ signal interactions occur, the dominant contributions come from the modeling of hadron interactions and of the material surrounding the target.

\section{Distributions of Reconstructed Vertex Position for Selected FGD2 Samples}
\label{sec:dis_vtx_pos_FGD2}

\begin{figure}
\centering
\subfloat[Signal sample]
{\includegraphics[width=0.45\textwidth]{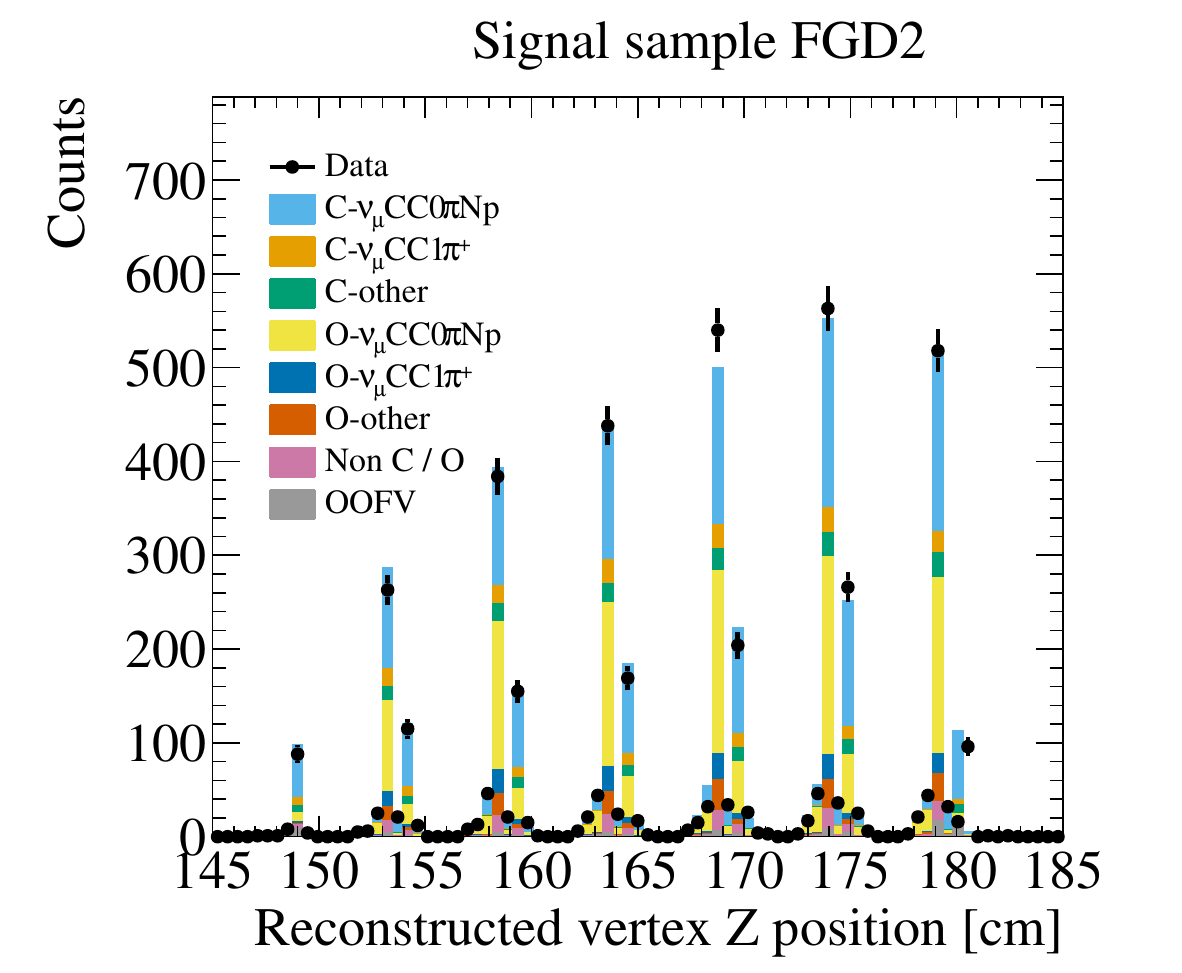}}
\\
\subfloat[Control sample]
{\includegraphics[width=0.45\textwidth]{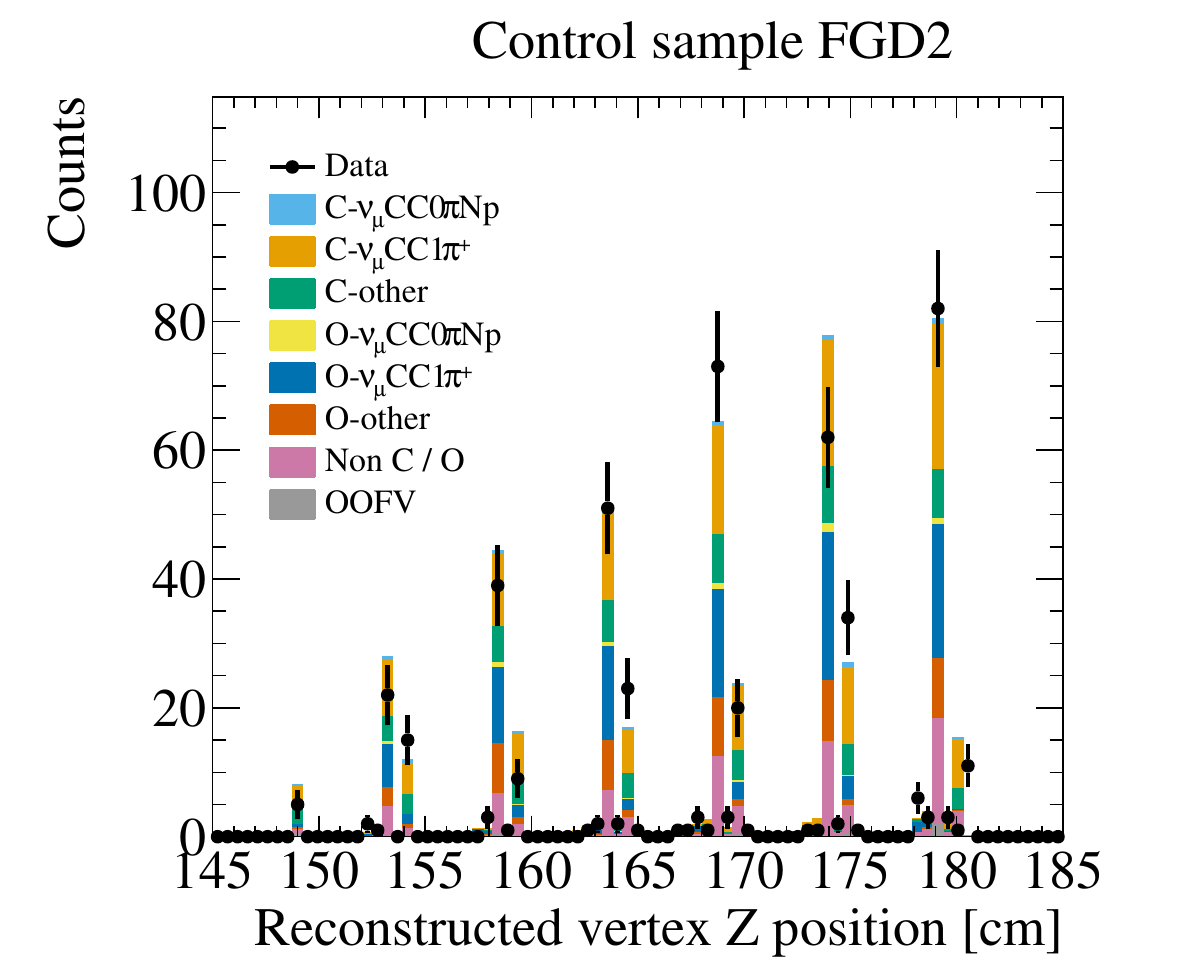}}
\caption{Distributions of the reconstructed vertex Z position for the selected FGD2 (a) signal and (b) control samples. The position is given in the ND280 coordinate system, in which the neutrino beam travels from left to right (i.e., in the direction of increasing Z). The MC distributions are normalized to the data POT and are decomposed into true topology categories, which also indicate the true interaction targets.}
\label{fig:dis_vtx_pos_FGD2}
\end{figure}

Figure~\ref{fig:dis_vtx_pos_FGD2} shows the distributions of the reconstructed vertex Z position for the selected FGD2 events, separately for the signal and control samples. This is the discriminating variable used to separate the FGD2X and FGD2Y samples: each selected event is assigned to the sample corresponding to the scintillator layer (X or Y) closest to its reconstructed vertex. Seven main peaks are visible, corresponding to the seven scintillator modules of FGD2, which are interleaved with six water layers. With the exception of the first module, whose X layer was removed, each main peak consists of two sub-peaks: the one at lower Z corresponds to the X layer, which lies immediately downstream of a water layer, and the one at higher Z corresponds to the Y layer. Overall, good agreement between data and MC is observed, with no significant differences in shape.

\end{document}